\documentclass[12pt]{article}
\usepackage[utf8]{inputenc}
\usepackage{amsmath,setspace,geometry}
\usepackage{amsthm}
\usepackage{amsfonts}
\usepackage[shortlabels]{enumitem}
\usepackage{rotating}
\usepackage{pdflscape}
\usepackage{graphicx}
\usepackage{bbm}
\usepackage[dvipsnames]{xcolor}
\usepackage{hyperref}
\hypersetup{colorlinks=true, linkcolor= BrickRed, citecolor = BrickRed, filecolor = BrickRed, urlcolor = BrickRed, hypertexnames = true}
\usepackage[]{natbib}
\bibpunct[:]{(}{)}{,}{a}{}{,}
\usepackage[english]{babel}
\usepackage{float}
\usepackage{caption}
\usepackage{subcaption}
\usepackage{booktabs}
\usepackage{pdfpages}
\usepackage{threeparttable}
\usepackage{lscape}
\usepackage{bm}
\begin{document}
\title{Nonparametric Estimation of Matching Efficiency and Mismatch: An Application to Japanese Labor Markets}
\author{Suguru Otani\thanks{\href{mailto:}{suguru.otani@e.u-tokyo.ac.jp}, Market Design Center, Department of Economics, University of Tokyo\\
I thank the Editor, Tito Boeri, and an anonymous referee for suggesting the use of more representative data, which substantially improved the empirical application. I thank Shoya Ishimaru, Keisuke Kawata, Fuhito Kojima, Akira Matsushita, Masaru Sasaki, Masayuki Sawada, Kazuhiro Teramoto, and Kosuke Uetake for their valuable advice. I am especially grateful to Yudai Higashi, Ryo Kambayashi, and Daiji Kawaguchi for their generous guidance on institutional context and invaluable help with data details. I also thank seminar and conference participants at Hitotsubashi University, Summer Workshop on Economic Theory (SWET), 27th Labor Economics Conference at University of Tokyo, Human Resource Allocation Workshop at Osaka University, and the ``Labor Shortages and Mismatch" invited session at the Sixth World Labor Conference SOLE-EALE-AASLE, Toronto. I thank Shunsuke Ishii and Hayato Kanayama for their excellent research assistance. An
earlier version of this paper was titled ``Nonparametric Estimation of Matching Efficiency and Mismatch in Labor Markets via Public Employment Security Offices in Japan, 1972--2024". This work was supported by JST ERATO Grant Number JPMJER2301 and JSPS KAKENHI 25K16620.
%, Keisuke Kawata
}
}
\date{
First version: July 31, 2024\\
Current version: \today
}
\maketitle

\begin{abstract}
\noindent
%150 words
I identify significant biases in the traditional Cobb-Douglas function under misspecification of nonadditive, time-varying matching efficiency, and evaluate the finite-sample performance of the nonparametric estimation method of \cite{lange2020beyond}. Additionally, I extend the mismatch index by \cite{csahin2014mismatch} to a nonparametric framework and develop a computational methodology.
Applying the method to Japanese labor market data from government surveys and Hello Work, I analyze changes in matching efficiency, elasticities, and mismatch. Both datasets show declining matching efficiency, aligned with decreasing job-finding and vacancy-filling rates. The unemployment elasticity is around 0.5--0.9 in the Hello Work data and lower in the nationally representative aggregate, while the vacancy elasticity is higher in the nationally representative aggregate. Furthermore, I demonstrate that occupational mismatch is more severe than geographic mismatch, with the Cobb-Douglas mismatch index significantly underestimating the true extent of mismatch.

%100 words AER
\end{abstract}

\newpage

\section{Introduction}

Understanding labor market matching efficiency and mismatch—defined as the misalignment between the distribution of vacant jobs and unemployed workers—is essential for analyzing labor market dynamics. The matching function, which maps hires to the number of job seekers and vacancies while incorporating matching efficiency and search effort, is central to understanding frictional labor markets. However, as \cite{lange2020beyond} highlight, a critical limitation of existing studies is their reliance on restrictive functional form assumptions, such as the Cobb-Douglas specification, which often lack empirical and theoretical justification. These assumptions can lead to biased estimates of matching efficiency and elasticity. To address this issue, \cite{lange2020beyond} propose a nonparametric approach to identify the matching function and recover matching efficiency without imposing functional form restrictions.

Despite this advancement, the gap between nonparametric and Cobb-Douglas approaches remains, particularly in measuring labor mismatch. This paper aims to bridge this gap by evaluating the finite-sample performance of the proposed estimator, documenting the bias of Cobb-Douglas approaches under misspecification, and extending the mismatch index proposed by \cite{csahin2014mismatch} to a nonparametric framework.

Several studies have assessed labor mismatch and its contribution to unemployment fluctuations using the Cobb-Douglas specification. \cite{csahin2014mismatch} provide a methodology to quantify mismatch as the deviation between observed labor market outcomes and socially optimal outcomes. However, the Cobb-Douglas specification oversimplifies the mismatch index by canceling out time-varying matching efficiency and elasticities. As a methodological contribution, this paper combines the nonparametric approach with the mismatch index, allowing for nonparametrically estimated matching efficiency and elasticities. Additionally, I demonstrate how to compute the index using the Mathematical Programming with Equilibrium Constraints (MPEC) framework proposed by \cite{su2012constrained}. To address practitioners' concerns, I highlight the bias of the traditional Cobb-Douglas matching function and mismatch index and evaluate the finite-sample performance of the proposed estimator via Monte Carlo simulation and empirical application.

My empirical application combines two complementary Japanese datasets. Dataset 1 (LFS + SET) merges the Labour Force Survey and the Survey on Employment Trends into a nationally representative annual panel at the country, region, and industry levels (country 1972--2024; region and industry 2012--2024). Its strength is broad coverage of observed hiring channels within the Survey on Employment Trends coverage---private-sector establishments with five or more regular employees in SET-covered industries, excluding sectors outside the SET such as government and primary industries---and statistical representativeness within that scope; its limitation is annual frequency and coarse cross-sectional resolution (13 regional blocks, 14 industries). Dataset 2 (Hello Work) draws on the Public Employment Security Office statistics, providing a monthly long-run series (1972--2023 for the aggregate estimation sample) at fine cross-sectional resolution (47 prefectures, 67 job categories), with full-time and part-time decompositions, for a clearly defined platform-level market; its limitation is partial coverage of the labor market, since Hello Work intermediates about 11--22\% of total hires nationwide (mean 18.5\% over 1991--2024). The matching function estimated on Dataset 1 therefore approximates an aggregate matching function for the Japanese labor market within SET coverage, while the matching function estimated on Dataset 2 represents a platform-level matching function for the Hello Work intermediation system. Hello Work statistics have been the de facto standard data source for Japanese matching function estimation \citep{kano2005estimating, kambayashi2006vacancy, sasaki2008matching, kambayashi2013role, higashi2018spatial, kawata2019}, but to my knowledge this is the first paper to complement the Hello Work-based platform matching function with a nationally representative aggregate matching function constructed from LFS + SET.

For Dataset 1, the nationally representative country-year matching efficiency, normalized to 1972, exhibits a long-run downward trend with notable fluctuations. The implied unemployment elasticity is centered around 0.3--0.5, and the vacancy elasticity is correspondingly centered around 0.5--0.7. At the region-year and industry-year level (2012--2024), the nonparametric mismatch index ranges between roughly 0.07 and 0.29 with a mean around 0.20--0.21, consistently exceeding the corresponding Cobb-Douglas indices and confirming the downward bias of the latter even at this more representative aggregate level.

For Dataset 2, the Hello Work matching efficiency, normalized to 1972, also exhibits a declining trend with notable fluctuations, aligning with downward trends in job-finding and vacancy-filling rates. This decline may reflect the increasing role of alternative matching opportunities outside the government-operated Hello Work platform. The implied unemployment elasticity ranges from 0.5 to 0.9, consistent with global findings such as \cite{petrongolo2001looking} (0.5--0.7) and previous Japanese studies. The vacancy elasticity has an interquartile range of 0.18--0.42 with a mean around 0.29, broadly consistent with estimates by \cite{lange2020beyond} (0.15--0.3) and Japanese studies (0.15--0.6). Substantial mismatches are observed at both the prefecture-month and occupation-month levels, with an upward trend over time: the average mismatch across prefectures is about 0.25, while mismatch across occupations averages about 0.52 and ranges from about 0.39 to 0.65---significantly larger than reported by \cite{shibata2020labor} and \cite{higashi2025did}, because their Cobb-Douglas index, by construction, ignores the time variation of matching efficiency and elasticity that the nonparametric estimator captures, thereby attenuating the measured deviation between actual and socially optimal allocations.

Across both datasets, the proposed nonparametric mismatch index reveals the limitations of the Cobb-Douglas mismatch index, demonstrating its downward bias through both Monte Carlo simulations and empirical application. Comparing the geographic and the sectoral or occupational dimensions, sectoral or occupational mismatch is consistently larger than geographic mismatch, with the largest values arising in the Hello Work occupation panel (mean about 0.52 with peaks reaching about 0.65), followed by the Hello Work prefecture panel (mean about 0.25); LFS + SET region and industry mismatches both lie in the 0.1--0.3 range. Part of this gap reflects the granularity of cross-sectional resolution: LFS + SET aggregates to 13 regional blocks and 14 broad industries, which mechanically washes out within-aggregate misallocation, whereas Hello Work resolves 47 prefectures and 67 job categories, which preserves and exposes finer skill- and location-specific frictions. The cross-dataset comparison further shows that the platform-level (Hello Work) matching dynamics are qualitatively informative about wider Japanese labor market trends, but with level differences that warrant caution when extrapolating Hello Work-based magnitudes (e.g.\ tightness $V/U$ or mismatch levels) to the aggregate. These findings underscore the importance of accounting for time-varying matching efficiency and elasticity in labor market analysis, as well as for the cross-sectional resolution at which mismatch is measured.

\subsection{Related Literature}
This paper contributes to three strands of literature.
First, I provide new evidence on matching efficiency in Japanese labor markets using the nonparametric approach of \cite{lange2020beyond}. Their approach identifies the matching function and recovers unobserved matching efficiency without imposing the usual independence assumption between matching efficiency and search on either side of the market, and it can accommodate multiple types of job seekers.
\cite{lange2020beyond} highlight positive correlations between efficiency and market structure, such as tightness, which induce a positive bias in estimates of the vacancy elasticity whenever unobserved matching efficiency is not controlled for, as in the traditional Cobb-Douglas matching function with constant elasticity parameters.\footnote{The approach is applicable to a wide range of matching markets: not only labor markets such as full-time labor, spot labor, and marriage markets \citep{otani2025nonparametric, kanayama2024nonparametric, otani2025marriage}, but also team formation \citep{nishihata2025team} and transportation markets such as global bulk shipping \citep{brancaccio2020geography,brancaccio2023search}. For reference, \cite{brancaccio2020guide} summarize practical issues.}
I investigate the finite-sample performance of the nonparametric approach to address practitioners' concerns regarding sample size, specification, and endogeneity issues, as \cite{lange2020beyond} provides only the theoretical identification proof. I highlight the bias inherent in the Cobb-Douglas matching function when nonadditive matching efficiency is misspecified and confirm the potential advantages of the nonparametric approach. By implementing this method, I contribute updated results derived from the more flexible approach, complementing the existing findings of \cite{kano2005estimating}, \cite{kambayashi2006vacancy}, \cite{sasaki2008matching}, and \cite{higashi2018spatial}, which relied on the traditional Cobb-Douglas matching function with geographic and occupational category fixed effects to capture regional and occupational heterogeneity using Hello Work data. Table \ref{tb:previous_literature} summarizes the previous findings.
My findings are also useful for comparison with other countries' results reported in \cite{bernstein2022matching} and \cite{petrongolo2001looking}.

\begin{table}[!htbp]
  \begin{center}
      \caption{Estimation results of the aggregate matching function in Japan}
      \begin{tabular}{|lcccccc|} \hline
   Paper &  CRS & FE & Estimation & Sample & $\frac{d \log H}{d \log U}$ & $\frac{d \log H}{d \log V}$\\
   \hline
   \cite{kano2005estimating} &  yes & yes & OLS & 1973--1999 & 0.56-0.59& 0.29-0.3\\
\cite{kambayashi2006vacancy} &  no & yes & OLS & 1996--2001 & 0.81& 0.30\\
\cite{sasaki2008matching} &  yes & no & OLS & 1982--2016 & & 0.38-0.45\\
\cite{shibata2013labor} &  yes & no & GMM & 2000--2013 & & 0.4\\
   \cite{shibata2020labor} &  yes & no & GMM & 2000--2019 & & 0.34-0.4\\
\cite{kawata2016multi} &  yes & yes & IV & 2000--2009 & & 0.51-0.6\\
\cite{kawata2019} &  yes & yes & OLS & 2012--2016 & & 0.52\\
\cite{higashi2018spatial} &  yes & no & OLS & 2008--2016 & 0.37-0.46& 0.17-0.24\\
\cite{higashi2020effects} &  yes & no & OLS & 2006--2016 & 0.48& 0.17\\
\cite{higashi2021agglomeration} &  yes & no & OLS, 2SLS & 2008--2018 & 0.22-0.46& 0.15-0.31\\
\cite{higashi2025did} &  yes & yes & OLS & 2016--2021& Calibrate & Calibrate \\
This paper (Hello Work) &  yes & no & Nonparametric & 1972--2023& 0.5--0.9 & 0.18--0.42 \\
This paper (LFS + SET) &  yes & no & Nonparametric & 1972--2024 & 0.3--0.5 & 0.5--0.7 \\
   \hline
 \end{tabular}
 \label{tb:previous_literature}
  \end{center}
  \footnotesize
  Note: All papers except for this paper use a Cobb-Douglas matching function. Constant returns to scale (CRS) imposes a restriction on matching elasticity parameters in estimation. Fixed effects (FE) capture geographic and time heterogeneity. Instrumental Variables (IV), Two-Stage Least Squares (2SLS), and Generalized Method of Moments (GMM) based on \cite{borowczyk2013accounting} use instruments in specific contexts. Blank cells mean no reported results. See the survey of \cite{miyamoto2025macroeconomic} for an overview of labor markets in Japan.
\end{table}

Second, this paper contributes to the literature on labor market mismatch, building on the foundational work of \cite{csahin2014mismatch}, which extends the earlier analysis by \cite{jackman1987structural}. In the static matching model with multiple sectors, \cite{jackman1987structural} demonstrate that aggregate hires are maximized when unemployment is distributed across sectors in a way that equalizes sectoral labor-market tightness. \cite{csahin2014mismatch} extend this framework by incorporating a dynamic, stochastic environment with multiple sources of heterogeneity, providing a methodology to construct counterfactual measures of unemployment in the absence of mismatch. Their focus is on mismatch unemployment, defined as unemployed workers searching in the ``wrong" sector.\footnote{A distinct body of literature, such as \cite{eeckhout2011identifying}, uses the term ``mismatch" to describe employed individuals in suboptimal jobs, where the alignment between worker skills and firm capital is not ideal. \cite{eeckhout2018sorting} provides an overview of this literature, identifying three primary causes of mismatch: search frictions \citep{eeckhout2011identifying}, stochastic sorting \citep{chade2014stochastic}, and multidimensional types \citep{lindenlaub2017sorting}. In addition, \cite{guvenen2020multidimensional} explores skill mismatch, which quantifies the gap between the skills required by an occupation and the abilities a worker possesses. Another approach, as in \cite{shimer2012reassessing}, takes a flow perspective by decomposing unemployment volatility into transition rates between four labor force states: (i) permanent-contract employment, (ii) temporary-contract employment, (iii) unemployment, and (iv) non-participation, providing a foundation for calculating mismatch measures based on employment status.}

The approach is widely used in a variety of contexts, for example, the United Kingdom in \cite{patterson2016working}.
In the context of Japanese labor markets, several papers estimate mismatch at geographic and occupational levels using a Cobb-Douglas specification (\cite{shibata2013labor,shibata2020labor},  \cite{kawata2016multi,higashi2018spatial,kawata2019}, \cite{higashi2020effects,higashi2021agglomeration}, \cite{higashi2025did}).
This paper points out that the Cobb-Douglas specification cancels out time variation in matching efficiency, which can bias the mismatch index. I then propose a nonparametric version of the mismatch index and apply a computational method known as Mathematical Programming with Equilibrium Constraints (MPEC), proposed by \cite{su2012constrained} and \cite{dube2012improving}, to calculate geographic and occupational mismatch indices.

Third, this paper contributes to the literature on the impact of COVID-19 on the matching efficiency of the government-operated worker-vacancy platform.
Because the related literature is extensive, I focus on the papers most closely connected to this setting.
The most closely related papers in my context are \cite{higashi2025did} and \cite{fukai2021describing}.
\cite{higashi2025did} estimate mismatch indices for local labor markets clustered by occupations vulnerable and not vulnerable to COVID-19, assuming a Cobb-Douglas matching function with calibrated matching elasticity parameters.
\cite{fukai2021describing} utilize the Labor Force Survey, a large-scale government statistic, to estimate group average treatment effects of COVID-19 on employment status for each month from January to June 2020.
To the best of my knowledge, this is the first paper to describe changes in matching efficiency before and after COVID-19 using a novel nonparametric approach.

% Third, I point out technical issues about scale normalization of matching efficiency on \cite{lange2020beyond}, which induces difficulty in comparing the estimated elasticities with respect to unemployed interacted with match efficiency across different datasets.

\section{Model}
\subsection{Nonparametric Aggregate Matching Function}
The main objects of interest are matching efficiency and matching elasticities with respect to the number of unemployed workers and vacancies in the labor market.
A matching function based on search models plays a central role in labor economics.\footnote{See \cite{pissarides2000equilibrium,petrongolo2001looking}, and \cite{rogerson2005search} for reference.}
The matching function relies on random search from both sides of the market, that is, individuals seeking jobs represent the supply of labor and recruiters represent the demand for labor.
To estimate the matching function and recover matching efficiency, I follow the novel approach proposed by \cite{lange2020beyond}.\footnote{\cite{lange2020beyond} additionally incorporate search effort \citep{mukoyama2018job} and a recruitment index \citep{davis2013establishment}. Unfortunately, my data do not contain this information.}
Their paper identifies the endogeneity problem of matching efficiency \citep{borowczyk2013accounting} and the restrictive nature of a Cobb-Douglas matching function with fixed matching elasticities, then proposes nonparametric identification and estimation of matching efficiency under the conditions introduced below.

Let unscripted capital letters $(A, U, V)$ denote random variables while realizations are subscripted by time $t$.
I consider the matching function $m(\cdot,\cdot)$ that maps period-$t$ unemployed workers $U_t$, per-capita matching efficiency of unemployed workers $A_t$, and vacancies $V_t$ into hires $H_t$.
I assume that the underlying data generating process is stationary and that I observe a long enough time-series so that I can treat the joint distribution $G: \mathbb{R}_{+}^3 \rightarrow[0,1]$ of $\left(H_t, U_t, V_t\right)$ as observed.
Also, denote by $F(A, U)$ the joint distribution of $A$ and $U$.

I identify the matching function as well as unobserved, time-varying matching efficiency, $A$.
First, I assume that $V$ and $A$ are independent conditional on $U$, that is, $A \perp V \mid U$.
Second, I assume that the matching function $m(AU,V):\mathbb{R}_{+}^2 \rightarrow \mathbb{R}$ has constant returns to scale (CRS).\footnote{To follow the original model of \cite{matzkin2003nonparametric}, I can rewrite $H=m(AU,V)$ into $H/U=m(A,V/U)$ under CRS. Here, $H/U$ and $V/U$ are known as a job-finding rate and market tightness. It is known that increasing returns to scale in the matching process can create multiple equilibria and the subsequent literature has found scant evidence for increasing returns, as in \cite{petrongolo2001looking}.}
Then, by applying nonparametric identification results of \cite{matzkin2003nonparametric}, Proposition 1 of \cite{lange2020beyond} proves that $G(H, U, V)$ identifies $F(A, U)$ and $m(A U, V): \mathbb{R}_{+}^2 \rightarrow \mathbb{R}_{+}$ up to a normalization of $A$ at one point, denoted $A_0$, in the support of $(A, U, V)$.\footnote{In Section \ref{sec:monte_carlo}, I investigate finite-sample performance with Monte Carlo simulation. Based on simulation results with sample size $T=50$, the sample size in this paper is enough to recover matching efficiency well. Also, see \cite{brancaccio2020guide} for practical issues. The approach is used to estimate a matching function in a trade model \citep{brancaccio2020geography,brancaccio2023search}.}

\subsection{Nonparametric Mismatch}
I construct a nonparametric version of a mismatch index which is a measure of the fraction of hires foregone due to mismatch.
Suppose there are $R$ local markets at time $t$ with a given number of vacancies, unemployed workers, and aggregate matching efficiencies in the economy.
The number of hires $H_{r t}$ in market $r$ at time $t$ is determined by a matching function $ m(A_{r t}U_{r t},V_{r t})$.
Note that, unlike \cite{csahin2014mismatch}, I allow any specification of $m$ satisfying CRS.
Given matching efficiency $A_{r t}$ and the number of vacancies $V_{r t}$, a social planner maximizes the total hires in the
economy by distributing a given number of unemployed workers to each labor market.

Following \cite{csahin2014mismatch}, I define mismatch as the deviation of the hires in the data from the efficient allocation chosen by the social planner.
Under the assumptions of homogeneous productivity and job separation rates across labor markets, the optimal allocation of unemployed workers satisfies the following equilibrium conditions:
\begin{align}
    \frac{\partial m}{\partial U}\left(A_{1t}U_{1t}^{*},V_{1t}\right)=\cdots=\frac{\partial m}{\partial U}\left(A_{Rt}U_{Rt}^{*},V_{Rt}\right) \label{eq:equilibrium_condition}
\end{align}
where $*$ denotes the planner's allocation.
The equilibrium condition means that, holding the total number of unemployed workers $U_{t}=\sum_{r=1}^{R}U_{r t}$ fixed, the planner allocates more unemployed workers to more effective labor markets, that is, markets with more vacancies and higher matching efficiency, until their marginal contribution to hires is equalized across markets.
The main difference from \cite{csahin2014mismatch} is that equilibrium condition \eqref{eq:equilibrium_condition} does not have an analytical formula like Cobb-Douglas specification.
In Section \ref{sec:mismatch_computation}, I explain how to compute the optimal allocation of unemployed workers numerically.

Using the optimal allocation of unemployed workers $U_{r t}^{*}$, the aggregate actual and optimal number of new hires can be expressed as
\begin{align*}
    H_{t}&=\sum_{r=1}^{R}m(A_{r t}U_{r t},V_{r t}),\\
    H_{t}^{*}&=\sum_{r=1}^{R}m(A_{r t}U_{r t}^{*},V_{r t}).
\end{align*}
Using these expressions, I define the mismatch index as
\begin{align}
    \mathcal{M}_{t}=1-\frac{H_{t}}{H_{t}^{*}}\label{eq:mismatch_index}
\end{align}
where $\mathcal{M}_{t}$ measures the fraction of hires lost in period $t$ because of misallocation.
This index accounts for the heterogeneity of matching efficiencies across labor markets. I explicitly refer to the mismatch index \eqref{eq:mismatch_index} under the Cobb-Douglas matching function as the Cobb-Douglas mismatch index.\footnote{See Appendix \ref{sec:cobb_douglas_mismatch_index} for construction of the Cobb-Douglas mismatch index.}

\subsection{Why Is the Nonparametric Mismatch Index Superior to the Cobb-Douglas Mismatch?}

To illustrate the potential benefit of the nonparametric mismatch index, consider \(R\) markets observed over \(T\) years. Estimating the Cobb-Douglas mismatch requires specifying a Cobb-Douglas matching function with additive matching efficiency, $H_{r t}=A_{r}A_{t}U_{r t}^{1-\gamma}V_{r t}^{\gamma}$, with \(R + T + 1\) parameters: \(R\) market fixed effects (\(A_r\)), \(T\) year fixed effects (\(A_t\)), and a constant vacancy share parameter (\(\gamma\)). The equilibrium condition \eqref{eq:equilibrium_condition} can be expressed as:

\[
    A_{r}A_{t}\left(\frac{V_{r t}}{U_{r t}^{\star}}\right)^\gamma (1-\gamma)=A_{r'}A_{t}\left(\frac{V_{r' t}}{U_{r' t}^{\star}}\right)^\gamma (1-\gamma).
\]

This formulation highlights a critical limitation of the Cobb-Douglas mismatch: the time variations in \(A_t\) and the elasticity \(\gamma\) are canceled out in the equilibrium condition.
In contrast, the proposed nonparametric approach estimates \(T \times R\) time- and market-specific efficiencies and corresponding time-varying elasticities, thereby preserving the temporal variations of \(A_t\) and elasticity. This allows for greater flexibility and precision in capturing the dynamics of matching efficiency and mismatch across both time and markets.

While the nonparametric approach offers more flexibility, it also requires a larger sample size to ensure precise estimates. The finite-sample performance of this approach is evaluated in Section \ref{sec:monte_carlo}.

\subsection{Extension: Inclusion of Heterogeneity}

\begingroup
\color{black}
The baseline mismatch index above follows the benchmark case in \cite{csahin2014mismatch} and assumes homogeneous productivity and job separation rates across markets. The nonparametric framework can incorporate these sources of heterogeneity without changing the estimated matching function. Let \(z_{r t}\) denote market-specific labor productivity and let \(\delta_{r t}\) denote the market-specific exogenous match destruction rate. The planner's problem can be generalized by maximizing the weighted value of matches,
\[
    \max_{\{U_{r t}^{*}\}_{r=1}^{R}}
    \sum_{r=1}^{R}\omega_{r t}m(A_{r t}U_{r t}^{*},V_{r t})
    \quad \text{subject to} \quad
    \sum_{r=1}^{R}U_{r t}^{*}=U_t,
\]
where \(\omega_{r t}\) is the value of a match in market \(r\). In a static extension, \(\omega_{r t}=z_{r t}\). In the dynamic extension of \cite{csahin2014mismatch}, a match in a high-destruction market has a shorter expected duration, so a natural weight is proportional to \(z_{r t}/[1-\beta(1-\Delta_t)(1-\delta_{r t})]\), up to the aggregate terms common across markets. The planner therefore equalizes \(\omega_{r t}\partial m(A_{r t}U_{r t}^{*},V_{r t})/\partial U\) across markets, rather than equalizing the unweighted marginal contribution to hires in \eqref{eq:equilibrium_condition}.

Operationally, this extension only changes the objective function and the equilibrium constraints in the MPEC computation. The nonparametric estimates of \(A_{r t}\), \(m(\cdot,\cdot)\), and the local derivatives can be used as before, while the optimal allocation gives more weight to markets where a new match is more productive or expected to last longer. The corresponding output-weighted mismatch index would be
\[
    \mathcal{M}_{t}^{\omega}
    =
    1-
    \frac{\sum_{r=1}^{R}\omega_{r t}m(A_{r t}U_{r t},V_{r t})}
    {\sum_{r=1}^{R}\omega_{r t}m(A_{r t}U_{r t}^{*},V_{r t})}.
\]
Thus, the current index is the special case \(\omega_{r t}=1\) for all markets. Implementing this richer index would require external measures of market-level productivity and destruction or separation rates, such as region- or occupation-specific wages, value added, or job separation rates. Because sufficiently detailed and consistent measures are not available for the markets in this paper, we view this heterogeneity-adjusted mismatch index as a natural but out-of-scope extension.
\endgroup

\section{Estimation}
\subsection{Matching Efficiency and Elasticities}
Following \cite{lange2020beyond}, I begin by estimating $F(A_0|U)$ across the support of $U$. To this end, I utilize the distribution of hires conditional on unemployed, $U$, and observed vacancies, $V$.
Specifically, I have
\begin{align*}
    F(A_0|\psi U_0) &= G_{H|U,V}(\psi H_0|\psi U_0, \psi V_0) \quad \text{for any arbitrary scalar } \psi.\\
    F(\psi A_0|\lambda U_0) &= G_{H|U,V}(\psi H_0|\lambda U_0, \psi V_0) \quad \text{where } \lambda > 0 \text{ is a scaling factor}
\end{align*}
where $F(A_0|\psi U_0)$ and $ G_{H|U,V}$ are conditional distributions.
By varying $(\psi, \lambda)$, I can therefore trace out $F(A|U)$ across the entire support of $(A, U)$.

Given finite data, I rely on an estimate of $G_{H|U,V}$ for the constructive estimator. Consider an arbitrary point $(H_\tau, U_\tau, V_\tau)$. To obtain $G(H_\tau|U_\tau, V_\tau)$, I compute the proportion of observations with observed hires below $H_\tau$ among observations proximate to $(U_\tau, V_\tau)$ in $(U, V)$-space. In practice, this is done by averaging across all observations in the data, using a kernel to downweight observations far from $(U_\tau, V_\tau)$. Consequently, the estimate is given by
\[
F(\psi A_0|\lambda U_0) = G_{H|U,V}(\psi H_0|\lambda U_0, \psi V_0)
\]
\[
\hat{F}(\psi A_0|\lambda U_0) = \sum 1(H_t < \psi H_0) \kappa(U_t, V_t, \lambda U_0, \psi V_0)
\]
where $\kappa(\cdot)$ denotes normalized bivariate normal kernel weights with baseline bandwidth 0.10.

Having recovered the distribution function $F(A|U)$, I invert $F(A_t|U_t)$ to derive $A_t$. This is achieved by
\[
A_t = F^{-1}(G(H_t|U_t, V_t)|U_t)
\]
for all observations $(H_t, U_t, V_t)$ in the dataset. Finally, I recover the matching function as
\[
m(A_t U_t, V_t) = G^{-1}_{H|U,V}(F(A_t|U_t)|U_t,V_t).
\]

To calculate matching elasticities, I run a LASSO regression projecting hires on vacancies, efficiency-weighted unemployment, their interaction, and their squared terms.
The estimates approximate the derivatives of the matching function with respect to vacancies and efficiency-weighted unemployment, yielding the elasticities of the matching function.
I denote the unemployment elasticity by $\varepsilon_U \equiv \partial \log m(AU,V)/\partial \log U = \partial \log m(AU,V)/\partial \log (AU)$ and the vacancy elasticity by $\varepsilon_V \equiv \partial \log m(AU,V)/\partial \log V$.
Under CRS, I report $\varepsilon_V = 1-\varepsilon_U$ in the main results, which imposes the adding-up restriction but does not manually censor local estimates that fall outside the unit interval; Appendix \ref{sec:robustness} reports an illustrative diagnostic of the unconstrained LASSO derivative.\footnote{The unemployment elasticity $\frac{d \log m(AU,V)}{d \log U}=\frac{d m(AU,V)}{d U}\frac{U}{H}=\frac{d m(AU,V)}{d AU}\frac{d AU}{dU}\frac{U}{H}=\frac{d m(AU,V)}{d AU}\frac{AU}{H}=\frac{d \log m(AU,V)}{d \log AU}$ is obtained from the regression coefficient of $H$ on $AU$ and multiplying it by $\frac{AU}{H}$. Concretely, I approximate $m$ by the second order polynomial $m=\beta_1 (AU) + \beta_2(AU)V+\beta_3 V + \beta_4 (AU)^2 + \beta_5 V^2$ and get $\frac{d m(AU,V)}{d AU}=\beta_1 + \beta_2V + 2\beta_4 (AU) $ and $\frac{d\log m}{d\log U}=\frac{d\log m}{d\log AU}=(\beta_1 + \beta_2V + 2\beta_4 (AU))\frac{AU}{H}$. For calculating mismatch, I need $\frac{dm}{dU}=\frac{d \log m(AU,V)}{d \log AU}\frac{H}{U}$ at $U=U^*$ and $H=H^*=m(AU^*,V)$.}

\subsection{Mismatch}\label{sec:mismatch_computation}

Unlike the original Cobb-Douglas specification of \cite{csahin2014mismatch}, equilibrium condition \eqref{eq:equilibrium_condition} does not have an analytical formula.
Instead, I employ the Mathematical Programming with Equilibrium Constraints (MPEC) approach proposed by \cite{su2012constrained} and \cite{dube2012improving}.
The constrained optimization problem in time $t$ is defined as follows.
\begin{align}
    \max_{U_{1t},\cdots,U_{Rt}} & \sum_{r=1}^{R}m\left(A_{r t}U_{r t},V_{r t}\right),\\
    \text{subject to }\frac{\partial m}{\partial U}\left(A_{1t}U_{1t}^{*},V_{1t}\right)&=\cdots=\frac{\partial m}{\partial U}\left(A_{Rt}U_{Rt}^{*},V_{Rt}\right) ,\nonumber\\
    U_{t}&=\sum_{r=1}^{R}U_{r t}.\nonumber
\end{align}
Maximizing the objective function is equivalent to solving equilibrium conditions \eqref{eq:equilibrium_condition} so that the central planner's problem represents a nonlinear system of equations with linear equilibrium constraints.
I solve the model using \texttt{Ipopt.jl} and \texttt{JuMP.jl}, which are standard tools for MPEC problems.
The formulation can incorporate additional equilibrium constraints, such as the government's cost of reallocating unemployed workers across markets in calibration exercises, but such extensions are outside the scope of this paper.

\section{Simulation}\label{sec:monte_carlo}
To address practitioners' concerns, I investigate the finite-sample performance of the estimator, extending Section 4 of \cite{matzkin2003nonparametric} to the aggregate labor market setting. The primary focus is on sample size, robustness to model specification, and sensitivity to endogeneity in estimating time-varying matching efficiency. I simulate 100 datasets with sample sizes \(T=50\), \(T=100\), and \(T=200\), representing aggregate-level data over 50, 100, and 200 months, respectively, and consider correlations between \(V\) and \(A\), \(\rho^{AV}=0, 0.1, 0.2, 0.5\). Positive \(\rho^{AV}\) induces upward bias. A stationary AR(1) process is generated for the inputs as \(U_t = \rho^{U} U_{t-1} + \varepsilon_{t}^{U}\), \(V_t = \rho^{V} V_{t-1} + \varepsilon_{t}^{V}\), and \(A_t = \rho^{A} A_{t-1} + \varepsilon_{t}^{A}\), with \((\rho^U, \rho^V, \rho^A) = (0.2, 0.2, 0.2)\), and \((\varepsilon_{t}^{U}, \varepsilon_{t}^{V}, \varepsilon_{t}^{A})\) drawn i.i.d. from the standard normal distribution. If \(\rho^{AV} > 0\), \(A_t\) is recomputed via Cholesky decomposition to achieve the desired correlation with \(V_t\). Inputs are rescaled to mimic actual data by multiplying \(U_t\) and \(V_t\) by 100 and \(A_t\) by 10, normalizing \(A_t\) such that \(A_1 = 100\).

Restricting the analysis to the CRS class, I specify matching function \(m(A_t U_t, V_t)\) as a Cobb-Douglas function \(((A_t U_t)^{\gamma} V_t^{1-\gamma})/10\) and a perfect substitute function \(((\gamma A_t U_t + \gamma V_t)/10)\), where \(\gamma=0.3\). The number of matches \(H_t\) is generated using the true \(m(A_t U_t, V_t)\). For nonparametric estimation, a Gaussian kernel with bandwidth \(h=0.1\) is applied. For comparison, a misspecified Cobb-Douglas model is estimated using OLS, i.e., \(\log(10 H_t) = \hat{A} + \hat{\gamma} \log(U_t) + (1-\hat{\gamma}) \log(V_t)\), recovering \(A = \exp\left(\frac{\hat{A}}{\hat{\gamma}}\right)\) as the normalized efficiency term.

Bias and root-mean-squared error (RMSE) of matching efficiency \(A_t\) are calculated at 50 randomly chosen fixed grid points. The averages of these results, across the 50 points and 100 simulations, are reported in Table \ref{tb:simulation_result_cobb_douglas_0.3_AR1_I0}. The results highlight differences between the nonparametric (NP) and ordinary least squares (OLS) approaches. When the sample size is \(T=200\), the NP method demonstrates robust performance with minimal bias (consistently below 1.5\%) and relatively low RMSE, even under varying levels of \(\rho^{AV}\). By contrast, the OLS approach exhibits substantial bias and significantly larger RMSE across all scenarios due to the misspecification of \(A_t\) as time-invariant. Furthermore, the inaccuracies in \(\hat{A}\) and \(\hat{\gamma}\) amplify errors in recovering \(A_t\) via log-transformation. These findings underscore the reliability of the NP method in capturing time-varying matching efficiency, particularly for larger sample sizes, and its modest robustness to endogeneity, contrasting with the limitations of the OLS approach.\footnote{The endogeneity problem can be resolved by applying the instrumental variable approach proposed by \cite{imbens2009identification}, although finding the instruments is usually difficult in the aggregate labor market setting.}

\begin{table}[htbp]
  \begin{center}
      \caption{Monte Carlo simulation results of matching efficiency $A_t$ for $t=1,\cdots,T$}
       \label{tb:simulation_result_cobb_douglas_0.3_AR1_I0}
      \subfloat[Cobb-Douglas]{
      \begin{tabular}[t]{rrrrrr}
      \toprule
      Sample size & $\rho^{AV}$ & Bias (NP) & RMSE (NP) & Bias (OLS) & RMSE (OLS)\\
      \midrule
      50 & 0.0 & 1.28 & 14.50 & 2300.12 & 2370.36\\
      50 & 0.1 & 1.96 & 14.73 & 917.18 & 1008.53\\
      50 & 0.2 & 1.14 & 14.51 & 181.41 & 305.90\\
      50 & 0.5 & 1.98 & 13.91 & -86.73 & 93.84\\
      100 & 0.0 & -0.43 & 13.13 & 406.54 & 456.63\\
      100 & 0.1 & -0.92 & 12.97 & 68.66 & 143.96\\
      100 & 0.2 & -0.28 & 14.24 & -29.17 & 105.86\\
      100 & 0.5 & -0.84 & 14.67 & -94.58 & 94.93\\
      200 & 0.0 & -0.25 & 12.20 & 79.84 & 118.95\\
      200 & 0.1 & -1.01 & 11.93 & -20.09 & 77.27\\
      200 & 0.2 & -0.82 & 12.93 & -65.47 & 71.56\\
      200 & 0.5 & -1.13 & 10.88 & -98.39 & 98.72\\
      \bottomrule
      \end{tabular}
      }\\
      \subfloat[Perfect substitute]{
      \begin{tabular}[t]{rrrrrr}
      \toprule
      Sample size & $\rho^{AV}$ & Bias (NP) & RMSE (NP) & Bias (OLS) & RMSE (OLS)\\
      \midrule
      50 & 0.0 & 3.51 & 9.33 & 269.67 & 357.98\\
      50 & 0.1 & 2.88 & 11.27 & 310.58 & 441.84\\
      50 & 0.2 & 2.93 & 11.58 & -51.56 & 105.06\\
      50 & 0.5 & 3.16 & 14.00 & -98.41 & 98.75\\
      100 & 0.0 & 1.78 & 7.78 & -24.58 & 86.31\\
      100 & 0.1 & 1.75 & 7.54 & -58.95 & 85.38\\
      100 & 0.2 & 1.51 & 8.92 & -79.45 & 86.61\\
      100 & 0.5 & 1.49 & 13.78 & -98.69 & 99.03\\
      200 & 0.0 & 0.90 & 6.97 & -60.63 & 69.08\\
      200 & 0.1 & 0.90 & 6.58 & -69.45 & 77.17\\
      200 & 0.2 & 1.41 & 7.73 & -87.99 & 88.41\\
      200 & 0.5 & -0.77 & 13.37 & -98.76 & 99.06\\
      \bottomrule
      \end{tabular}
      }
  \end{center}\footnotesize
  Note: NP = nonparametric approach, OLS = ordinary least squares. The true value of matching efficiency $A_t$ is normalized to 100 at $t=1$. I calculate bias and root-mean-squared-error (RMSE) of matching efficiency $A_t$ at 50 randomly chosen fixed grid points. The averages of these results over the 50 points and 100 simulations are reported. The replication code is available on the author's GitHub.
\end{table}

% Second, I report bias and root-mean-squared-error (RMSE) of the mismatch index. I compare my nonparametric mismatch index with the classical mismatch index \citep{jackman1987structural} and Cobb-Douglas mismatch index \citep{csahin2014mismatch}.

% \begin{itemize}
%     \item How large is the bias of mismatch from the Cobb Douglass specification?
% \end{itemize}

\section{Data}

I use two datasets for complementary goals: (i) nationally representative evidence for external validity and (ii) a clearly defined platform-based market for nonparametric identification of matching dynamics. Dataset~1 (Labour Force Survey combined with the Survey on Employment Trends) covers observed hiring channels within the Survey on Employment Trends coverage---private-sector establishments with five or more regular employees in SET-covered industries, excluding sectors outside the SET such as government and primary industries---but is annual and limited to industry- and region-level analysis, and the matching function estimated on it approximates an aggregate matching function for the Japanese labor market within that scope.\footnote{I am grateful to the Editor, Tito Boeri, for suggesting that I complement the Hello Work-based estimates with a nationally representative aggregate matching function, and to Daiji Kawaguchi and Ryo Kambayashi for guidance on the data availability needed to construct this aggregate from the Labour Force Survey and the Survey on Employment Trends.} Dataset~2 (Hello Work) covers only the Public Employment Security Office channel but provides a long monthly time series with finer cross-sectional resolution and a full-time / part-time decomposition, and the matching function estimated on it is interpreted as a platform-level matching function for the Hello Work intermediation system.

\subsection{Dataset 1: Nationally Representative Aggregate Data}
For the first dataset, I combine two government surveys to construct annual matching panels at three different resolutions---country, industry, and region: the Survey on Employment Trends (SET, \textit{Koyou Doukou Chousa}), an establishment survey published by the Ministry of Health, Labour and Welfare (MHLW), and the Labour Force Survey (LFS, \textit{Roudouryoku Chousa}), a household survey published by the Ministry of Internal Affairs and Communications (MIC).
The SET provides hires ($H$) and unfilled vacancies ($V$), and the LFS provides unemployment ($U$); all variables are measured in thousands of persons.
Unlike the Hello Work data in Dataset~2, these series are nationally representative within the SET's coverage of private-sector establishments with five or more regular employees, and they capture matches realized through observed channels in that coverage rather than only through the Public Employment Security Office.

I define the three matching variables as follows.
For the region and industry panels, let $r$ index a cross-sectional market, corresponding to a region or an industry. Hires $H_{r t}$ are the annual flow of \emph{hires from non-employment} only (\textit{mishuugyou nyuushokusha} in the SET), explicitly excluding job-to-job transitions, and are taken from the SET table on accessions by prefecture, industry, enterprise size, sex, and job history. Restricting $H$ to hires from non-employment is required for stock--flow consistency with the LFS unemployment stock $U_{r t}$ (which excludes job changers) and avoids the mechanical $H/U > 1$ that would otherwise arise from Japan's high inter-firm mobility; see Appendix~\ref{sec:data_construction} for full discussion.
Vacancies $V_{r t}$ are the end-of-June stock of unfilled positions, including both full-time and part-time openings, from the SET table on unfilled vacancies by industry, enterprise size, region, and occupation.
Unemployment $U_{r t}$ is the annual average of completely unemployed persons from the LFS, classified by industry of previous job for the industry panel, by prefecture for the region panel, and as the national total for the country panel.
Because $V$ is an end-of-June stock, $U$ is an annual average, and $H$ is an annual flow, the three variables are not aligned at a single point in time within a year; this is a standard limitation of annual matching function estimation.

I construct three panels with different geographic and industry resolutions; each panel has a different sample period determined by data availability and classification consistency.

\paragraph{Country panel (year, 1972--2024).}
At the national level, I observe $T = 53$ annual observations spanning 1972--2024. Hires $H_t$, vacancies $V_t$, unemployment $U_t$, and the labor force are taken from the SET and the LFS over the same span. Because the publicly available SET vacancy and hire tabulations do not extend back to 1972 in their currently published form, the country-level series for 1972 to the SET-publication start year are reconstructed from longer-running SET rate series and LFS employee counts; the full reconstruction details are in Appendix~\ref{sec:data_construction}.

\paragraph{Industry panel (year $\times$ industry, 2012--2024).}
The industry panel comprises the 16 major industry groups common to the SET (workplace establishment classification) and the LFS (industry of previous job) over 2012--2024, yielding $N = 208$ industry-year observations. I use the industry-of-previous-job classification because the LFS does not provide unemployment by destination occupation. At the estimation stage I drop two industries (Mining and Electricity/Gas/Water) for which $U_{r t}$ takes too few distinct values to identify the local nonparametric regression, leaving $N = 182$ effective observations; details are reported in Appendix~\ref{sec:data_construction}.

\paragraph{Region panel (year $\times$ region, 2012--2024).}
The region panel aggregates prefecture-level hires and unemployment to the 13 SET regional blocks---Hokkaido, Tohoku, Kita-Kanto, Minami-Kanto, Hokuriku, Tokai, Kinki, Keihanshin, San'in, San'yo, Shikoku, Kita-Kyushu, and Minami-Kyushu---over 2012--2024, yielding $N = 169$ region-year observations. The sample starts in 2012 because of the SET's post-2012 industry classification revision and the availability of LFS prefecture-level unemployment estimates on a consistent basis. Small-cell suppressions in the SET are treated as missing rather than zero, producing an unbalanced panel.

Two coverage caveats are worth noting in the main text.
First, the LFS classifies unemployment by industry of \emph{previous} job, so $U_{r t}$ for an industry cell includes workers seeking employment in other industries; the estimated matching function should therefore be read as a relationship that incorporates inter-industry mobility.
Second, the geographic units of $U$ (place of residence in the LFS) and $V$ (workplace location in the SET) differ, although the use of broad regional blocks rather than prefectures attenuates the resulting commuting-flow measurement error.
Detailed source-table mappings, the JSIC industry crosswalk, and the full set of measurement caveats are reported in Appendix~\ref{sec:data_construction}.

\subsection{Dataset 2: Hello Work Data with a Clear Market Definition}
For the second dataset, I use the Report on Employment Service (\textit{Shokugyo Antei Gyomu Tokei}), published by the Ministry of Health, Labour and Welfare (MHLW), which records job seekers, vacancies, and successful placements through the Public Employment Security Office, commonly known as Hello Work. Hello Work is a government-operated employment service that provides job counseling, placement services, and vocational training. Because posted wages in Hello Work vacancies tend to be lower than the national average, and because the service is provided free of charge by the government, Hello Work should also be understood as an employment safety net rather than as a private matching platform. Because these statistics cover the Public Employment Security Office channel rather than all matching channels in Japan, the estimates are interpreted as outcomes in the Hello Work segment. The data are widely used in Japanese matching-function studies \citep{kano2005estimating, kambayashi2006vacancy, sasaki2008matching, kambayashi2013role, higashi2018spatial}, and \cite{kawata2021first} use recent Hello Work data to measure the welfare effects of the COVID-19 shock on unemployed workers.\footnote{\url{https://www.crepe.e.u-tokyo.ac.jp/material/crepecl12.html}: Accessed 2024 June 6.} I use three Hello Work panels.

\paragraph{Country panel (month, 1972--2023).}
The aggregate country panel covers the full-time + part-time Hello Work market at the monthly level from January 1972 to December 2023, yielding $T = 624$ observations; the full-time and part-time series each also contain 624 monthly observations. Hires $H_t$ are successful job placements through Hello Work, vacancies $V_t$ are active job openings registered at Hello Work, and unemployment $U_t$ is measured by active job seekers registered at Hello Work. New graduates and vacancies targeting them are excluded; operationally, I use the ``general worker'' (\textit{ippan}) series in the published tables. In the Hello Work system, part-time workers are defined as workers who work fewer hours than regular employees at the same establishment, so the full-time and part-time decomposition reflects the platform's administrative classification. This panel is used to estimate the long-run platform-level matching function and to compare its aggregate dynamics with the nationally representative LFS + SET country-year panel. Appendix \ref{sec:additional_figure} reports additional analysis using year-level aggregate data covering 1963--2023 and decomposes the monthly Hello Work aggregate into full-time and part-time series.

\paragraph{Prefecture panel (month $\times$ prefecture, 2013--2024).}
The prefecture panel covers 47 prefectures at the monthly level from January 2013 to April 2024, yielding $N = 6{,}392$ prefecture-month observations in the nonparametric estimation output. It provides the main geographic cross-sectional panel for the Hello Work market. Hires, vacancies, and job seekers are all measured within the Hello Work system, so the panel has a consistent platform-level market definition across prefectures.

\paragraph{Occupation panel (month $\times$ job category, 2012--2023).}
The occupation panel covers 67 Hello Work job categories at the monthly level from March 2012 to March 2023, yielding $N = 8{,}911$ job-category-month observations in the nonparametric estimation output. Unemployed workers are tabulated by registered desired occupation (\textit{yuukou kyuushokusha}), and vacancies are tabulated by occupation-specific openings (\textit{yuukou kyuujin}) under the MHLW occupational classification. This panel is used to measure occupational mismatch within the Hello Work platform. The sample period is constrained by the revision of job classifications before and after 2012, as in \cite{kawata2019} and \cite{higashi2018spatial}. Because occupation classifications and publication formats are not consistently available jointly with prefectures over the full period, I analyze prefecture and occupation panels separately rather than estimating a combined prefecture-by-occupation panel. Finer market definitions are conceptually attractive because they expand the set of across-market reallocations available to the planner; mechanically, mismatch indices tend to be weakly larger as markets become more disaggregated. \cite{csahin2014mismatch} make a related choice in their appendix: because CPS unemployment microdata become too thin for fully disaggregated county-by-occupation markets, they use the interaction of two-digit occupations and nine Census divisions at the quarterly frequency to reduce small-cell noise. The same trade-off applies here: occupation-region cells would be useful, but stable measurement requires a worker-side panel that is not publicly available with consistent hires, job seekers, and vacancies over the relevant period.

\subsection{Hello Work Coverage and Cross-Dataset Comparison}\label{sec:hellowork_coverage}

\begin{figure}[!ht]
  \begin{center}
  \includegraphics[width = 0.9\textwidth]{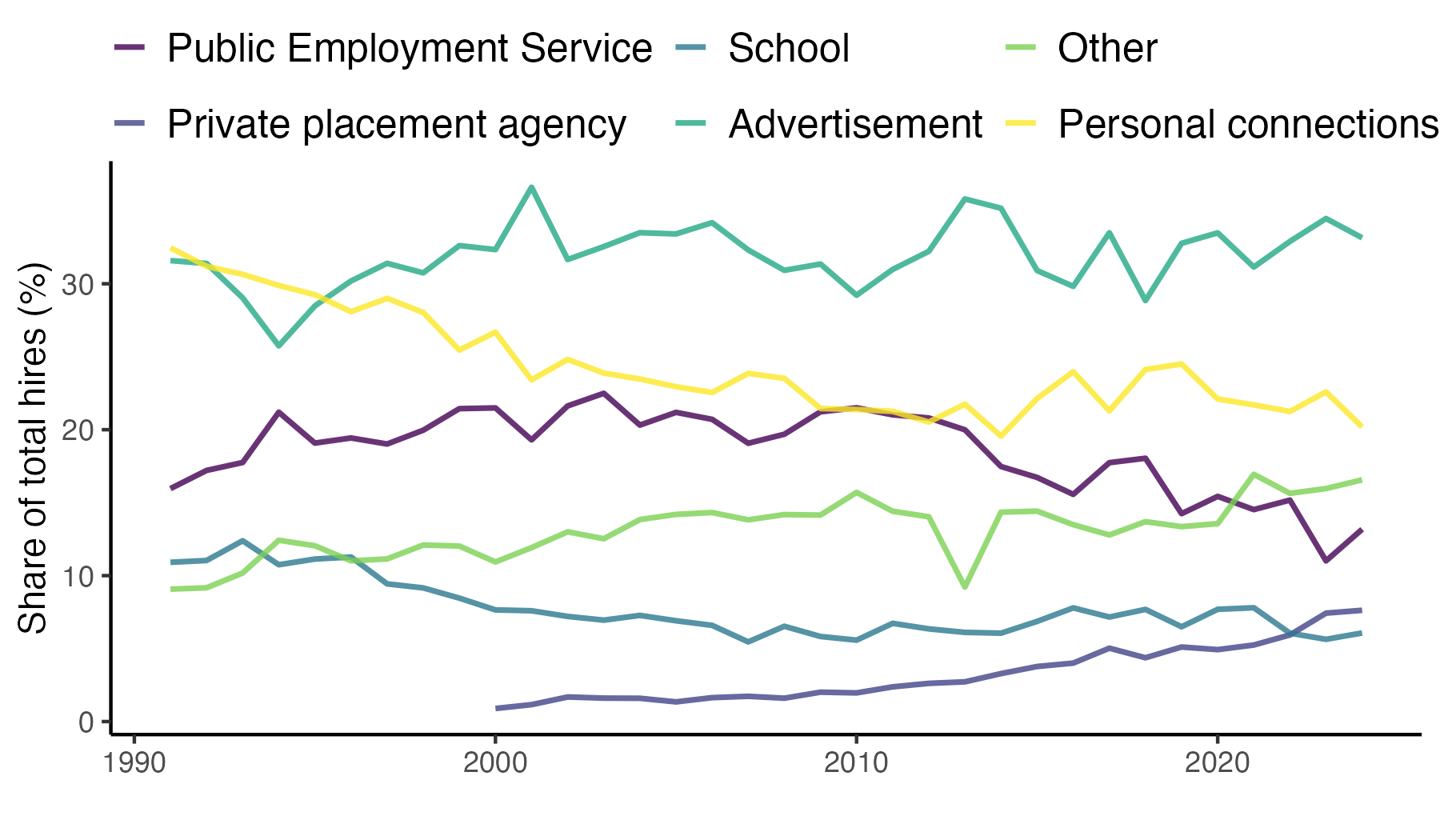}
  \caption{Share of annual hires by hiring route, 1991--2024}
  \label{fg:hire_route_share}
  \end{center}
  \footnotesize
  Note: Computed from the Survey on Employment Trends as route-specific hires divided by total hires. ``Other'' aggregates the original ``other'' category, secondment, and return from secondment. Private placement agencies are surveyed only from 2000.
\end{figure}

Figure~\ref{fg:hire_route_share} reports the share of annual hires by hiring route, computed from the Survey on Employment Trends for 1991--2024. The Public Employment Service (Hello Work) accounts for about 11--22\% of total hires across the sample (mean 18.5\% over 1991--2024), while advertisement-driven hires are consistently the largest single channel at about 30--35\%, followed by personal connections (which decline from 32\% in 1991 to 20\% in 2024). Private placement agencies, surveyed only from 2000, expand from below 1\% to about 8\%, and school referrals decline from 11\% to 6\%. Hello Work therefore covers one segment of the Japanese hiring market rather than the aggregate; it remains, however, the official statistic with the longest monthly history and the finest cross-sectional resolution by prefecture and occupation, which explains its prominent use in the matching function literature \citep{kano2005estimating, higashi2018spatial, kawata2019}.

\begin{figure}[!ht]
  \begin{center}
  \subfloat[Hello Work (1973--2023)]{\includegraphics[width = 0.45\textwidth]{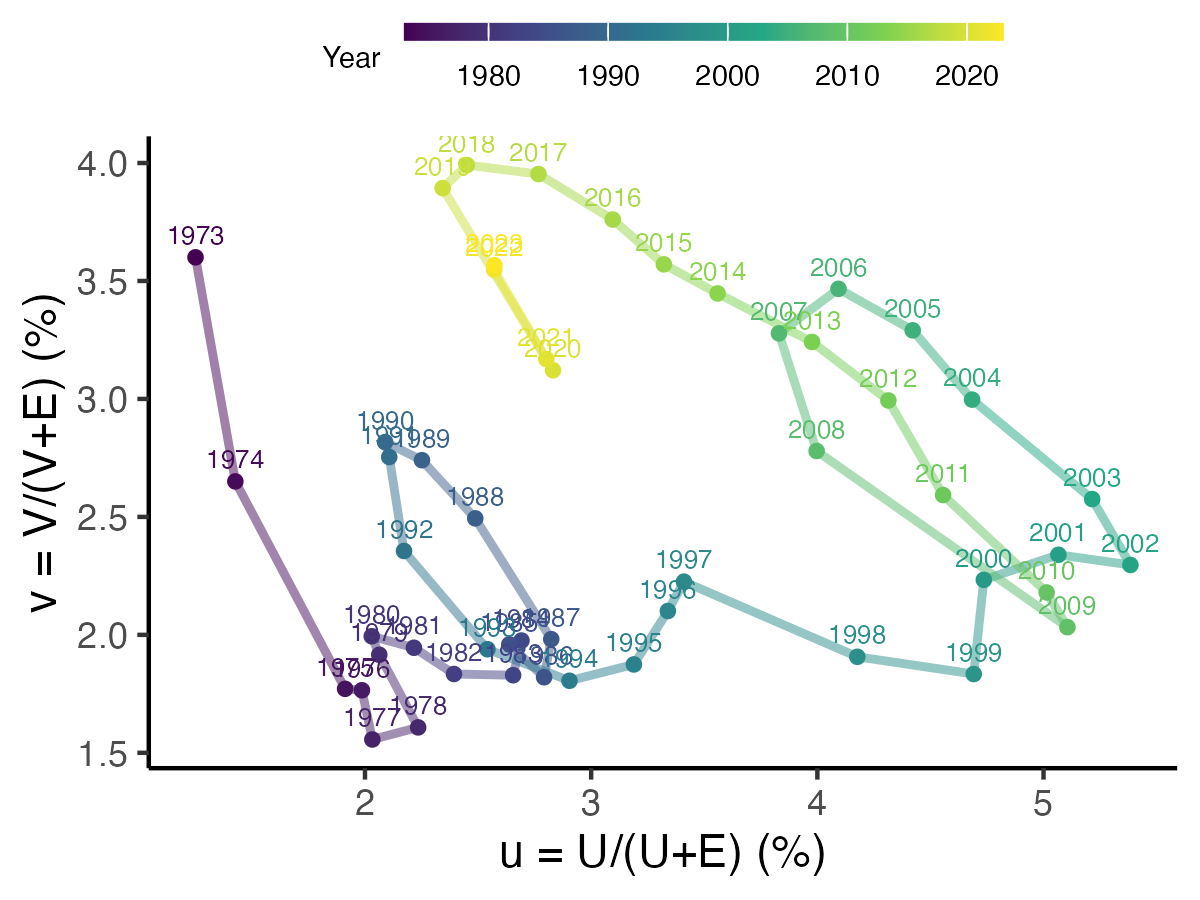}}
  \subfloat[Nationally Representative LFS + SET (1972--2024)]{\includegraphics[width = 0.45\textwidth]{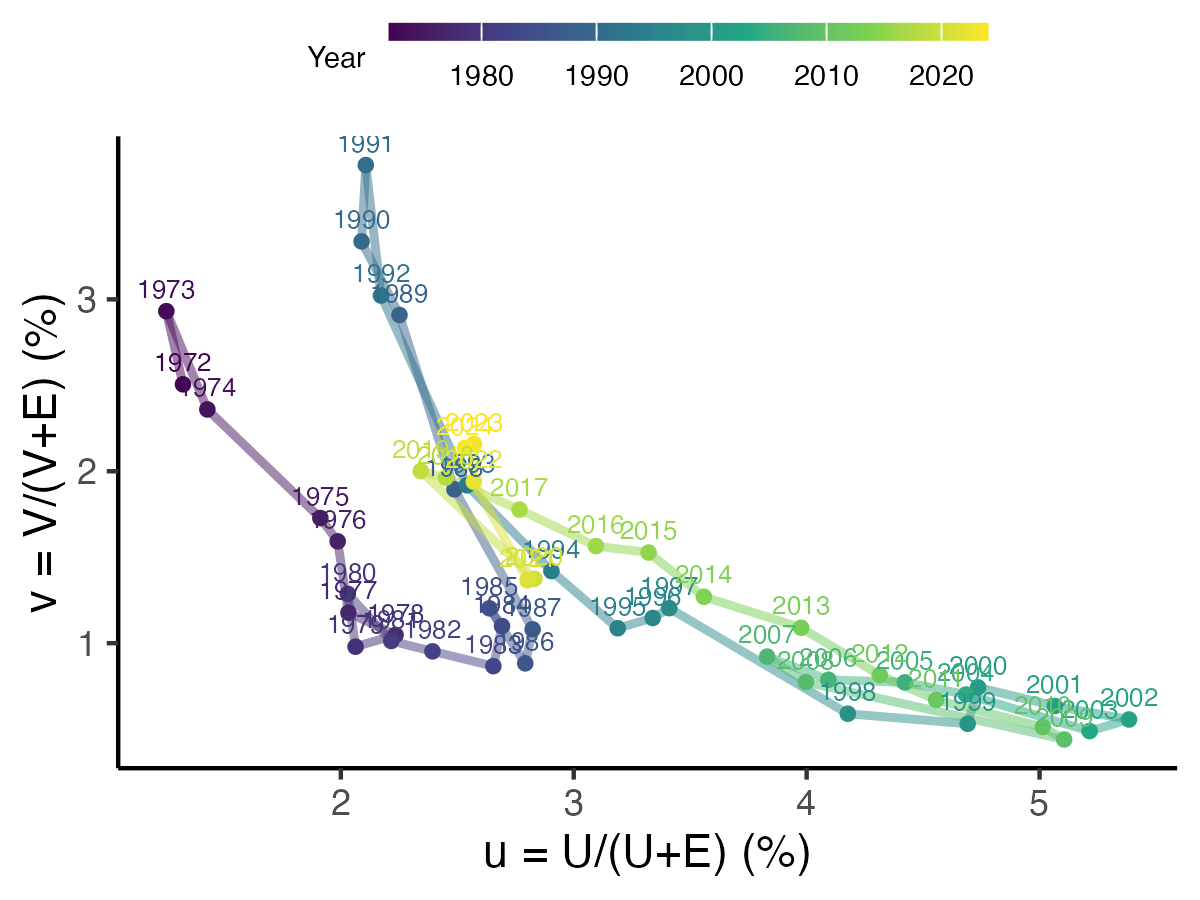}}
  \caption{Beveridge curve: Hello Work vs.\ nationally representative data}
  \label{fg:beveridge_comparison}
  \end{center}
  \footnotesize
  Note: Annual Beveridge curves with $u_t = U_t/(U_t+E_t)$ and $v_t = V_t/(V_t+E_t)$, where $U_t$ is unemployment, $L_t$ is the labor force, and $E_t = L_t - U_t$ is employment from the Labour Force Survey. Vacancy $V_t$ in panel~(a) is from Hello Work (active openings, full-time and part-time), and in panel~(b) from the Survey on Employment Trends. Years are labeled and color-coded by viridis (early years dark, recent years bright) for traceability of the cyclical path. Detailed construction of the vacancy series is in Appendix~\ref{sec:data_construction}.
\end{figure}

Figure~\ref{fg:beveridge_comparison} compares annual Beveridge curves constructed under the \cite{miyamoto2025macroeconomic} definition $u_t = U_t/(U_t+E_t)$ and $v_t = V_t/(V_t+E_t)$, where $U_t$ is unemployment, $L_t$ is the labor force, and $E_t = L_t - U_t$ is employment from the Labour Force Survey in both panels; only the vacancy source differs. Panel~(a) pairs the LFS unemployment and employment series with Hello Work active vacancies (full-time and part-time, 1973--2023), which cover only the Public Employment Security Office channel. Panel~(b) pairs the same LFS unemployment and employment series with the SET vacancy series for SET-covered establishments (1972--2024), so it does not use Hello Work vacancies; the construction of $V$ for the SET series is detailed in Appendix~\ref{sec:data_construction}. The two curves trace qualitatively similar dynamics: a downward-sloping locus with outward shifts during the 1990s and 2000s and inward shifts in recent years, consistent with \cite{miyamoto2025macroeconomic}. Vacancy levels in panel~(a) are lower than in panel~(b), reflecting Hello Work's partial coverage, and the recent inward shift is more pronounced in panel~(a), consistent with the post-2015 decline in the Hello Work share documented in Figure~\ref{fg:hire_route_share}. The qualitative agreement supports interpreting Hello Work-based matching dynamics as informative about the wider Japanese labor market, while the level differences caution against extrapolating Hello Work-based tightness ($V/U$) to the aggregate.

In summary, the two datasets are complementary rather than substitutable. Dataset~1 (LFS + SET) is nationally representative within the SET's coverage of private-sector establishments with five or more regular employees in SET-covered industries, excluding sectors outside the SET such as government and primary industries, but it is annual and limited to industry- and region-level resolutions over 2012--2024 (with country totals reconstructed back to 1972). Dataset~2 (Hello Work) sacrifices population coverage---it captures only the Public Employment Security Office channel---but provides a long monthly estimation sample (1972--2023 at the aggregate level), a clearly defined platform-level market, and finer cross-sectional resolution by prefecture and occupation. The matching function estimated on Dataset~2 should therefore be read as a platform-level matching function for the Hello Work intermediation system, whereas the matching function estimated on Dataset~1 approximates an aggregate matching function for the Japanese labor market within SET coverage. The Beveridge curve comparison in Figure~\ref{fg:beveridge_comparison} suggests that the two move in qualitatively similar directions, so the platform-level dynamics in Dataset~2 are informative about the aggregate, but their levels are not directly comparable.

\section{Empirical Results}
I report empirical applications based on two datasets.
The empirical objective is to balance representativeness and market-definition clarity: Dataset 1 provides a nationally representative benchmark, while Dataset 2 identifies matching dynamics in the clearly defined Hello Work segment over a long time horizon. I report aggregate, disaggregated, and mismatch results below, with additional diagnostics in the Online Appendix.

Before presenting the aggregate trend results, I test the assumption that vacancies are independent of matching efficiency, conditional on overall labor supply, i.e., \( V \perp A \mid U \).
Specifically, I use the residuals from a regression of vacancies \( V \) on the unemployed \( U \), and similarly, the residuals of implied matching efficiency \( A \) on \( U \).
For the aggregate data, the correlation between these two residuals is close to zero (0.04), indicating no systematic relationship between them.
For the Hello Work full-time and part-time data, the corresponding correlations are 0.14 and -0.14.
As shown in Figure \ref{fg:residual_plots} in the Appendix, visible deviations are driven by a few outliers, so any violation of the independence assumption appears limited.

\subsection{Dataset 1: Nationally Representative Aggregate Data}

\subsubsection{Aggregate Trends in 1972--2024}\label{sec:dataset1_aggregate}

\begin{figure}[!ht]
  \begin{center}
  \subfloat[Unemployed ($U$), Vacancy ($V$), and Tightness ($\frac{V}{U}$)]{\includegraphics[width = 0.39\textwidth]
  {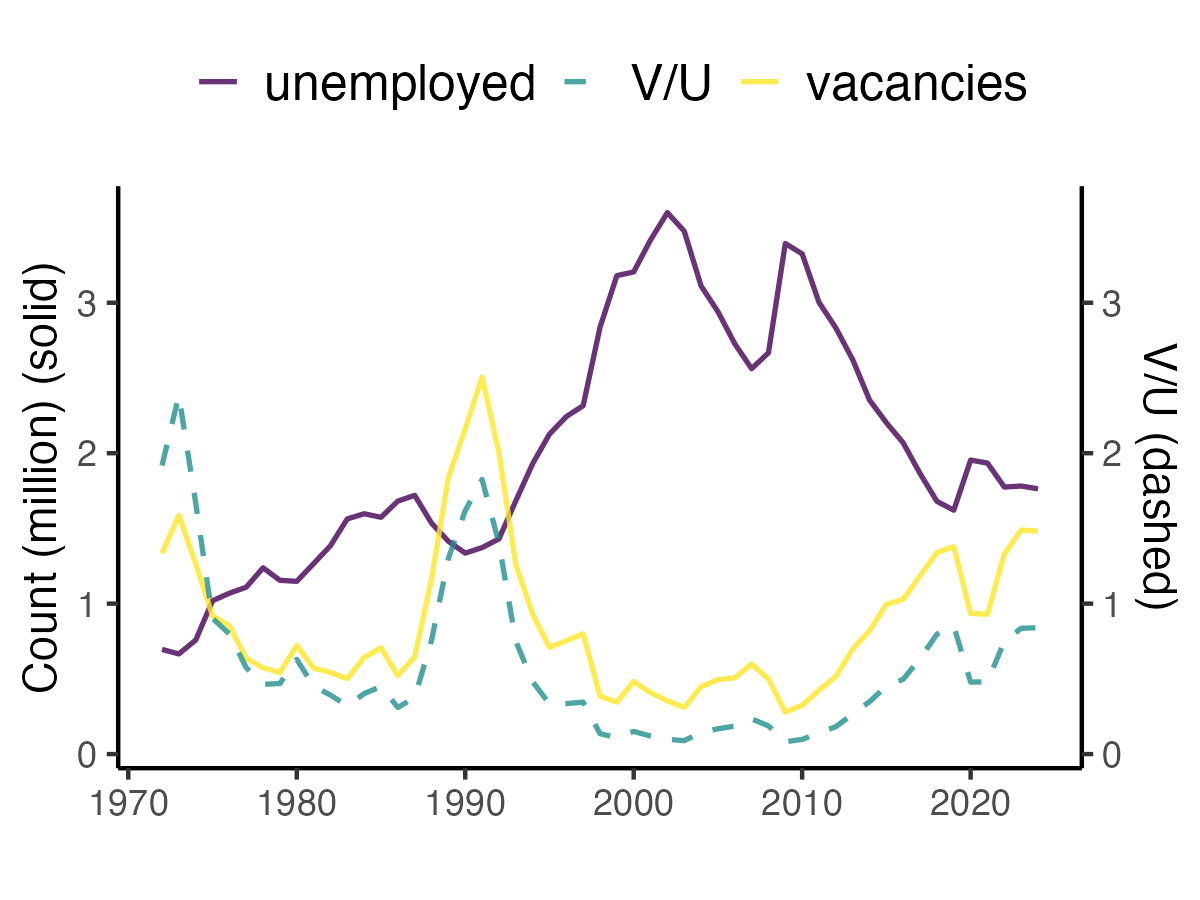}}\hspace{0.15\textwidth}
  \subfloat[Hire ($H$)]{\includegraphics[width = 0.39\textwidth]
  {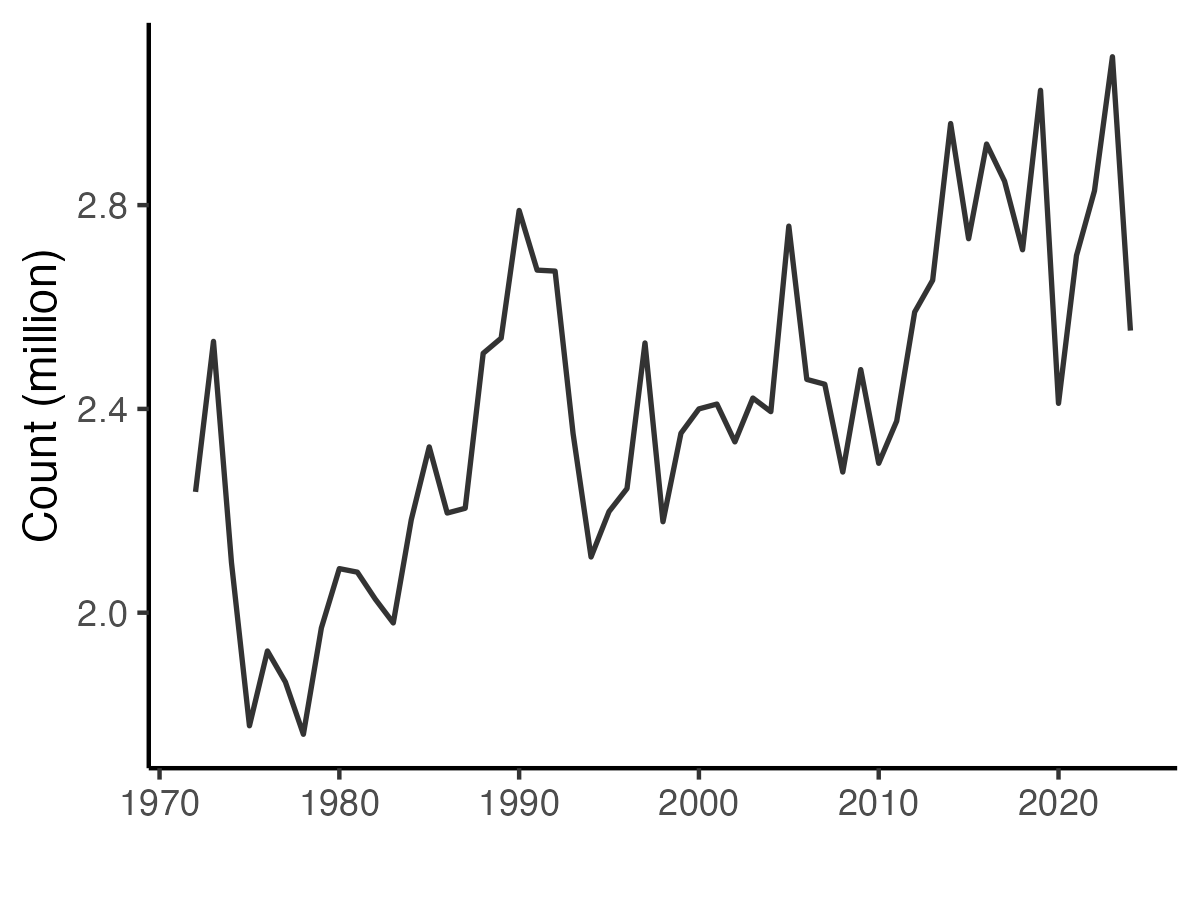}}\\
  \subfloat[Job-finding rate ($\frac{H/U}{12}$) and vacancy-filling rate ($\frac{H/V}{12}$)]{\includegraphics[width = 0.39\textwidth]
  {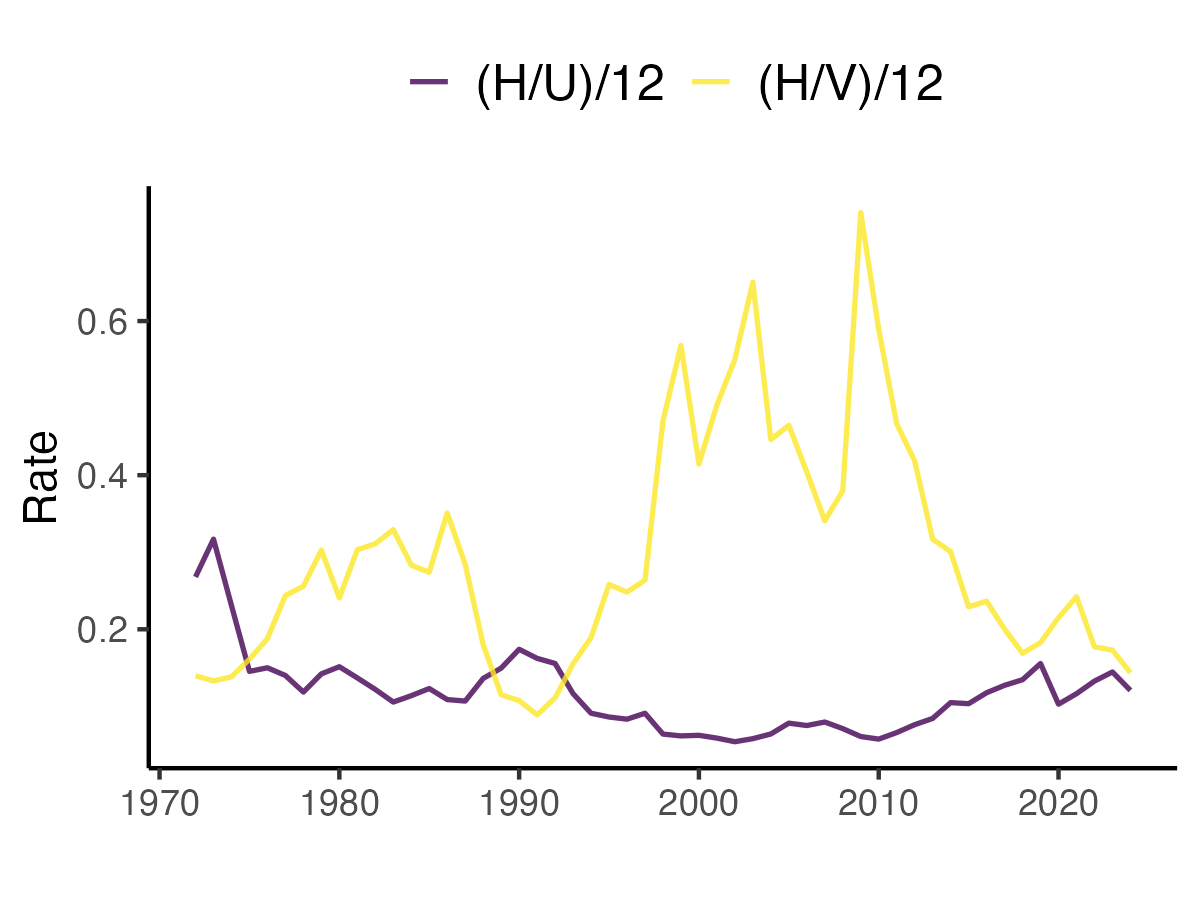}}\hspace{0.15\textwidth}
  \subfloat[Matching Efficiency ($A$)]{\includegraphics[width = 0.39\textwidth]
  {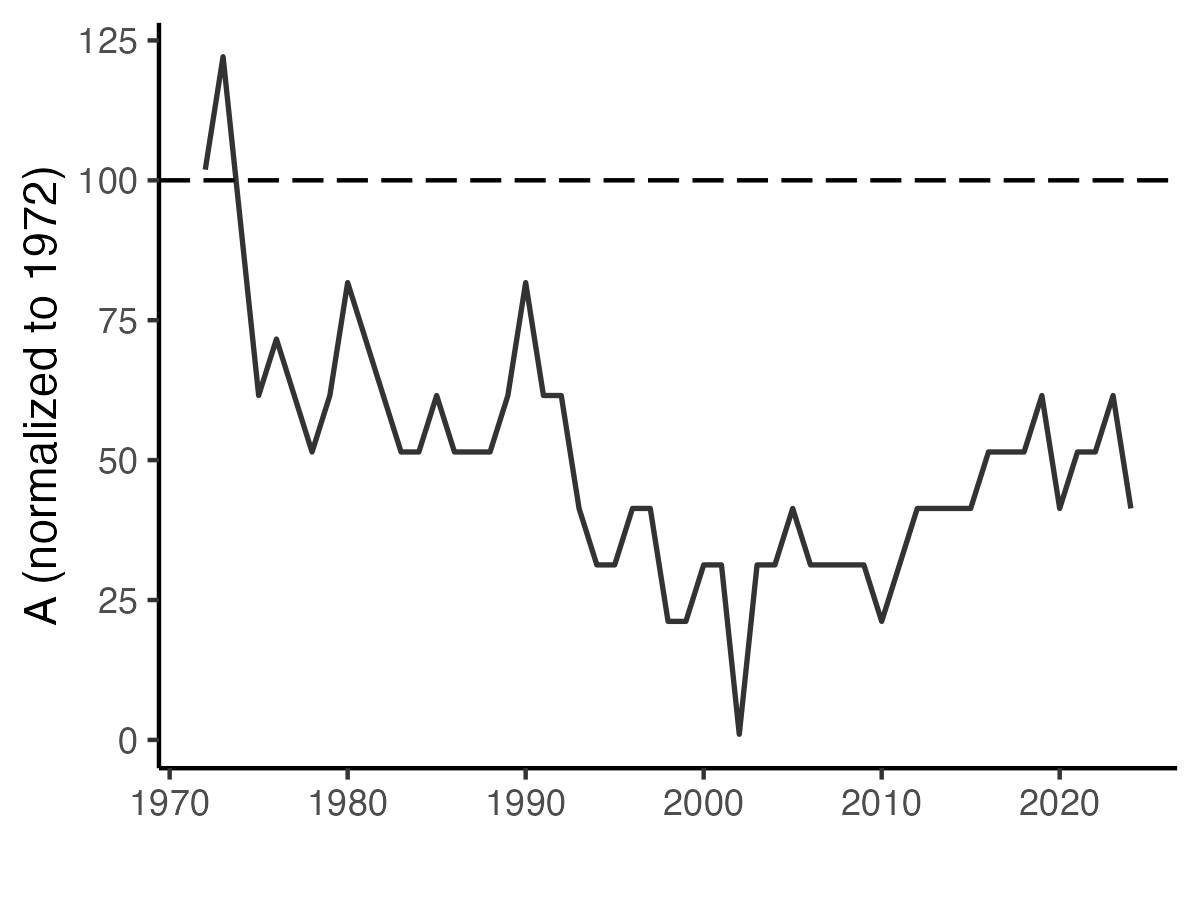}}
  \\
  \subfloat[Matching Elasticity ($\frac{d\ln m}{d \ln AU}$, $\frac{d\ln m}{d\ln V}$)]{\includegraphics[width = 0.39\textwidth]
  {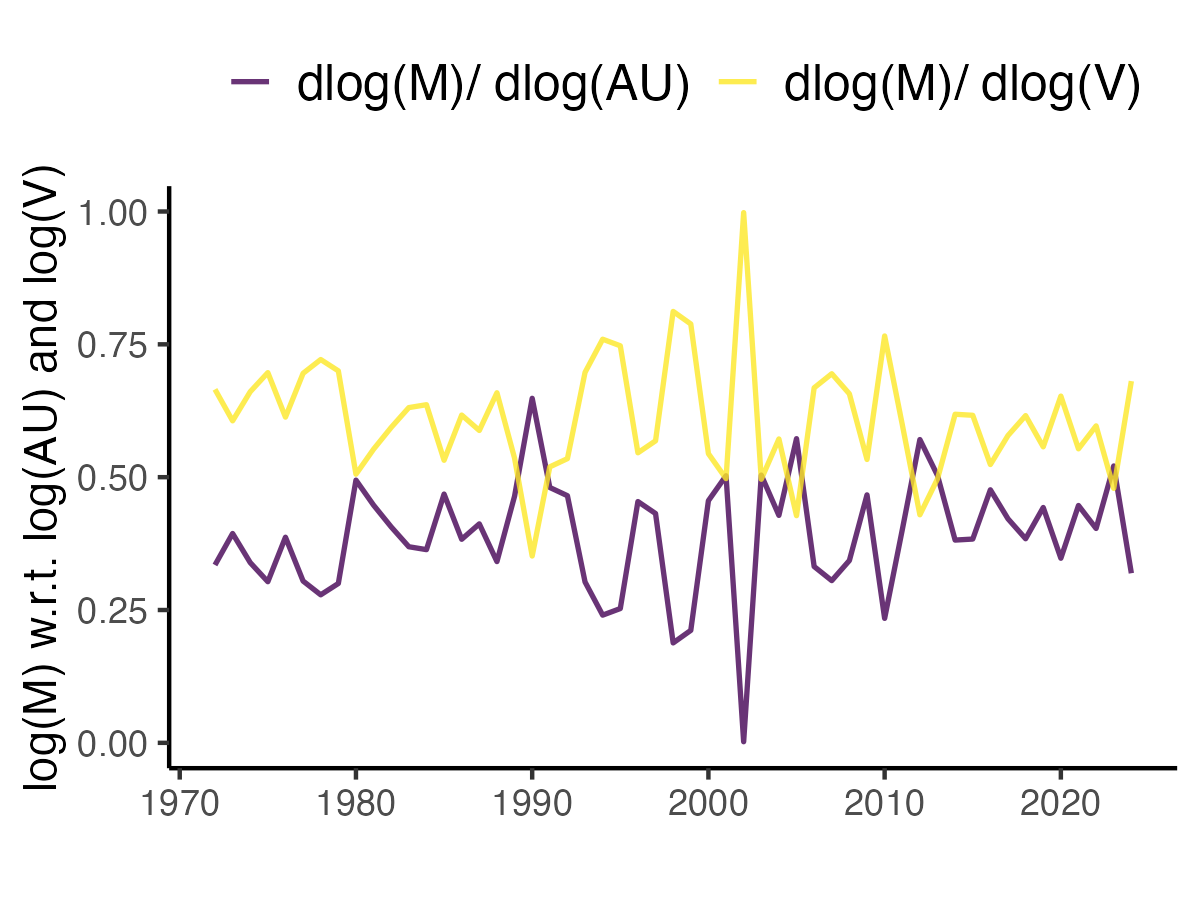}}\hspace{0.15\textwidth}
  \subfloat[Efficiency ($A$) and ($\ln\frac{V}{U}$, $\ln\frac{H}{U}$, $\ln\frac{H}{V}$)]{\includegraphics[width = 0.39\textwidth]
  {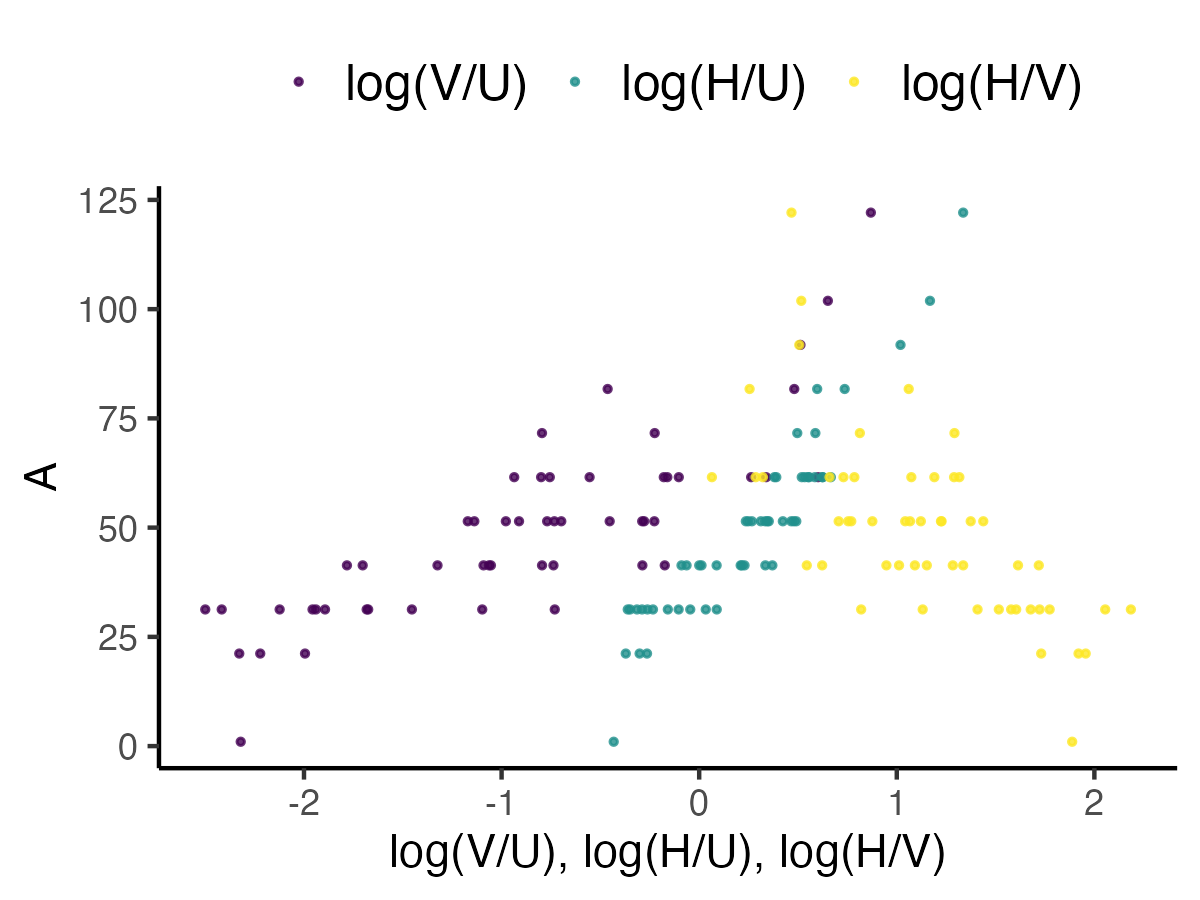}}
  \caption{Country-year-level aggregate results 1972--2024 (LFS + SET data)}
  \label{fg:lfs_set_country_year_results}
  \end{center}
  \footnotesize
  Note: $U$ is the annual average stock of completely unemployed persons from the Labour Force Survey (LFS); $V$ is the stock of unfilled vacancies from the Survey on Employment Trends (SET); $H$ is the annual flow of hires from non-employment from the SET. The finding rates $(H/U)/12$ and $(H/V)/12$ are divided by 12 because $H$ is a 12-month cumulated flow whereas $U$ and $V$ are stocks, so the division yields a monthly-equivalent rate comparable to the Hello Work monthly series in Figure \ref{fg:month_part_and_full_time_results}. Detailed data construction is in Appendix~\ref{sec:data_construction}.
\end{figure}

Figures \ref{fg:lfs_set_country_year_results} (a)-(c) provide annual data patterns of unemployed individuals, vacancies, labor market tightness ($V/U$), hires from non-employment, and the monthly-equivalent job-finding and vacancy-filling rates ($H/(12U)$ and $H/(12V)$) constructed from LFS + SET over 1972--2024. The finding rates are scaled to a monthly basis because $H$ is a 12-month cumulated flow whereas $U$ and $V$ are stocks.
In summary, the number of unemployed fluctuates between roughly 0.7 and 3.6 million while vacancies range between roughly 0.3 and 2.5 million, both tracking business cycle variations; tightness peaks in the late 1980s and again in the late 2010s. The late-1980s vacancy boom corresponds to Japan's asset-price bubble, while the rise in unemployment and fall in vacancies around 2008--2010 correspond to the global financial crisis. Hires from non-employment fluctuate between roughly 1.8 and 3.1 million per year and recover toward recent years, while the monthly-equivalent finding rates trace standard countercyclical and procyclical comovements (job-finding 0.05--0.32 per month, vacancy-filling 0.09--0.74 per month).
Figures \ref{fg:lfs_set_country_year_results} (d)-(f) present the estimation results of the matching function. Matching efficiency (normalized to 1972) shows a long-run downward trend with notable fluctuations, qualitatively consistent with the Hello Work-based estimates in Section \ref{sec:month_level}; the implied unemployment elasticity is centered around 0.3--0.5, and the vacancy elasticity is correspondingly centered around 0.5--0.7. Panel (f) plots efficiency against the three log market-structure variables ($\ln V/U$, $\ln H/U$, $\ln H/V$). The patterns differ noticeably from the Hello Work counterpart in Figure~\ref{fg:month_part_and_full_time_results}~(f): for LFS + SET, $\ln V/U$ and $\ln H/U$ comove positively with efficiency as predicted by \cite{lange2020beyond}, but $\ln H/V$ (yellow) shows a negative correlation with efficiency, in contrast to the more uniformly positive Hello Work pattern. \textcolor{black}{This difference is consistent with the fact that, at the nationally representative aggregate, $V$ reflects broader SET-covered vacancy demand, so periods of high efficiency are typically associated with $H$ rising less than proportionally with $V$ (i.e., $H/V$ stable or falling), whereas in the Hello Work segment a rise in efficiency translates into an unambiguous co-movement with vacancy-filling because the Hello Work platform captures only a small share of total vacancies and is not constrained by aggregate vacancy creation.}

\subsubsection{Region-Level Aggregate Results in 2012--2024}\label{sec:dataset1_area}

\begin{figure}[!ht]
  \begin{center}
  \subfloat[Efficiency ($A$), Area]{\includegraphics[width = 0.52\textwidth]
  {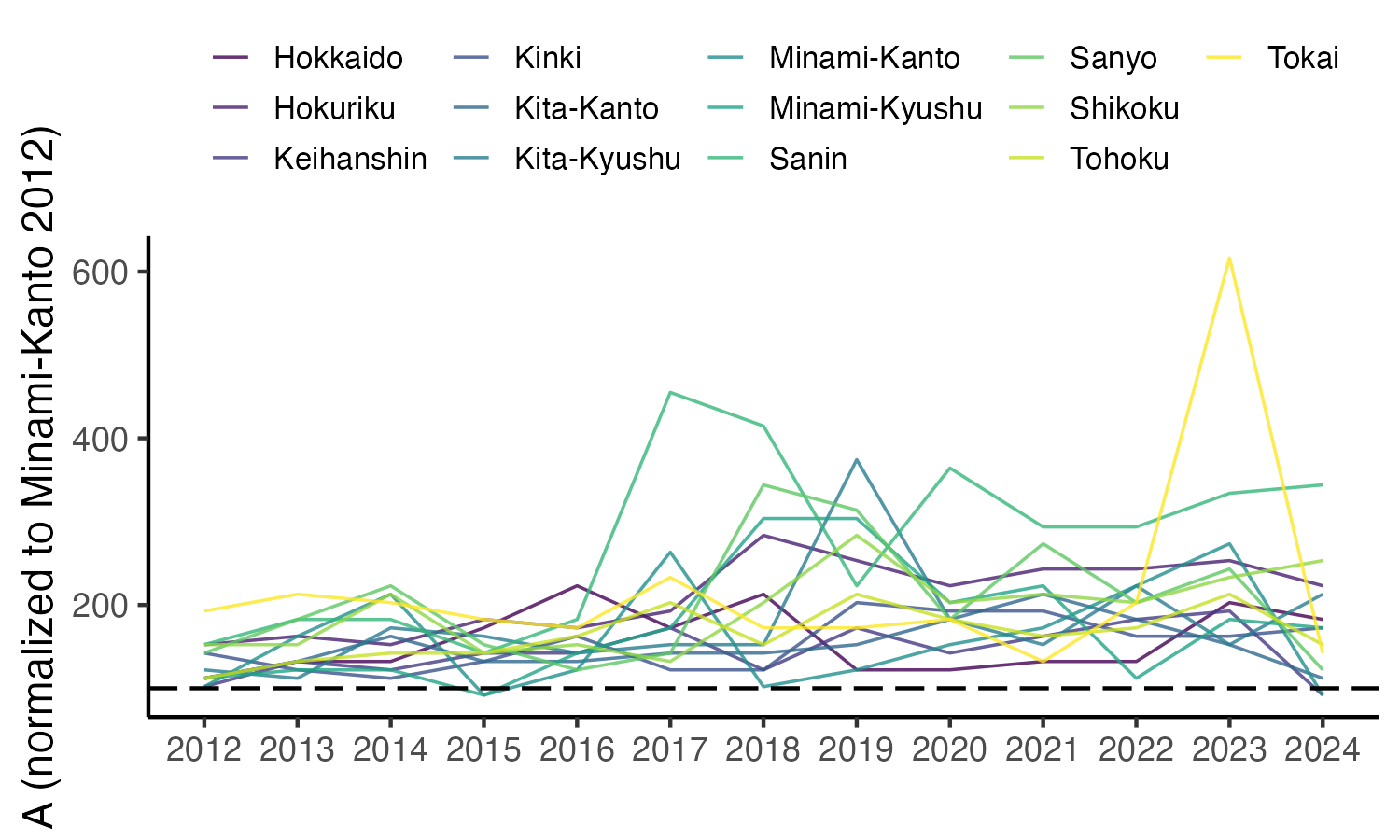}}
  \subfloat[Efficiency ($A$), Industry]{\includegraphics[width = 0.52\textwidth]
  {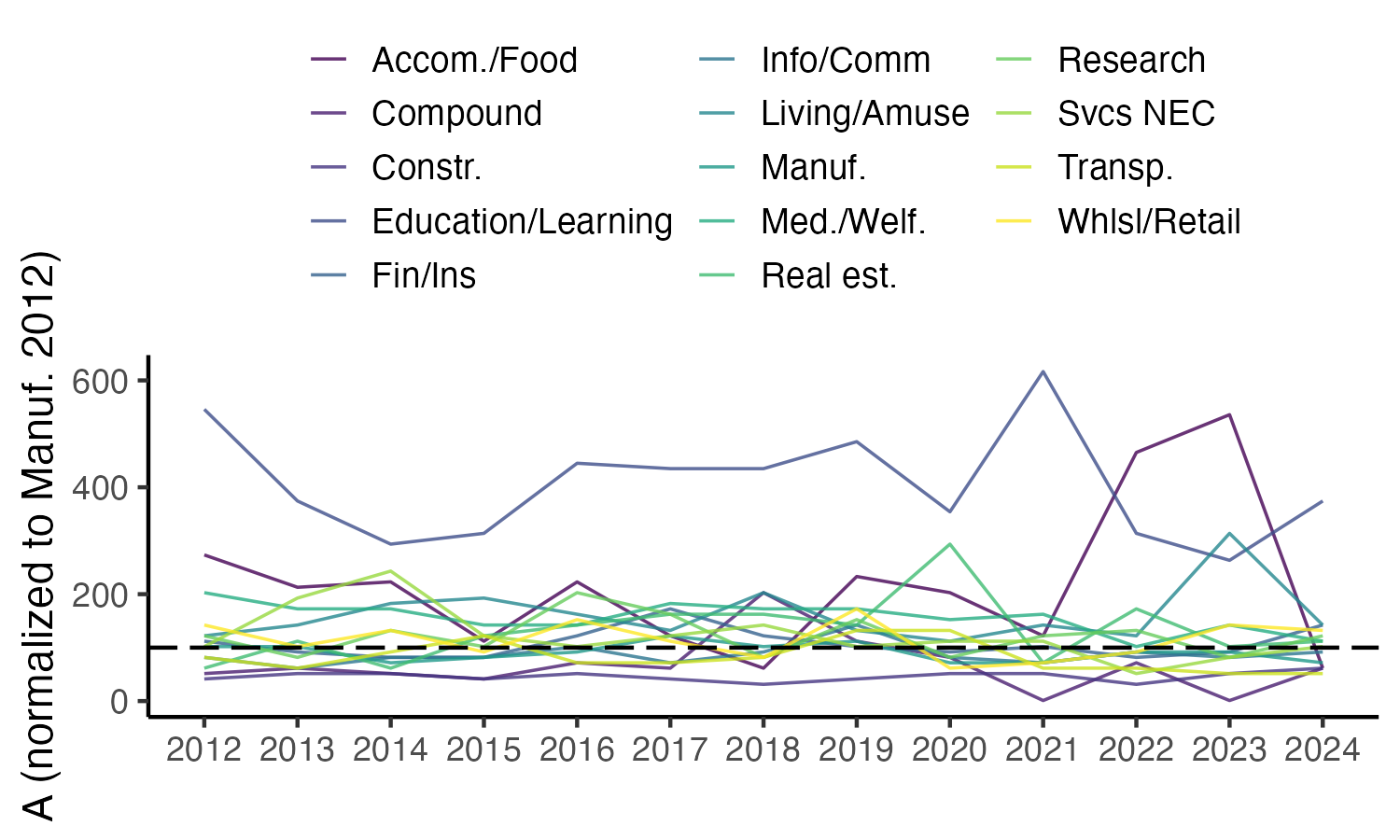}}\\
  \subfloat[$d \ln m / d \ln (AU)$, Area]{\includegraphics[width = 0.52\textwidth]
  {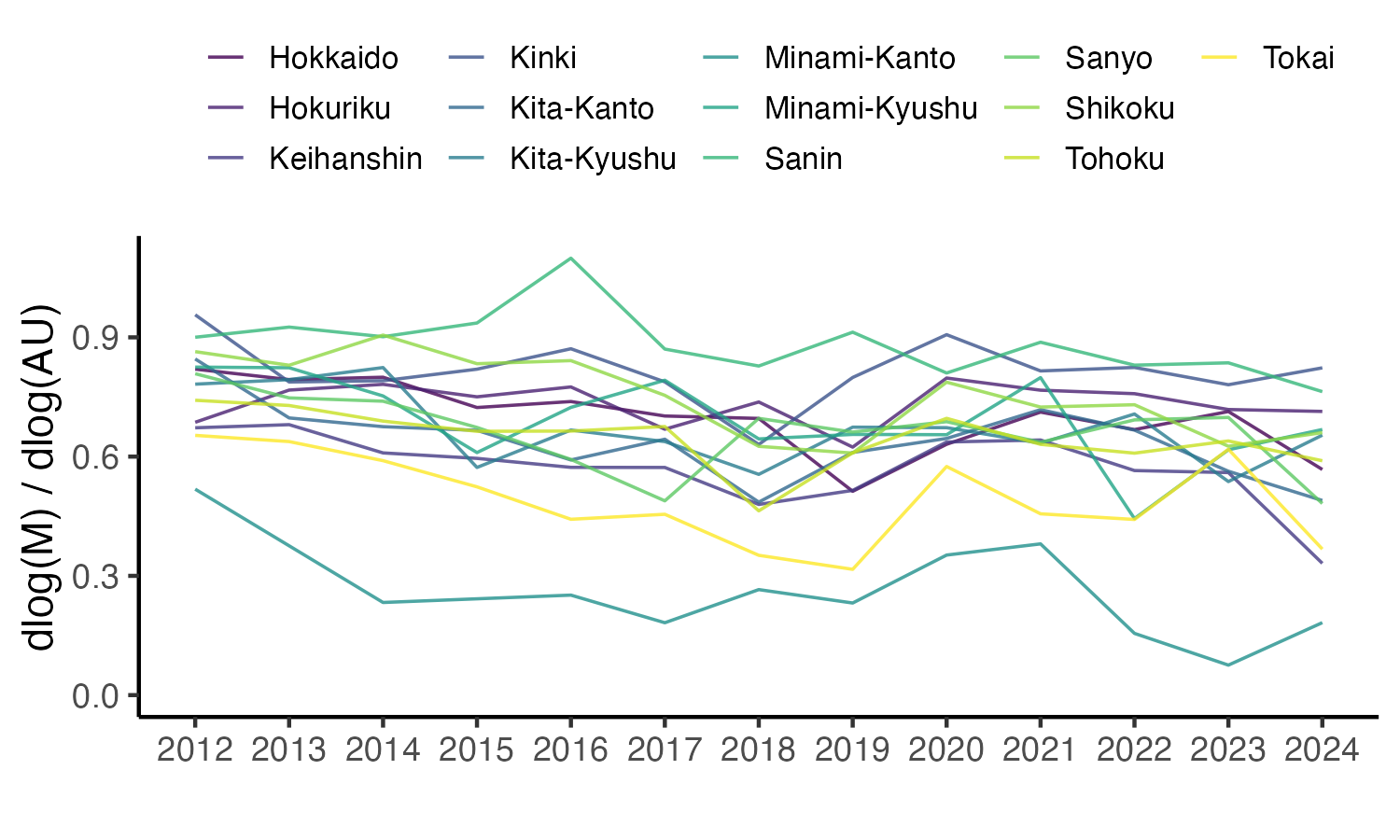}}
  \subfloat[$d \ln m / d \ln (AU)$, Industry]{\includegraphics[width = 0.52\textwidth]
  {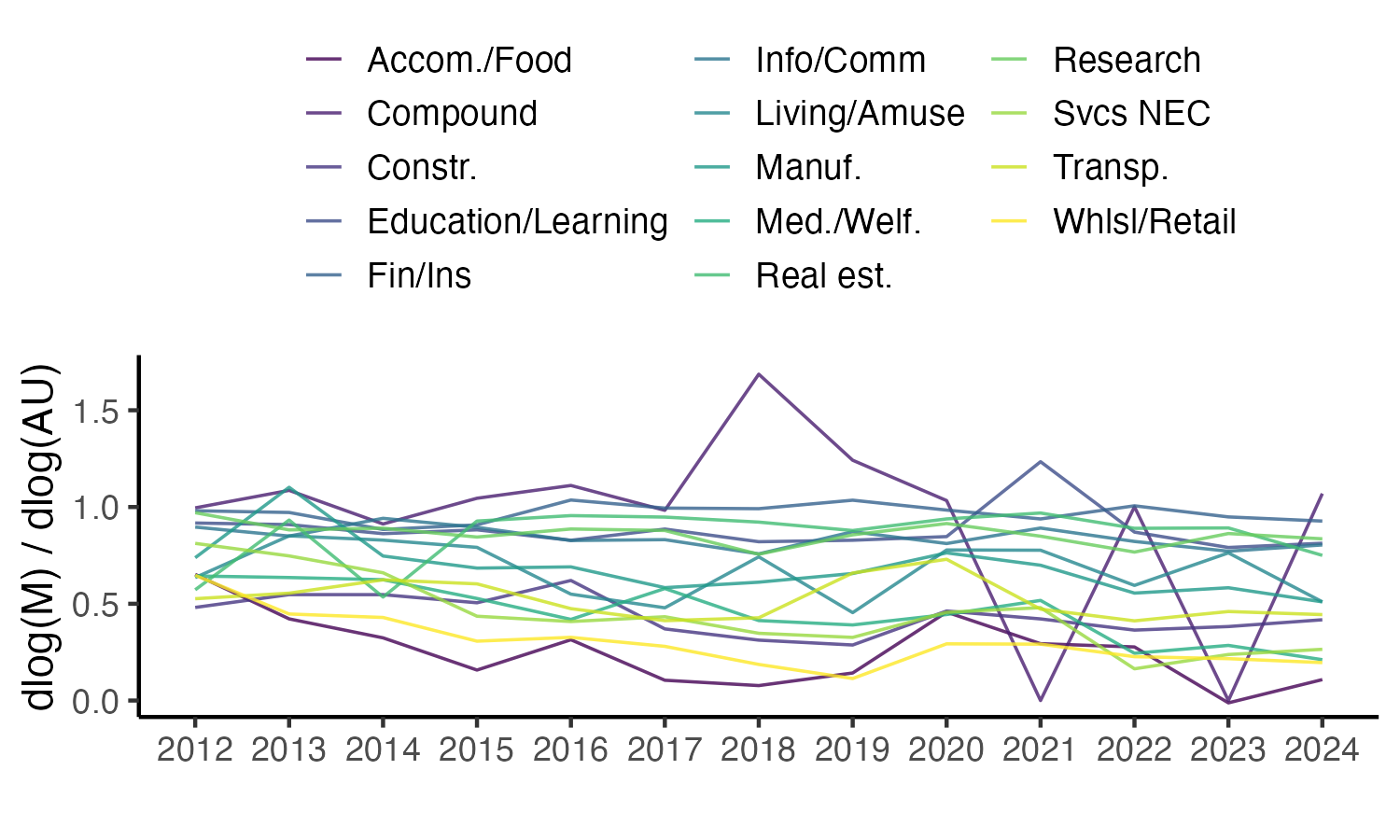}}\\
  \subfloat[$d \ln m / d \ln V$, Area]{\includegraphics[width = 0.52\textwidth]
  {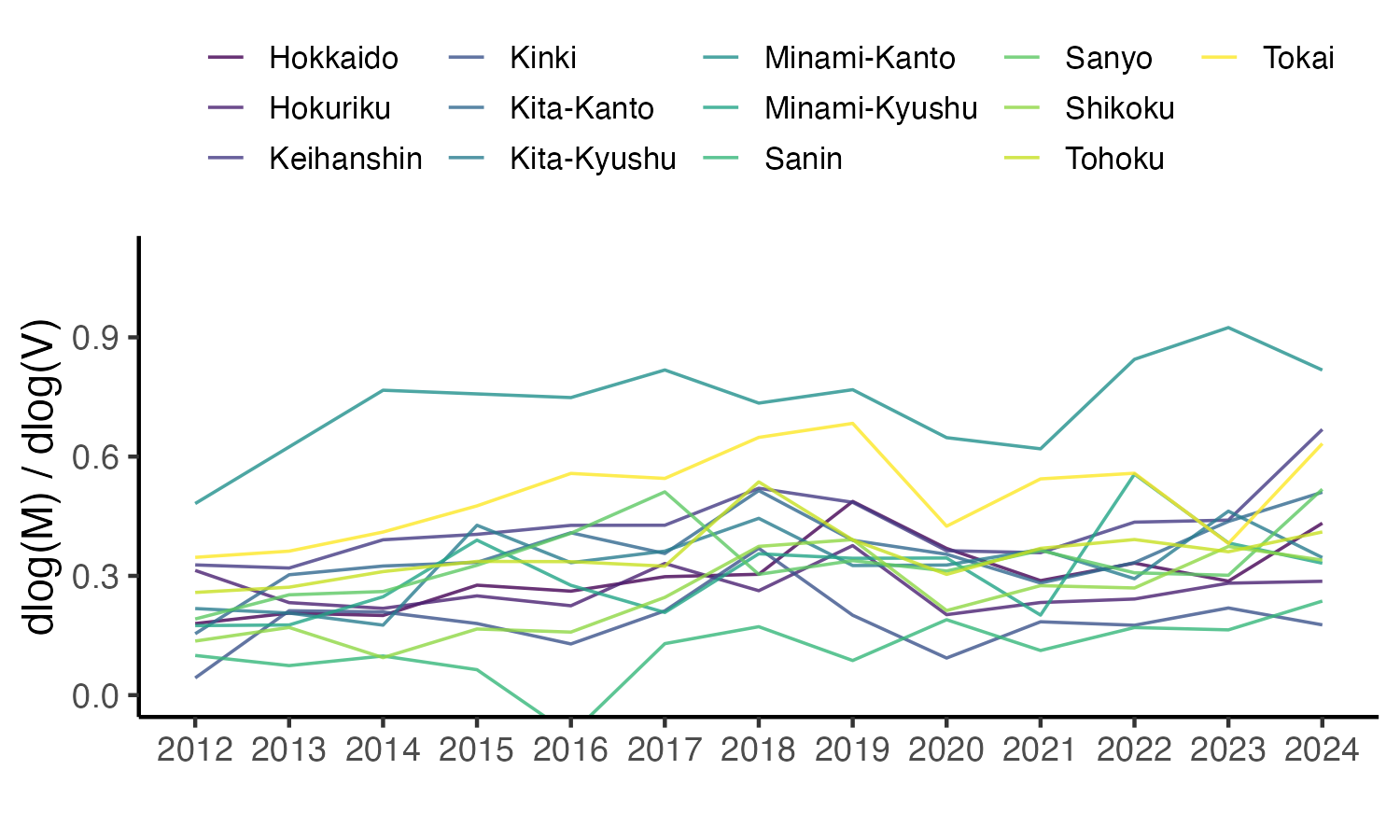}}
  \subfloat[$d \ln m / d \ln V$, Industry]{\includegraphics[width = 0.52\textwidth]
  {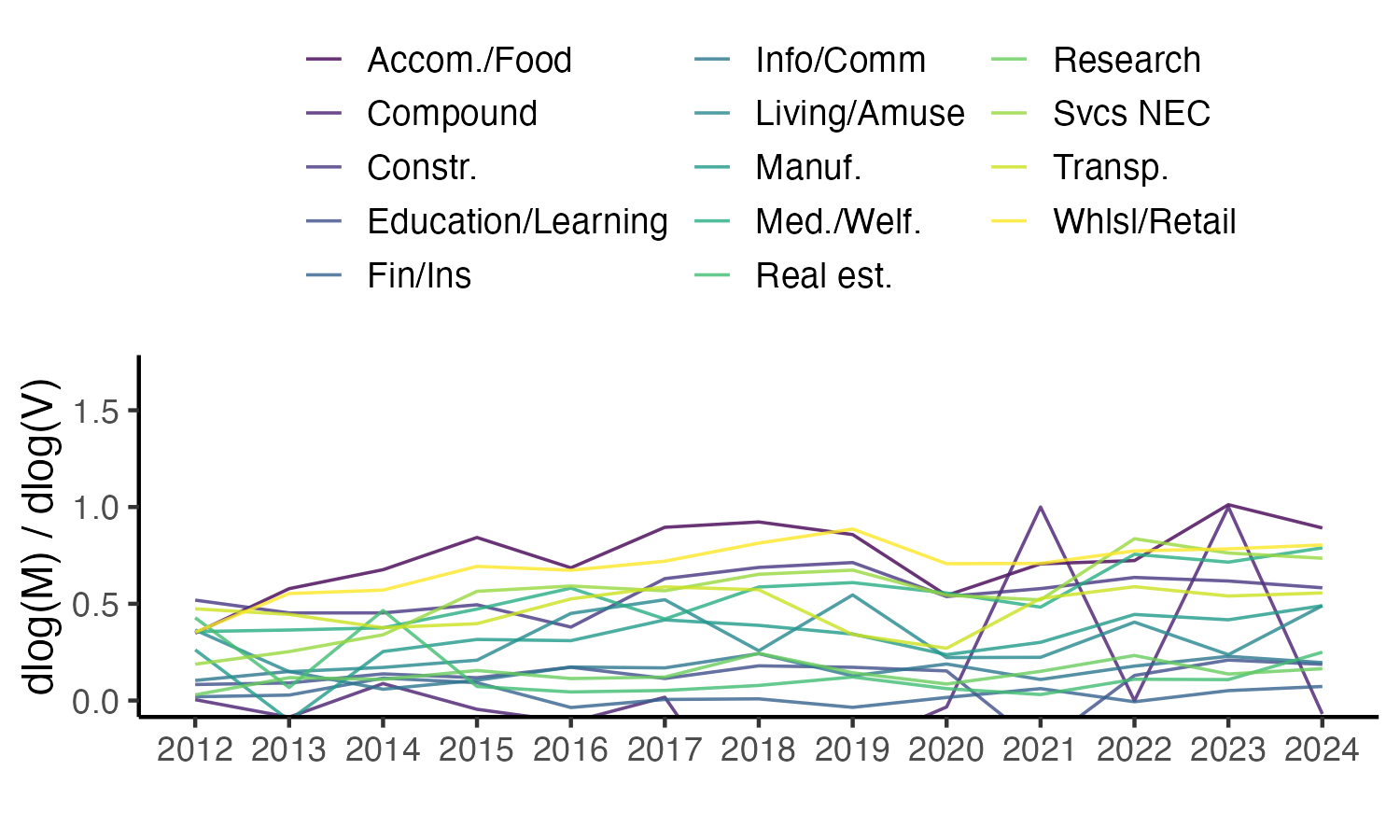}}
  \caption{Area-level and industry-level matching function estimates 2012--2024 (LFS + SET data)}
  \label{fg:lfs_set_area_occupation_results}
  \end{center}
  \footnotesize
  Note: Anchors: Minami-Kanto 2012 (Area) and Manufacturing 2012 (Industry) normalized to $A = 100$. Mining and Electricity/Gas/Water are dropped from the industry panel due to insufficient $U$ variation.
\end{figure}

Figure \ref{fg:lfs_set_area_occupation_results} (a), (c), and (e) display region-level estimates over 2012--2024 across the 13 regional blocks. Matching efficiency $A$, normalized to Minami-Kanto in 2012, has a regional median ranging from 132 (Hokkaido) to 294 (San'in), with rural blocks (San'in, Hokuriku, Shikoku) systematically more efficient than urban blocks (Minami-Kanto, Keihanshin, Kita-Kanto). The implied unemployment elasticity $d \ln m / d \ln (AU)$ has a mean of 0.66 and an interquartile range of 0.59--0.78 across region-year observations, comparable to existing Japanese estimates (\citealp{higashi2018spatial,kawata2019}); the vacancy elasticity $d \ln m / d \ln V$ has a mean of 0.34 and an interquartile range of 0.22--0.41. The reverse urban--rural ranking of $A$ is consistent with thinner but better-screened labor markets in lower-density regions, where employer--worker networks and re-employment information frictions may be smaller relative to the size of the unemployed pool.\footnote{A small number of disaggregated market-time estimates fall outside the unit interval. I interpret these values as diagnostics of the limits of the local approximation---possibly reflecting small-cell noise, local departures from the CRS approximation, or other economically implausible local slopes---rather than as observations to be manually censored. They are therefore retained in the figures and summary statistics to avoid an ad hoc trimming rule and to make visible the instability of the most granular estimates.}

\subsubsection{Industry-Level Results in 2012--2024}\label{sec:dataset1_occupation}

Figure \ref{fg:lfs_set_area_occupation_results} (b), (d), and (f) display industry-level estimates over 2012--2024 across the 14 industries common to the SET and LFS (Mining and Electricity/Gas/Water are dropped due to insufficient $U$ variation, see Appendix \ref{sec:data_construction}). Matching efficiency, normalized to Manufacturing in 2012, varies widely from a median of 52 (Construction) to 374 (Education/Learning), with service-sector industries---Education/Learning, Accommodation/Food, and Medical/Welfare---substantially more efficient than goods-producing and transport industries. The unemployment elasticity $d \ln m / d \ln (AU)$ has a broad cross-industry spread (medians 0.3--1.0), with the lowest values in Accommodation/Food and Wholesale/Retail and the highest in Compound services and Finance/Insurance; the vacancy elasticity $d \ln m / d \ln V$ has industry medians spanning roughly 0.0 (Compound services, Finance/Insurance) to 0.7 (Accommodation/Food, Wholesale/Retail). These cross-industry patterns should be interpreted with caution, because the LFS classifies unemployment by industry of \emph{previous} job, so a portion of $U_{r t}$ in any industry consists of workers seeking employment in other industries, attenuating within-industry matching frictions and confounding industry rankings of $A_{r t}$ with inter-industry mobility flows.

\subsection{Dataset 2: Hello Work Platform Results}
I next report results for the Public Employment Security Office (Hello Work) segment. This dataset does not cover all matching channels in Japan, but it provides a clear, consistent, granular, and high-frequency platform-level market definition.

\subsubsection{Aggregate Trends in 1972--2023}\label{sec:month_level}

\begin{figure}[!ht]
  \begin{center}
  \subfloat[Unemployed ($U$), Vacancy ($V$), and Tightness ($\frac{V}{U}$)]{\includegraphics[width = 0.39\textwidth]
  {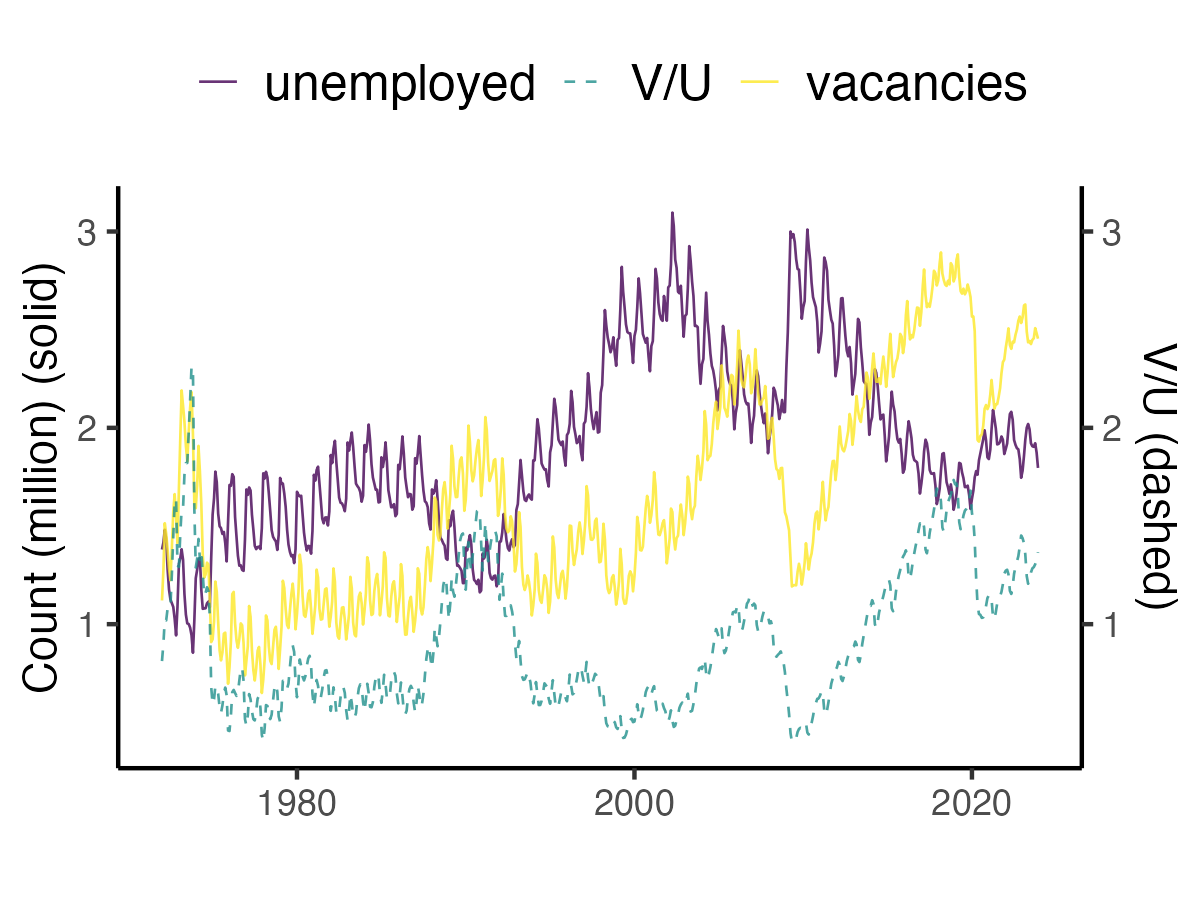}}\hspace{0.15\textwidth}
  \subfloat[Hire ($H$)]{\includegraphics[width = 0.39\textwidth]
  {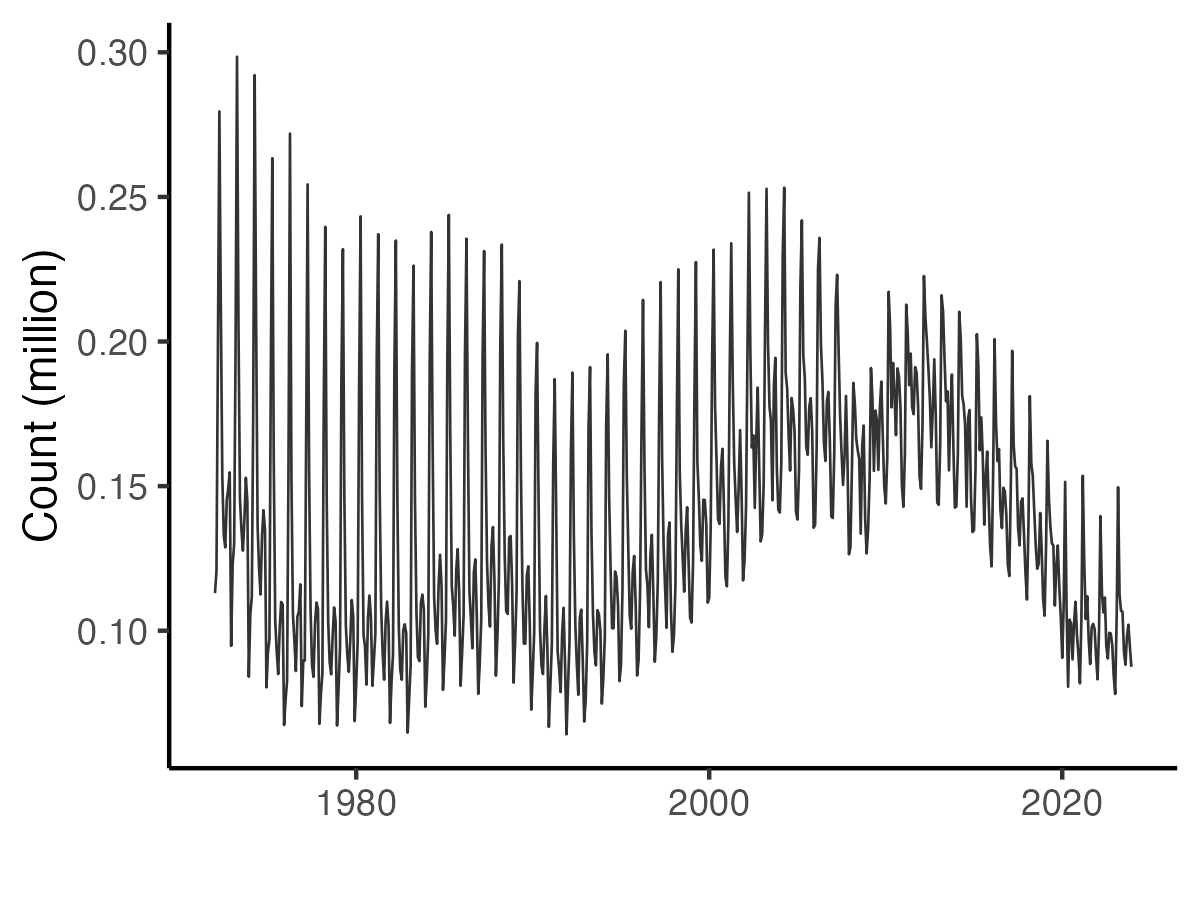}}\\
  \subfloat[Job-finding rate ($\frac{H}{U}$) and vacancy-filling rate ($\frac{H}{V}$)]{\includegraphics[width = 0.39\textwidth]
  {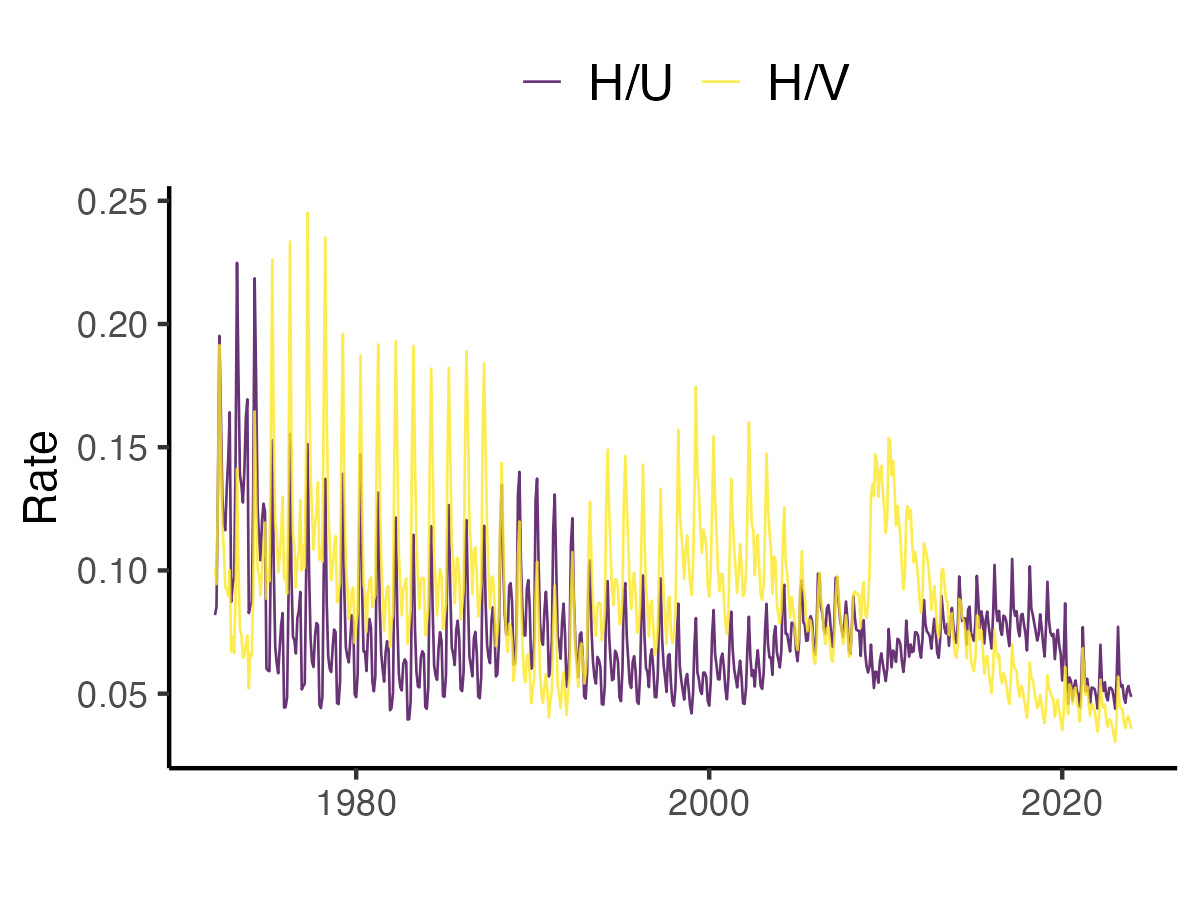}}\hspace{0.15\textwidth}
  \subfloat[Matching Efficiency ($A$)]{\includegraphics[width = 0.39\textwidth]
  {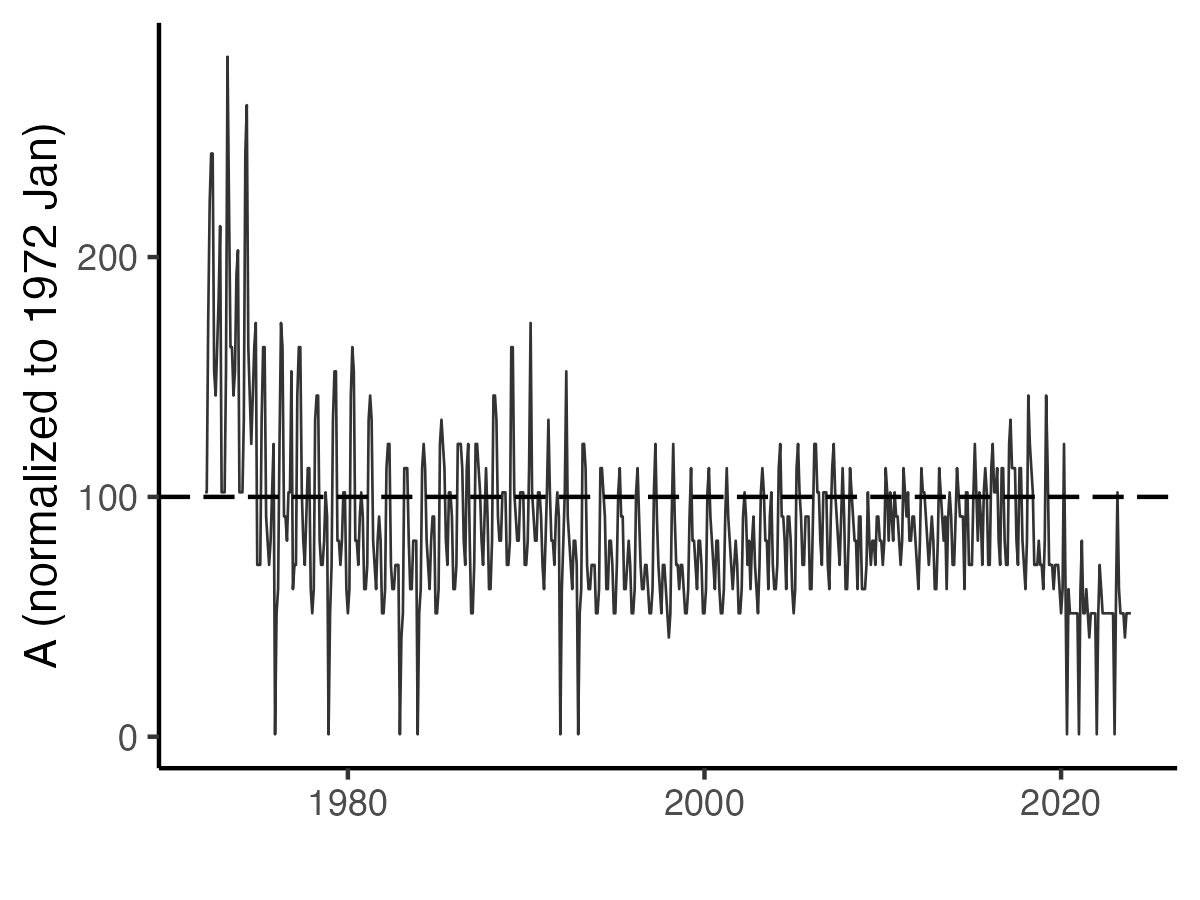}}
  \\
  \subfloat[Matching Elasticity ($\frac{d\ln m}{d \ln AU}$, $\frac{d\ln m}{d\ln V}$)]{\includegraphics[width = 0.39\textwidth]
  {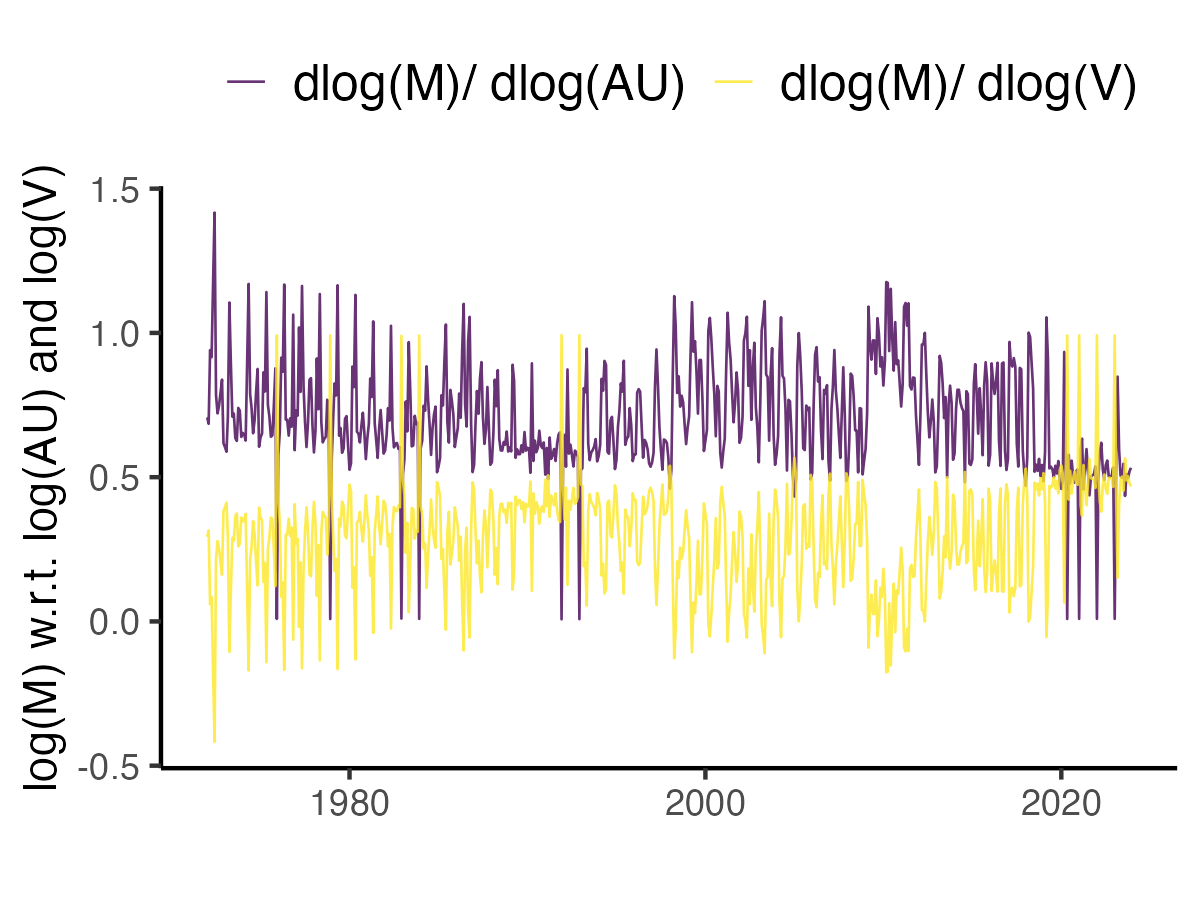}}\hspace{0.15\textwidth}
  \subfloat[Efficiency ($A$) and ($\ln\frac{V}{U}$, $\ln\frac{H}{U}$, $\ln\frac{H}{V}$)]{\includegraphics[width = 0.39\textwidth]
  {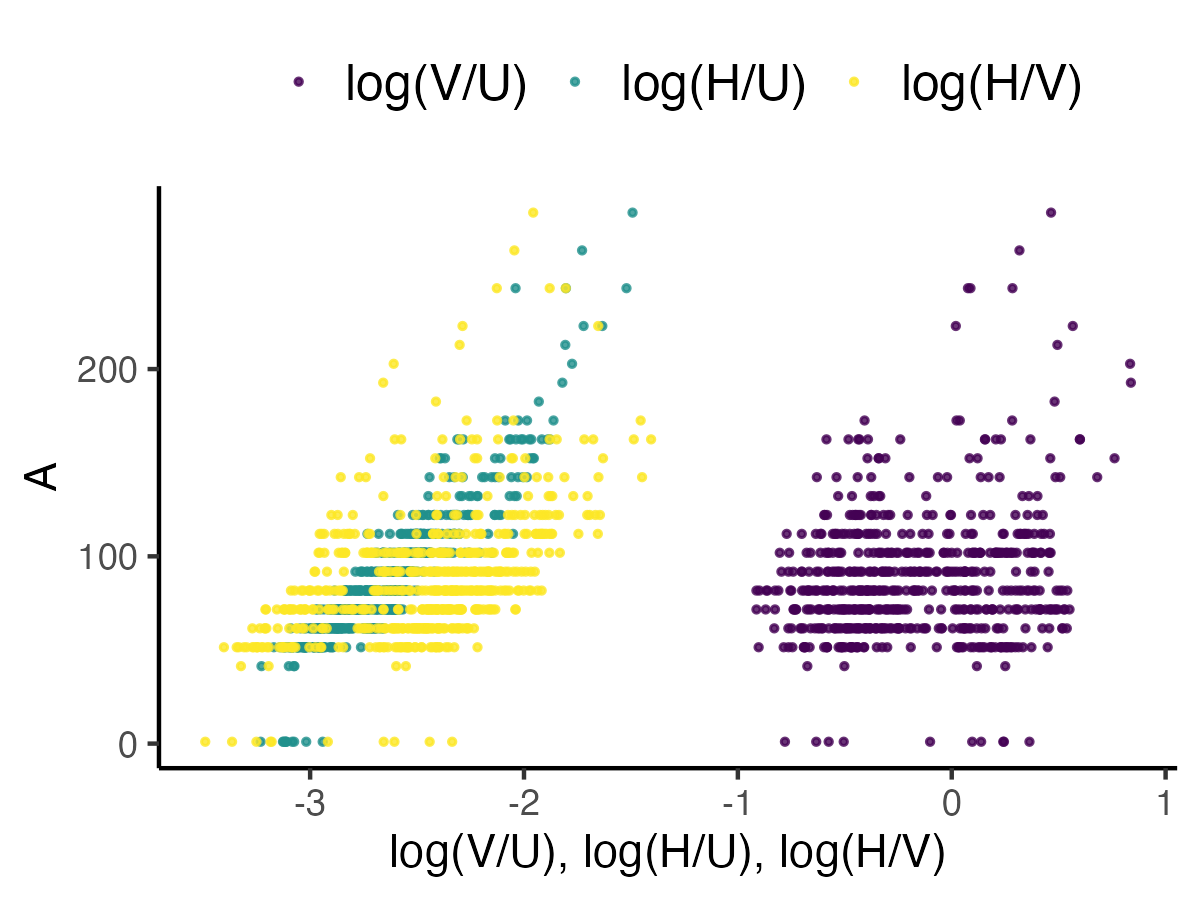}}
  \caption{Country-month-level aggregate results 1972--2023 (Hello Work data)}
  \label{fg:month_part_and_full_time_results}
  \end{center}
  \footnotesize
  %Note: $U$ is the monthly stock of registered job seekers (\textit{yuukou kyuushokusha}) at the Public Employment Security Office (Hello Work); $V$ is the monthly stock of registered vacancies (\textit{yuukou kyuujin}) at Hello Work; $H$ is the monthly flow of successful job placements (\textit{shuushoku kensuu}) through Hello Work. All three series cover the full-time + part-time aggregate (general workers, excluding new graduates) over 1972--2023. Detailed data construction is in Appendix~\ref{sec:data_construction}.
\end{figure}

Figures \ref{fg:month_part_and_full_time_results} (a)-(c) provide monthly data patterns of unemployed individuals, vacancies, labor market tightness (\(V/U\)), hires, and job-finding and vacancy-filling rates (\(H/U\) and \(H/V\)) in aggregate labor markets.\footnote{Vacancy-filling rates are also known as vacancy yields or worker-finding rates. \cite{davis2013establishment} define vacancy yields in month $t$ as the number of hires in month $t\left(h_t\right)$ divided by the number of vacancies in the previous month $\left(v_{t-1}\right)$ instead of the same month $\left(v_t\right)$, the two series are very similar.} Hires display strong seasonality, with the average monthly peak in April and a secondary high level in March, consistent with Japan's fiscal-year hiring cycle.
In summary, the numbers of unemployed individuals and vacancies increase with fluctuations, corresponding with market tightness, while the number of hires shows a noticeable decline around the late 1970s, peaks and troughs from the mid-1980s to the late 1990s, peaks around mid-2000s and 2010, and a sharp decline toward recent years.

Figures \ref{fg:month_part_and_full_time_results} (d) and (e) present the estimation results of the matching function, including matching efficiency and elasticities (\(\frac{d\ln m}{d\ln AU}\) and \(\frac{d\ln m}{d\ln V}\)). Matching efficiency, normalized to 1972, shows a declining trend with notable fluctuations, consistent with the downward trends in job-finding and vacancy-filling rates.
The post-2015 decline is especially pronounced.
This pattern may reflect an increase in matching opportunities outside the government-operated platform, as discussed in \cite{otani2025nonparametric} using proprietary on-the-job search platform data and \cite{kanayama2024nonparametric} using spot gig work platform data.

The implied unemployment elasticity is 0.5--0.9, which is comparable to previous worldwide findings such as \cite{petrongolo2001looking} (range: 0.5--0.7) and Japanese studies such as \cite{higashi2018spatial} (0.38 for 2000--2014 monthly), \cite{kawata2019} (0.48 for 2012--2017 prefecture-month-level), \cite{kano2005estimating} (0.56 for 1972--1999 prefecture-year-level), \cite{sasaki2007measuring} (about 0.6 for 1998--2007 prefecture-quarter-level), and \cite{kambayashi2006vacancy} (about 0.8 for 1996--2001 prefecture-month-level).\footnote{For reference, \cite{petrongolo2001looking} summarize early aggregate studies in many countries based on a Cobb-Douglas matching function with the flow of hires on the left-hand side and the stock of unemployment and job vacancies on the right-hand side. In short, the unemployment elasticity is in the range of 0.5--0.7. \cite{bernstein2022matching} review the recent empirical literature that estimates the matching elasticity in the U.S. by four types of specification; Cobb-Douglas, Cobb-Douglas with endogeneity correction, a constant elasticity of substitution (CES), and nonparametric \citep{lange2020beyond}. The broad range of estimates is partly due to variations in data choices and periods. Higher estimates are typically derived from JOLTS data \citep{borowczyk2013accounting, csahin2014mismatch}, while lower estimates are often obtained from CPS flows data \citep{barnichon2015labor} or occupation-level hires from CPS data \citep{csahin2014mismatch}.}
% However, I discuss the need for careful interpretation of the estimates due to scale normalization of matching efficiency in Section \ref{sec:month_level}.
The implied vacancy elasticity has a pre-2015 mean of 0.27 (interquartile range 0.16--0.39) and a post-2015 mean of 0.39, broadly comparable to \cite{lange2020beyond} (range: 0.15--0.3) and Japanese studies such as \cite{higashi2018spatial} (0.24 for 2000--2014 monthly), \cite{kawata2019} (0.52 for 2012--2017 prefecture-month-level), \cite{kano2005estimating} (0.3 for 1972--1999 prefecture-year-level), \cite{sasaki2007measuring} (about 0.2 for 1998--2007 prefecture-quarter-level), and \cite{kambayashi2006vacancy} (about 0.3 for 1996--2001 prefecture-month-level).

Figure \ref{fg:month_part_and_full_time_results} (f) illustrates some correlation patterns between matching efficiency and three log market-structure variables ($\ln V/U$, $\ln H/U$, $\ln H/V$) overlaid by color. Consistent with \cite{lange2020beyond}, these highlight positive correlations between efficiency and market structure, such as tightness, which induce a positive bias in the estimates of the vacancy elasticity whenever unobserved matching efficiency is not controlled for, as is the case in traditional estimators.

\subsubsection{Prefecture-Level Aggregate Results in 2013--2024}

\begin{figure}[!ht]
  \begin{center}
  \subfloat[Hokkaido]{\includegraphics[width = 0.3\textwidth]
  {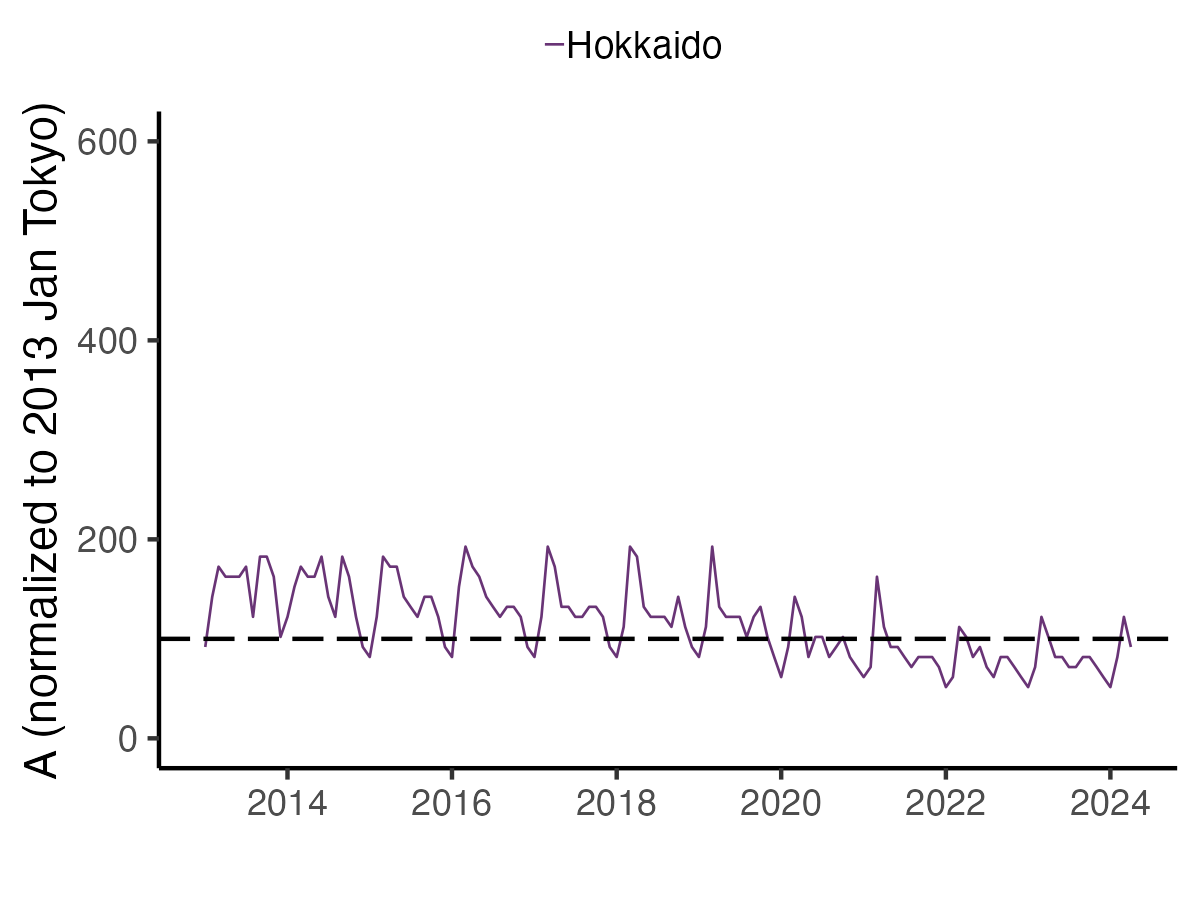}}
  \subfloat[Tohoku]{\includegraphics[width = 0.3\textwidth]
  {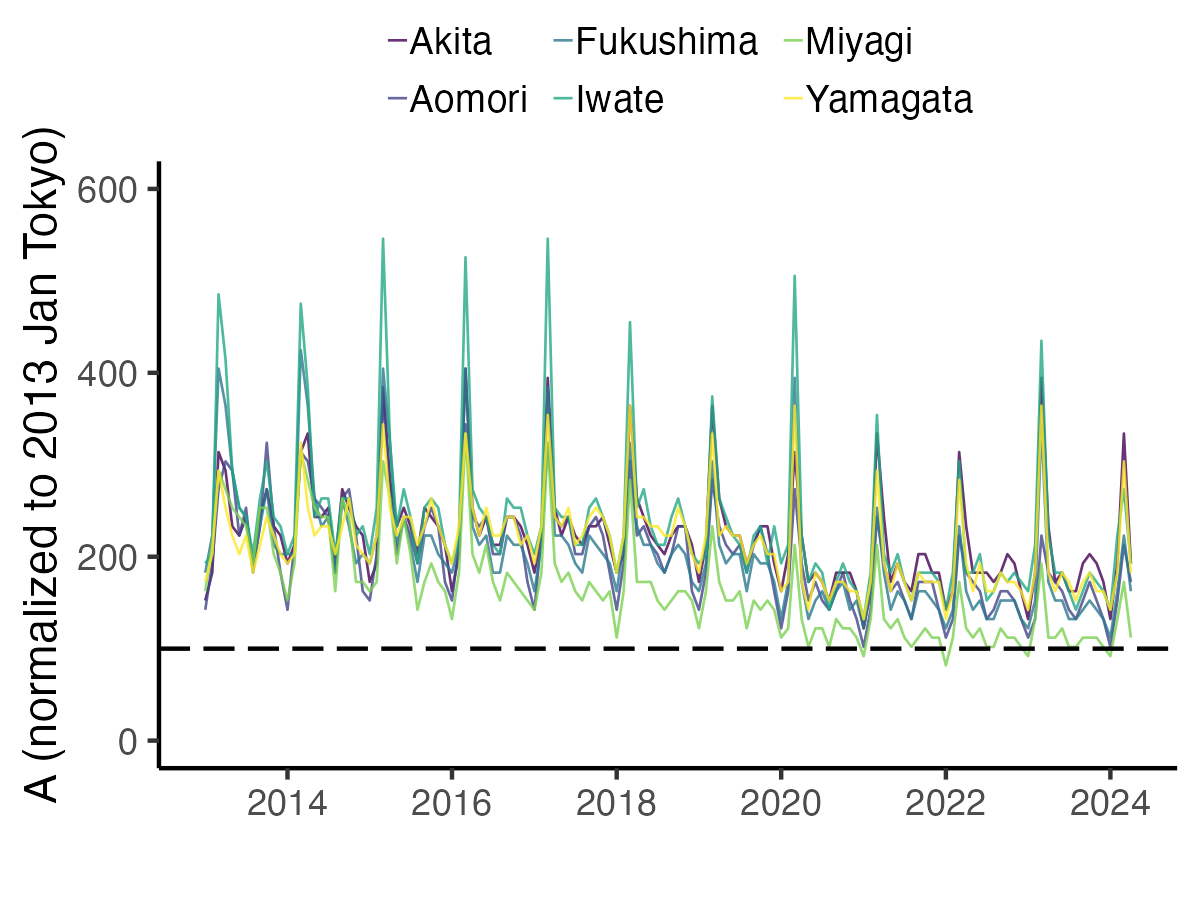}}\\
  \subfloat[Kanto]{\includegraphics[width = 0.3\textwidth]
  {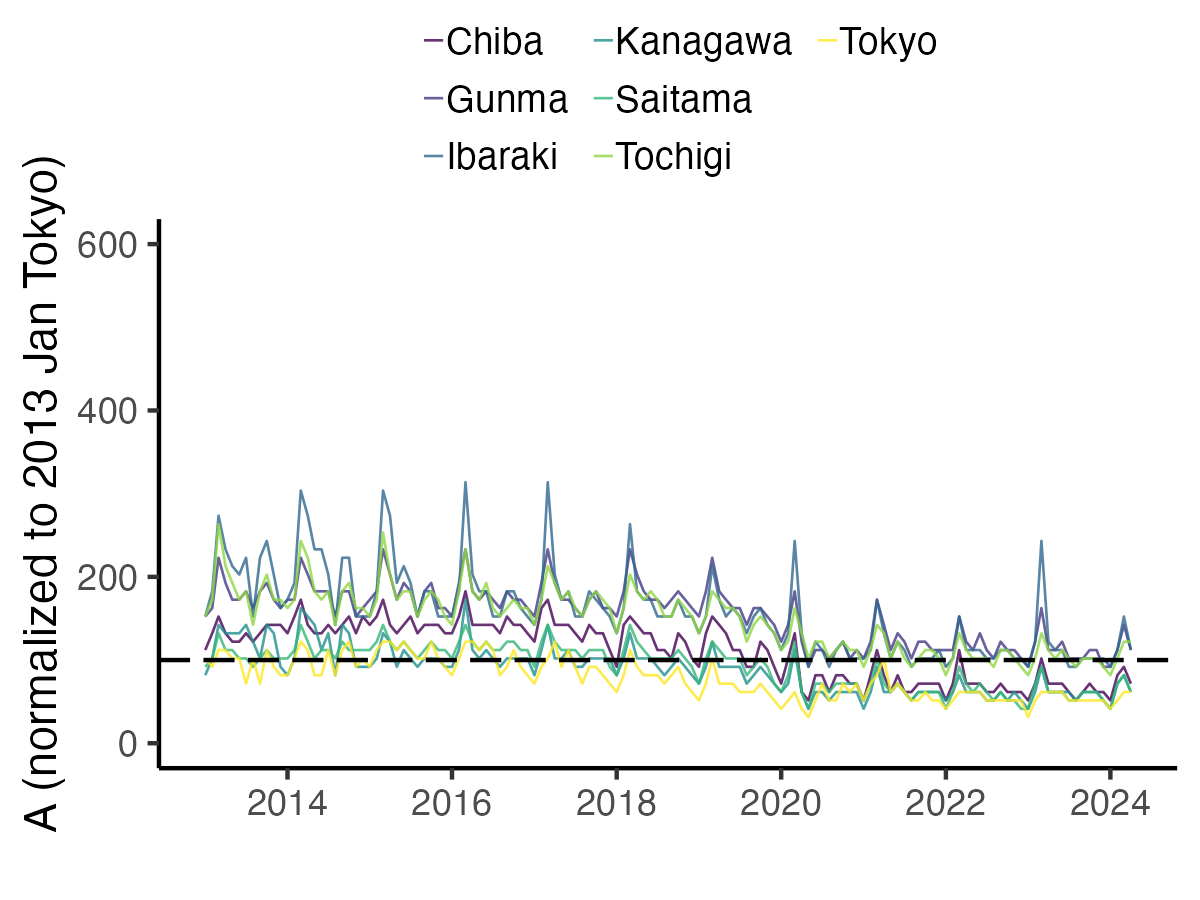}}
  \subfloat[Chubu]{\includegraphics[width = 0.3\textwidth]
  {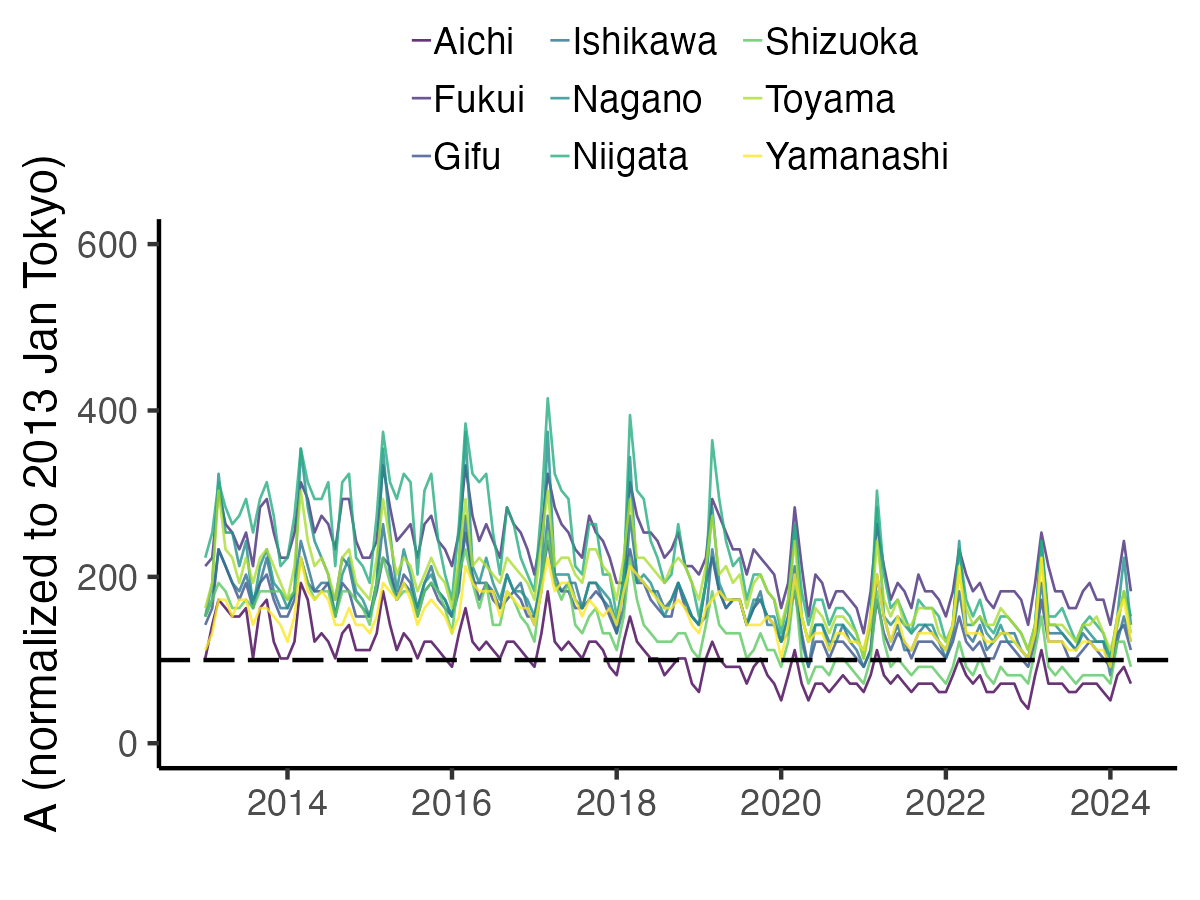}}
  \\
  \subfloat[Kansai]{\includegraphics[width = 0.3\textwidth]
  {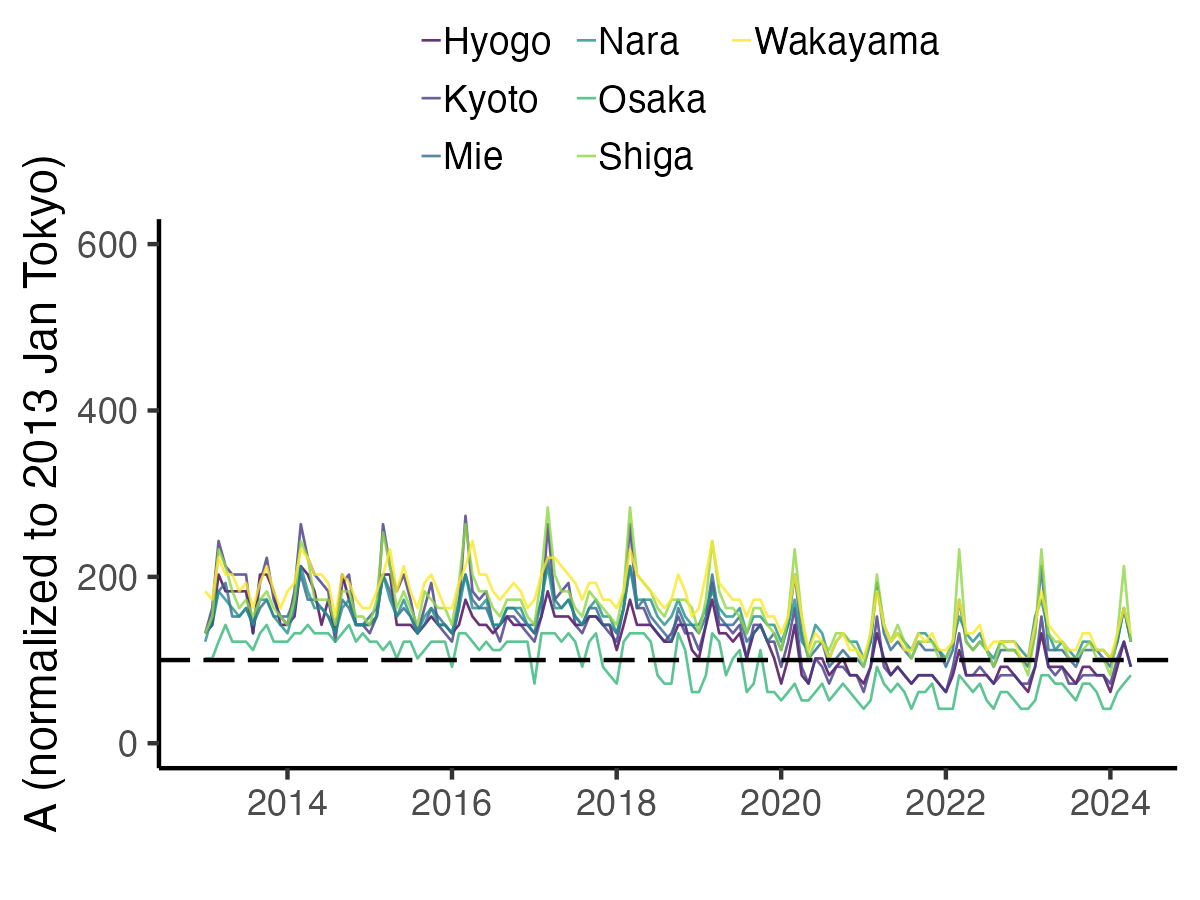}}
  \subfloat[Chugoku]{\includegraphics[width = 0.3\textwidth]
  {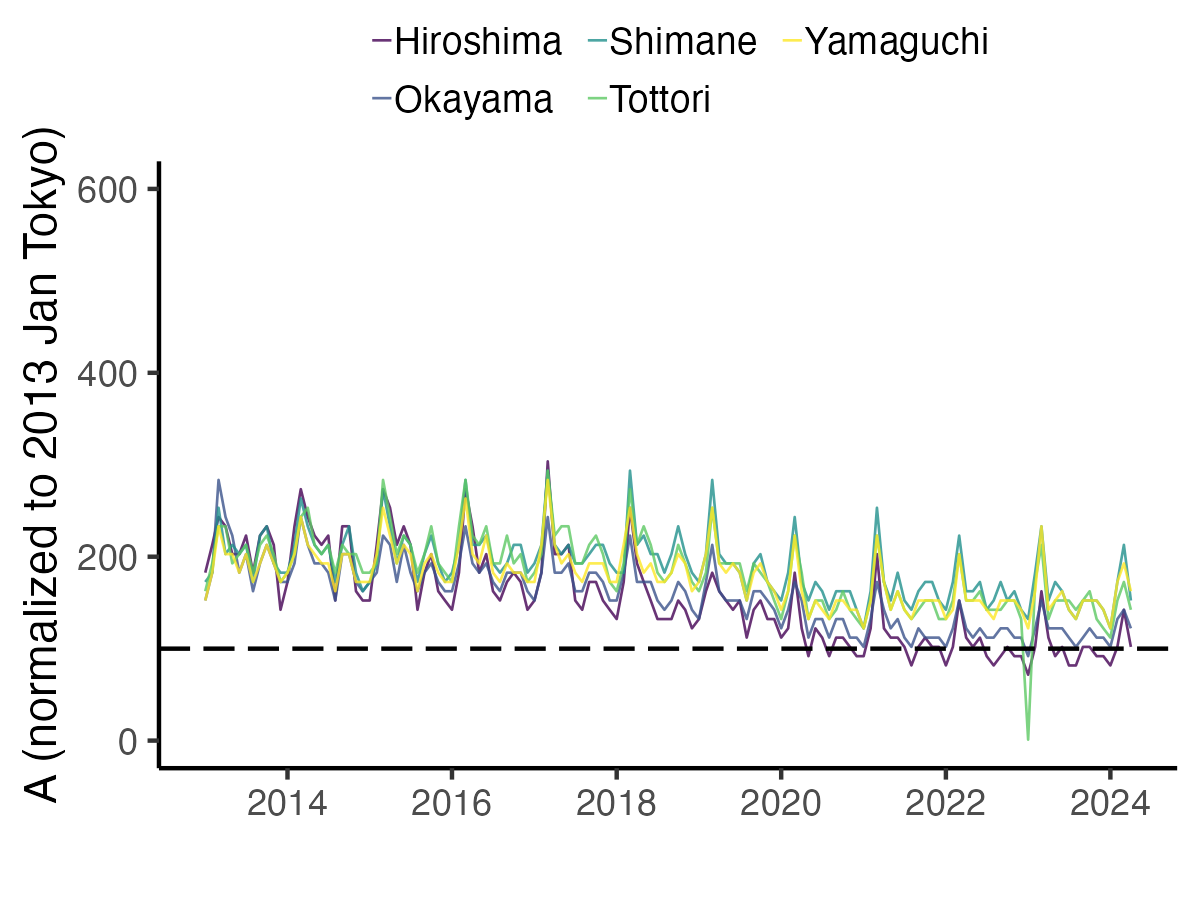}}\\
  \subfloat[Shikoku]{\includegraphics[width = 0.3\textwidth]
  {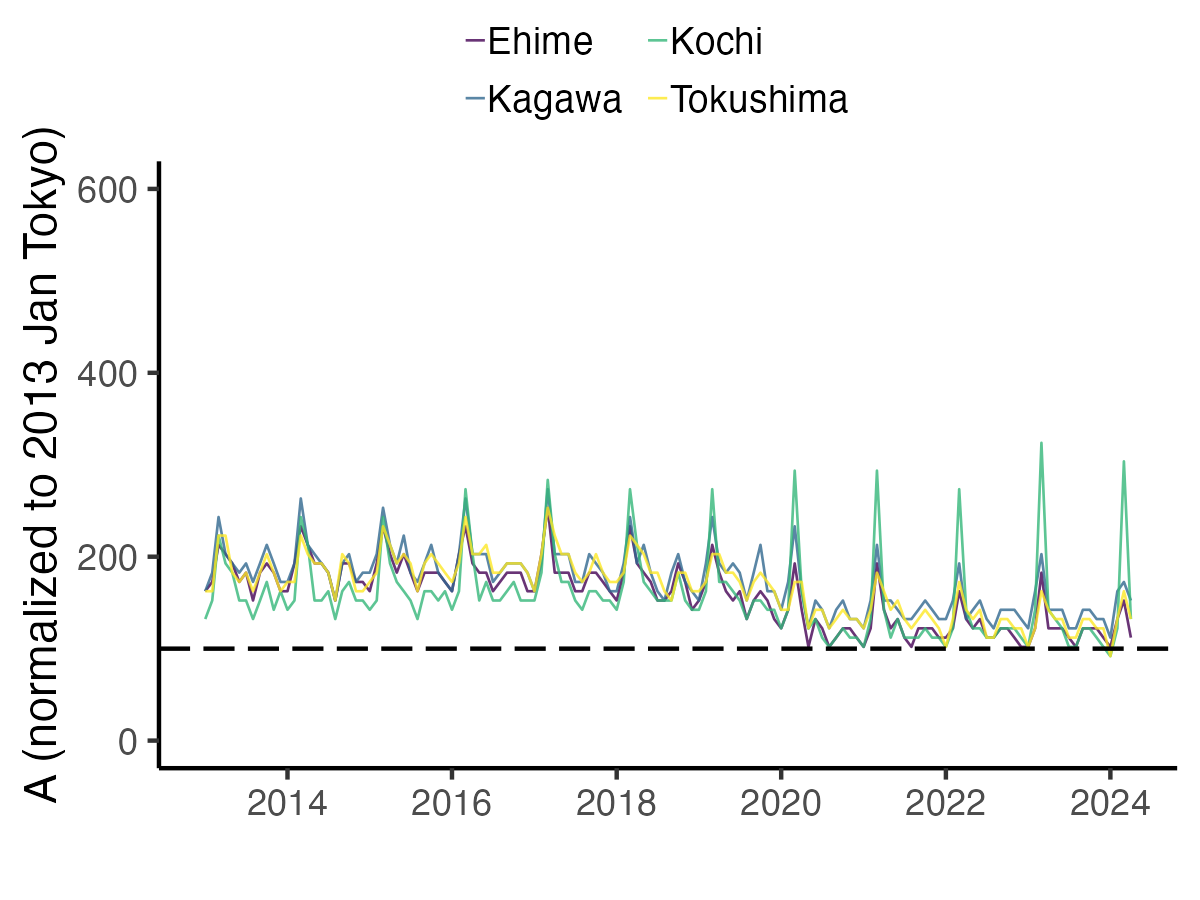}}
  \subfloat[Kyushu, Okinawa]{\includegraphics[width = 0.3\textwidth]
  {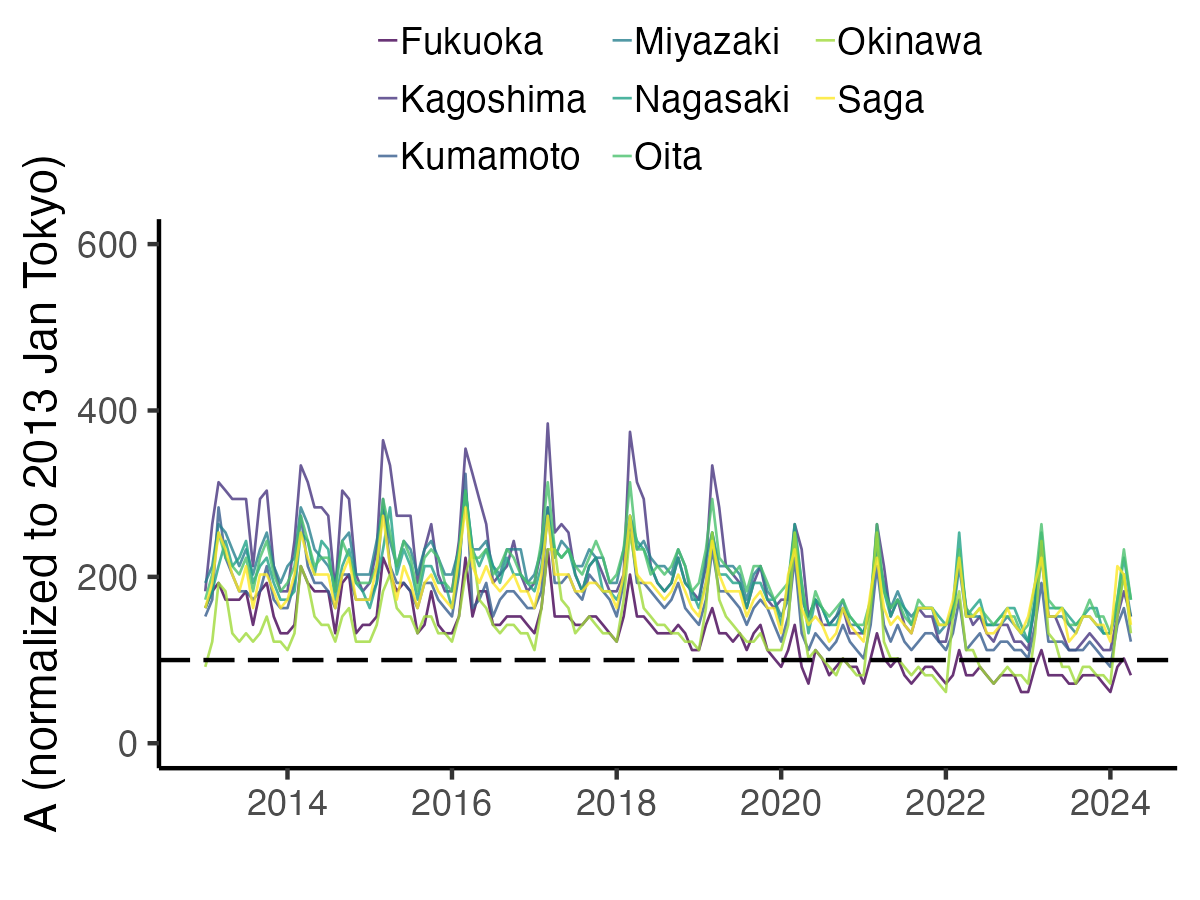}}
  \caption{Month-prefecture-level matching efficiency 2013--2024}
  \label{fg:month_part_and_full_time_matching_efficiency_prefecture_results}
  \end{center}
  \footnotesize
  %Note:
\end{figure}

Figure \ref{fg:month_part_and_full_time_matching_efficiency_prefecture_results} displays month-prefecture-level matching efficiency, normalized to Tokyo in January 2013, across regions of Japan from 2013 to 2024. Each panel shows efficiency trends within a region, highlighting notable regional variation. For instance, the Tohoku region (panel b) and Chubu region (panel d), which include many rural prefectures, exhibit relatively high volatility and generally higher efficiency values than other regions. In contrast, urban regions such as Kanto (panel c) and Kansai (panel e) show more stable and lower efficiency levels. The Tohoku region is approximately three times as efficient as the Tokyo region. These patterns suggest that regional economic factors and labor market conditions significantly influence matching efficiency across Japan.

\begin{figure}[!ht]
  \begin{center}
  \subfloat[Hokkaido]{\includegraphics[width = 0.3\textwidth]
  {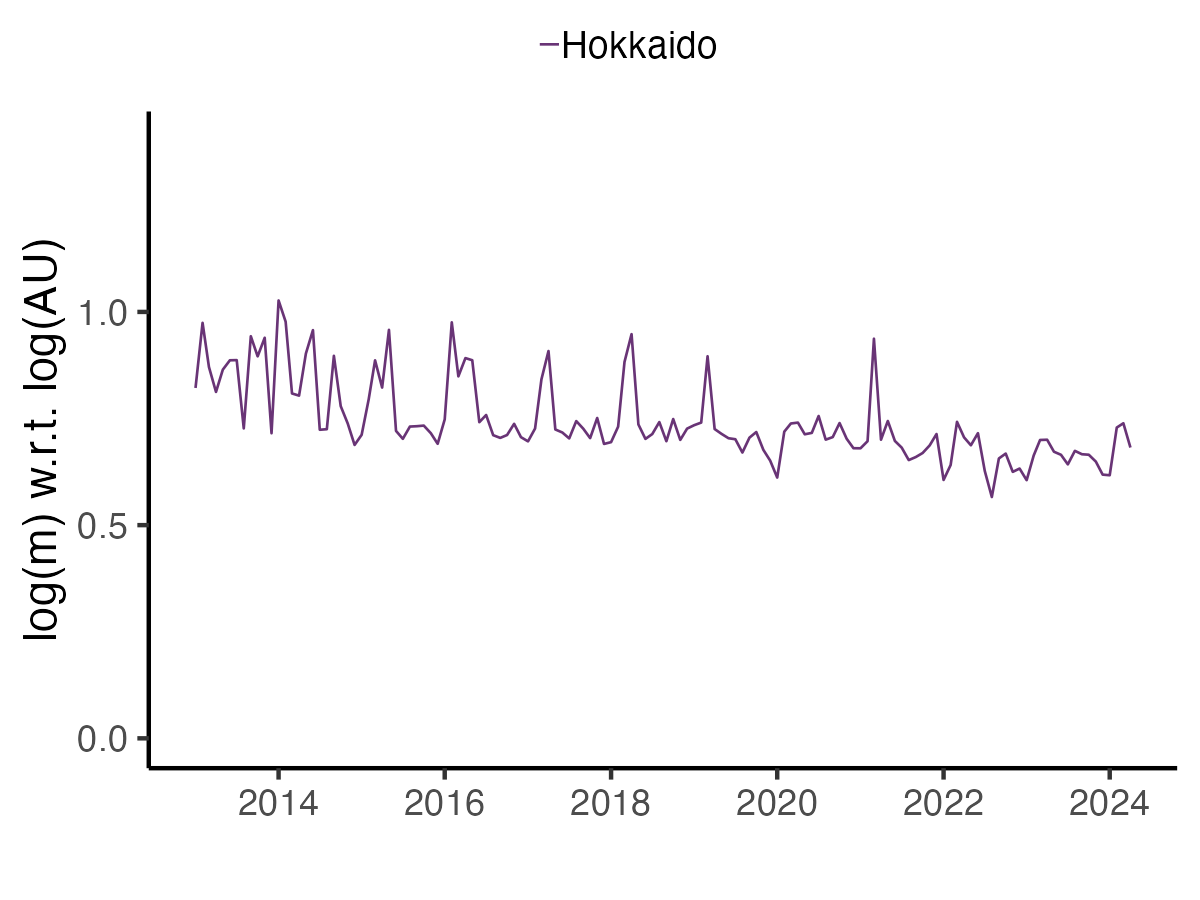}}
  \subfloat[Tohoku]{\includegraphics[width = 0.3\textwidth]
  {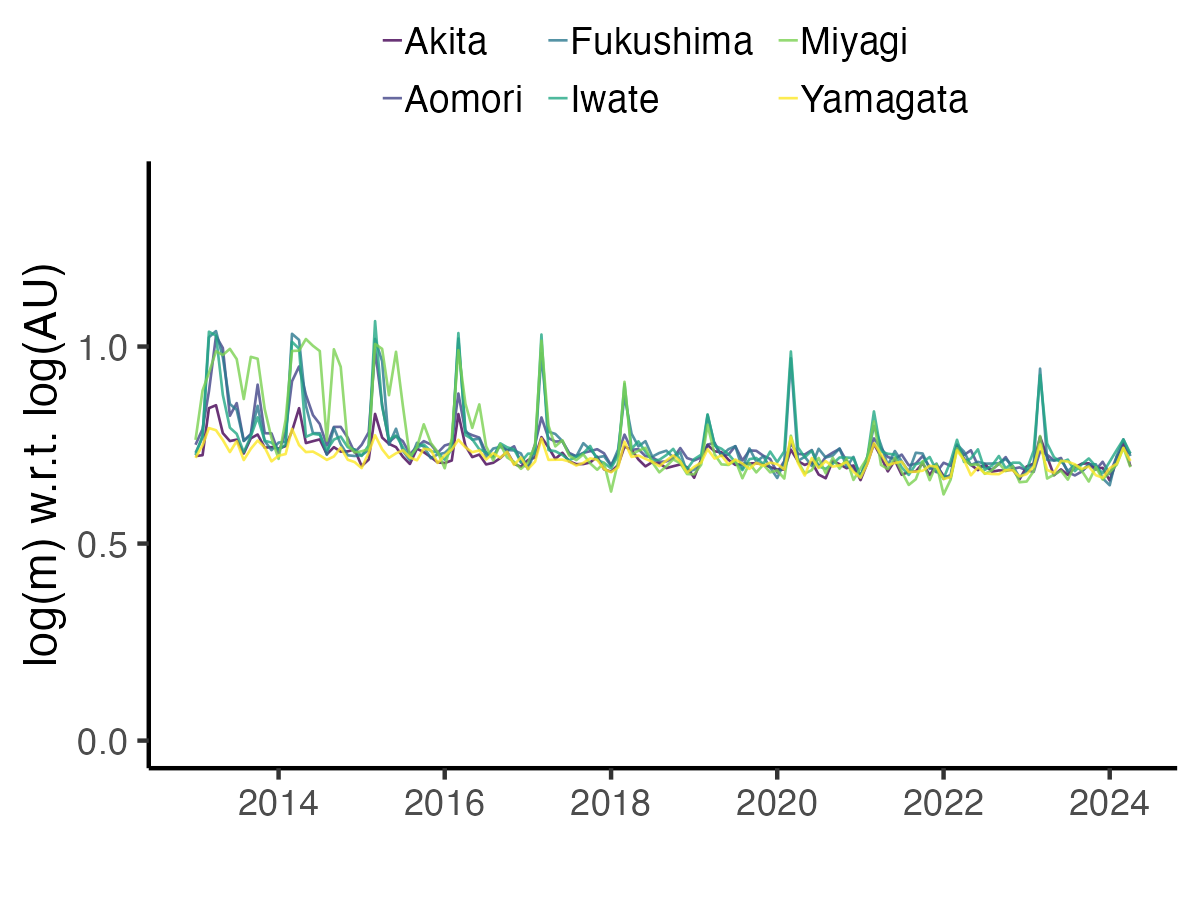}}\\
  \subfloat[Kanto]{\includegraphics[width = 0.3\textwidth]
  {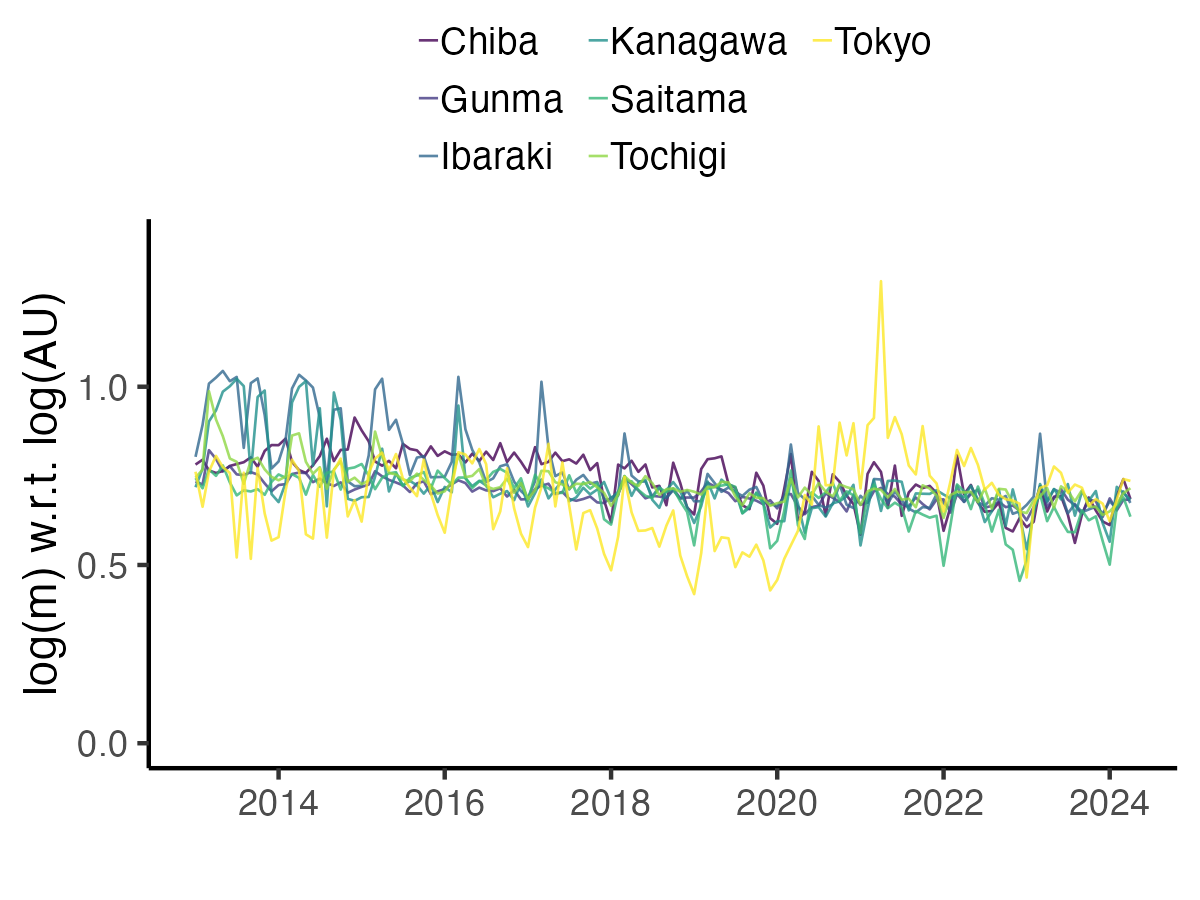}}
  \subfloat[Chubu]{\includegraphics[width = 0.3\textwidth]
  {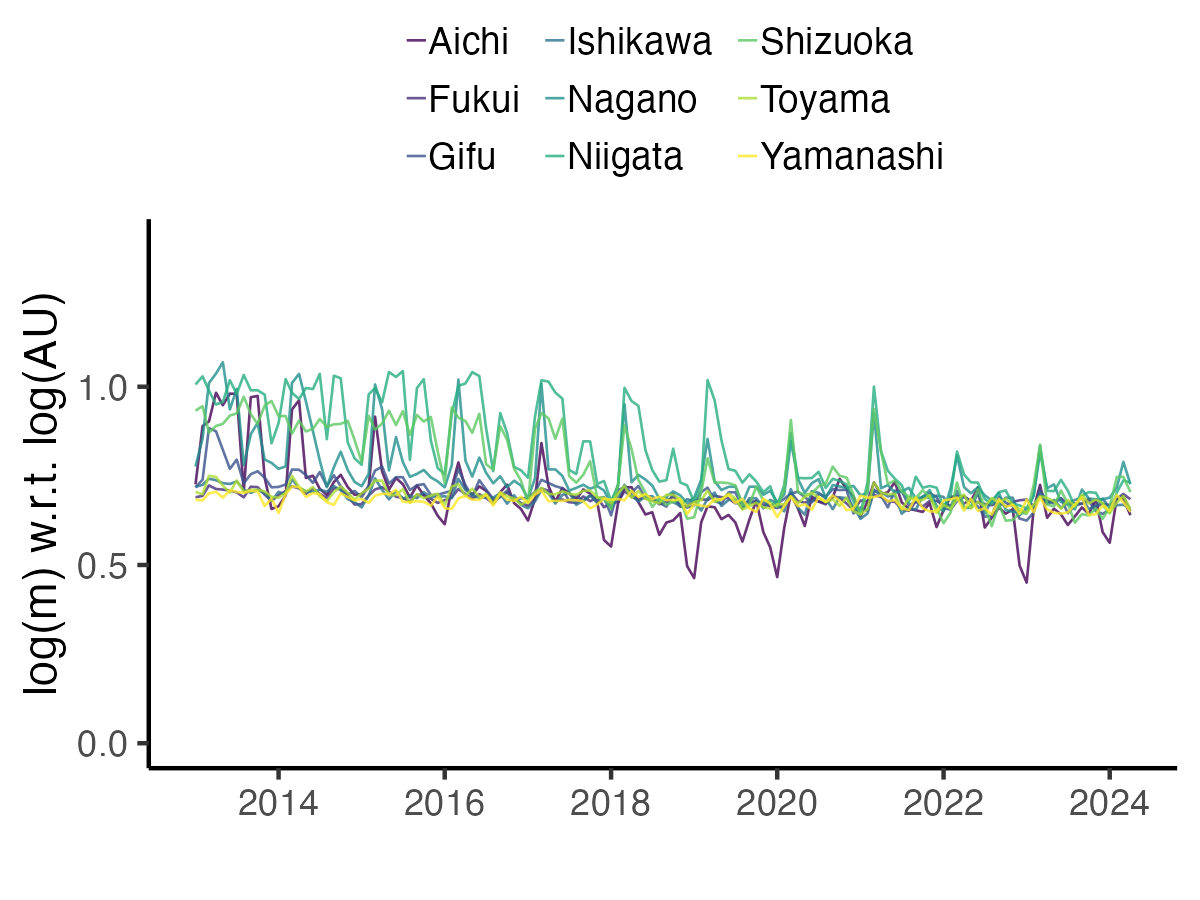}}
  \\
  \subfloat[Kansai]{\includegraphics[width = 0.3\textwidth]
  {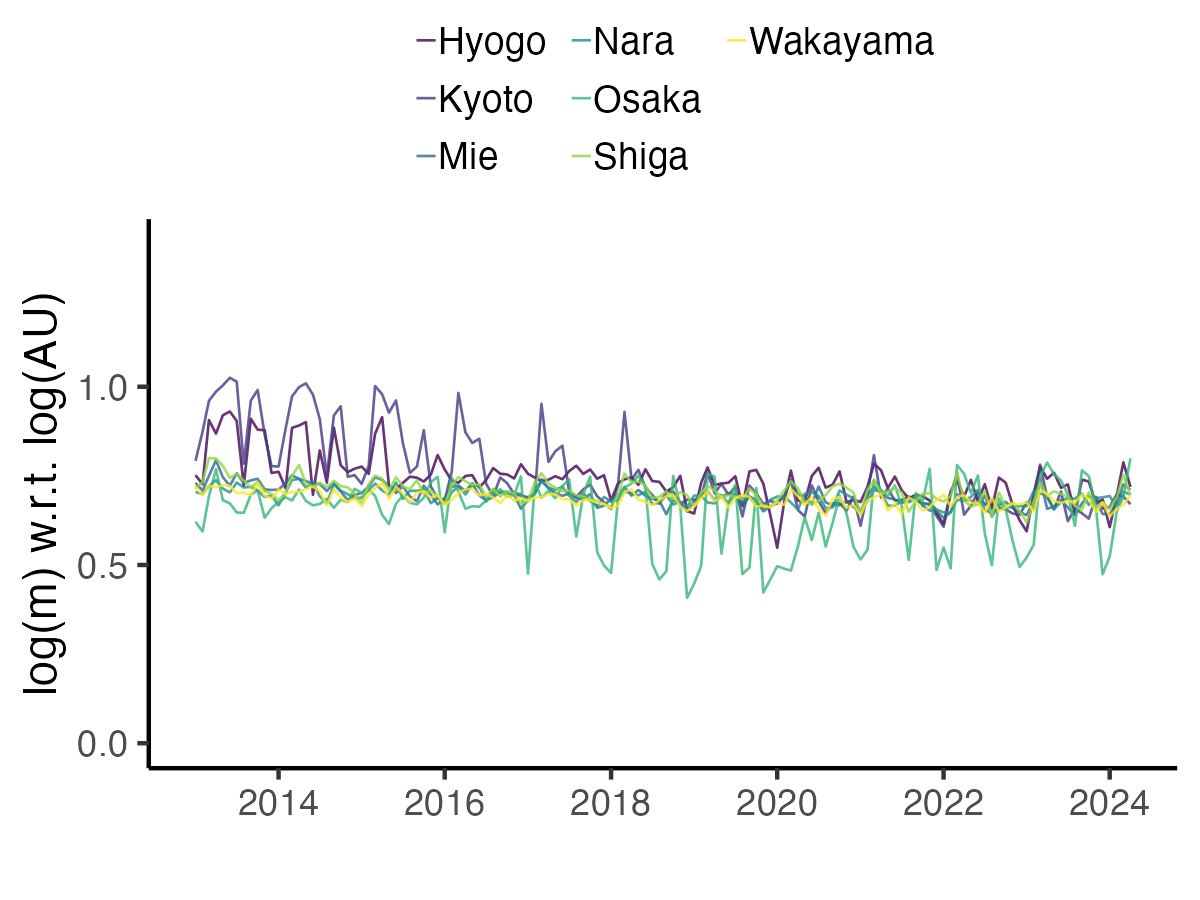}}
  \subfloat[Chugoku]{\includegraphics[width = 0.3\textwidth]
  {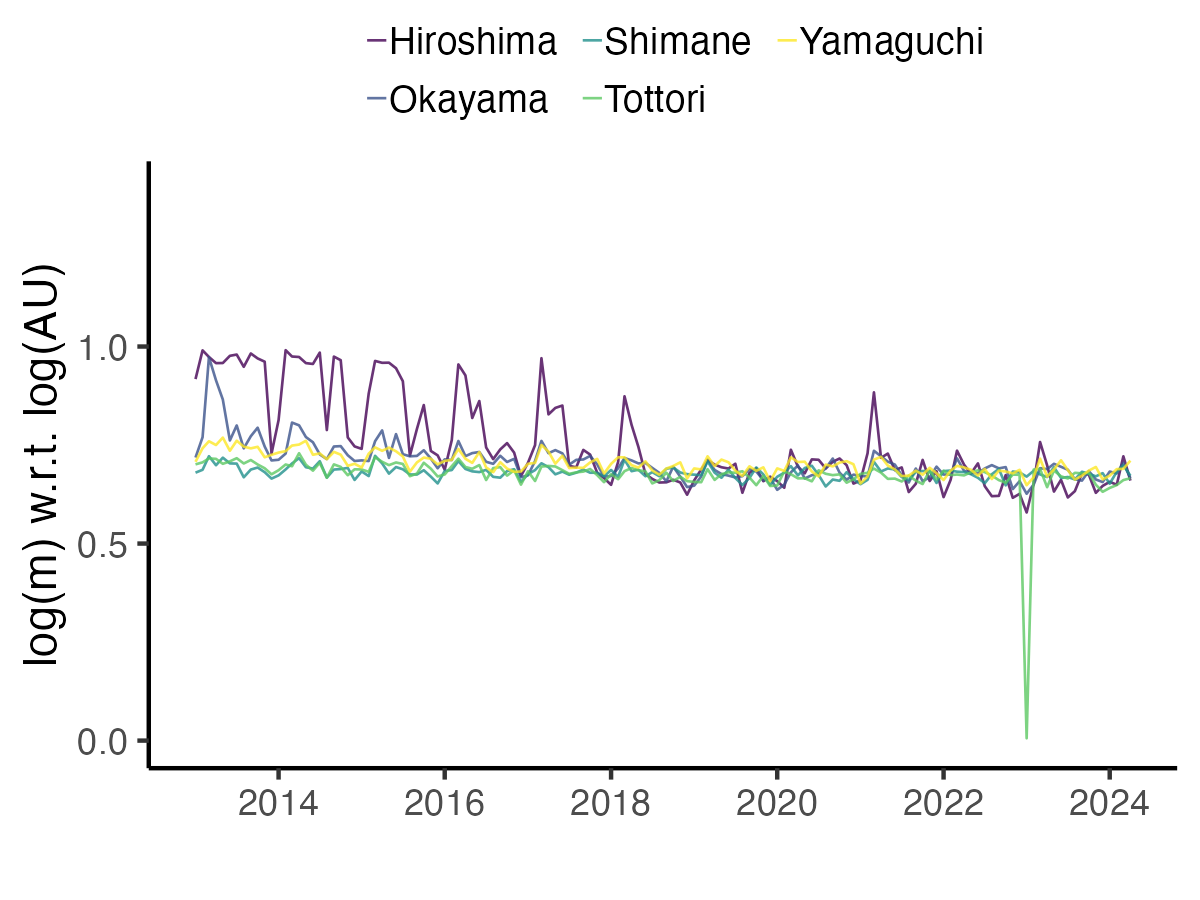}}\\
  \subfloat[Shikoku]{\includegraphics[width = 0.3\textwidth]
  {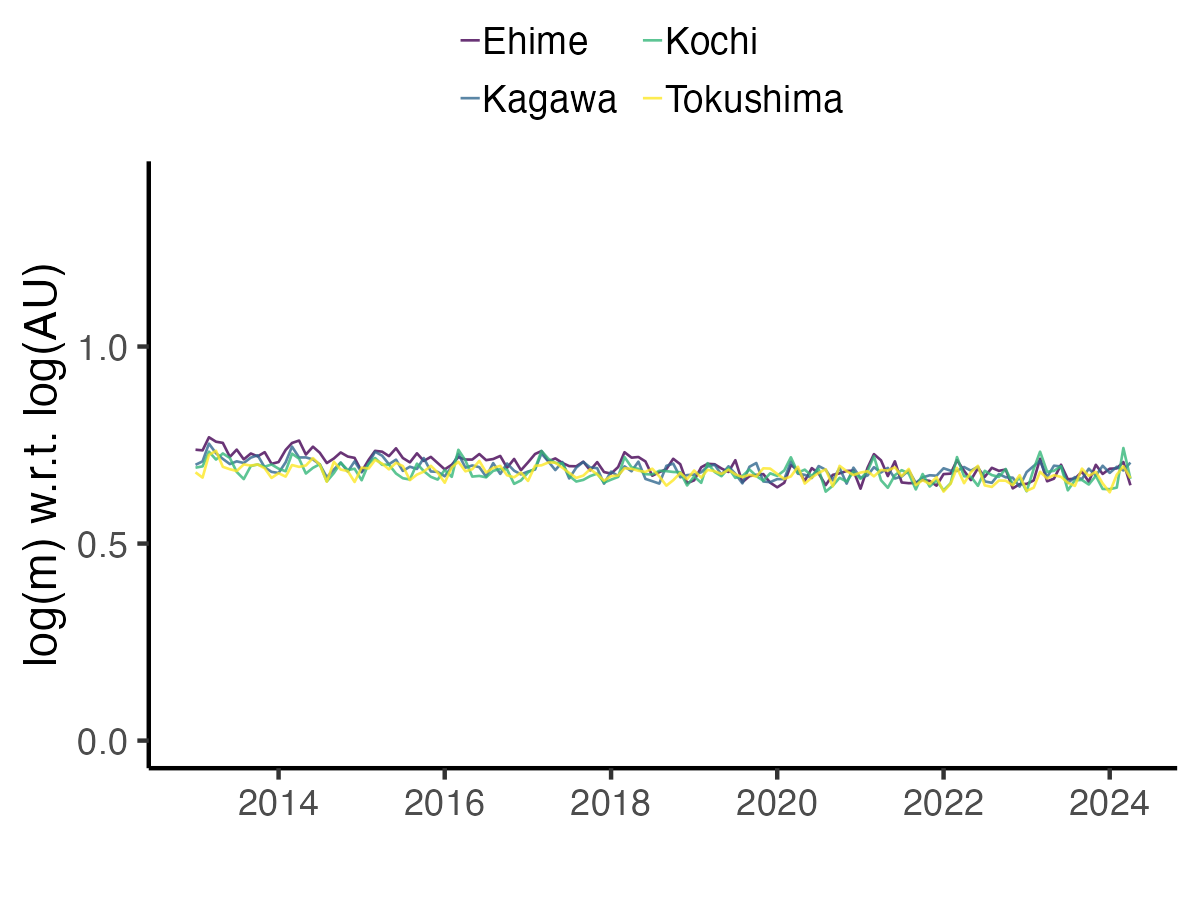}}
  \subfloat[Kyushu, Okinawa]{\includegraphics[width = 0.3\textwidth]
  {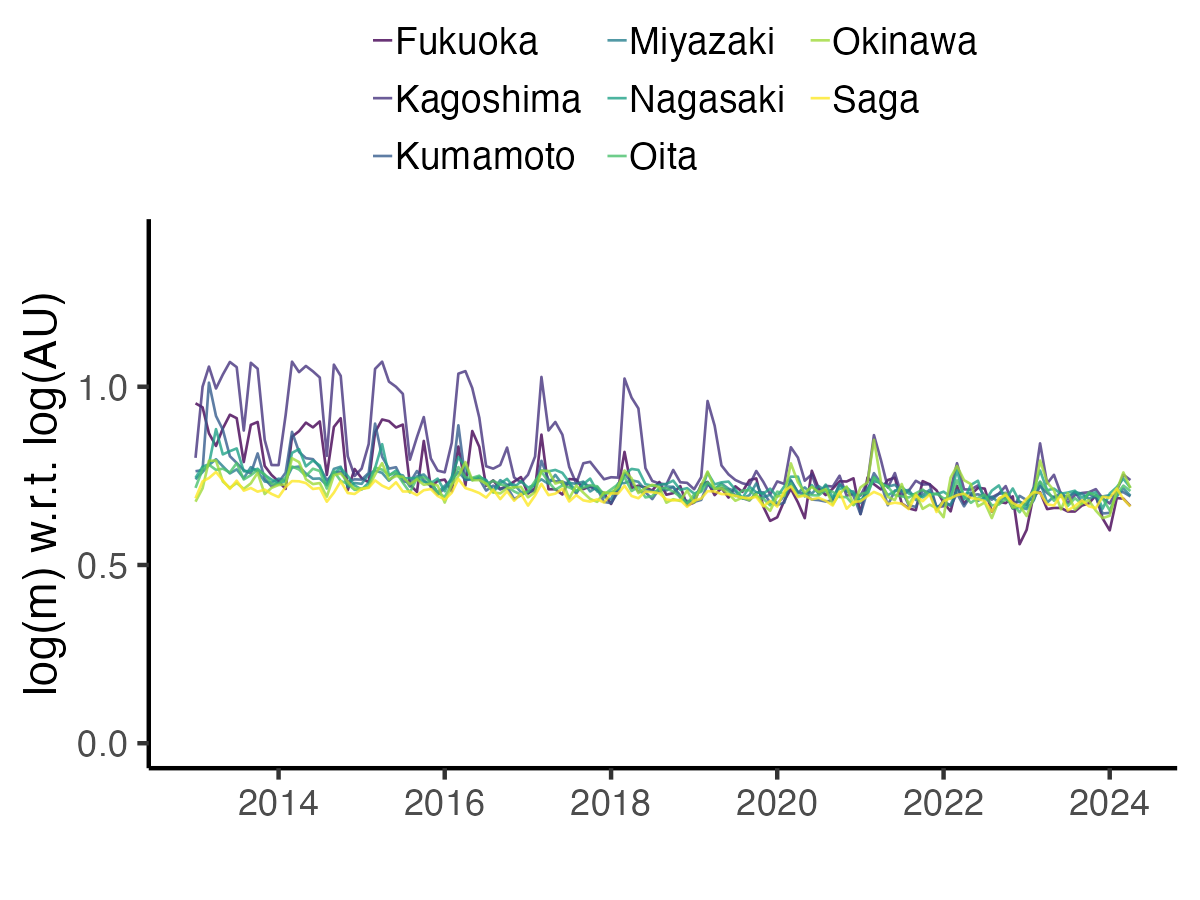}}
  \caption{Month-prefecture-level unemployment elasticity 2013--2024}
  \label{fg:month_part_and_full_time_elasticity_unemployed_month_aggregate_prefecture_results}
  \end{center}
  \footnotesize
  %Note:
\end{figure}

\begin{figure}[!ht]
  \begin{center}
  \subfloat[Hokkaido]{\includegraphics[width = 0.3\textwidth]
  {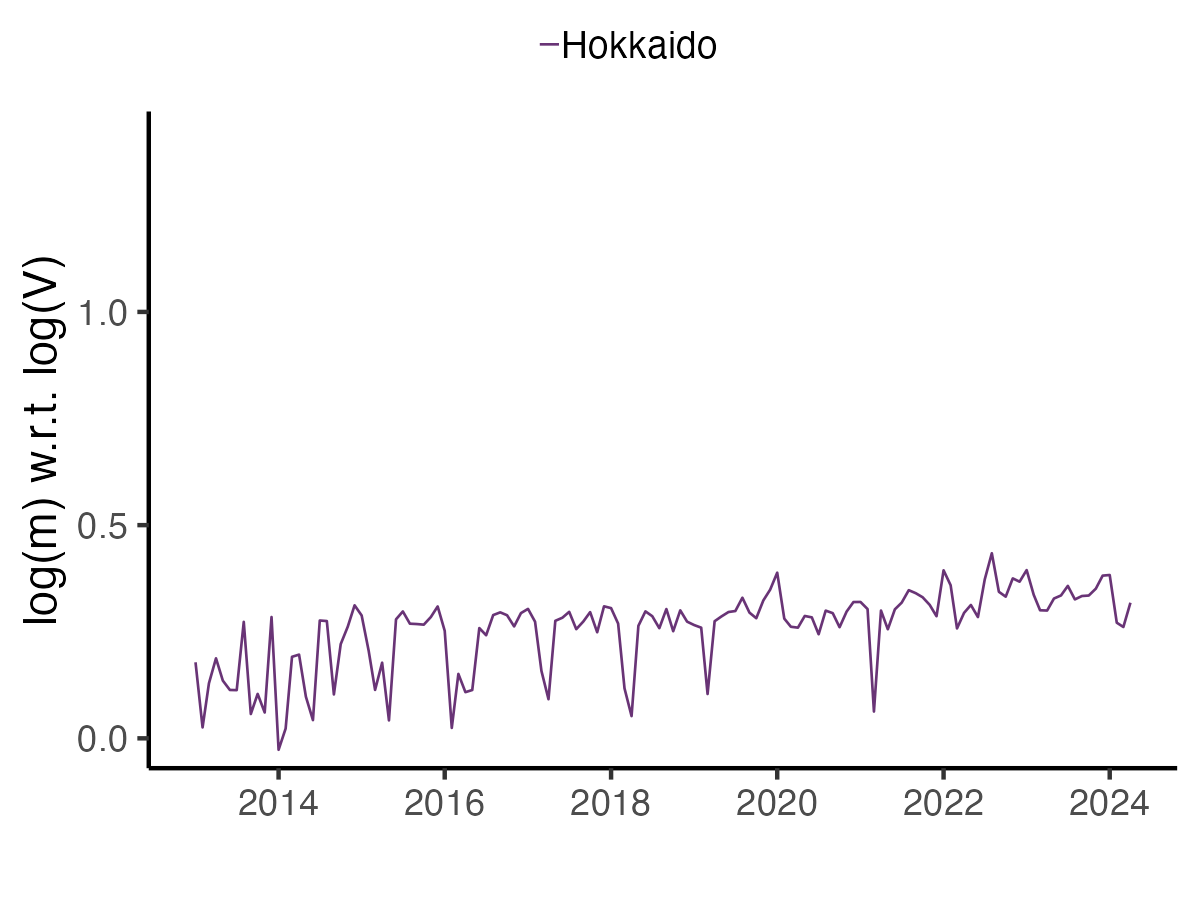}}
  \subfloat[Tohoku]{\includegraphics[width = 0.3\textwidth]
  {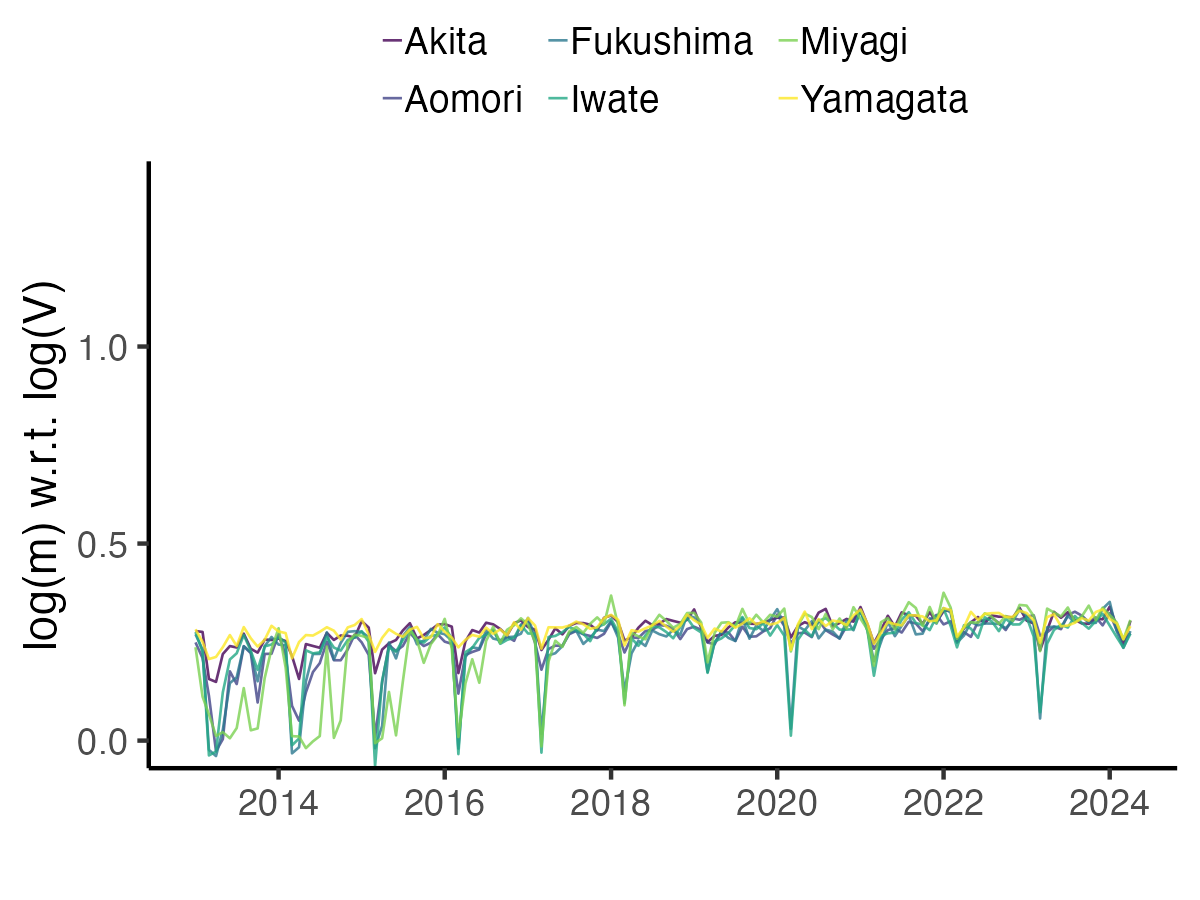}}\\
  \subfloat[Kanto]{\includegraphics[width = 0.3\textwidth]
  {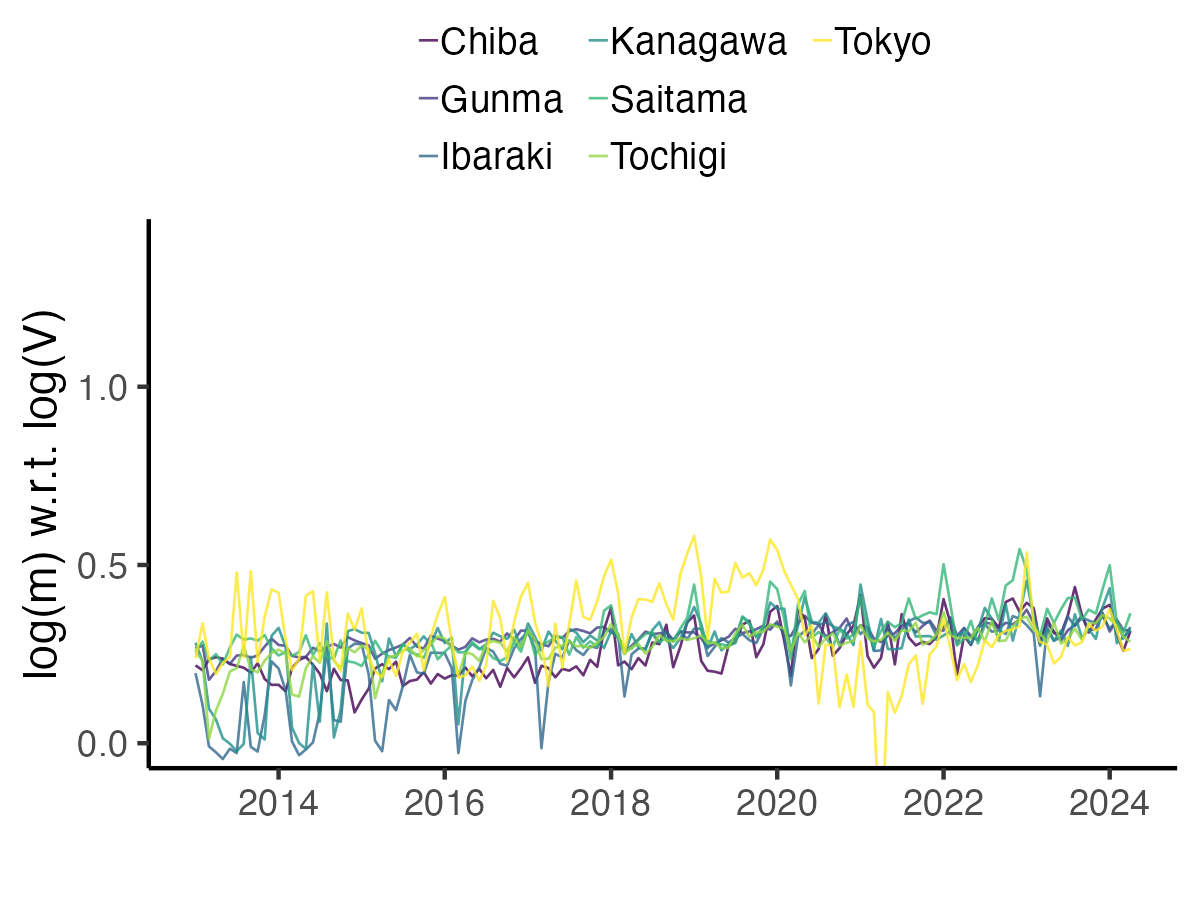}}
  \subfloat[Chubu]{\includegraphics[width = 0.3\textwidth]
  {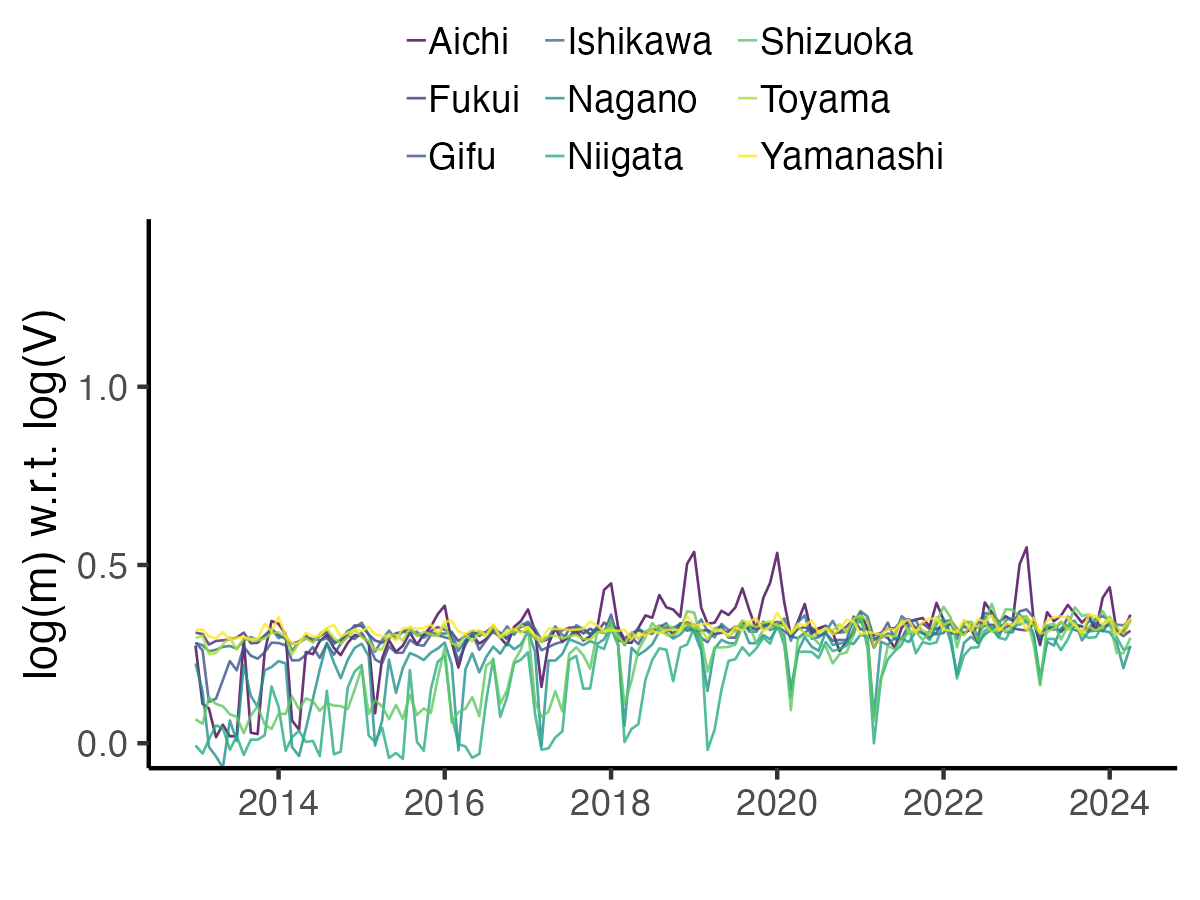}}
  \\
  \subfloat[Kansai]{\includegraphics[width = 0.3\textwidth]
  {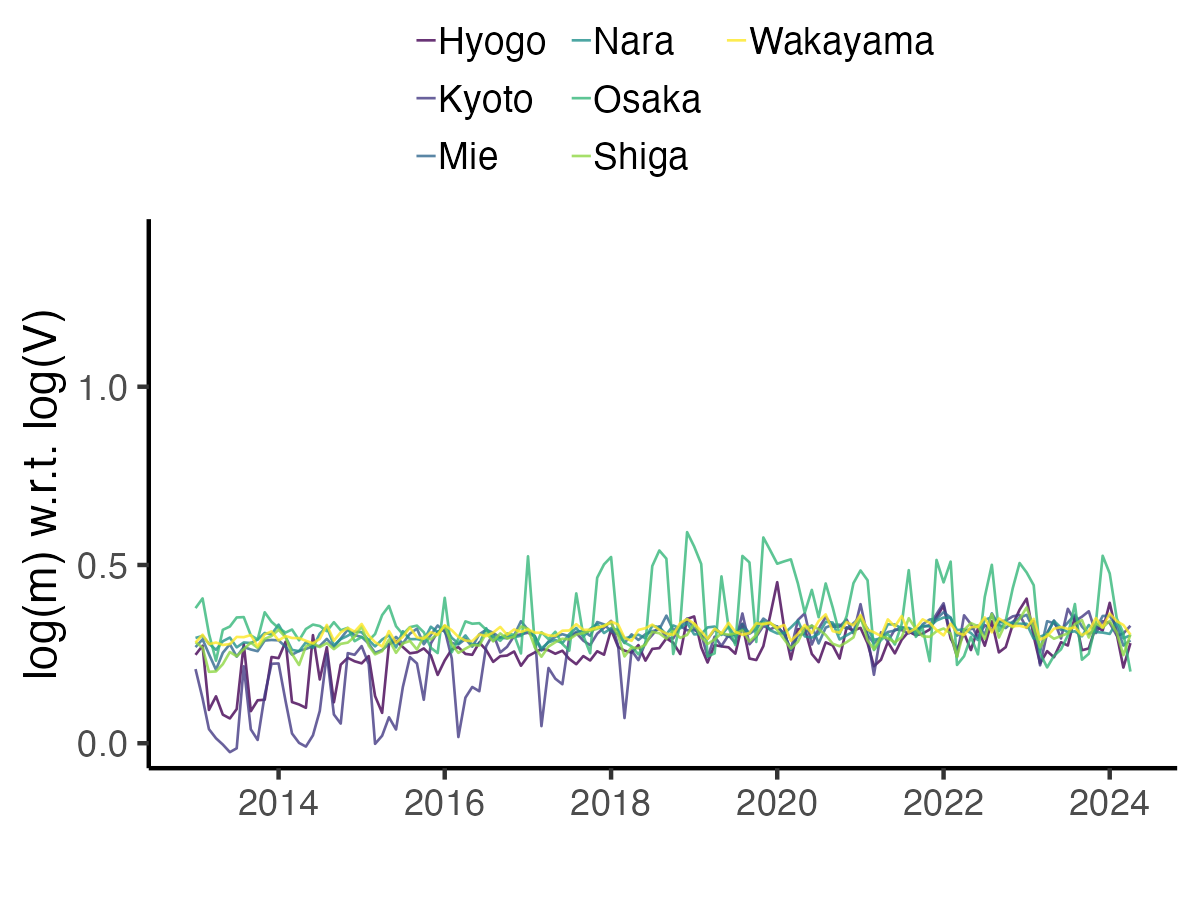}}
  \subfloat[Chugoku]{\includegraphics[width = 0.3\textwidth]
  {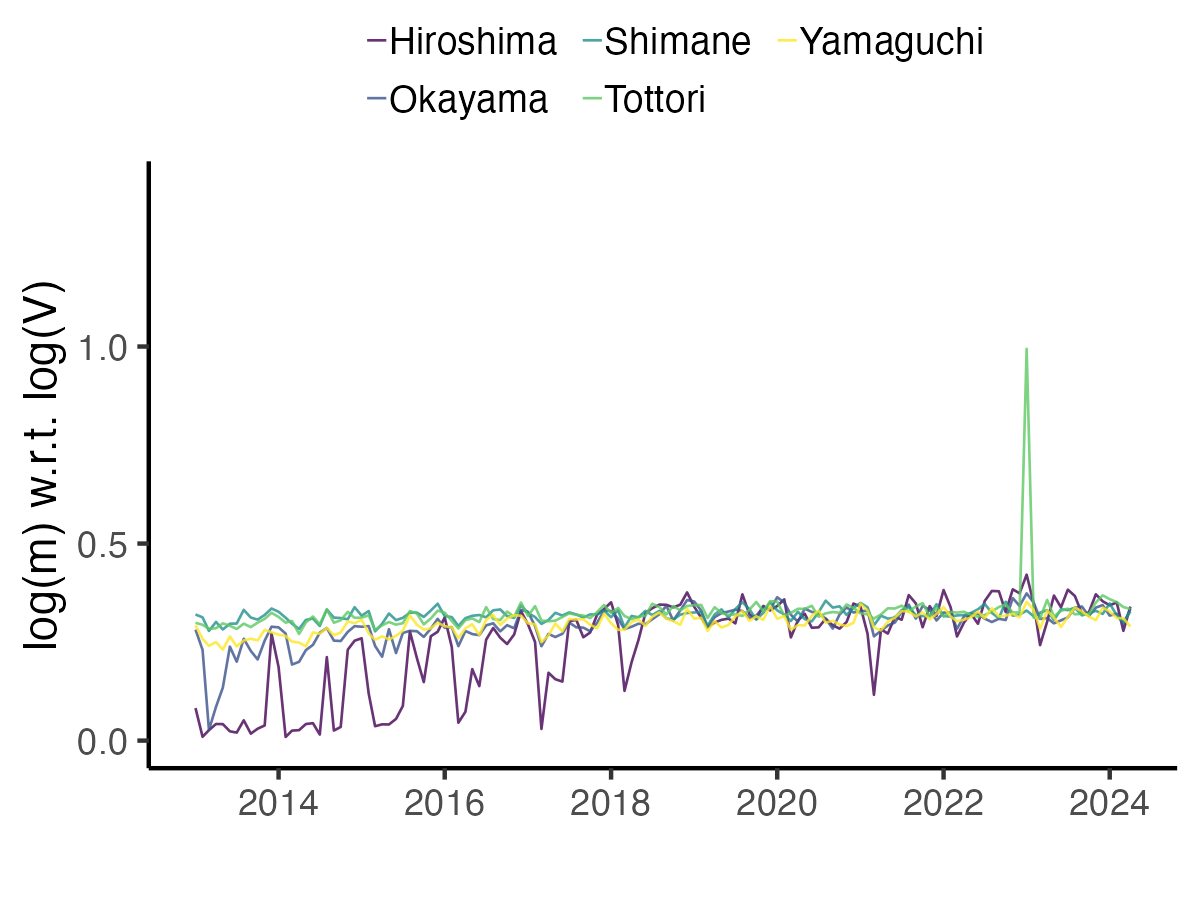}}\\
  \subfloat[Shikoku]{\includegraphics[width = 0.3\textwidth]
  {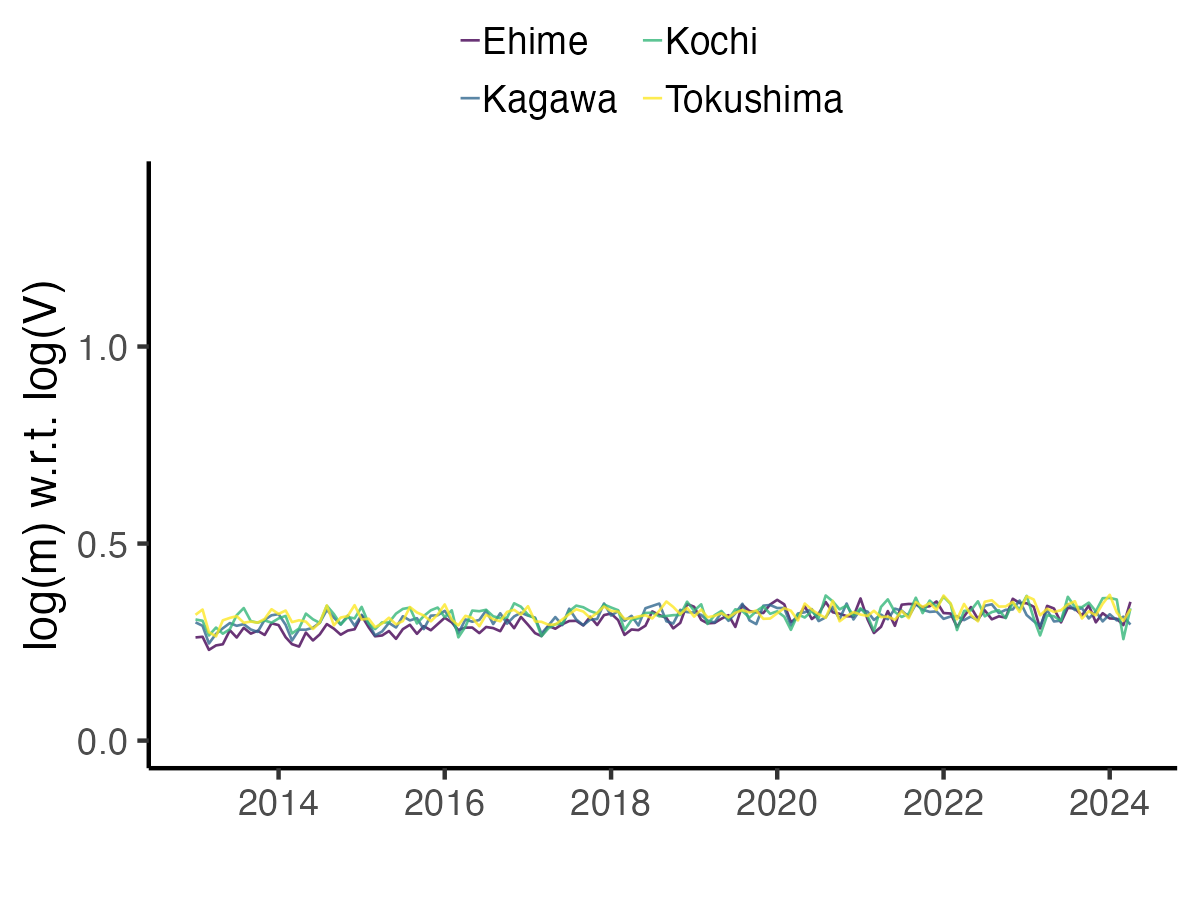}}
  \subfloat[Kyushu, Okinawa]{\includegraphics[width = 0.3\textwidth]
  {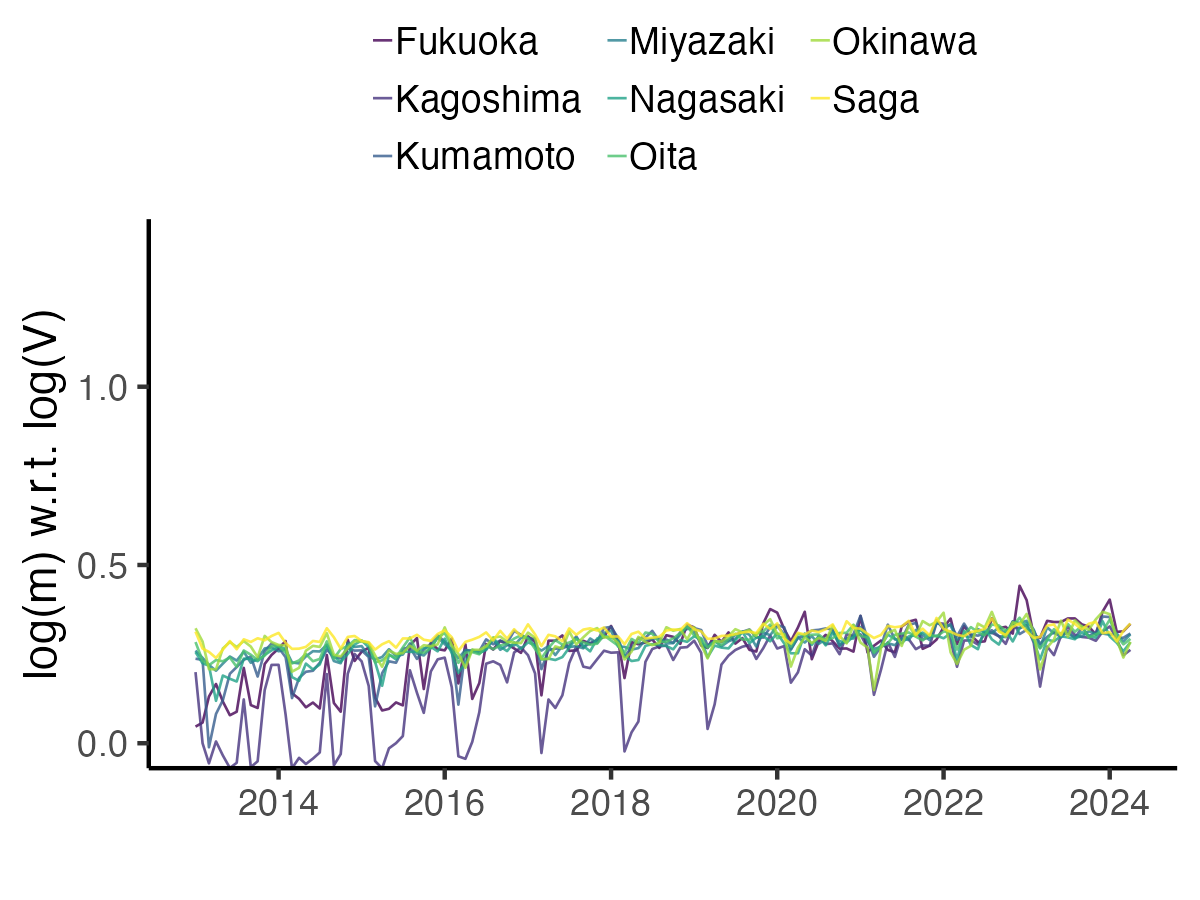}}
  \caption{Month-prefecture-level vacancy elasticity 2013--2024}
  \label{fg:month_part_and_full_time_elasticity_vacancy_month_aggregate_prefecture_results}
  \end{center}
  \footnotesize
  %Note:
\end{figure}

Figures \ref{fg:month_part_and_full_time_elasticity_unemployed_month_aggregate_prefecture_results} and \ref{fg:month_part_and_full_time_elasticity_vacancy_month_aggregate_prefecture_results} illustrate month-prefecture-level unemployment and vacancy elasticities in 2013--2024.
There is substantial heterogeneity across prefectures.
The unemployment elasticity is centered around 0.7, and the vacancy elasticity is centered around 0.3 (interquartile range 0.27--0.32) on average across prefectures.

\subsubsection{Occupation-Level Results in 2012--2023}

\begin{figure}[!ht]
  \begin{center}
  \subfloat[Managerial]{\includegraphics[width = 0.30\textwidth]
  {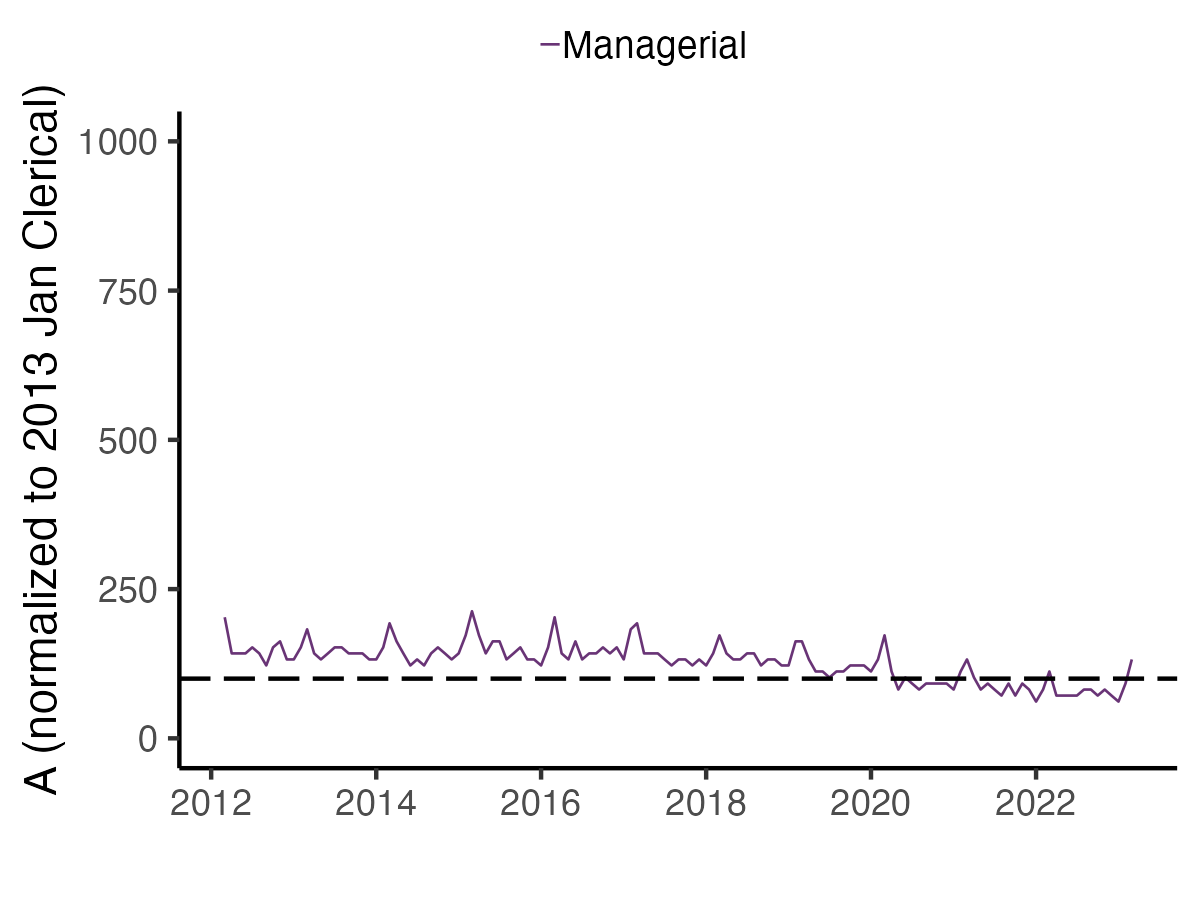}}
  \subfloat[Professional and Technical]{\includegraphics[width = 0.30\textwidth]
  {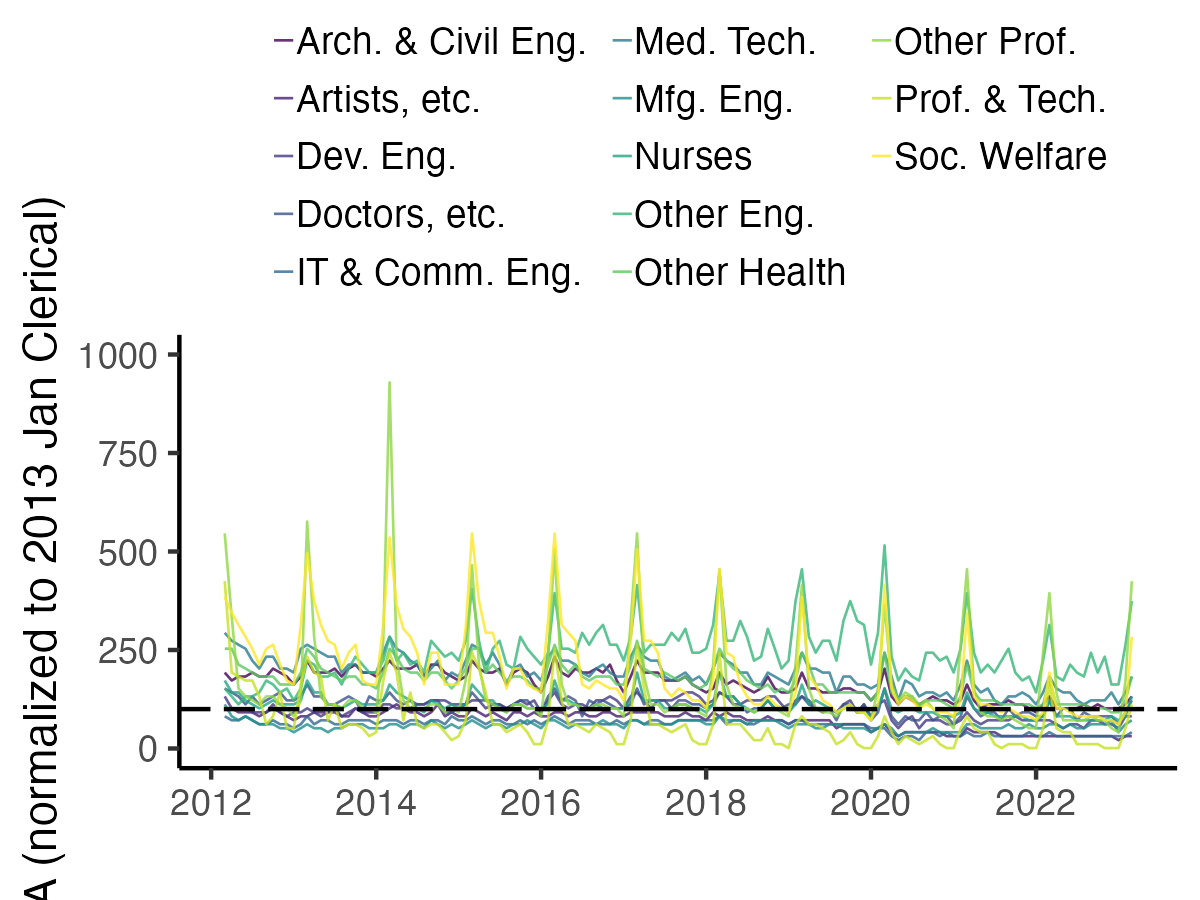}}
  \subfloat[Clerical]{\includegraphics[width = 0.30\textwidth]
  {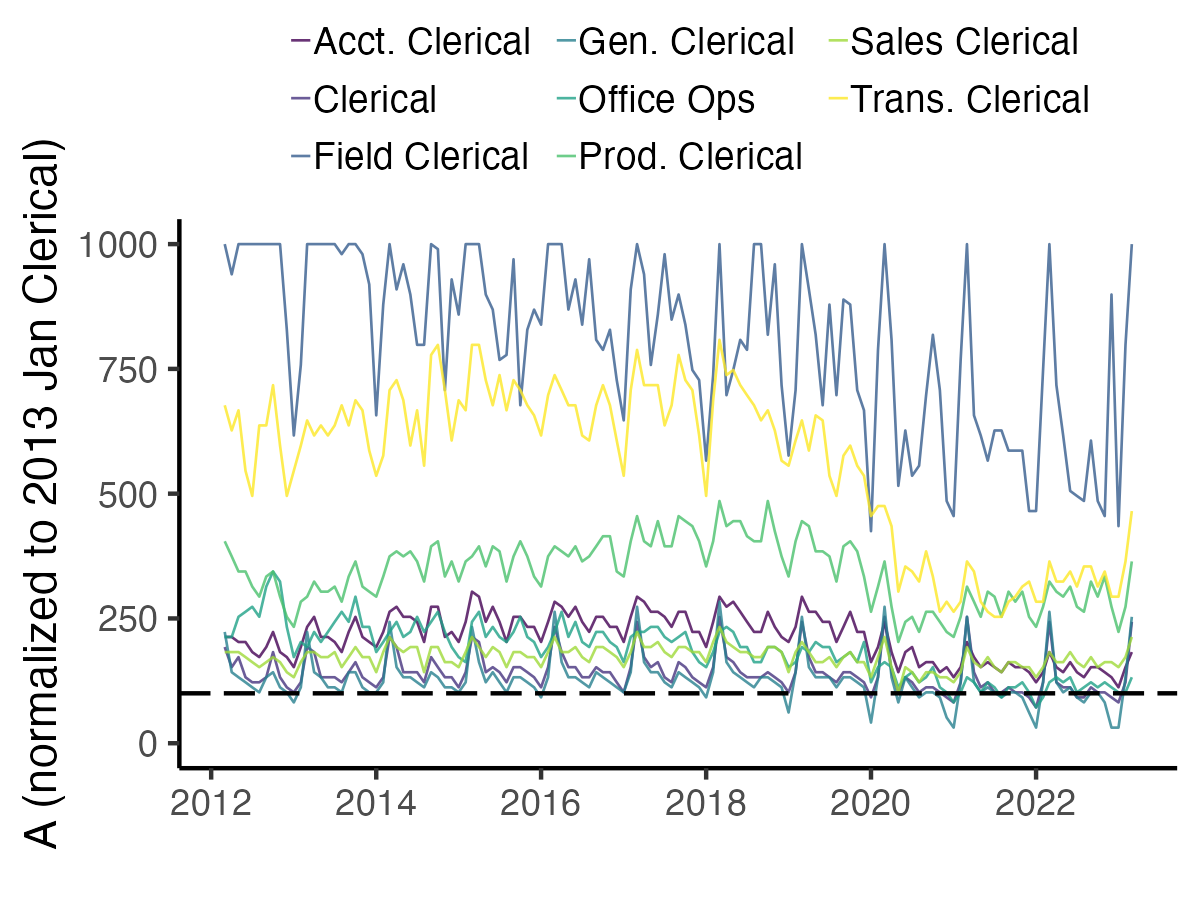}}\\
  \subfloat[Sales]{\includegraphics[width = 0.30\textwidth]
  {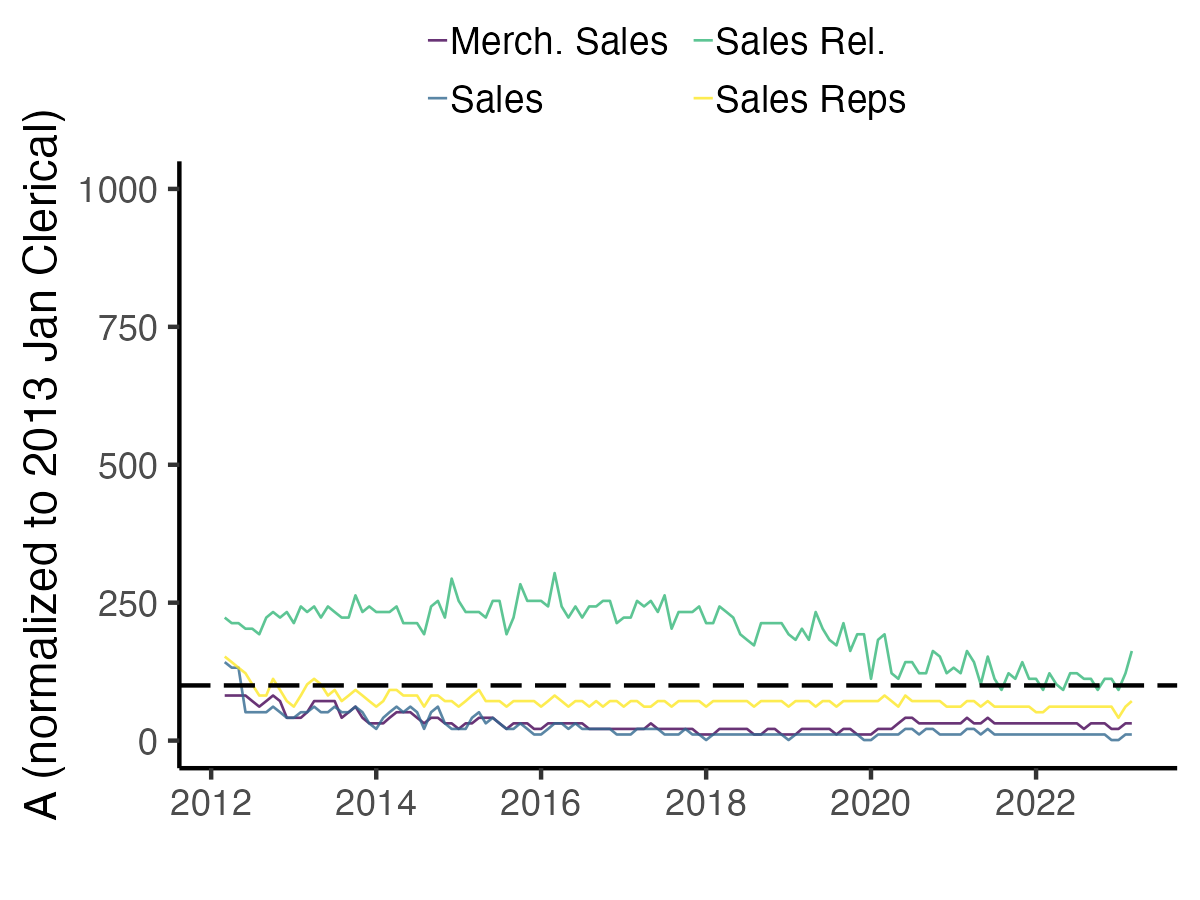}}
  \subfloat[Service]{\includegraphics[width = 0.30\textwidth]
  {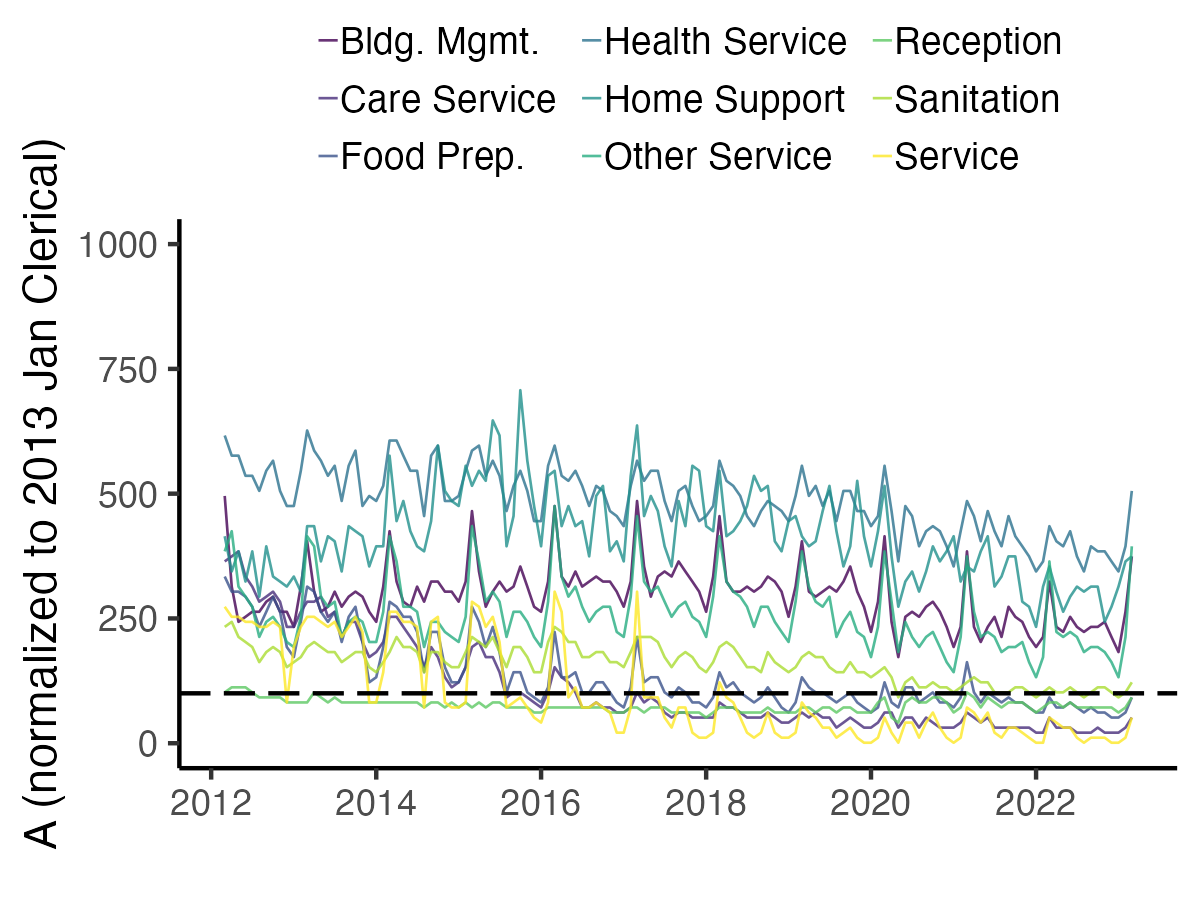}}
  \subfloat[Agriculture Forestry and Fishing]{\includegraphics[width = 0.30\textwidth]
  {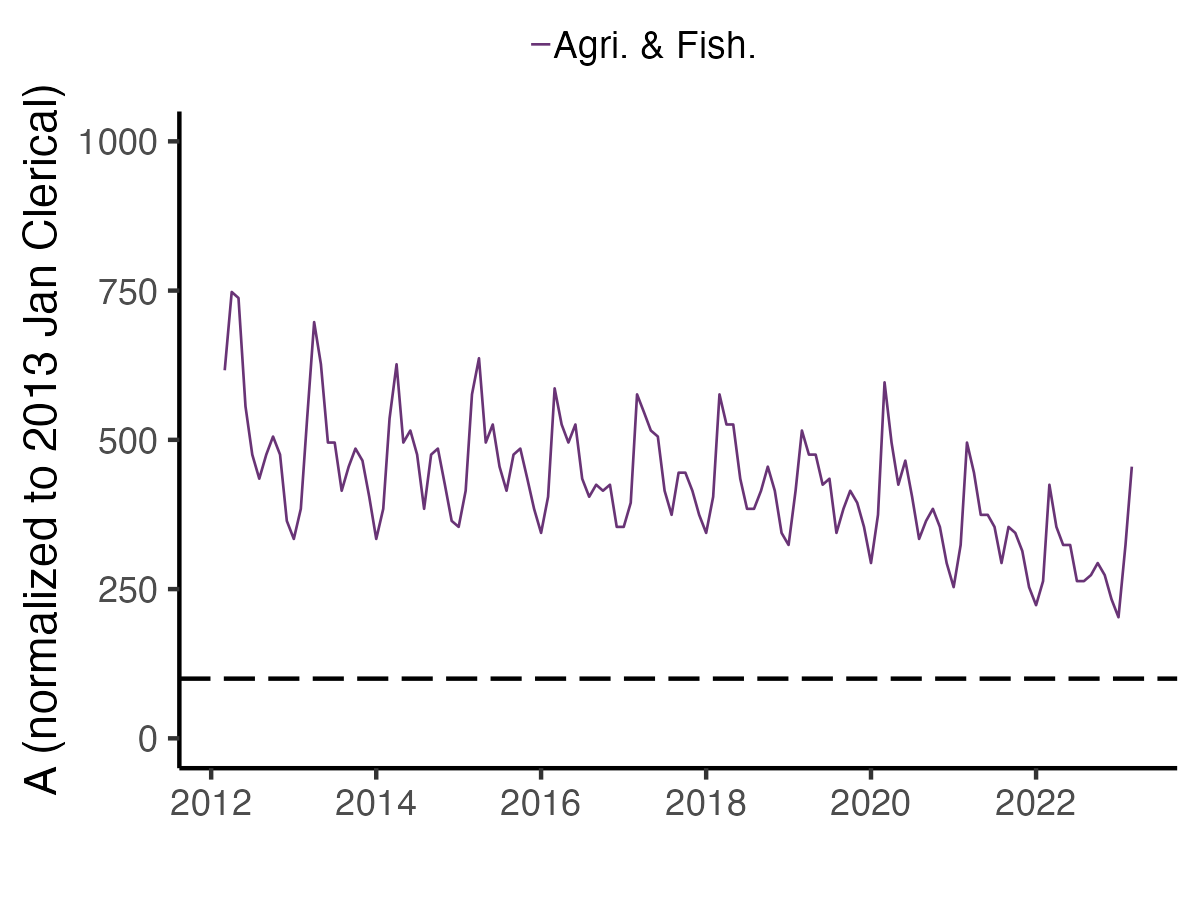}}\\
  \subfloat[Production]{\includegraphics[width = 0.30\textwidth]
  {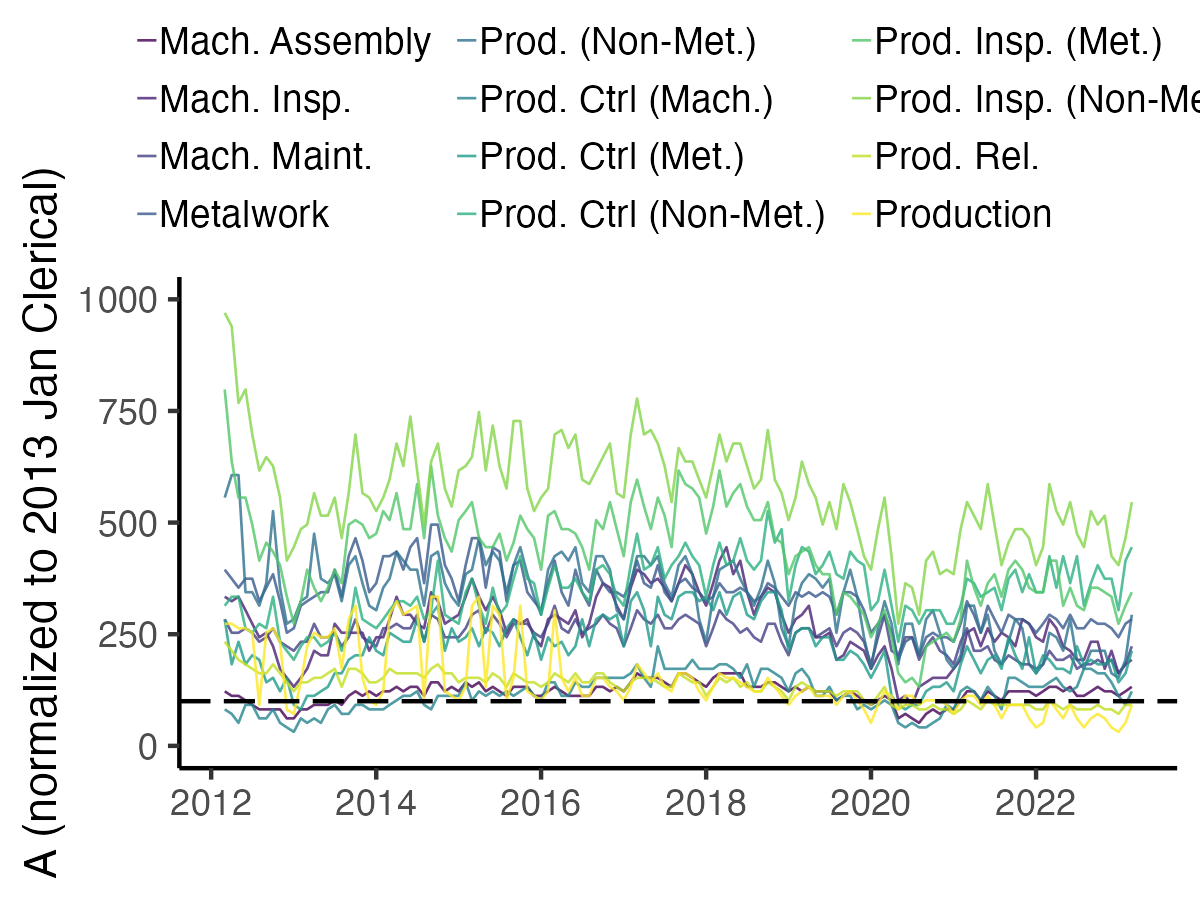}}
  \subfloat[Transportation and Machine Operation]{\includegraphics[width = 0.30\textwidth]
  {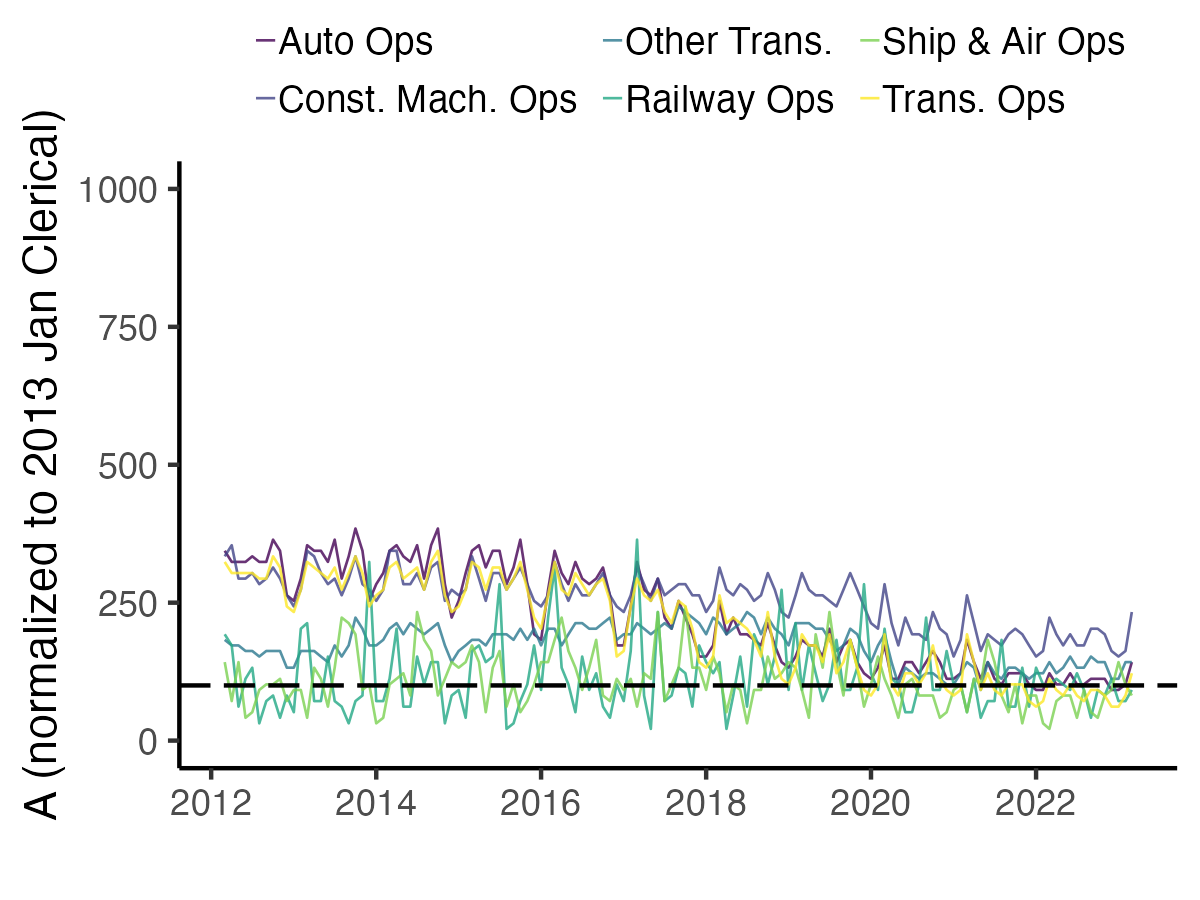}}
  \subfloat[Construction and Mining]{\includegraphics[width = 0.30\textwidth]
  {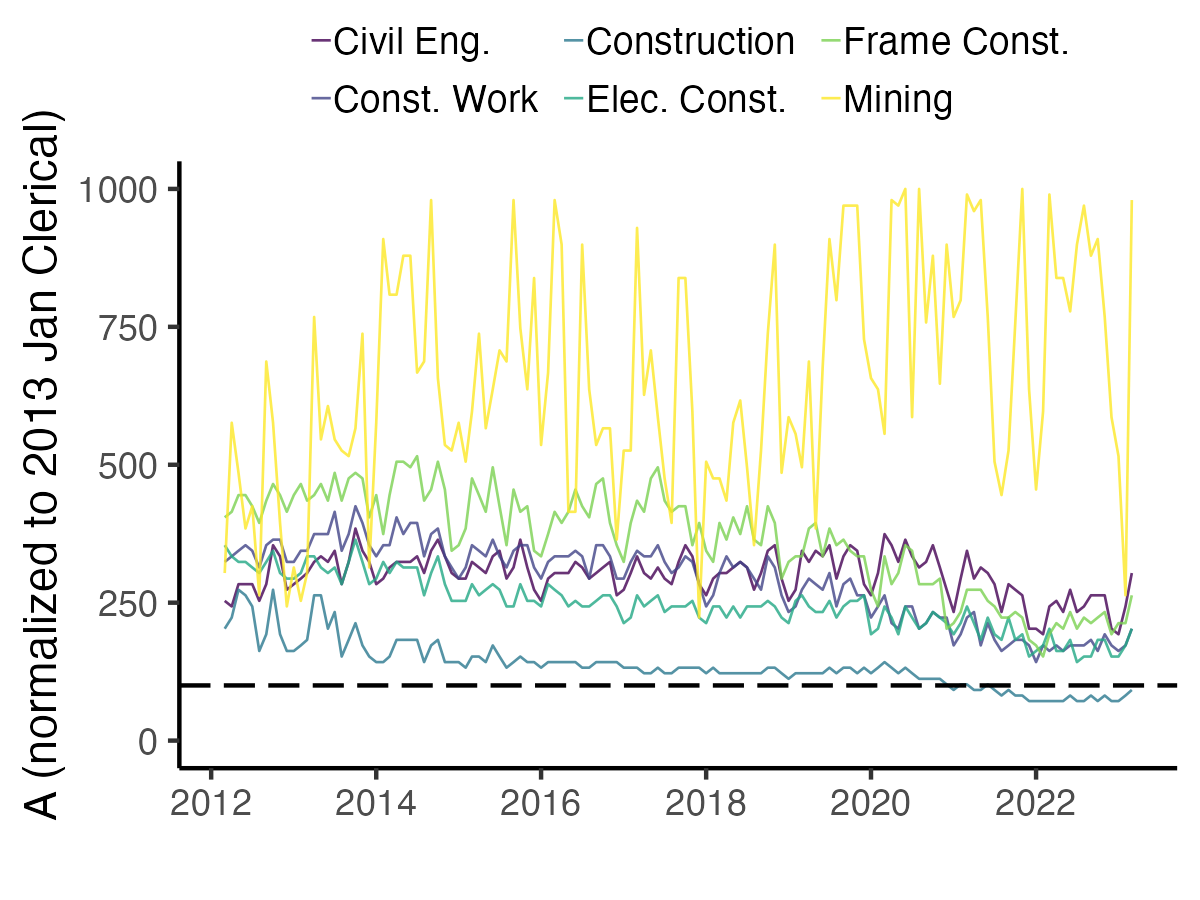}}\\
  \subfloat[Transportation Cleaning and Packaging]{\includegraphics[width = 0.30\textwidth]
  {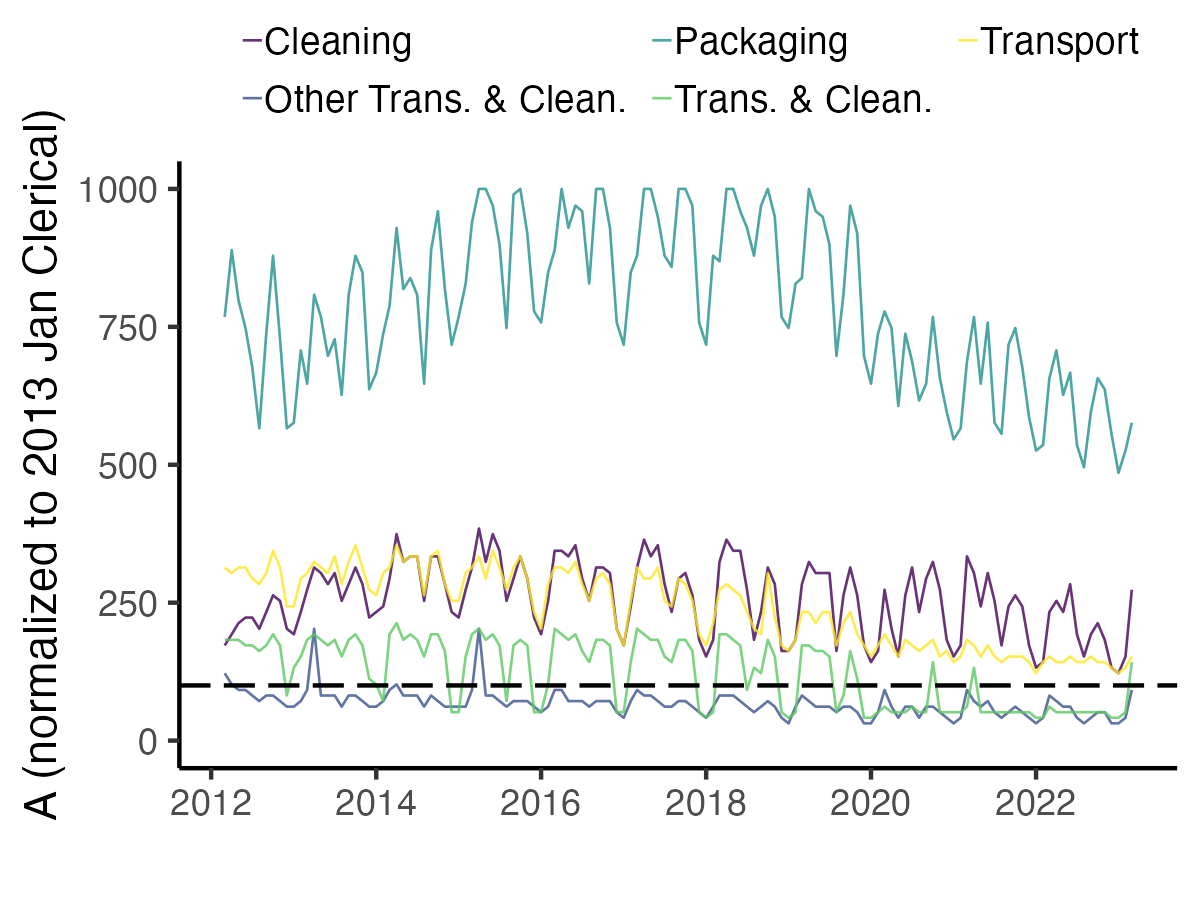}}
  \subfloat[Security]{\includegraphics[width = 0.30\textwidth]
  {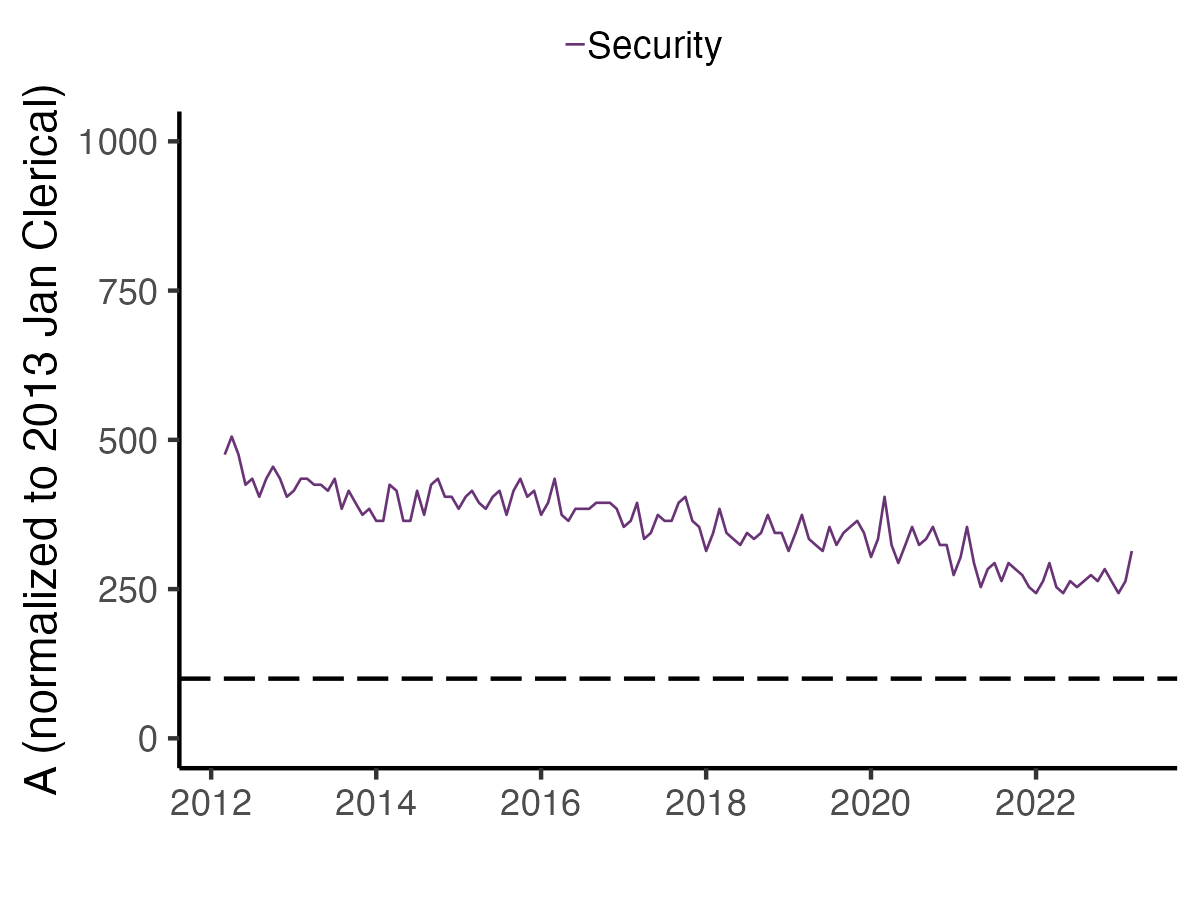}}
  \caption{Month-occupation-level matching efficiency results 2012--2023}
  \label{fg:month_part_and_full_time_matching_efficiency_job_category_results}
  \end{center}
  \footnotesize
  %Note:
\end{figure}

Figure \ref{fg:month_part_and_full_time_matching_efficiency_job_category_results} illustrates significant disparities in matching efficiency across occupations, normalized to Clerical jobs in January 2013.
Primary and secondary industries show relatively higher efficiency than tertiary industries.
For example, categories like ``Agriculture, Forestry, and Fishing" (panel f) and ``Construction and Mining" (panel i) in primary and secondary industries exhibit high volatility and generally higher efficiency levels. In contrast, occupations such as ``Managerial" (panel a) and ``Service" (panel e) display more stable and lower efficiency values. These variations highlight the diverse nature of labor market dynamics across different sectors, influenced by factors such as skill specialization, economic cycles, and industry-specific demand fluctuations.

\begin{figure}[!ht]
  \begin{center}
  \subfloat[Managerial]{\includegraphics[width = 0.30\textwidth]
  {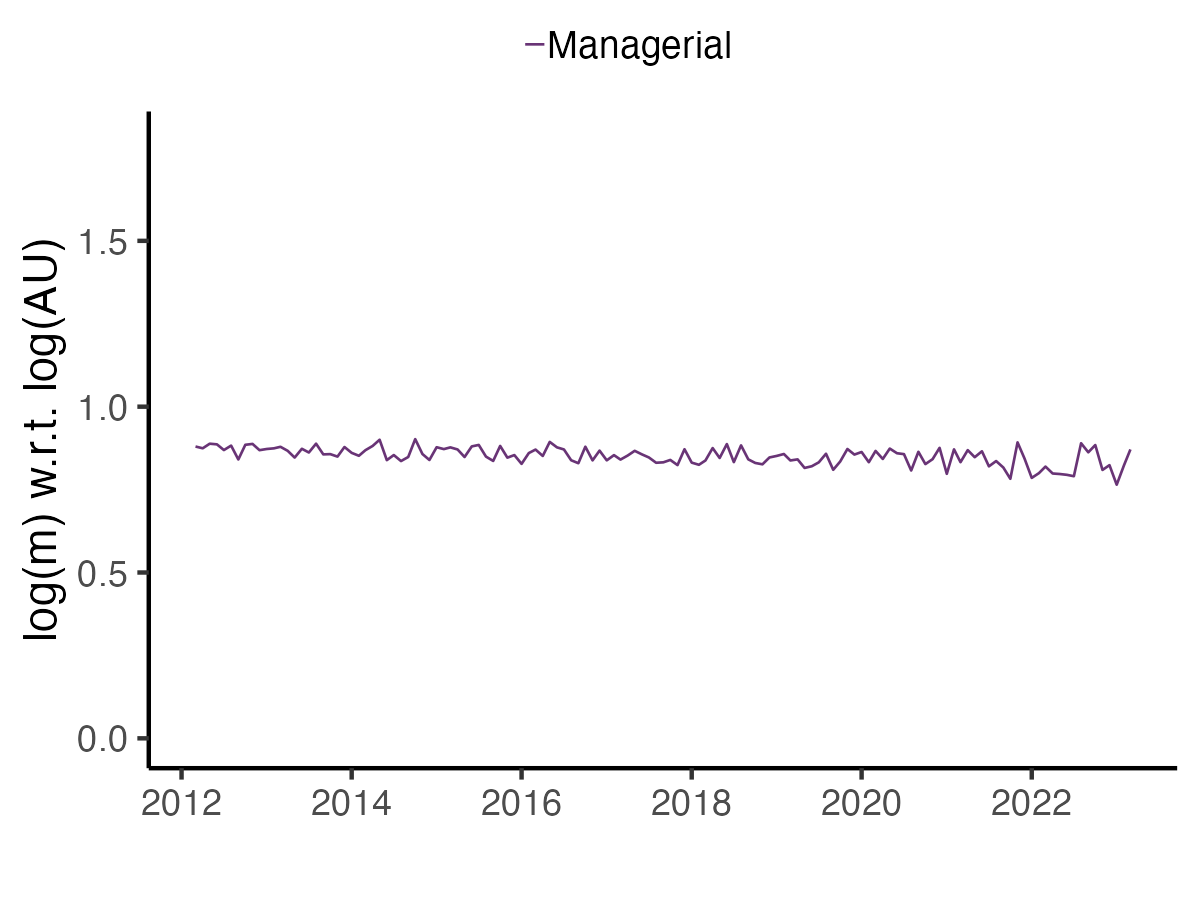}}
  \subfloat[Professional and Technical]{\includegraphics[width = 0.30\textwidth]
  {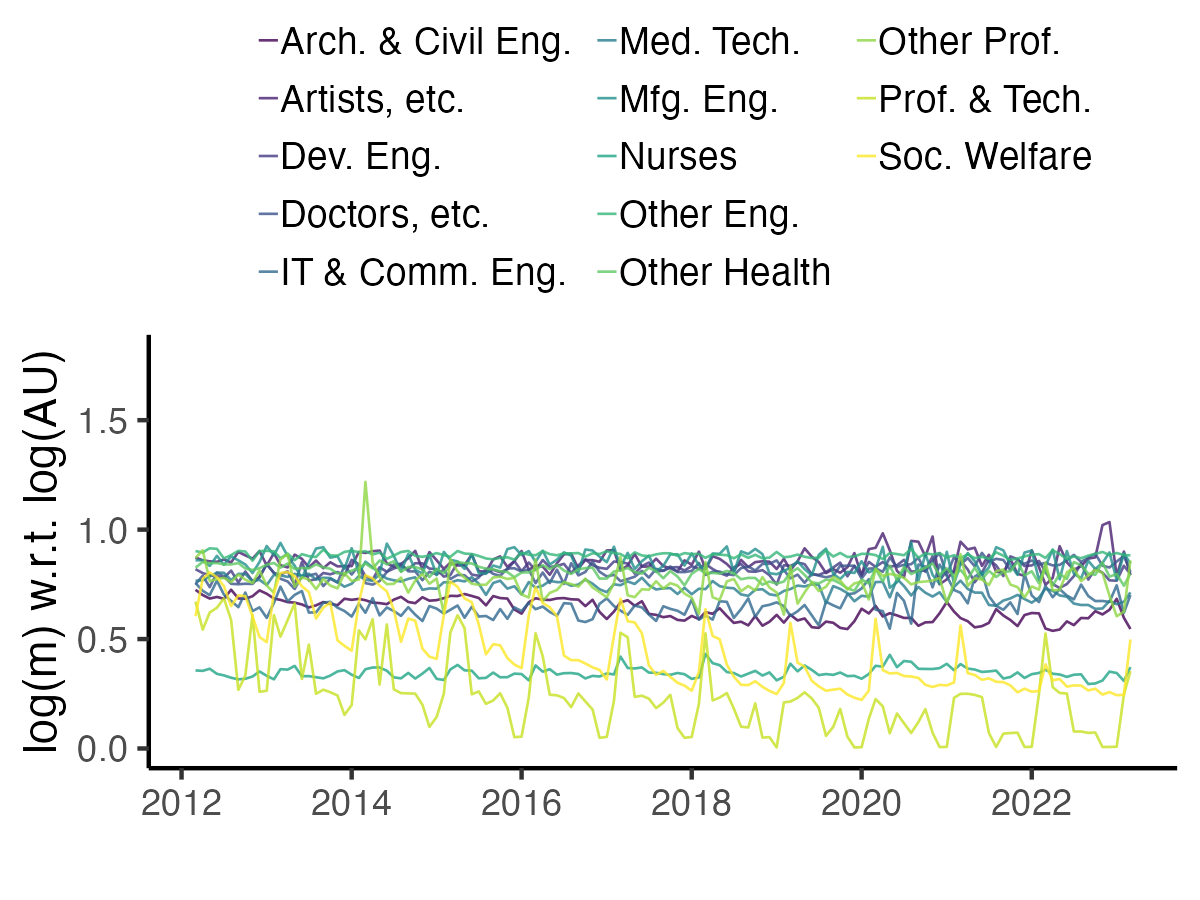}}
  \subfloat[Clerical]{\includegraphics[width = 0.30\textwidth]
  {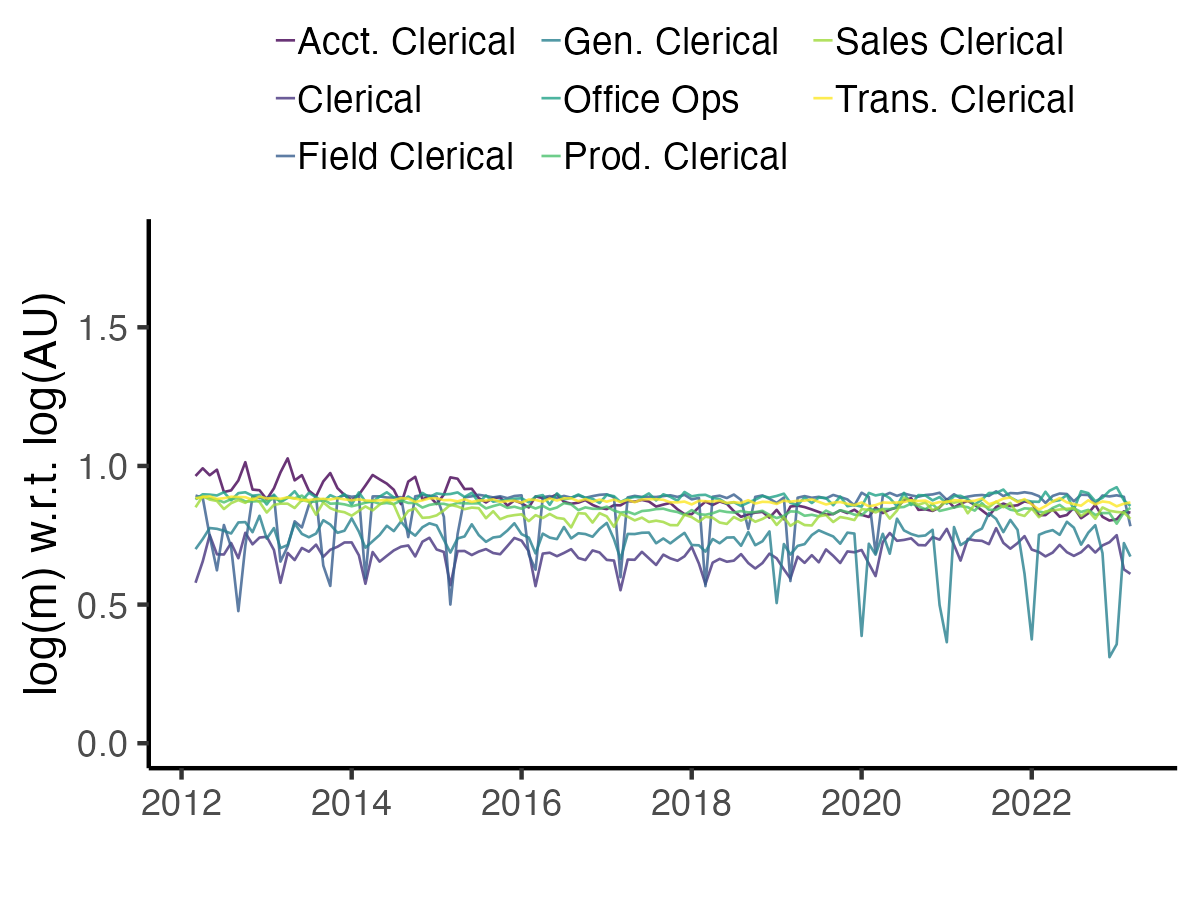}}\\
  \subfloat[Sales]{\includegraphics[width = 0.30\textwidth]
  {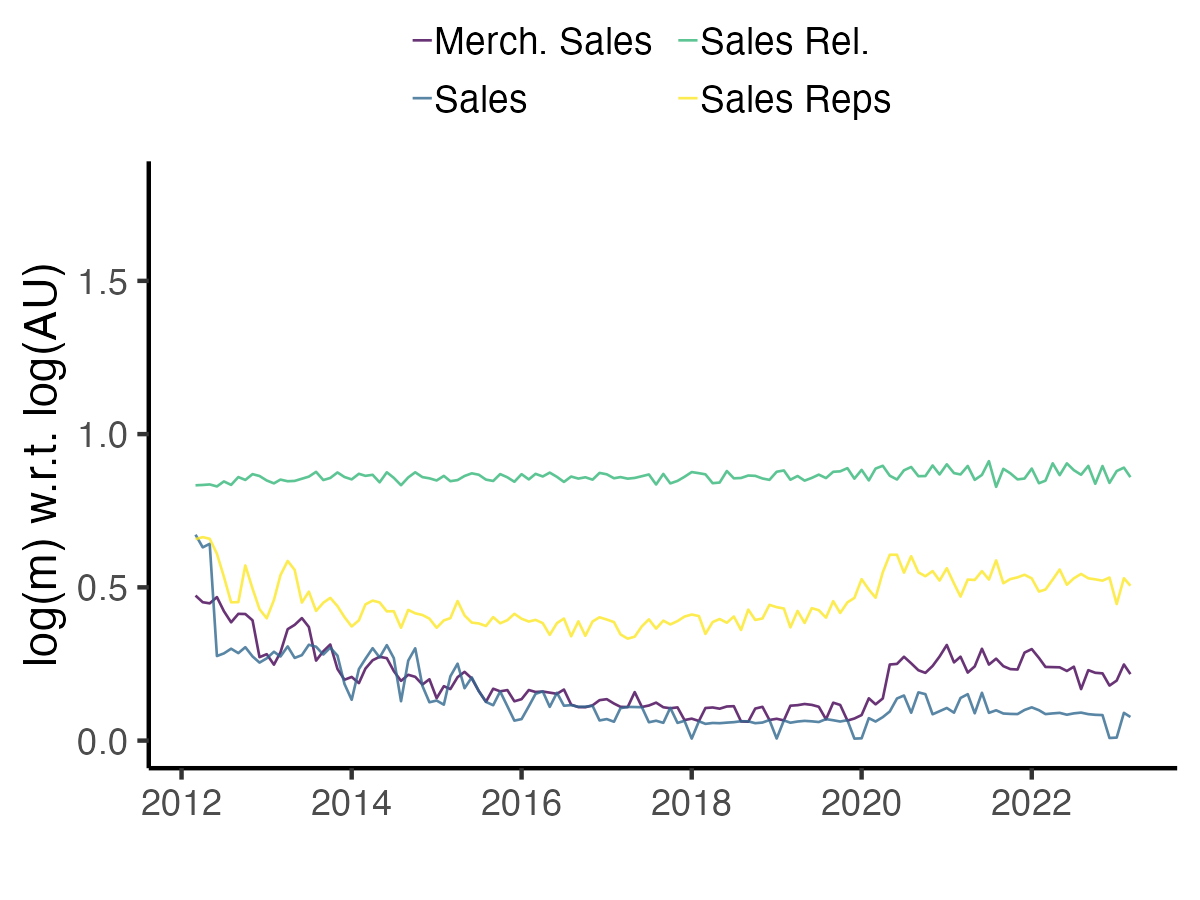}}
  \subfloat[Service]{\includegraphics[width = 0.30\textwidth]
  {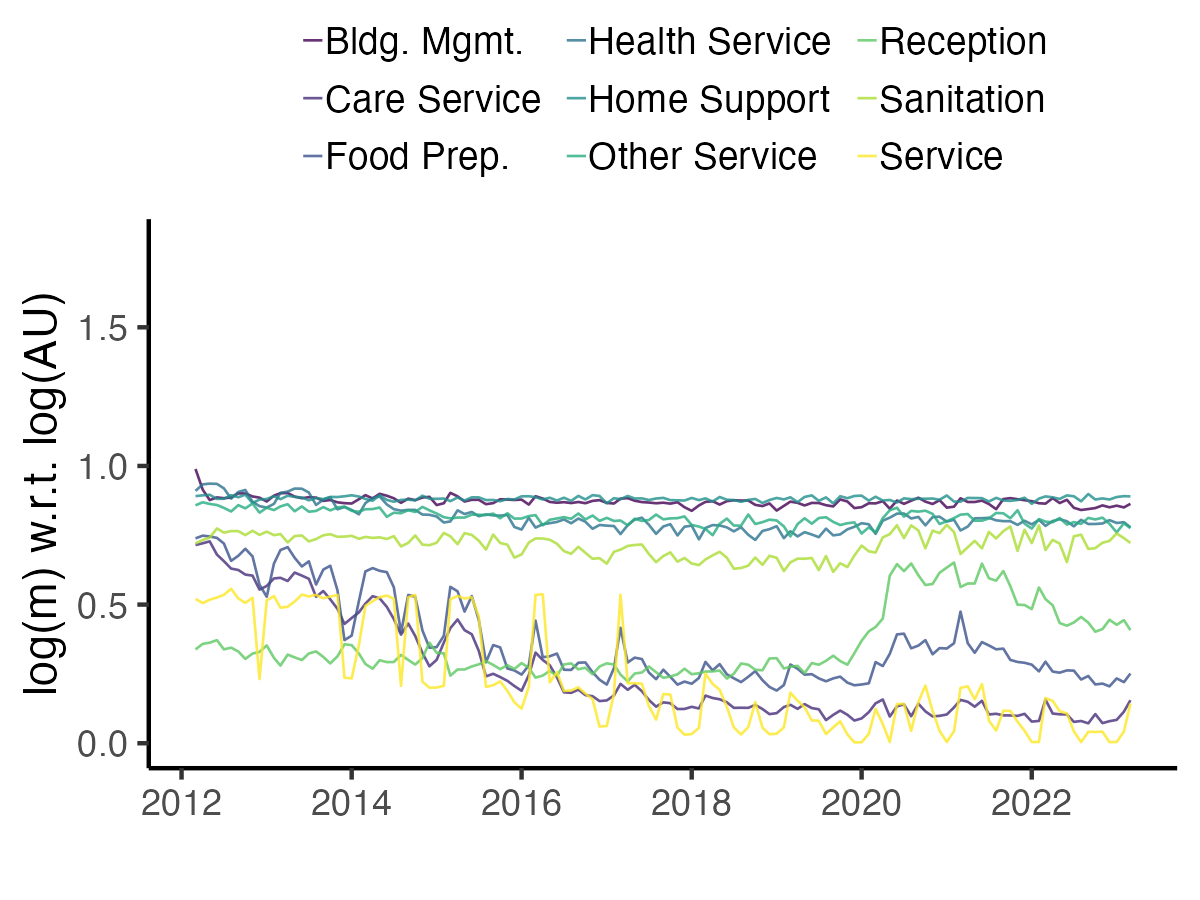}}
  \subfloat[Agriculture Forestry and Fishing]{\includegraphics[width = 0.30\textwidth]
  {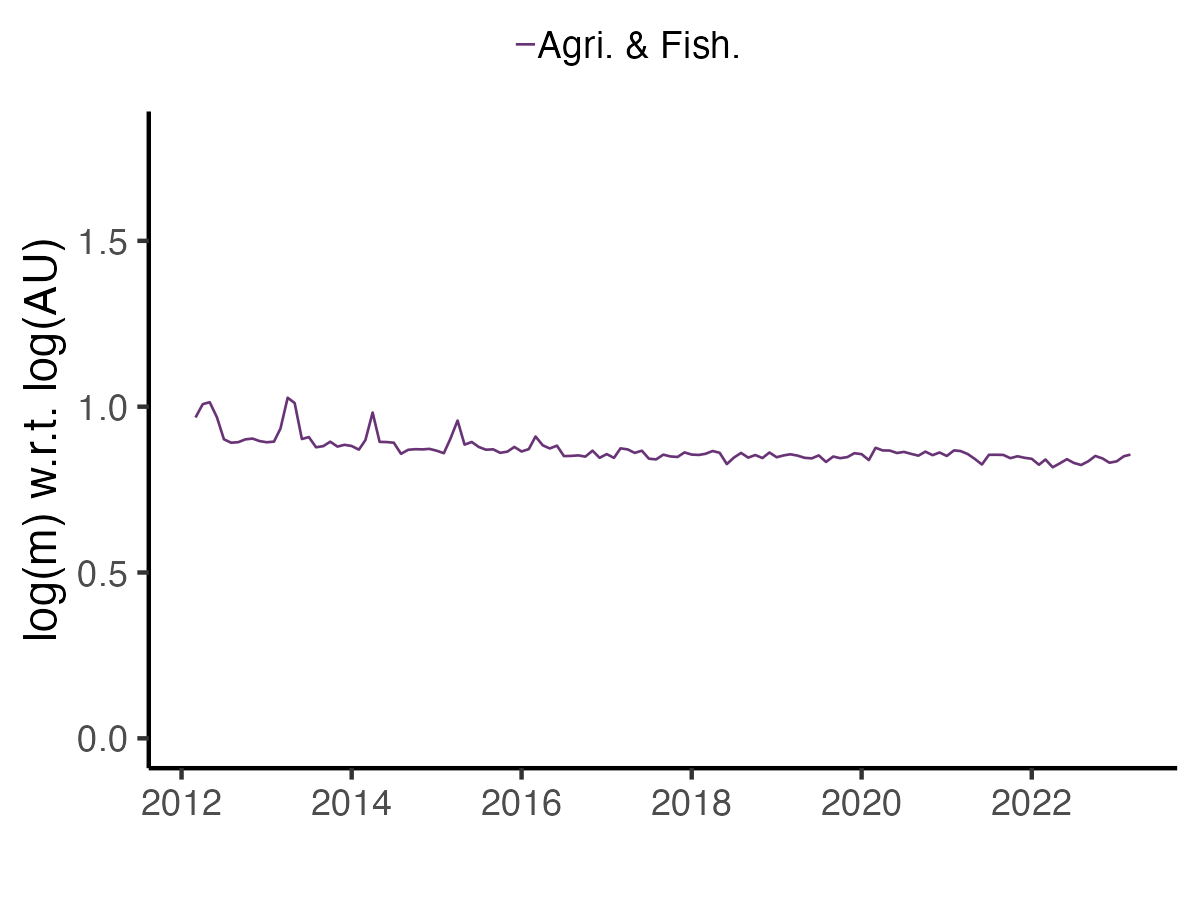}}\\
  \subfloat[Production]{\includegraphics[width = 0.30\textwidth]
  {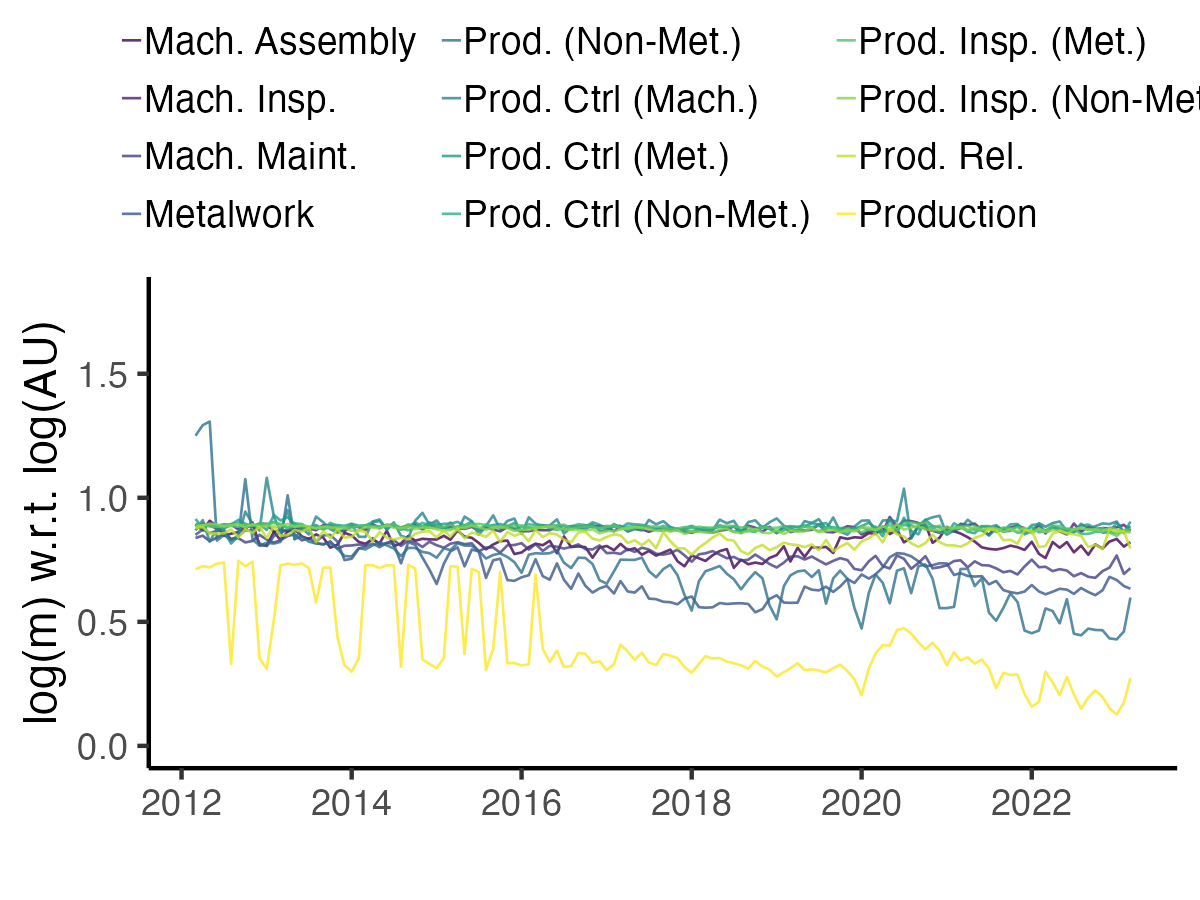}}
  \subfloat[Transportation and Machine Operation]{\includegraphics[width = 0.30\textwidth]
  {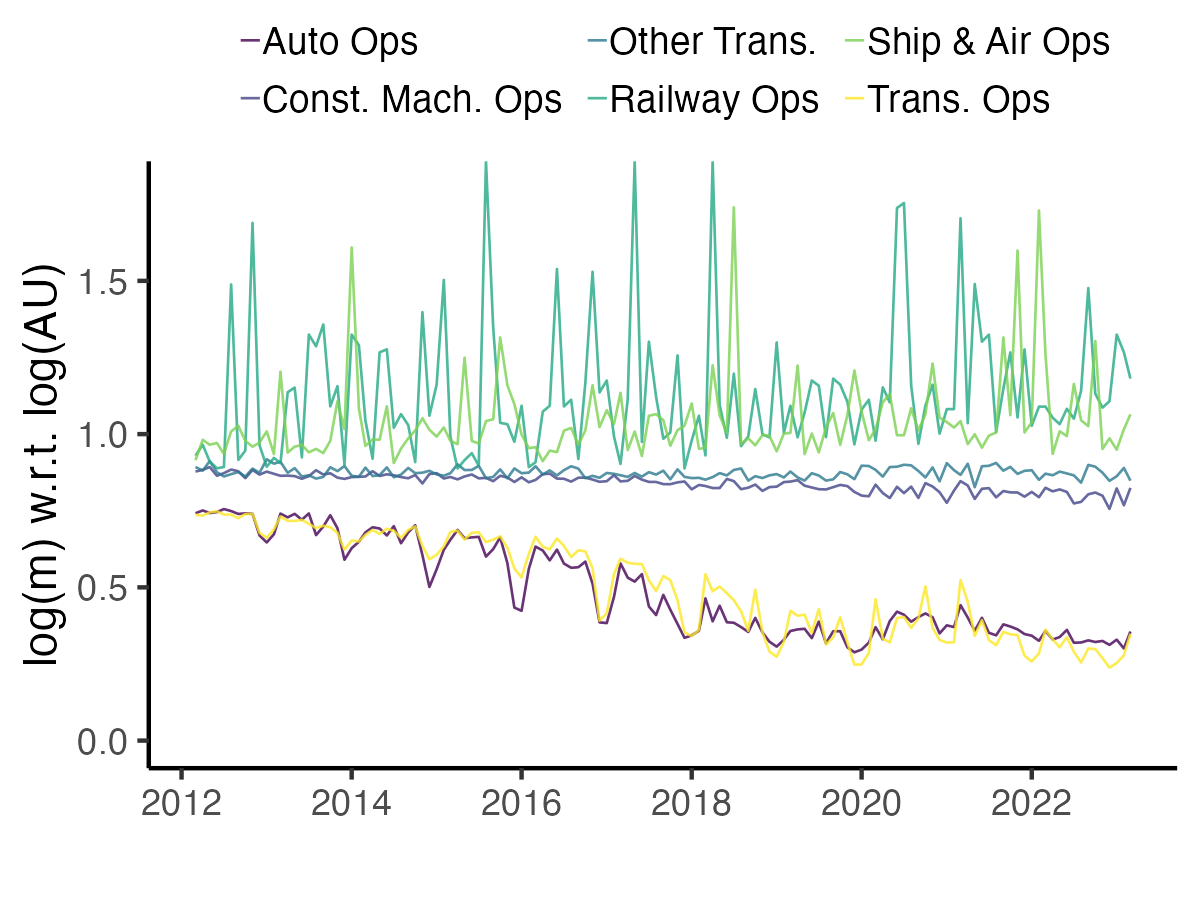}}
  \subfloat[Construction and Mining]{\includegraphics[width = 0.30\textwidth]
  {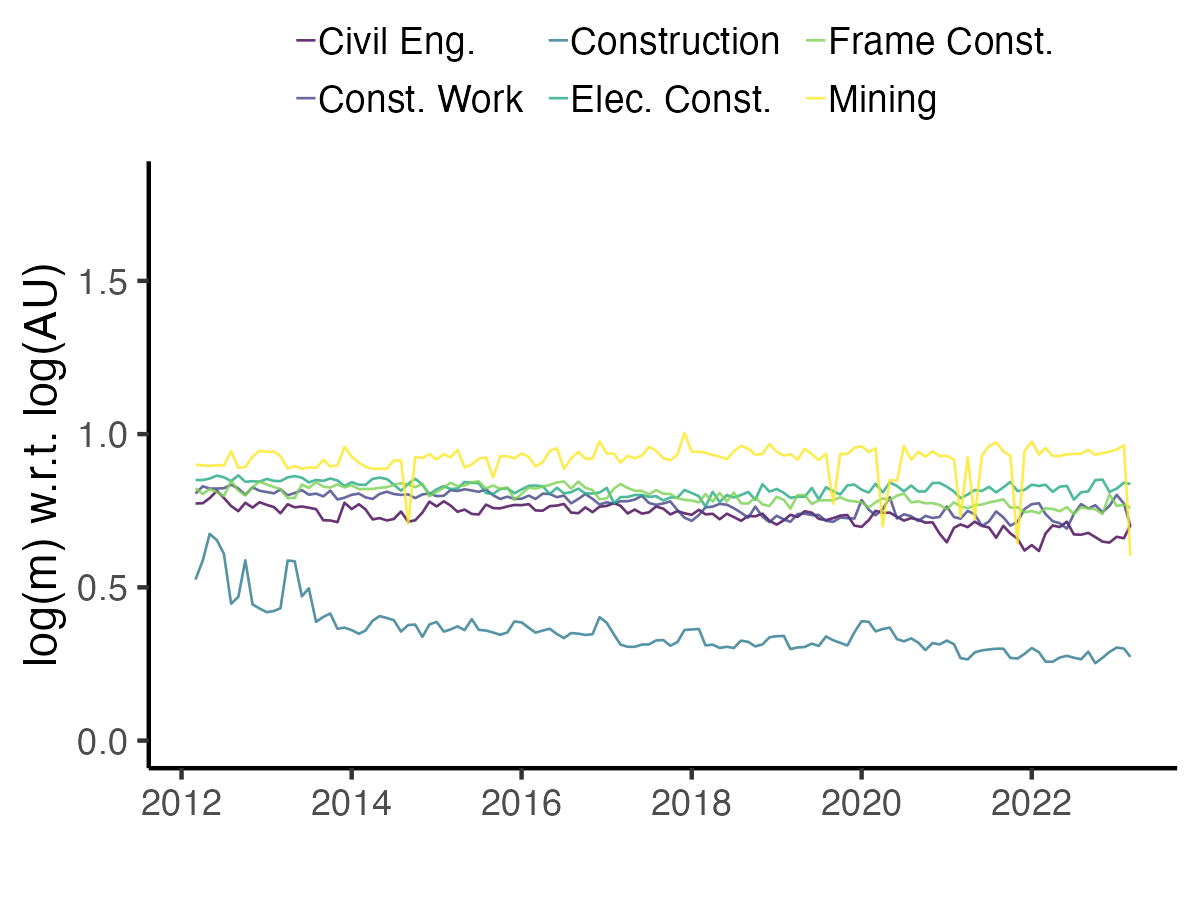}}\\
  \subfloat[Transportation Cleaning and Packaging]{\includegraphics[width = 0.30\textwidth]
  {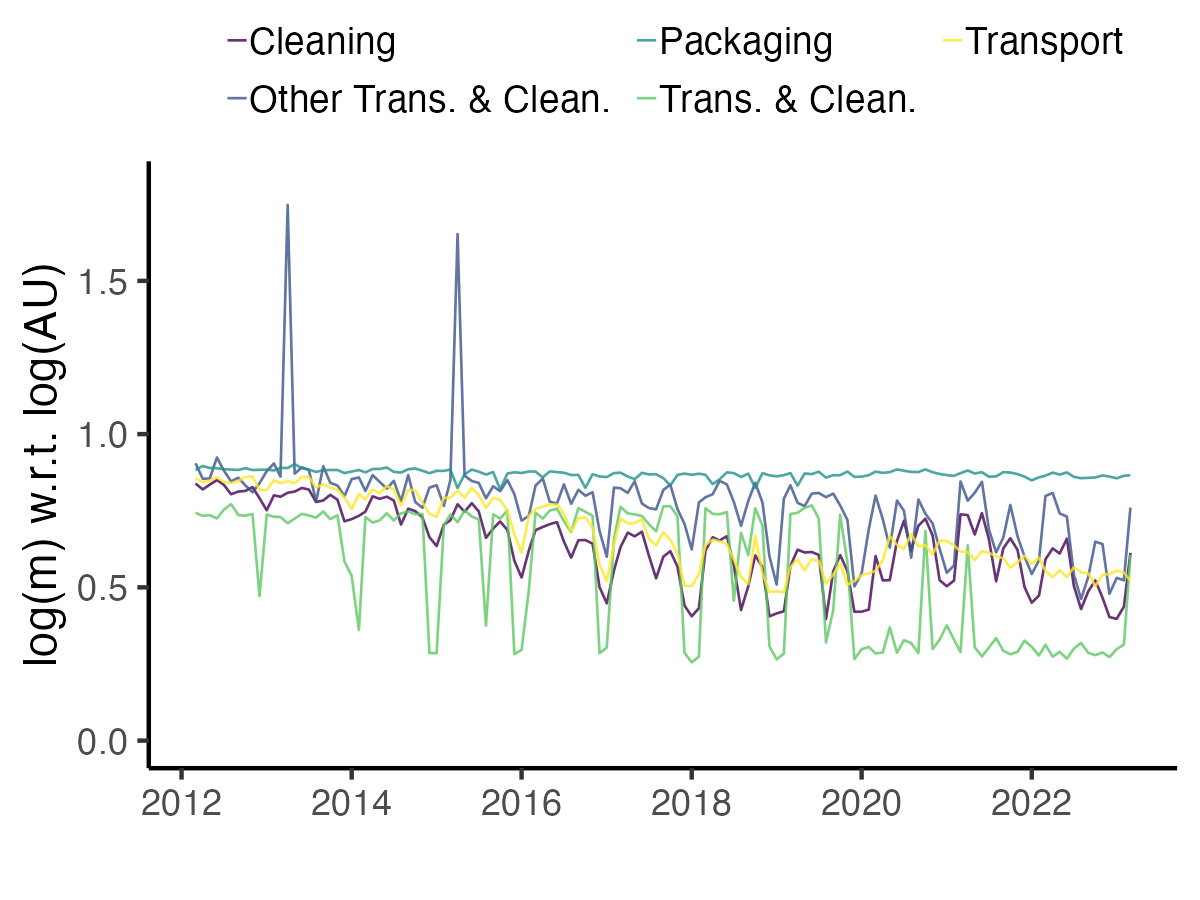}}
  \subfloat[Security]{\includegraphics[width = 0.30\textwidth]
  {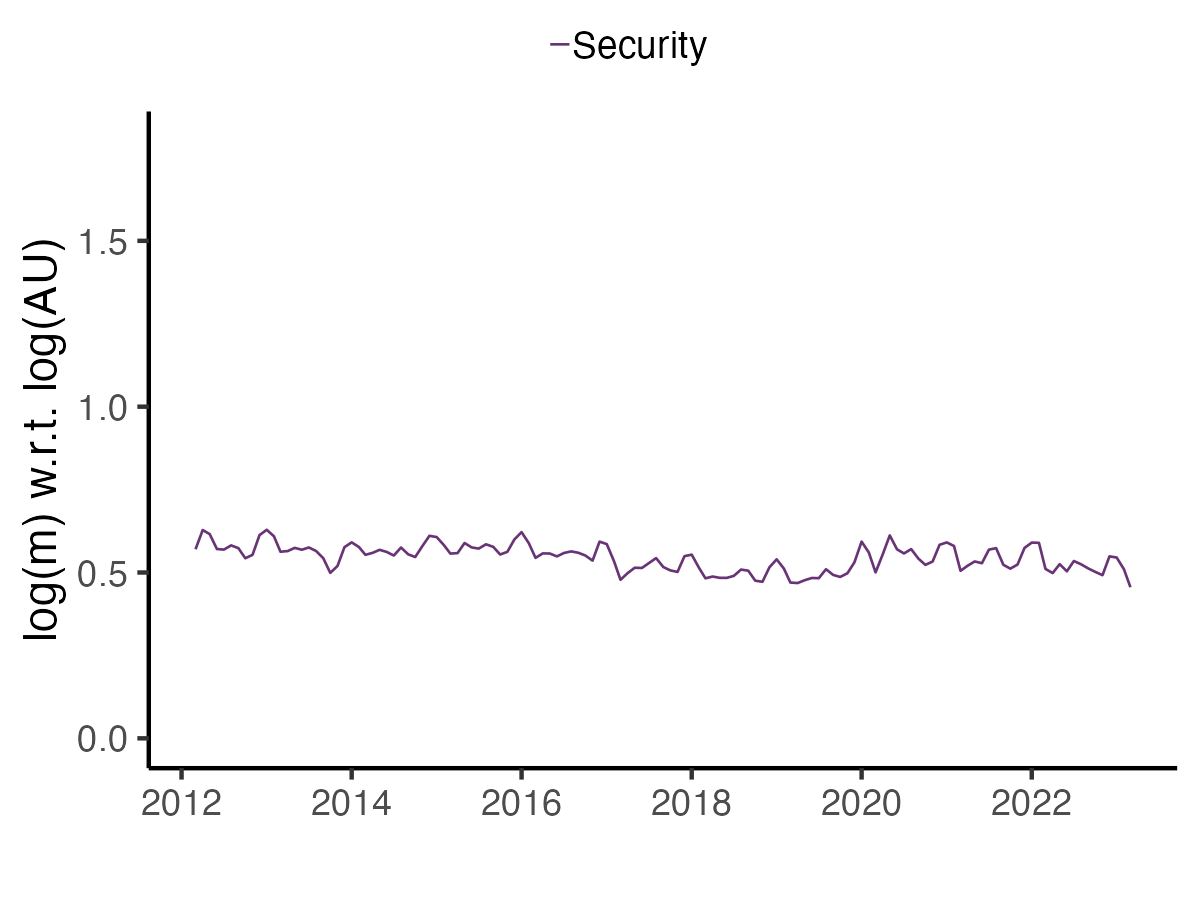}}
  \caption{Month-occupation-level unemployment elasticity 2012--2023}
  \label{fg:month_part_and_full_time_elasticity_unemployed_month_aggregate_job_category_results}
  \end{center}
  \footnotesize
  %Note:
\end{figure}

\begin{figure}[!ht]
  \begin{center}
  \subfloat[Managerial]{\includegraphics[width = 0.30\textwidth]
  {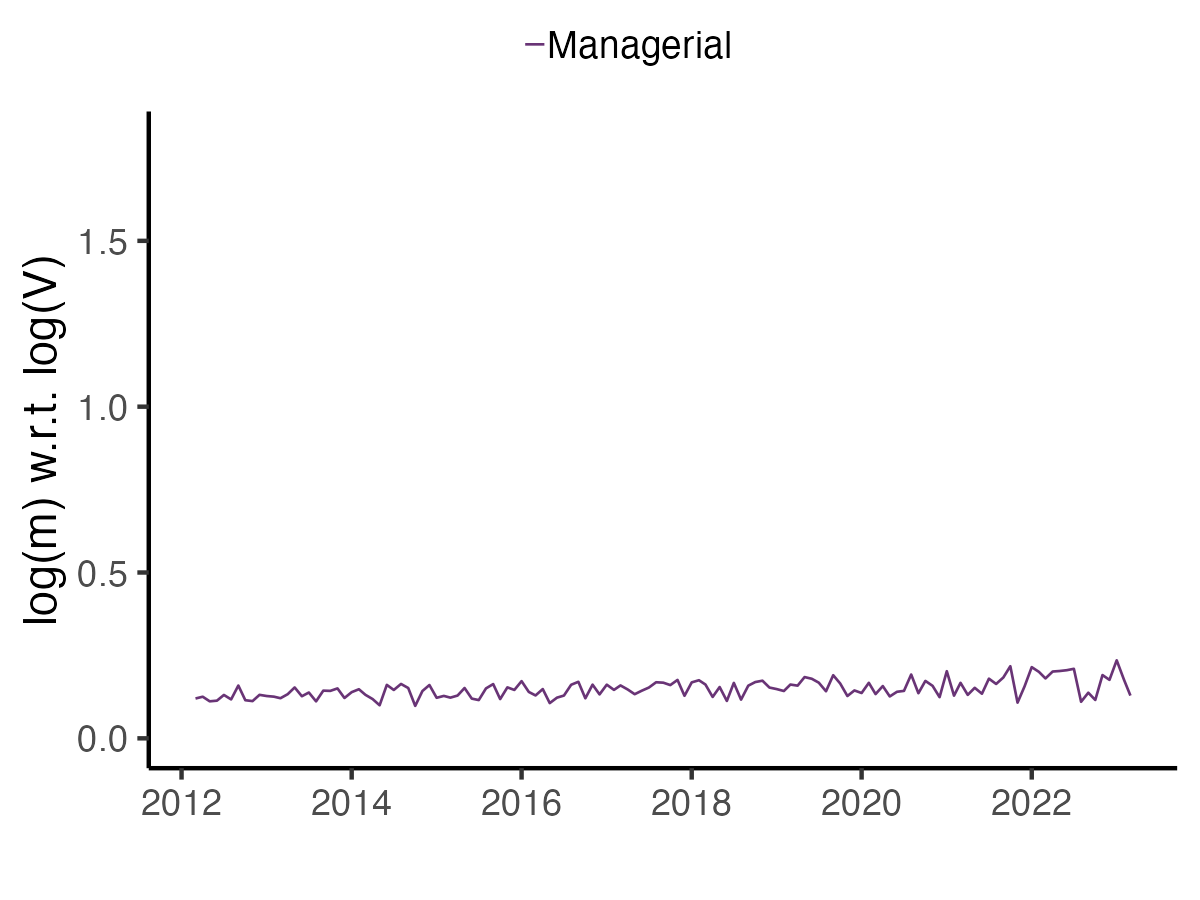}}
  \subfloat[Professional and Technical]{\includegraphics[width = 0.30\textwidth]
  {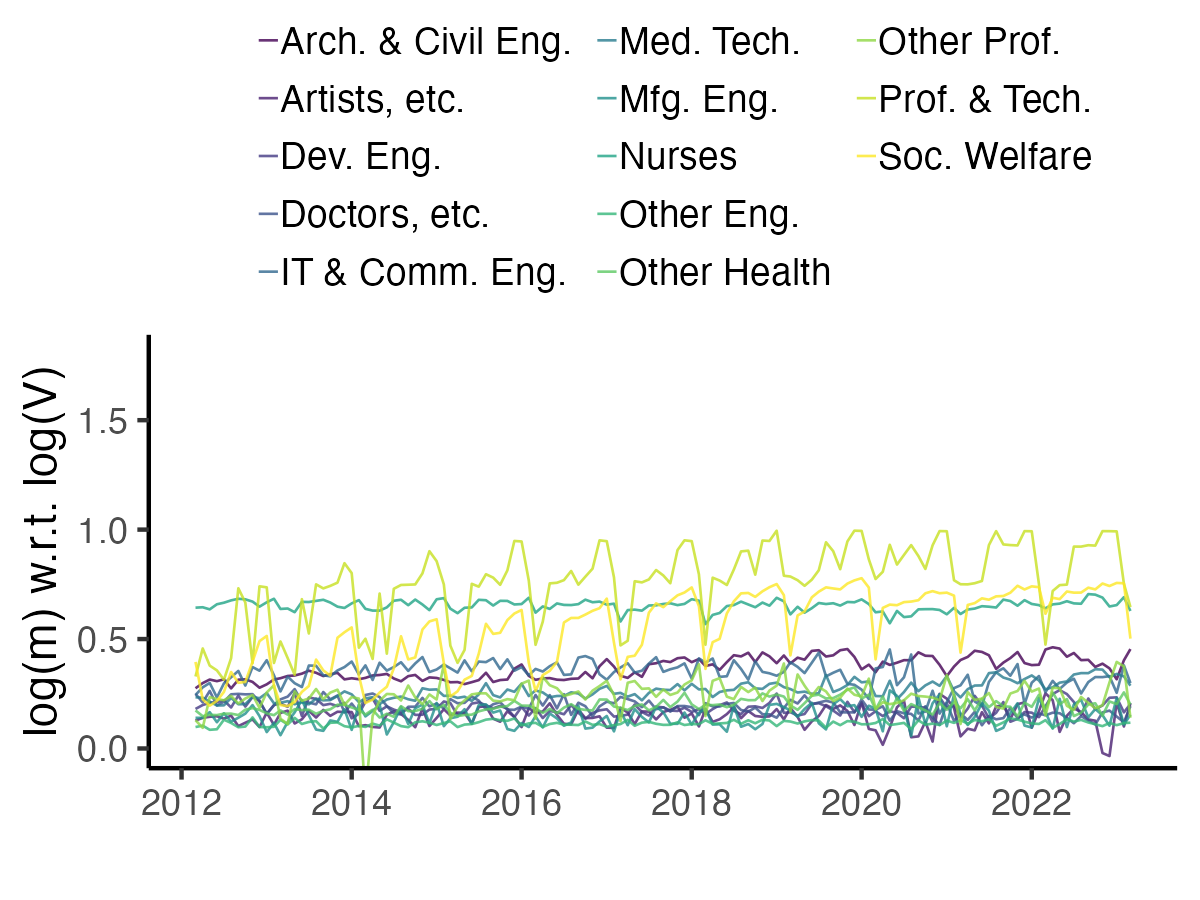}}
  \subfloat[Clerical]{\includegraphics[width = 0.30\textwidth]
  {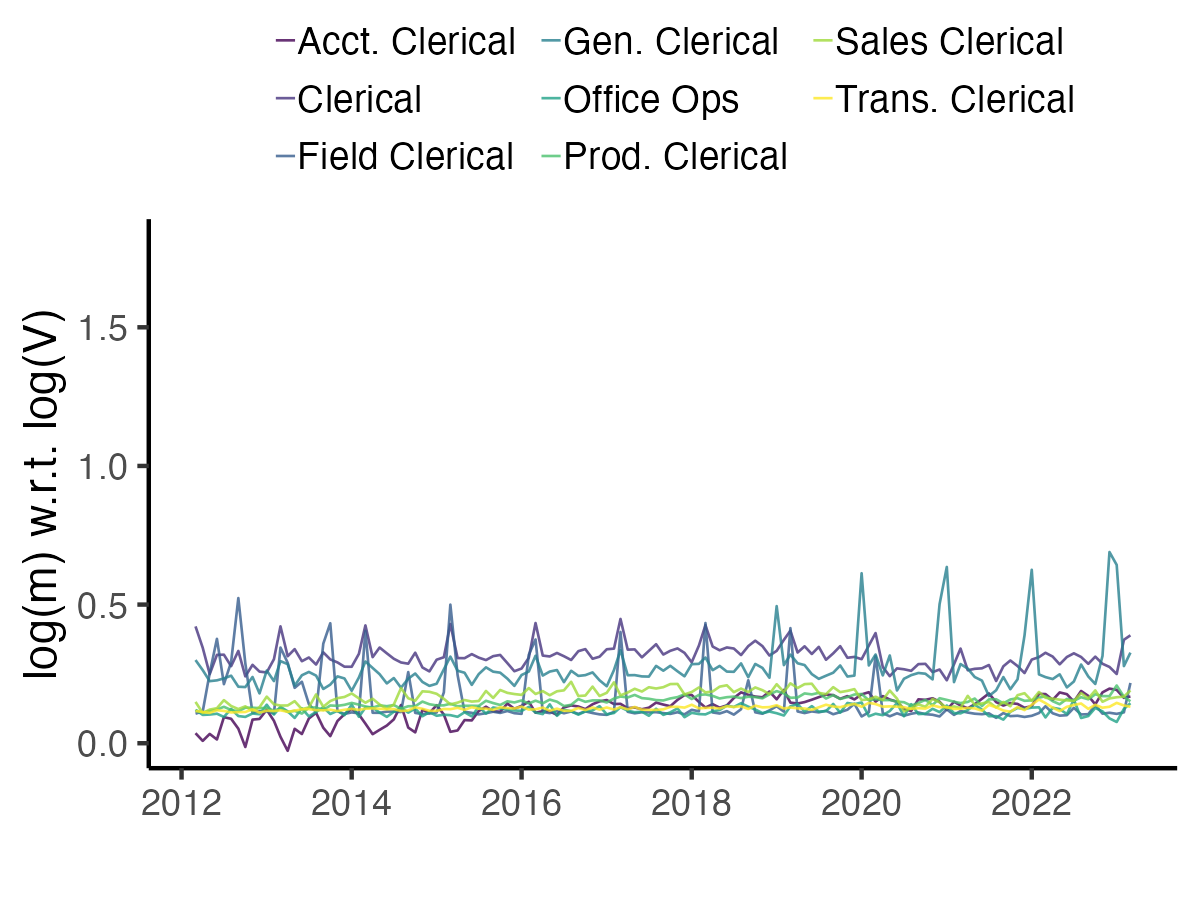}}\\
  \subfloat[Sales]{\includegraphics[width = 0.30\textwidth]
  {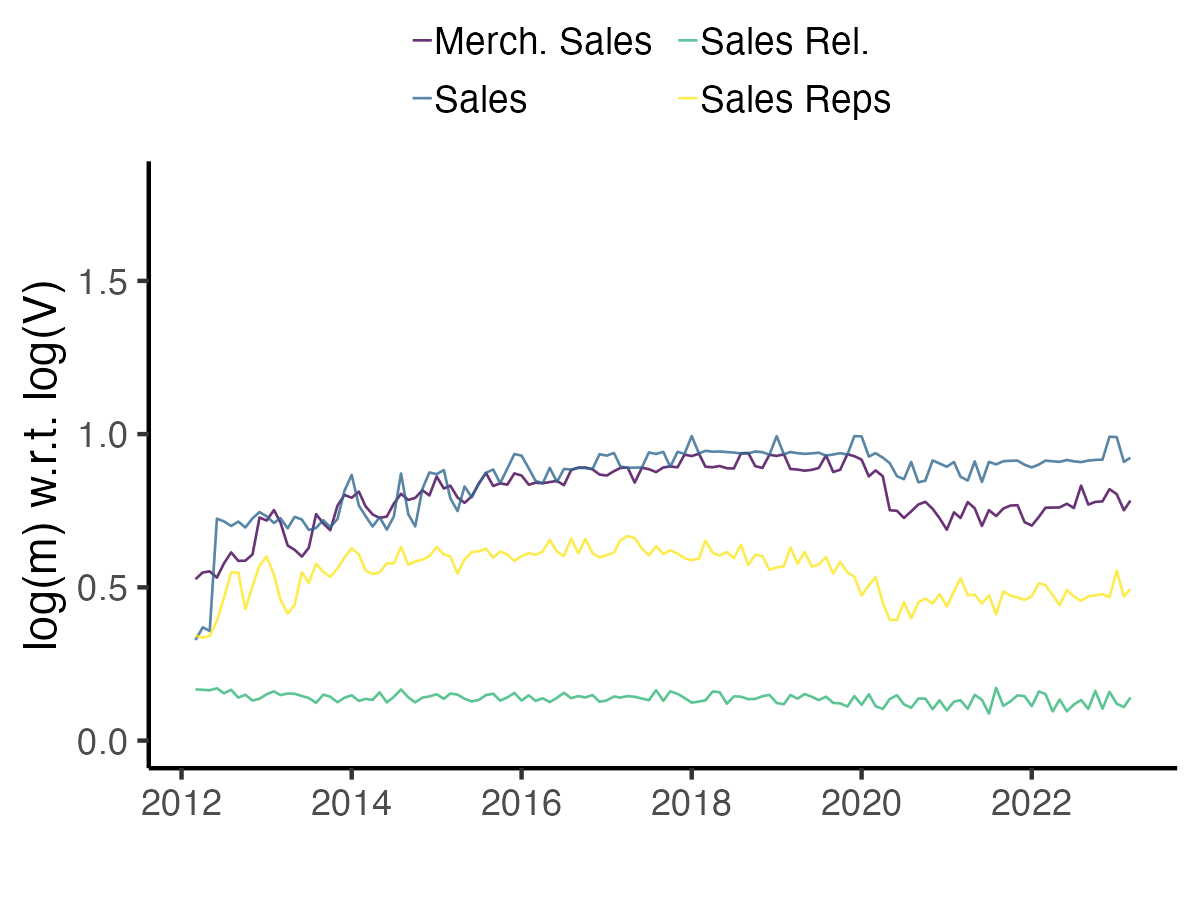}}
  \subfloat[Service]{\includegraphics[width = 0.30\textwidth]
  {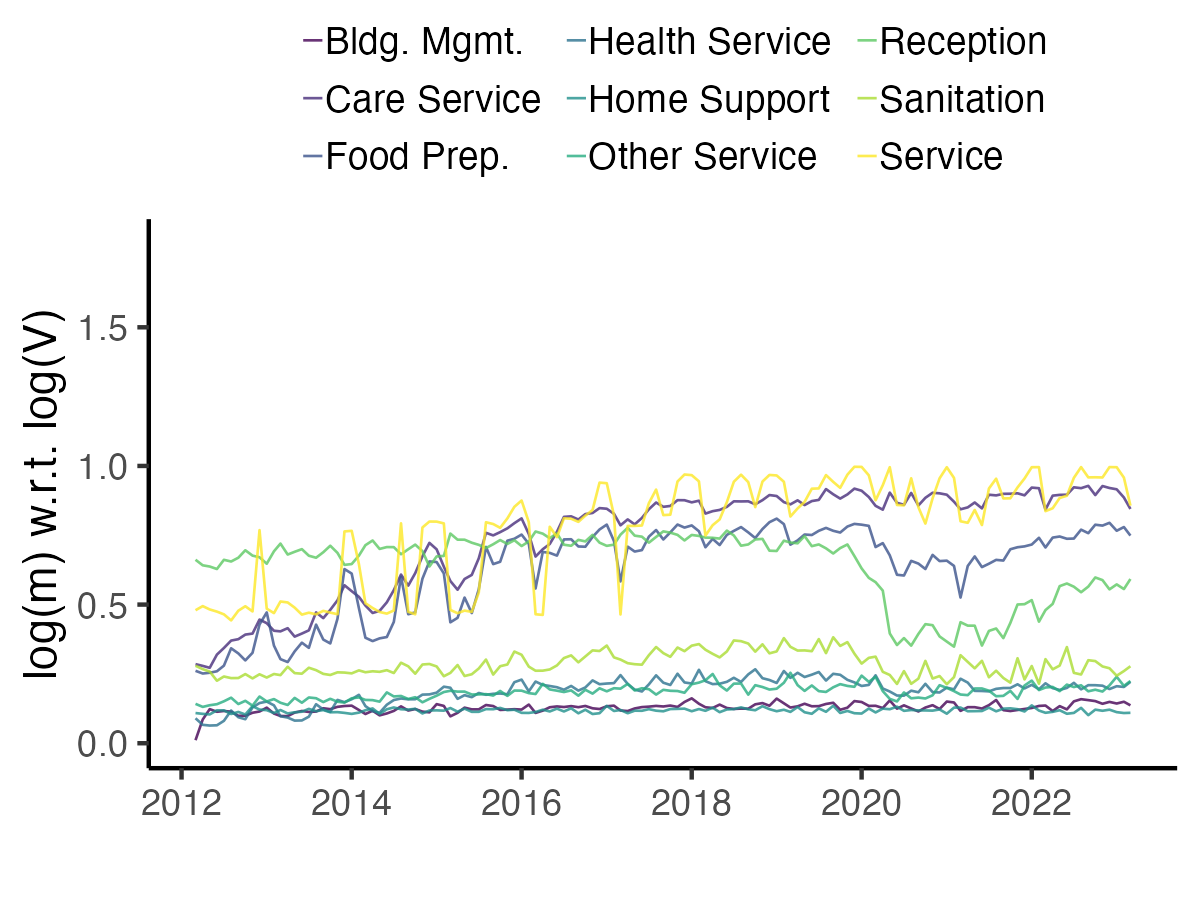}}
  \subfloat[Agriculture Forestry and Fishing]{\includegraphics[width = 0.30\textwidth]
  {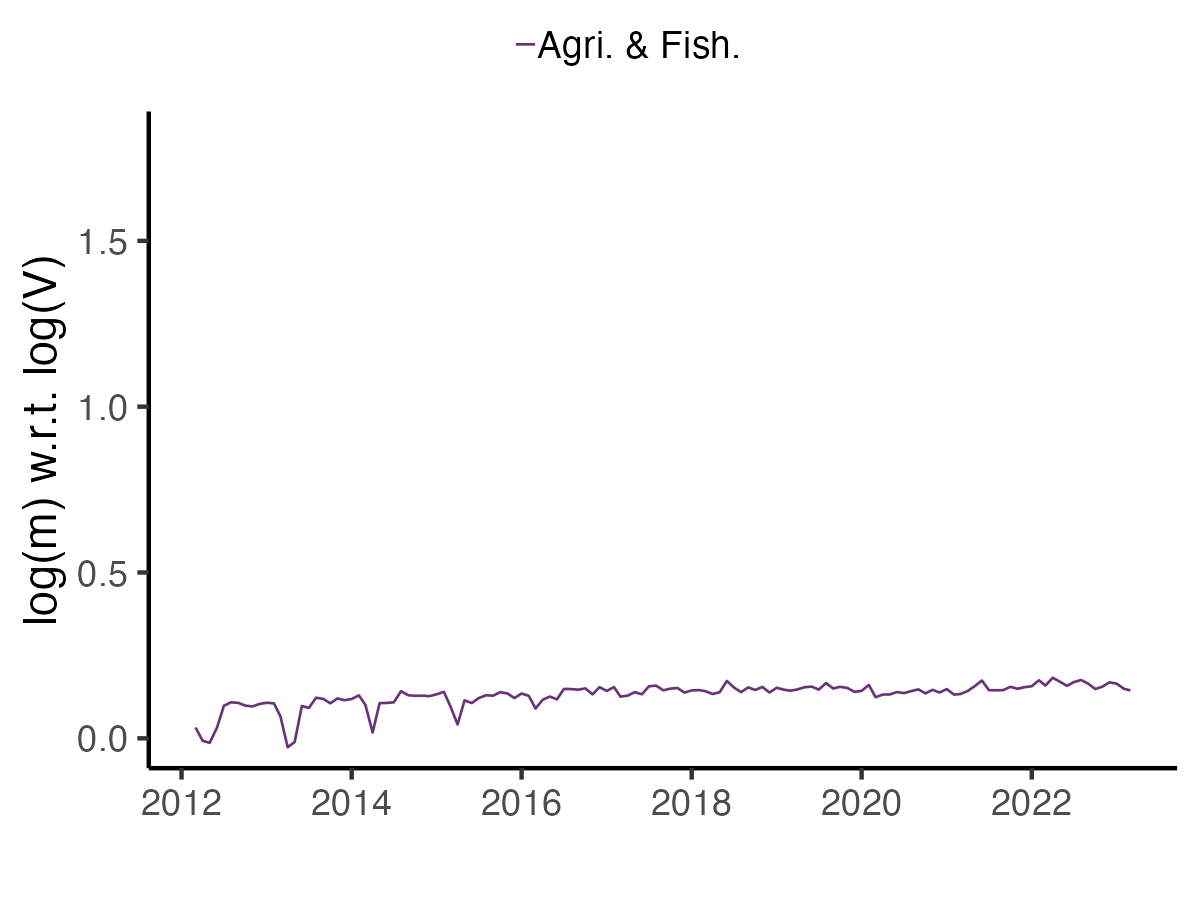}}\\
  \subfloat[Production]{\includegraphics[width = 0.30\textwidth]
  {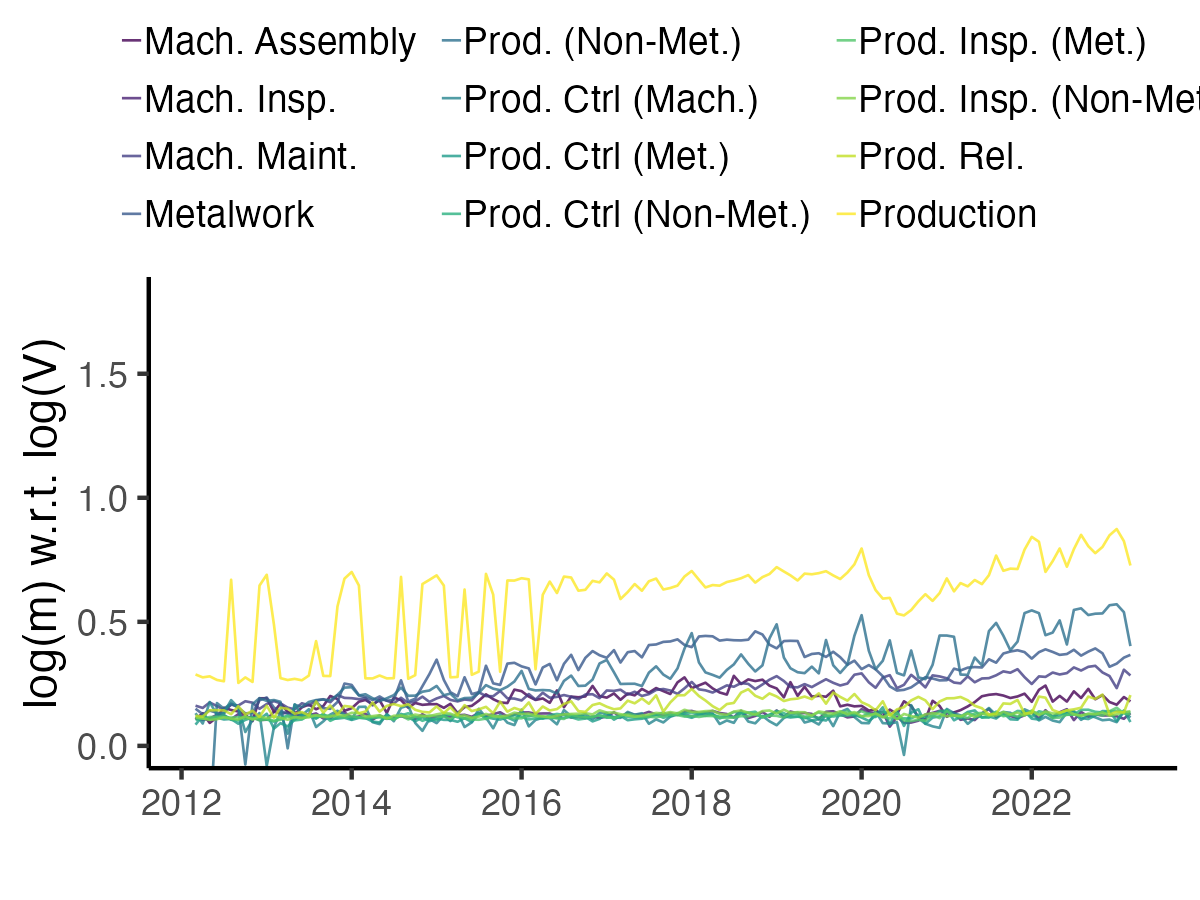}}
  \subfloat[Transportation and Machine Operation]{\includegraphics[width = 0.30\textwidth]
  {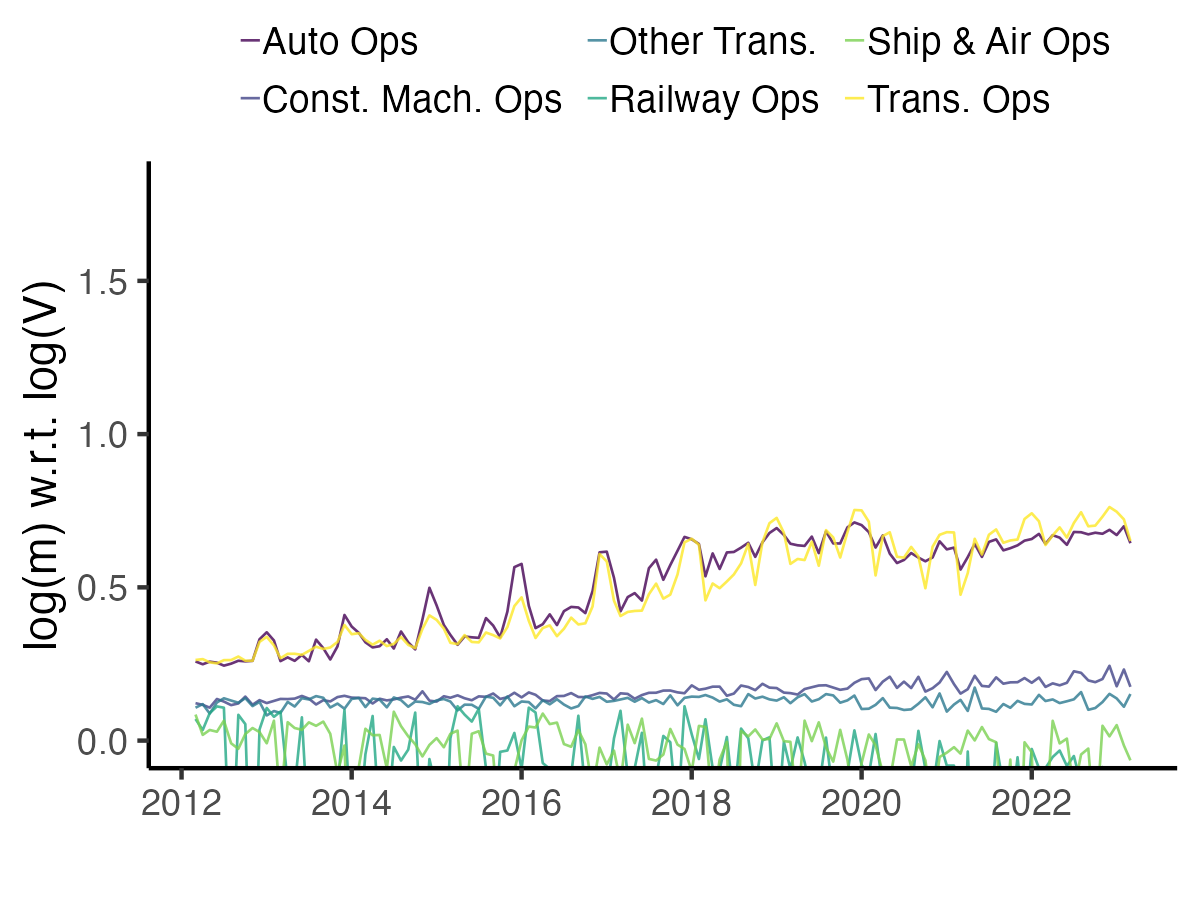}}
  \subfloat[Construction and Mining]{\includegraphics[width = 0.30\textwidth]
  {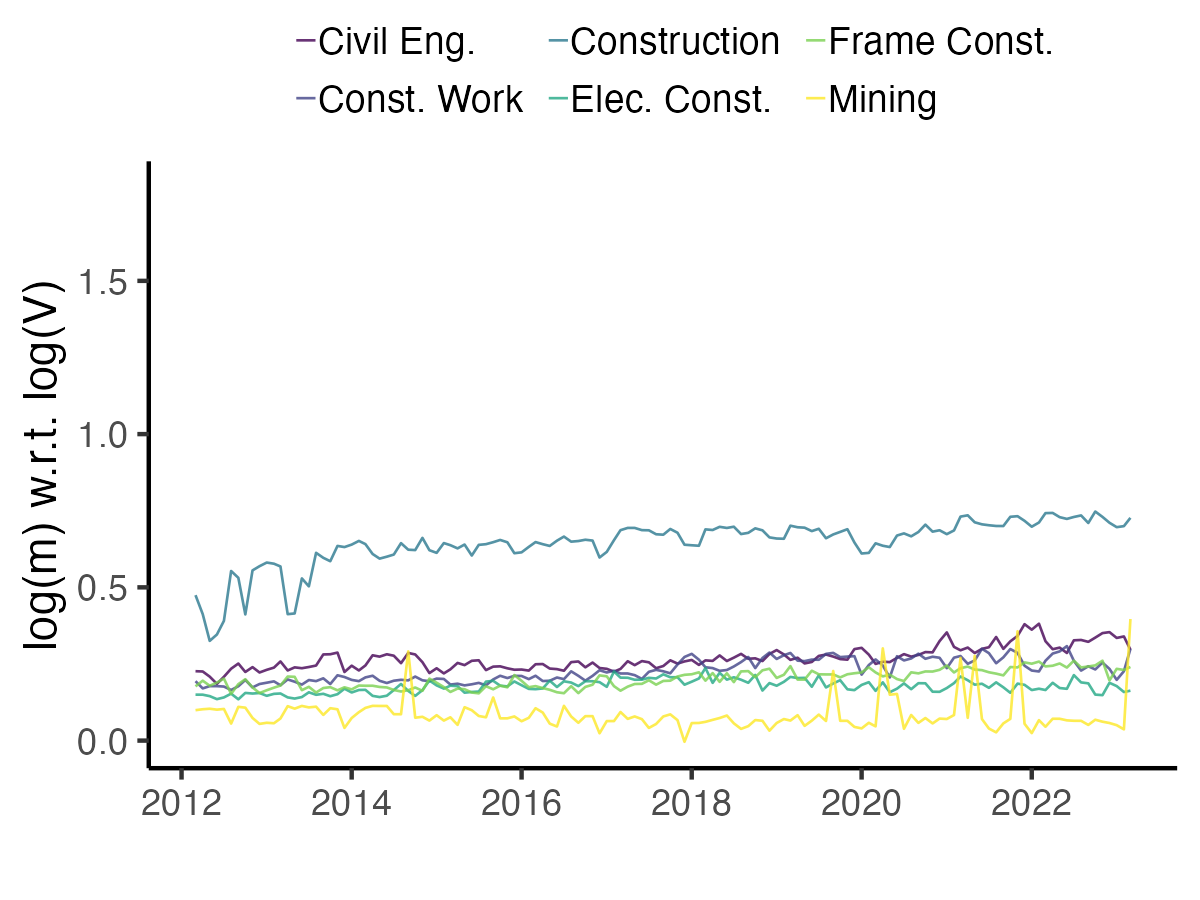}}\\
  \subfloat[Transportation Cleaning and Packaging]{\includegraphics[width = 0.30\textwidth]
  {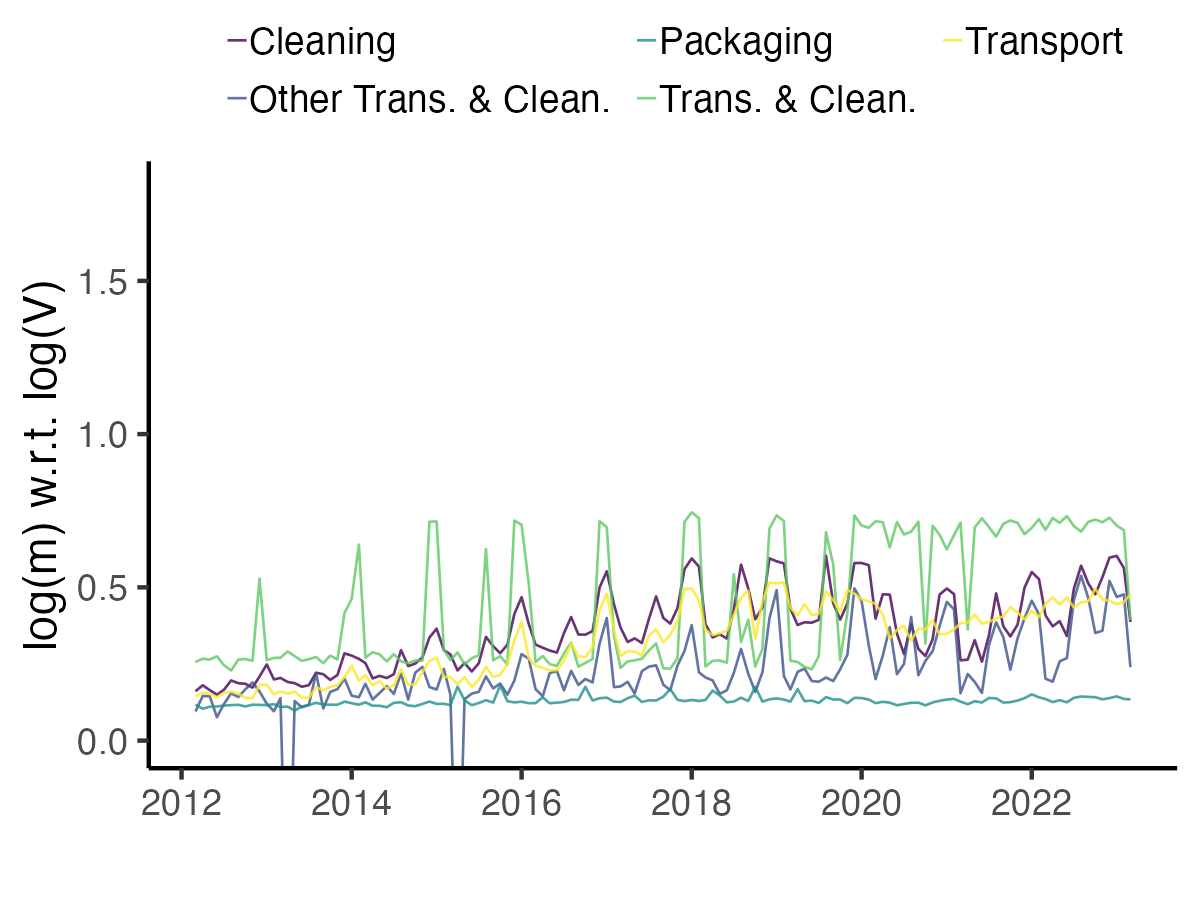}}
  \subfloat[Security]{\includegraphics[width = 0.30\textwidth]
  {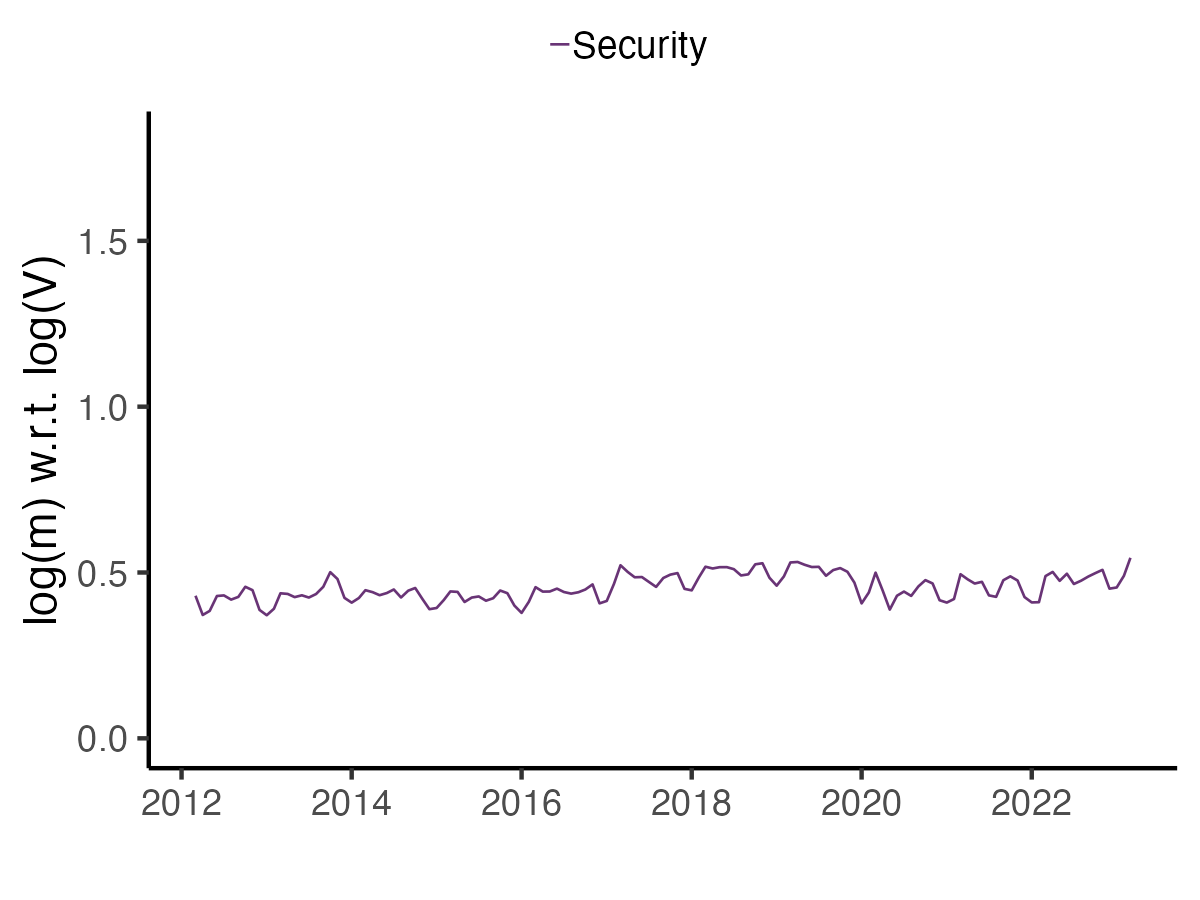}}
  \caption{Month-occupation-level vacancy elasticity 2012--2023}
  \label{fg:month_part_and_full_time_elasticity_vacancy_month_aggregate_job_category_results}
  \end{center}
  \footnotesize
  %Note:
\end{figure}

Figures \ref{fg:month_part_and_full_time_elasticity_unemployed_month_aggregate_job_category_results} and \ref{fg:month_part_and_full_time_elasticity_vacancy_month_aggregate_job_category_results} illustrate unemployment and vacancy elasticities in each occupation in 2012--2023.
As in the prefecture-level results, there is substantial heterogeneity across job categories.
The unemployment elasticity fluctuates with a range generally between 0.2 and 0.9 across job categories, and the vacancy elasticity is correspondingly between 0.1 and 0.8, with greater dispersion than the prefecture-level results.
This indicates that matching heterogeneity depends on the definition of markets and matters more at the occupational level than at the geographic level.

\subsection{Geographic, Industry, and Occupational Mismatch in 2012--2024}

\begin{figure}[!ht]
  \begin{center}
  \subfloat[Across regions (LFS + SET)]{\includegraphics[width = 0.48\textwidth]
  {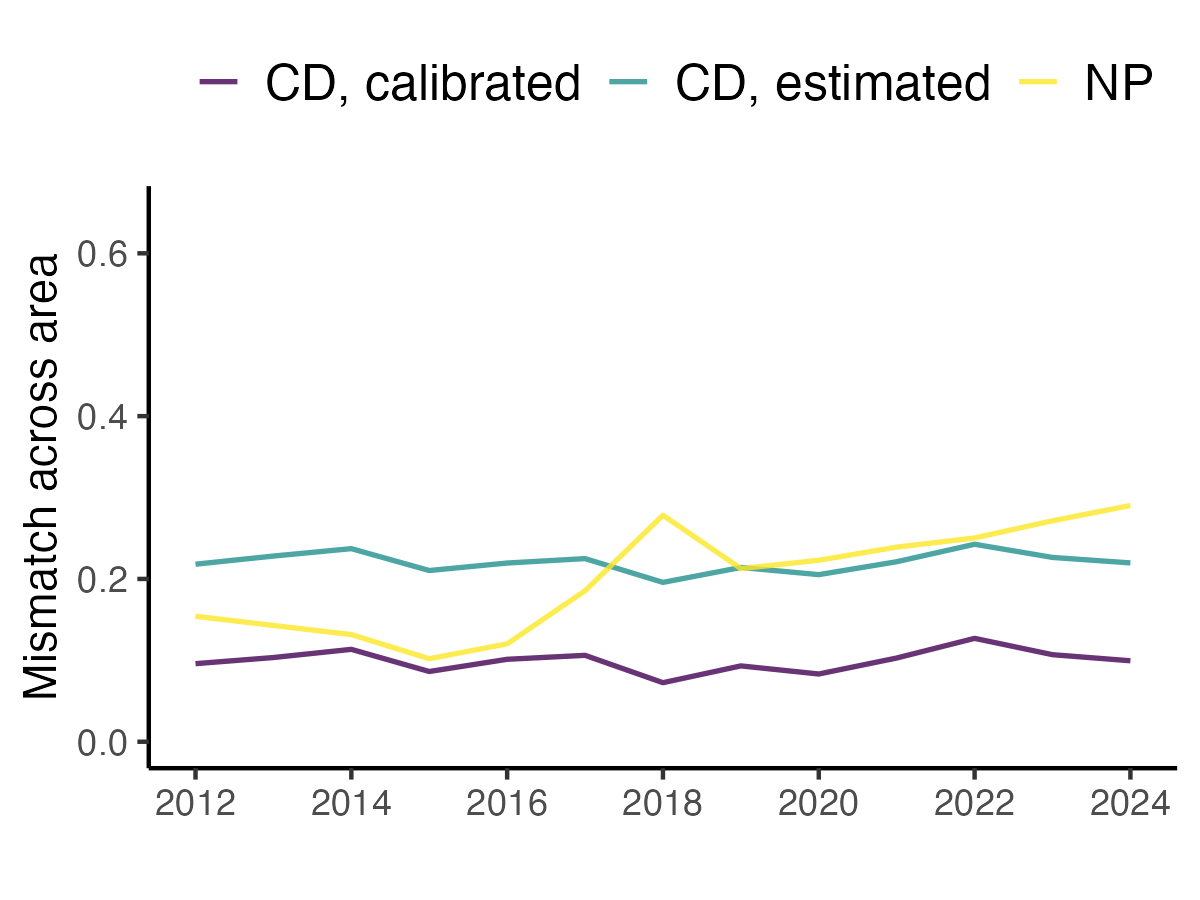}}
  \subfloat[Across prefectures (Hello Work)]{\includegraphics[width = 0.48\textwidth]
  {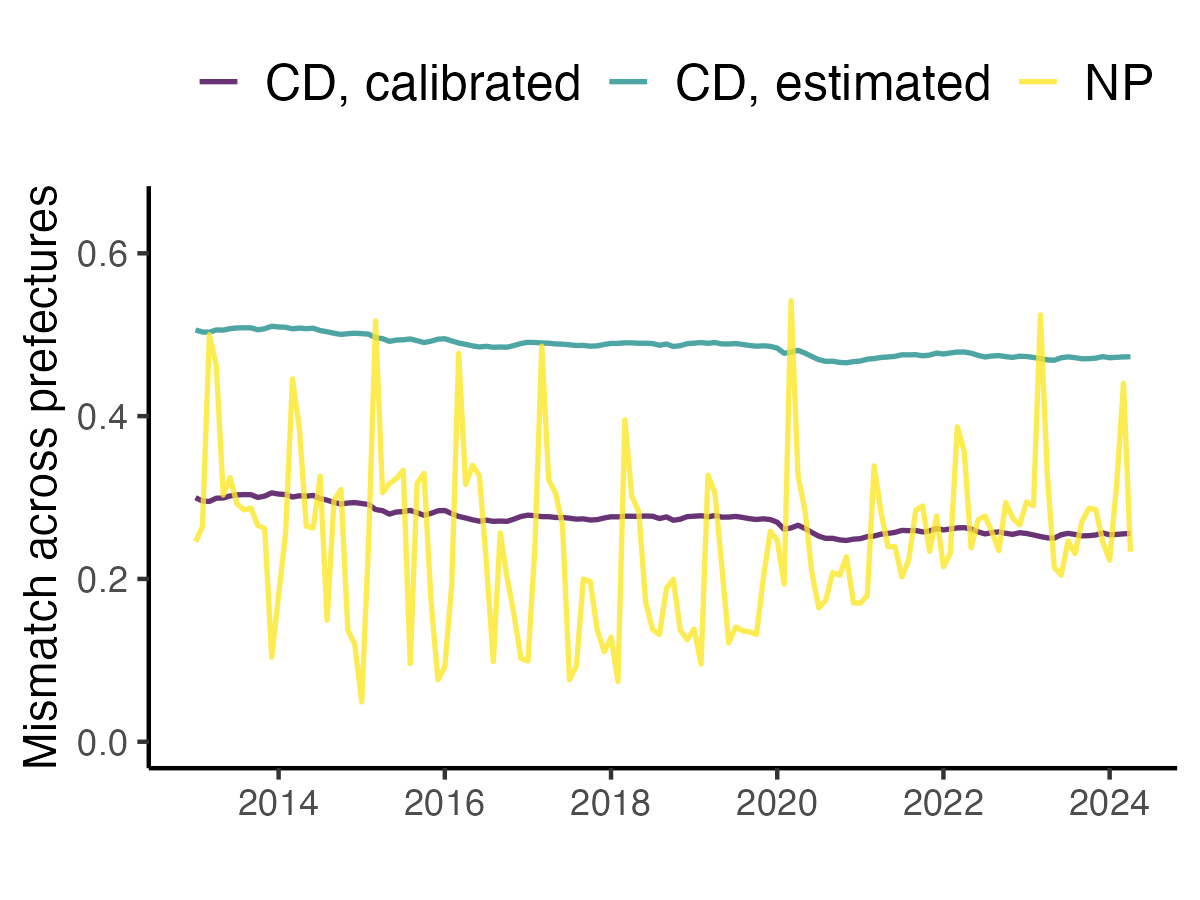}}\\\vspace{0.05\textwidth}
  \subfloat[Across industries (LFS + SET)]{\includegraphics[width = 0.48\textwidth]
  {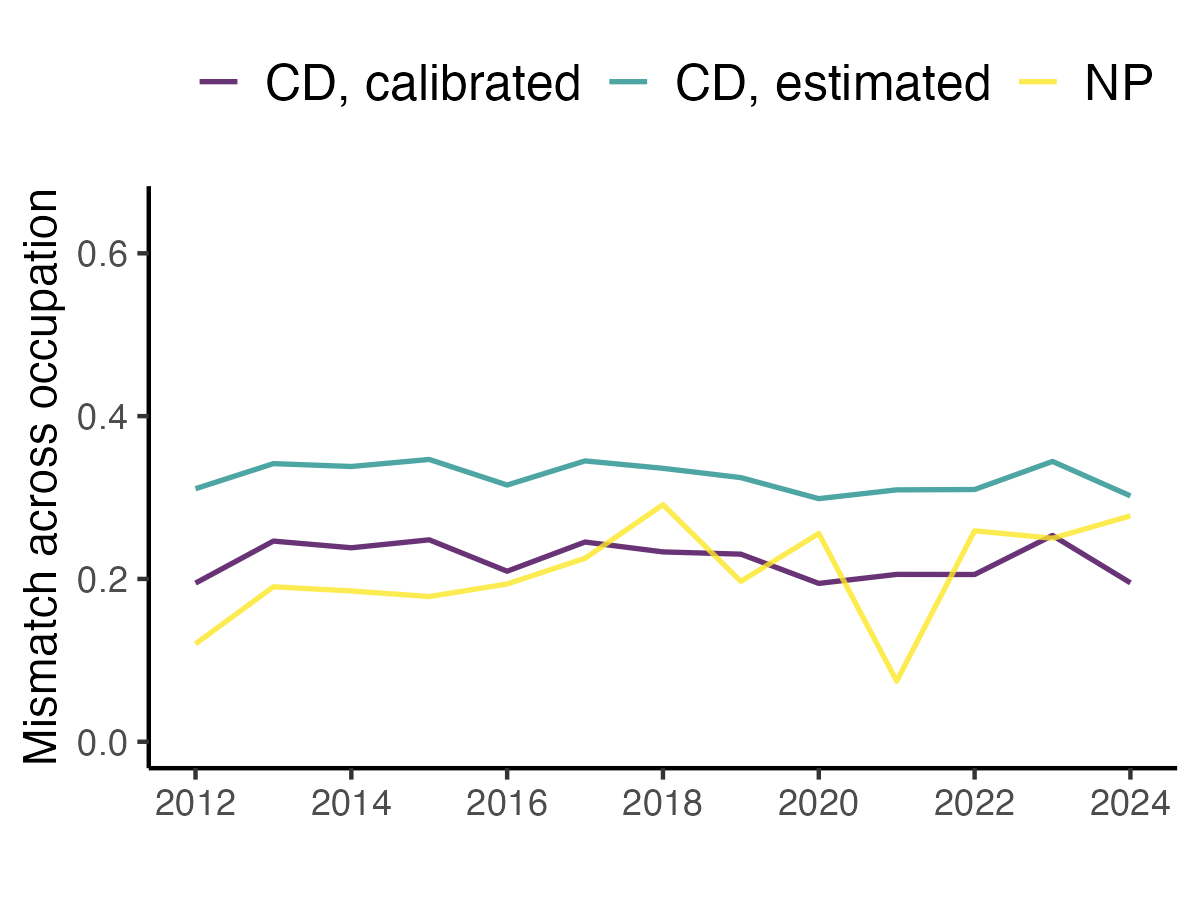}}
  \subfloat[Across job categories (Hello Work)]{\includegraphics[width = 0.48\textwidth]
  {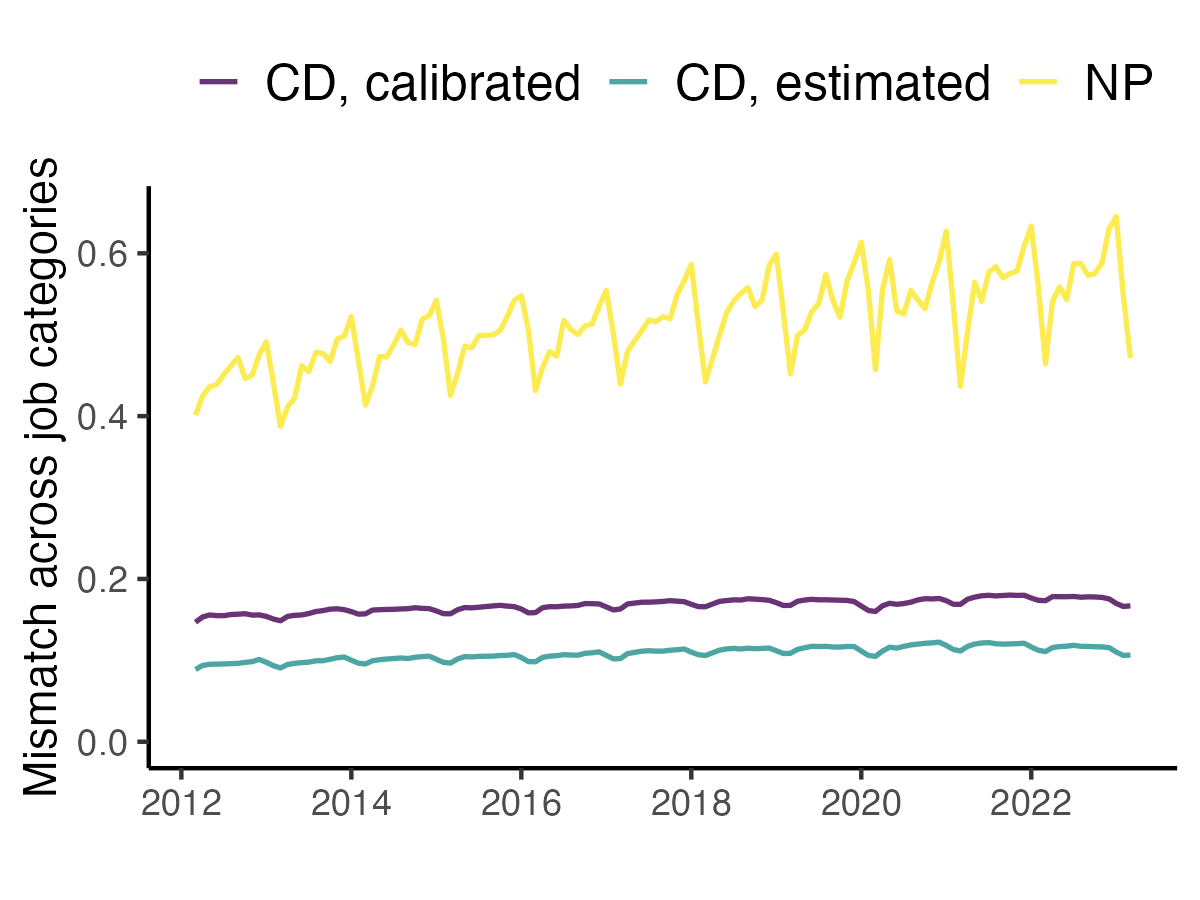}}
  \caption{Estimated nonparametric mismatch index}
  \label{fg:mismatch_part_and_full_time_monthly_prefecture}
  \end{center}
  \footnotesize
  Note: CD, estimated = Cobb-Douglas mismatch index with estimated vacancy elasticity, as in \cite{csahin2014mismatch}, \cite{kawata2019}, and \cite{shibata2020labor}. CD, calibrated = Cobb-Douglas mismatch index with calibrated vacancy elasticity set to 0.5 \citep{higashi2025did,kawata2019} for geographic panels (a, b) and 0.3 \citep{shibata2020labor} for industry and occupational panels (c, d) for comparison with existing studies. NP = proposed nonparametric mismatch index. Panels (a) and (c) (LFS + SET) cover 13 regional blocks and 14 industries respectively at the year level, 2012--2024. Panels (b) and (d) (Hello Work) cover 47 prefectures (2013--2024) and 67 job categories (2012--2023) at the month level. Mining and Electricity/Gas/Water are dropped from panel (c) due to insufficient $U$ variation.
\end{figure}

I compute the estimated nonparametric mismatch index and compare it with Cobb-Douglas (CD) benchmarks constructed using estimated and calibrated vacancy elasticities.\footnote{See Appendix \ref{sec:cobb_douglas_mismatch_index} for the Cobb-Douglas matching function estimation by OLS with fixed effects.} Figure \ref{fg:mismatch_part_and_full_time_monthly_prefecture} reports four panels: (a) and (c) use the nationally representative LFS + SET data at the region-year and industry-year levels (2012--2024), while (b) and (d) use the Hello Work data at the prefecture-month and job-category-month levels.

\paragraph{Nationally Representative Aggregate Data.}
Panel (a) reports the region-level mismatch from LFS + SET. The NP index ranges between approximately 0.10 and 0.29 with a mean around 0.20, exhibiting a gradual upward trend through 2017 and a partial inward shift toward recent years. The CD-estimated and CD-calibrated indices lie below the NP index, consistent with the downward bias documented for the Cobb-Douglas approximation when matching efficiency and elasticity are time-varying \citep{csahin2014mismatch}. Panel (c) reports the industry-level mismatch. The NP index ranges from about 0.07 to 0.29 with a mean around 0.21, again exceeding the CD counterparts and showing a similar mild upward drift. Both panels confirm that, even at the more representative LFS + SET aggregate, allowing for time-varying efficiency and elasticity raises the measured mismatch relative to the Cobb-Douglas benchmarks reported in \cite{shibata2020labor} and \cite{higashi2025did}.

\paragraph{Hello Work Data.}
Panel (b) presents the nonparametric mismatch index across 47 prefectures from 2013 to 2024 using Hello Work month-level data. The NP index, capturing the deviation of actual hires from the social planner's optimal level, ranges from about 0.05 to 0.54 with a mean around 0.25, significant seasonal fluctuations, and an overall mild upward trend toward 2024. It also displays a cyclical pattern, indicating the influence of seasonal factors or regional economic cycles on labor market dynamics. Notably, the mismatch level across prefectures is consistently higher than the Cobb-Douglas mismatch indices reported by \cite{shibata2020labor} (0.05--0.1) and \cite{higashi2025did} (0.15--0.25). While I can partially replicate their findings using their calibrated matching elasticity values, the results highlight that the Cobb-Douglas mismatch index underestimates fluctuations and is biased downward due to its inability to account for time-varying matching efficiency and elasticity across prefectures. Comparing panels (a) and (b), the LFS + SET region-level NP index lies in a similar range to the Hello Work prefecture-level NP index, suggesting that prefecture-level mismatch dynamics in the Hello Work platform are broadly representative of the wider Japanese labor market at the regional aggregation level.

Panel (d) illustrates the mismatch index across occupations during the same period using Hello Work month-level data. Unlike the prefecture-based index, it shows a steady upward trend with minimal seasonal variation, ranging from about 0.39 to 0.65 and averaging 0.52---substantially higher than the levels reported in \cite{shibata2020labor} and \cite{higashi2025did}. This discrepancy underscores the limitations of the Cobb-Douglas approach, which underestimates mismatch due to its exclusion of time-varying efficiency and elasticity across occupations. Even when I use their calibrated matching elasticity values, I cannot reproduce their reported levels. Although the fixed effects estimated from my more recent data differ from the old estimates in existing studies, panel (d) suggests that incorporating time-varying efficiency and elasticity plays a critical role in mitigating the downward bias of the Cobb-Douglas mismatch index, particularly at higher levels of disaggregation.\footnote{This property about the levels of disaggregation is further discussed in online Appendix A.A5 of \cite{csahin2014mismatch}.} Comparing panels (c) and (d), the Hello Work occupation-level NP index reaches considerably higher values (mean about 0.52) than the LFS + SET industry-level NP index (mean about 0.21), reflecting both the finer Hello Work occupation classification (67 job categories vs 14 industries) and the platform-specific composition of vacancies and job seekers.

The steady rise in the Hello Work occupational mismatch index in panel (d) reflects a growing misalignment between the skills provided by job seekers and those demanded by employers in the Hello Work segment. This trend likely stems from evolving industry demands, technological advancements, and a lag in workforce adaptation to new skill requirements. The sharper increase post-2020 may further indicate the exacerbation of skill mismatches due to the COVID-19 pandemic, driven by shifts such as accelerated digital transformation and uneven sectoral impacts. The corresponding LFS + SET industry-level index in panel (c) shows a milder upward trend over the same period, suggesting that the steepest occupation-level rises are concentrated in the Hello Work platform and may be attenuated when aggregating across the broader, nationally representative industry classification.

\section{Conclusion}

This study highlights significant biases in the traditional Cobb-Douglas function when nonadditive, time-varying matching efficiency is misspecified, and evaluates the finite-sample performance of the nonparametric estimation method proposed by \cite{lange2020beyond}. By extending the mismatch index of \cite{csahin2014mismatch} to a nonparametric framework and developing a computational methodology, I analyze Japan's labor market using two complementary datasets: a nationally representative aggregate panel constructed from the Labour Force Survey and the Survey on Employment Trends (LFS + SET, 1972--2024) and the Hello Work platform statistics (1972--2023 for the aggregate monthly estimation sample). Both datasets reveal a declining trend in matching efficiency, consistent with falling job-finding and vacancy-filling rates. Estimated unemployment elasticities are around 0.5 to 0.9 in the Hello Work data and lower in the LFS + SET country-year aggregate, with the vacancy elasticity correspondingly lower in the Hello Work data and higher in the LFS + SET aggregate. The results show that sectoral or occupational mismatch is more pronounced than geographic mismatch in both datasets, with the traditional Cobb-Douglas mismatch index significantly underestimating the extent of mismatch. The cross-dataset comparison further indicates that platform-level (Hello Work) matching dynamics are qualitatively informative about wider Japanese labor market trends, but level differences---driven in part by Hello Work's partial coverage and finer cross-sectional resolution (47 prefectures, 67 job categories) versus the broader LFS + SET aggregation (13 regional blocks, 14 industries)---warrant caution when extrapolating Hello Work-based magnitudes to the aggregate. These findings underscore the importance of adopting nonparametric approaches to better capture time-varying matching efficiency, elasticity, and mismatch, as well as accounting for the cross-sectional resolution at which mismatch is measured.

Future research could extend this analysis to other private platforms and to individual-level behavior. Recent studies of job-matching intermediaries \citep{kambayashi2025decomposing}, on-site spot gig-work platforms \citep{kanayama2026justminimumwagehikes,cullen2021outsourcing}, off-site gig-work platforms \citep{stanton2025benefits}, and scouting platforms \citep{roussille2025bidding} point to settings in which richer microdata could deepen our understanding of labor market dynamics. Another advantage of the proposed computational framework is that the equilibrium constraints can be modified flexibly without changing the nonparametric estimation of the matching function. This feature makes it possible to incorporate richer market heterogeneity, such as productivity differences, market-specific job destruction rates, or policy-relevant reallocation constraints, once sufficiently detailed data become available.

\bibliographystyle{ecca}
\bibliography{matching_function}

\newpage

\appendix
\section{Additional Results}\label{sec:additional_figure}

\subsection{Residual Plots}

\begin{figure}[!ht]
  \begin{center}
  \subfloat[Aggregate (Hello Work)]{\includegraphics[width = 0.40\textwidth]
  {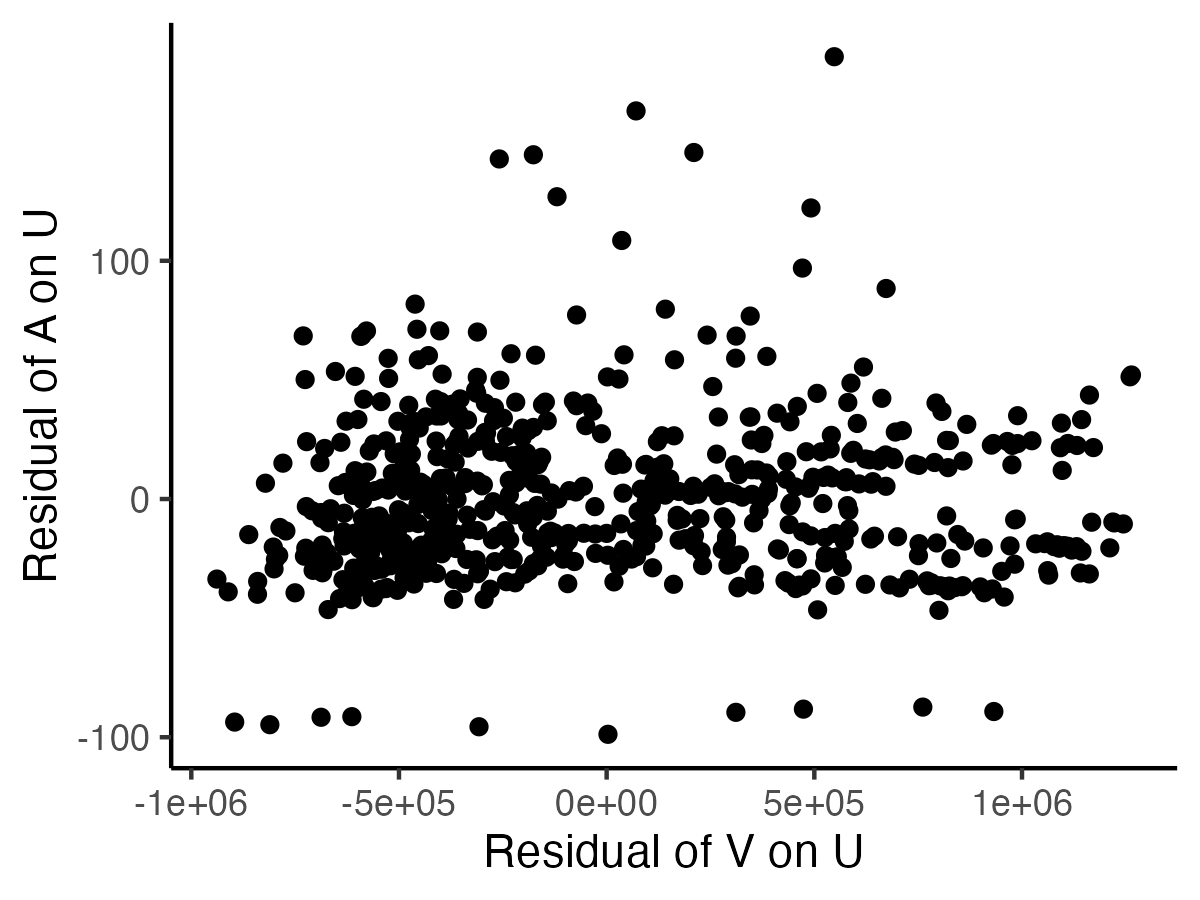}}
  \subfloat[Full-time (Hello Work)]{\includegraphics[width = 0.40\textwidth]
  {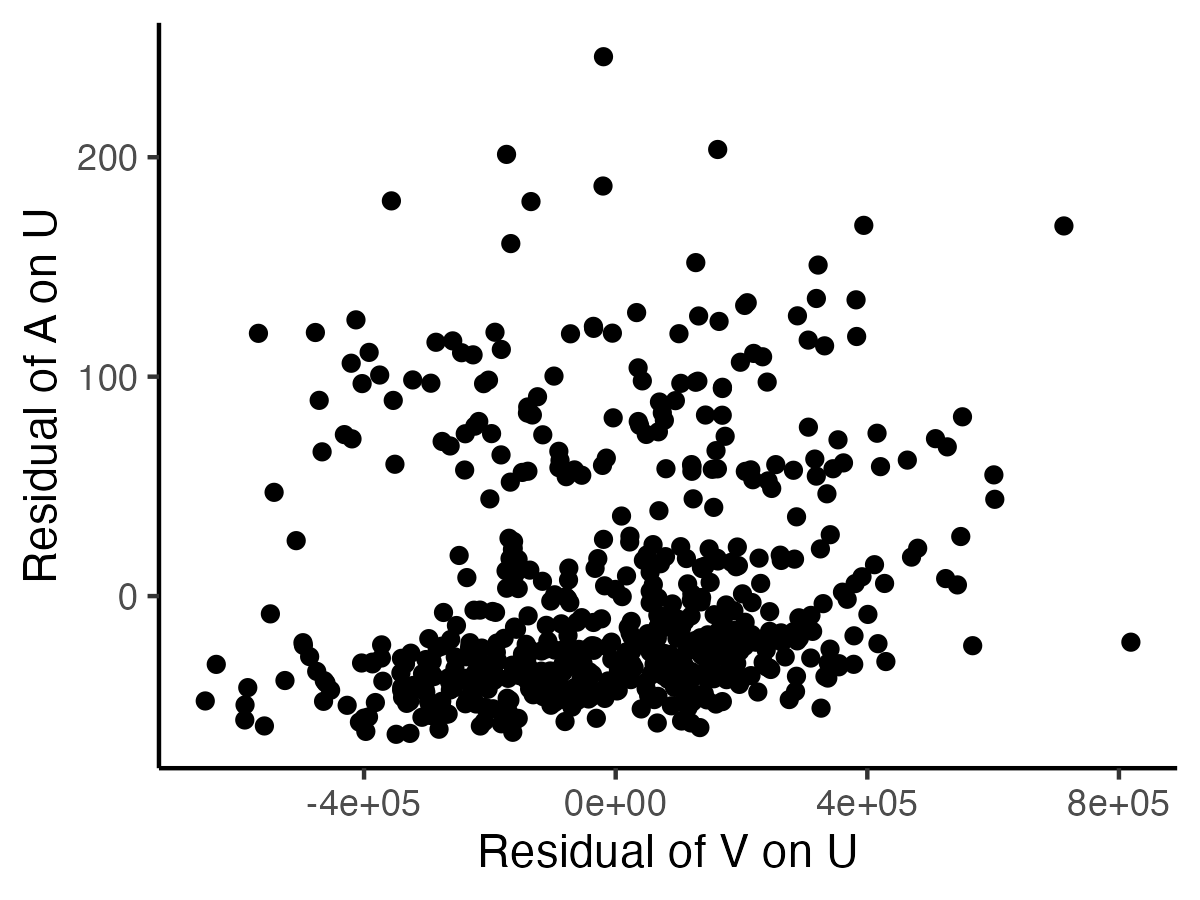}}\\\vspace{0.05\textwidth}
  \subfloat[Part-time (Hello Work)]{\includegraphics[width = 0.40\textwidth]
  {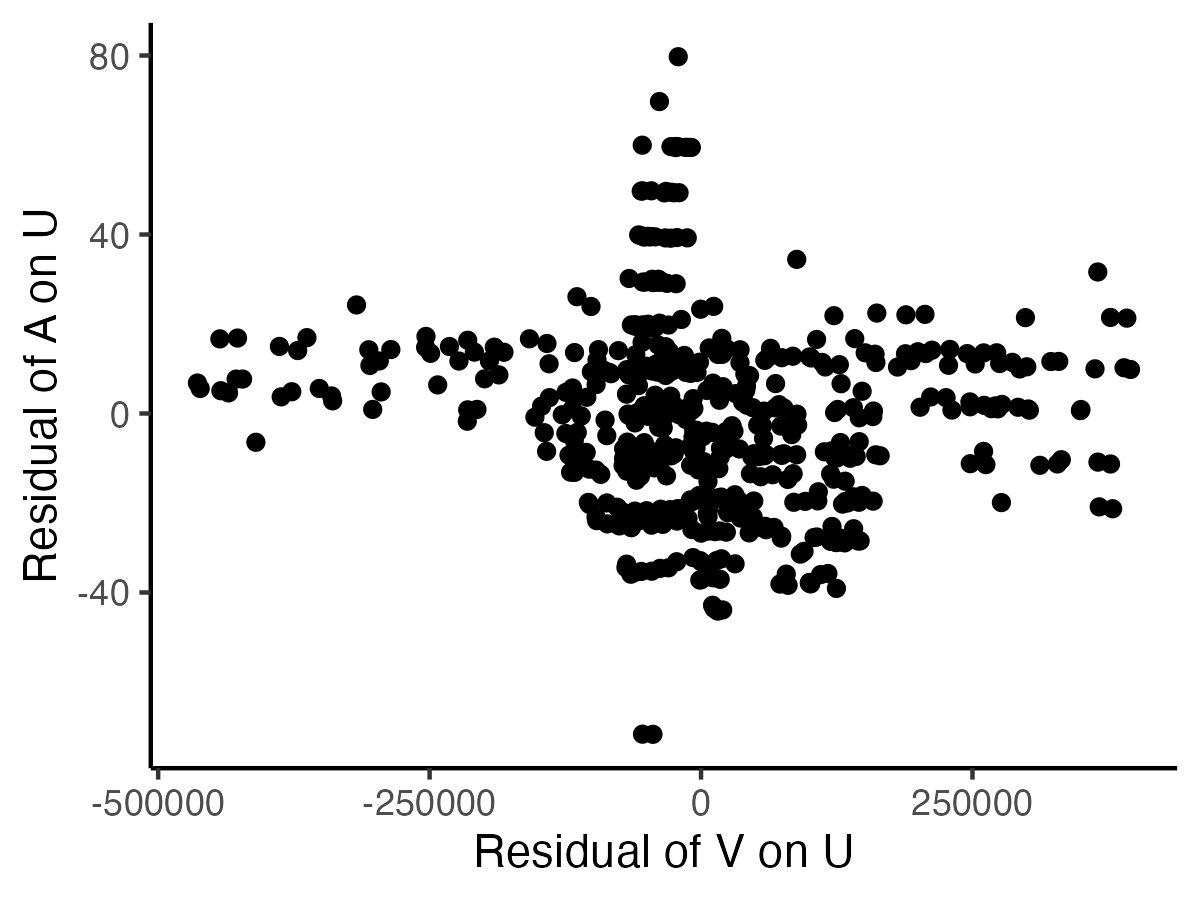}}
  \subfloat[Country-year (LFS + SET)]{\includegraphics[width = 0.40\textwidth]
  {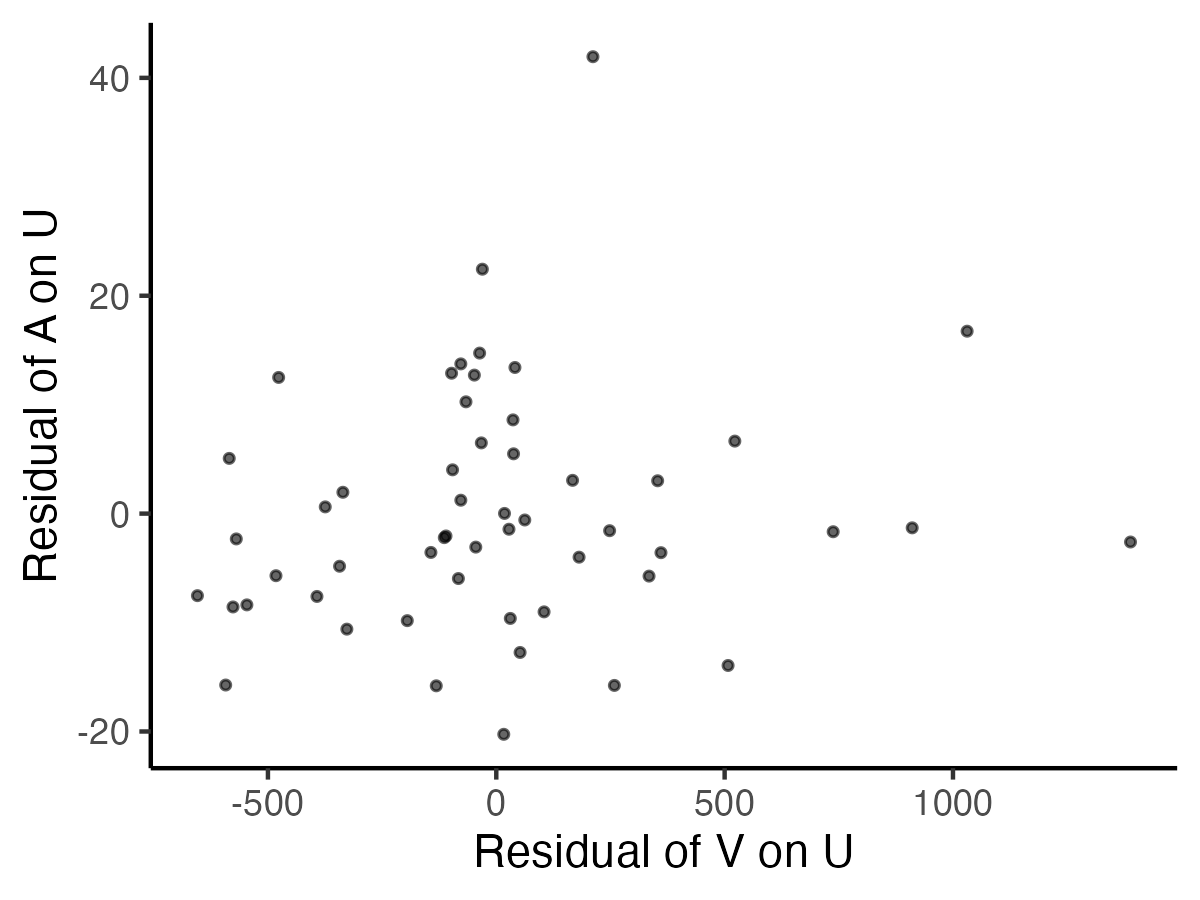}}
  \caption{Residual plot}
  \label{fg:residual_plots}
  \end{center}
  \footnotesize
  Note: Each panel plots the residual of $V$ on $U$ (horizontal axis) against the residual of $A$ on $U$ (vertical axis) to assess the conditional independence assumption $V \perp A \mid U$. Panels (a)--(c) use Hello Work data: (a) full-time + part-time aggregate (1972--2023 monthly), (b) full-time only, (c) part-time only. Panel (d) uses the nationally representative LFS + SET country-year panel (1972--2024 annual). Sample correlations are small in absolute value (Hello Work aggregate $+0.04$; full-time $+0.14$; part-time $-0.14$; LFS + SET country-year $+0.15$), indicating no systematic relationship between vacancies and matching efficiency conditional on unemployment.
\end{figure}

Figures \ref{fg:residual_plots} show the residual plots used to test the conditional independence assumption $V \perp A \mid U$ for both datasets. For Hello Work (panels a--c), the patterns are largely scattered with no systematic positive or negative comovement, and the impact of any visible outliers is limited. For the LFS + SET country-year panel (panel d), the same diagnostic is applied to the nationally representative aggregate; the residuals are similarly scattered, supporting the use of the nonparametric estimator under the independence assumption in both datasets. The effect of the violation of independence therefore appears limited across the prefecture and occupation panels at month level (Hello Work) and at the country-year level (LFS + SET).

\subsubsection{Year-Level Analysis in 1963--2023}
\begin{figure}[!ht]
  \begin{center}
 \subfloat[Unemployed ($U$), Vacancy ($V$), and Tightness ($\frac{V}{U}$)]{\includegraphics[width = 0.42\textwidth]
  {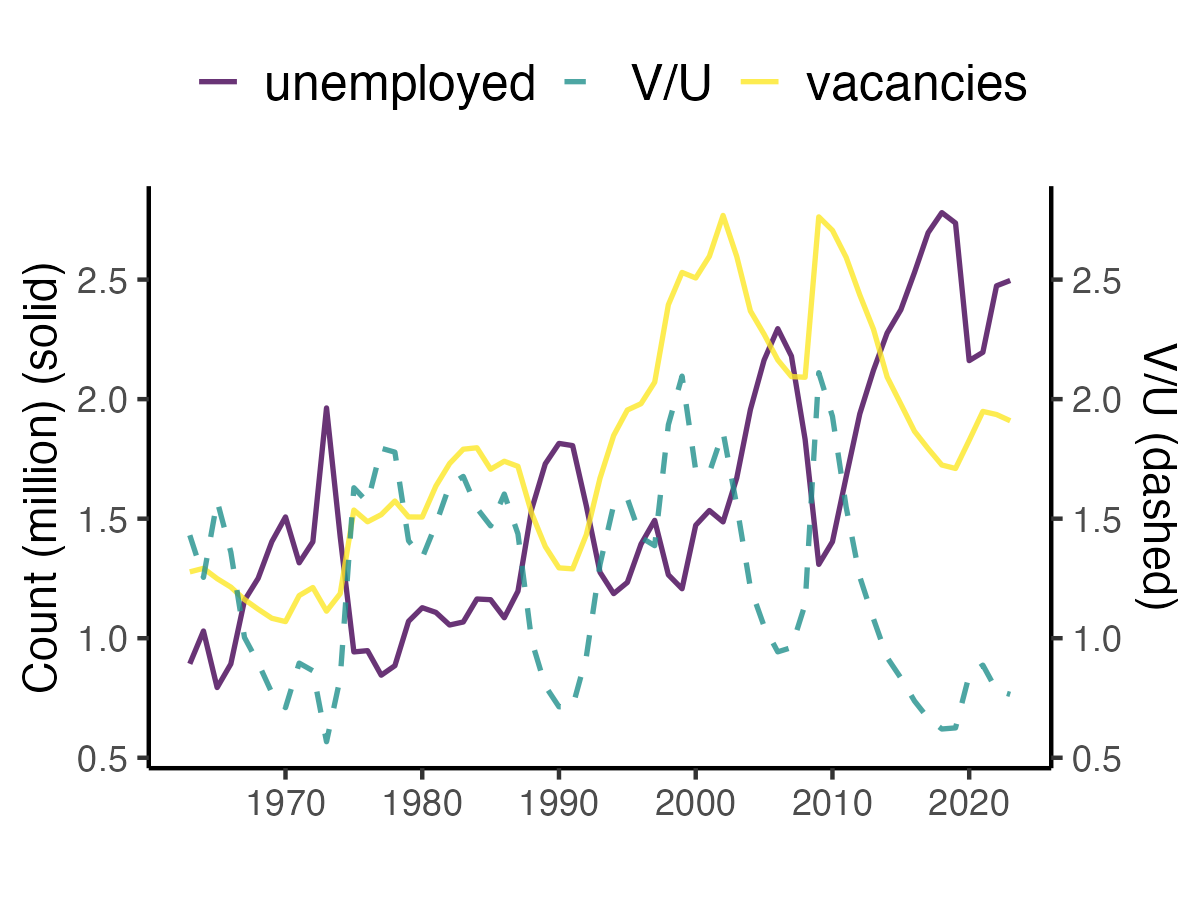}}\hspace{0.08\textwidth}
  \subfloat[Hire ($H$)]{\includegraphics[width = 0.42\textwidth]
  {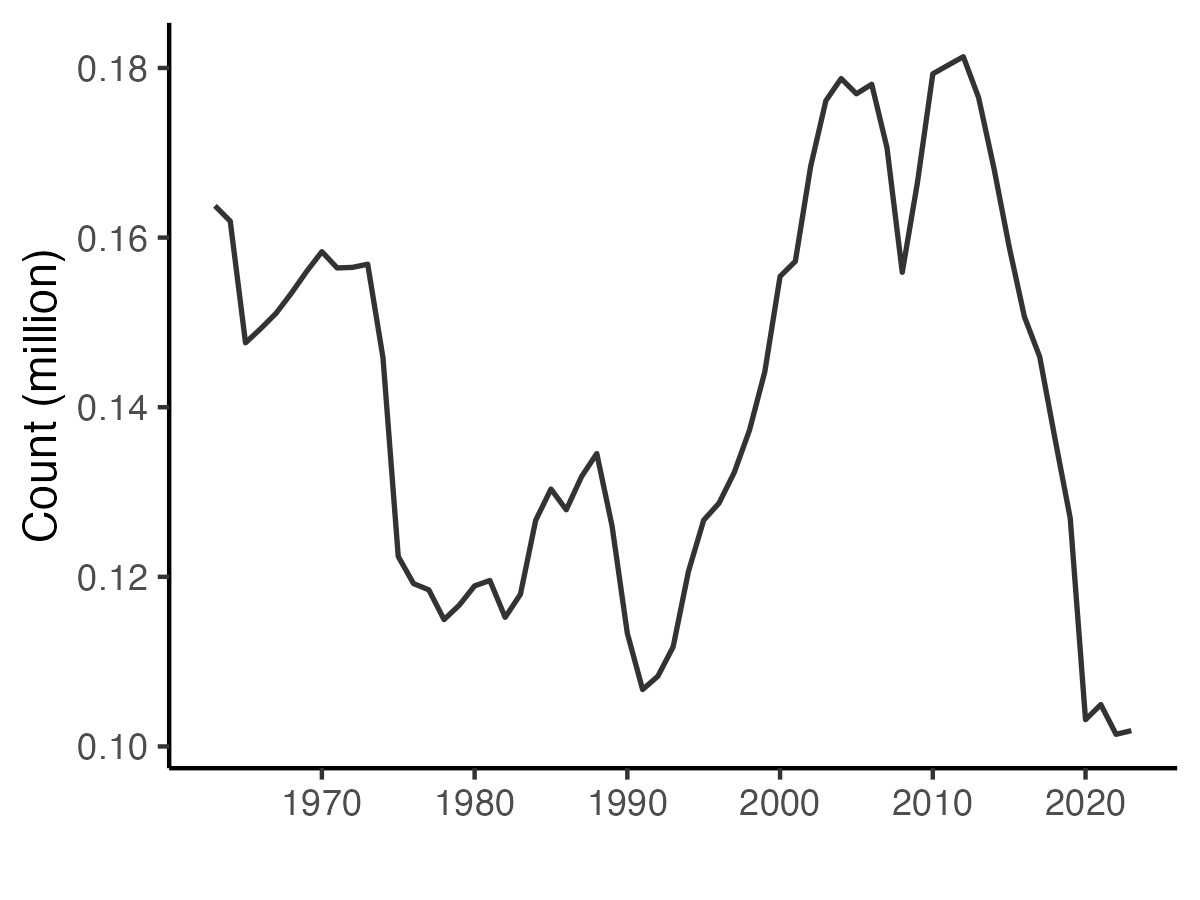}}\\
  \subfloat[Job-finding rate ($\frac{H}{U}$) and vacancy-filling rate ($\frac{H}{V}$)]{\includegraphics[width = 0.42\textwidth]
  {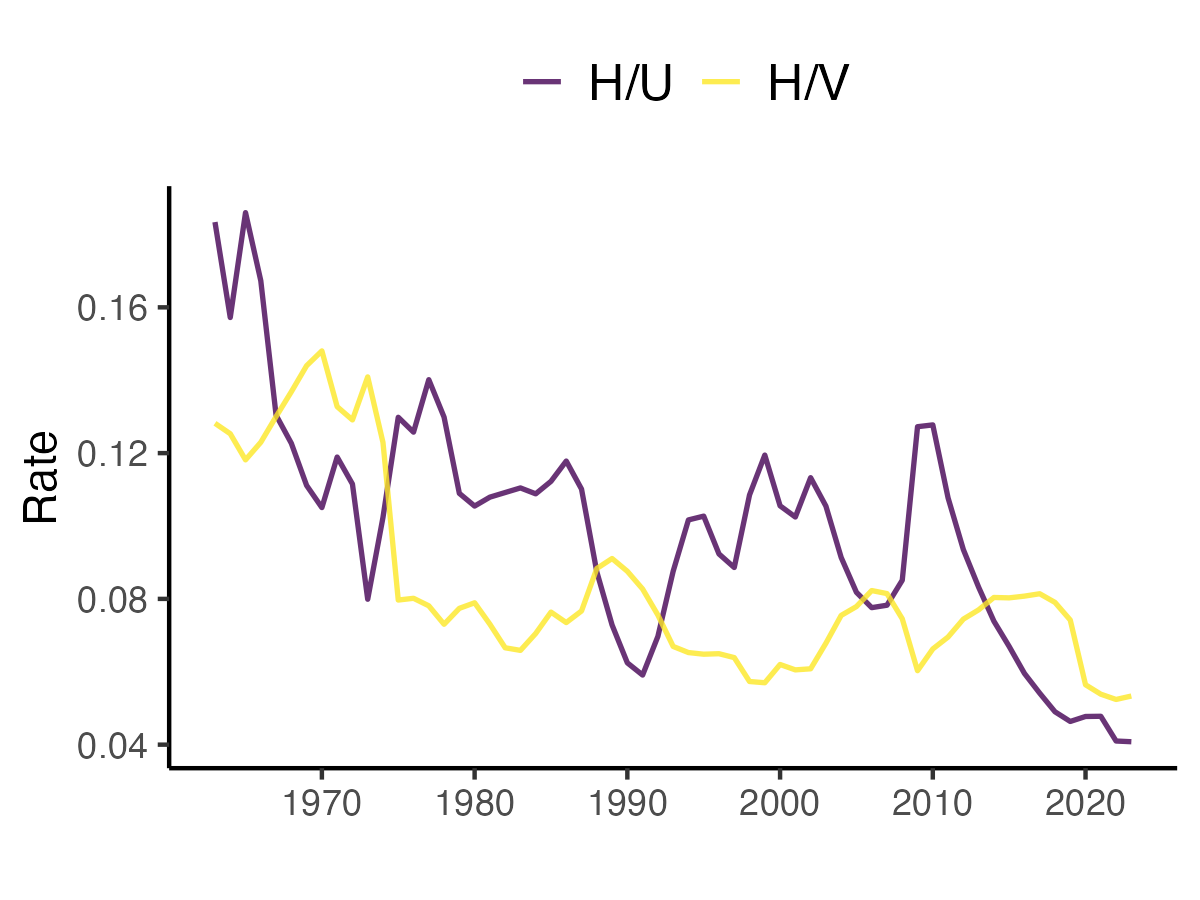}}\hspace{0.08\textwidth}
  \subfloat[Matching Efficiency ($A$)]{\includegraphics[width = 0.42\textwidth]
  {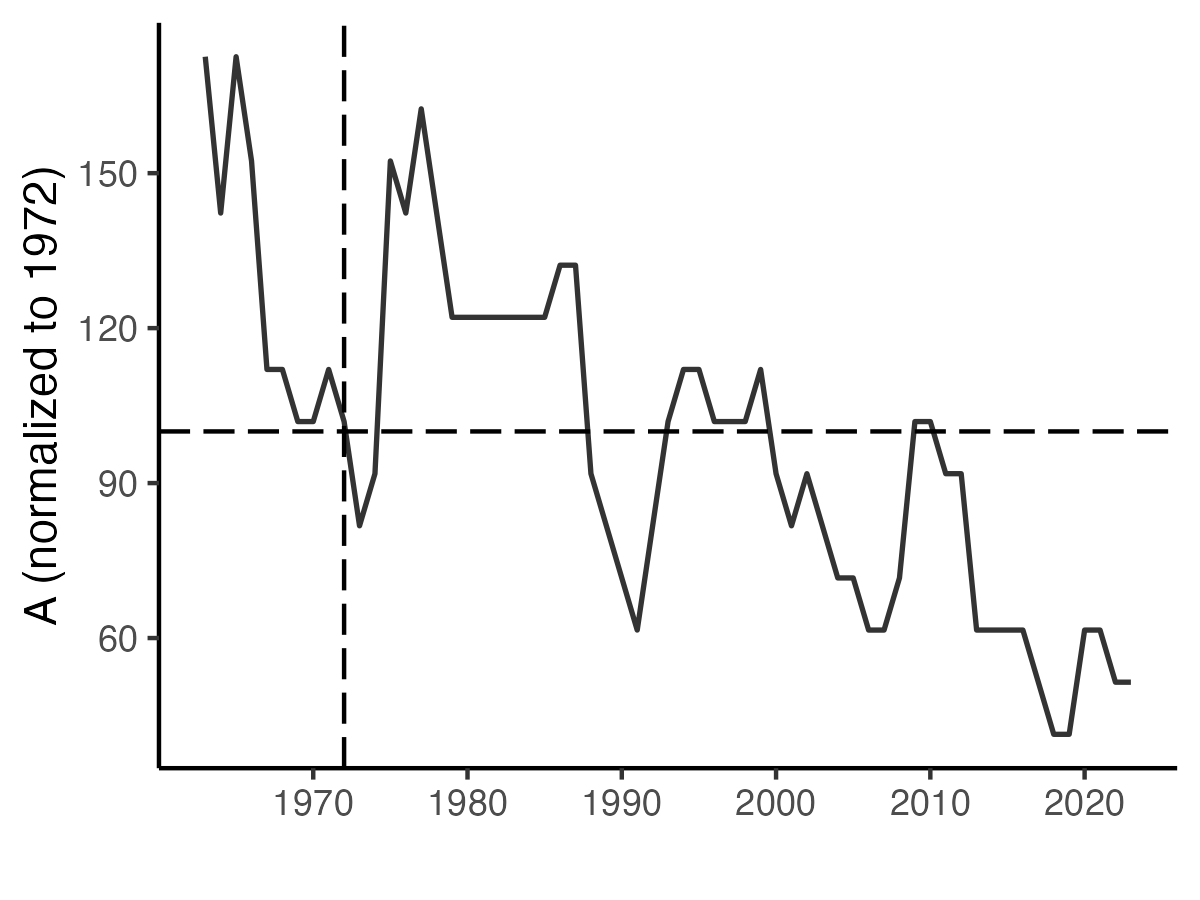}}
  \\
  \subfloat[Matching Elasticity ($\frac{d\ln m}{d\ln U}$, $\frac{d\ln m}{d\ln V}$)]{\includegraphics[width = 0.42\textwidth]
  {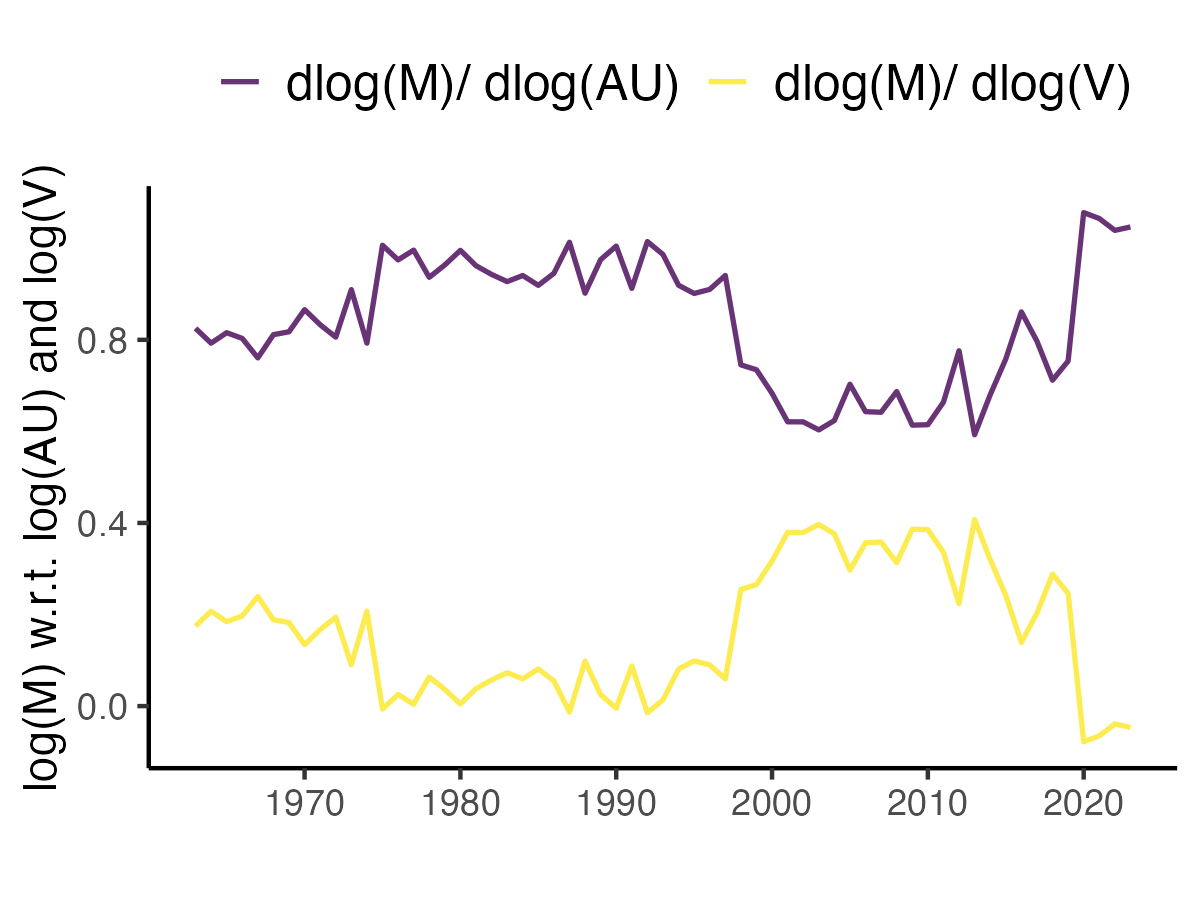}}\hspace{0.08\textwidth}
  \subfloat[Efficiency ($A$) and ($\ln \frac{V}{U}$, $\ln \frac{H}{U}$, $\ln \frac{H}{V}$)]{\includegraphics[width = 0.42\textwidth]
  {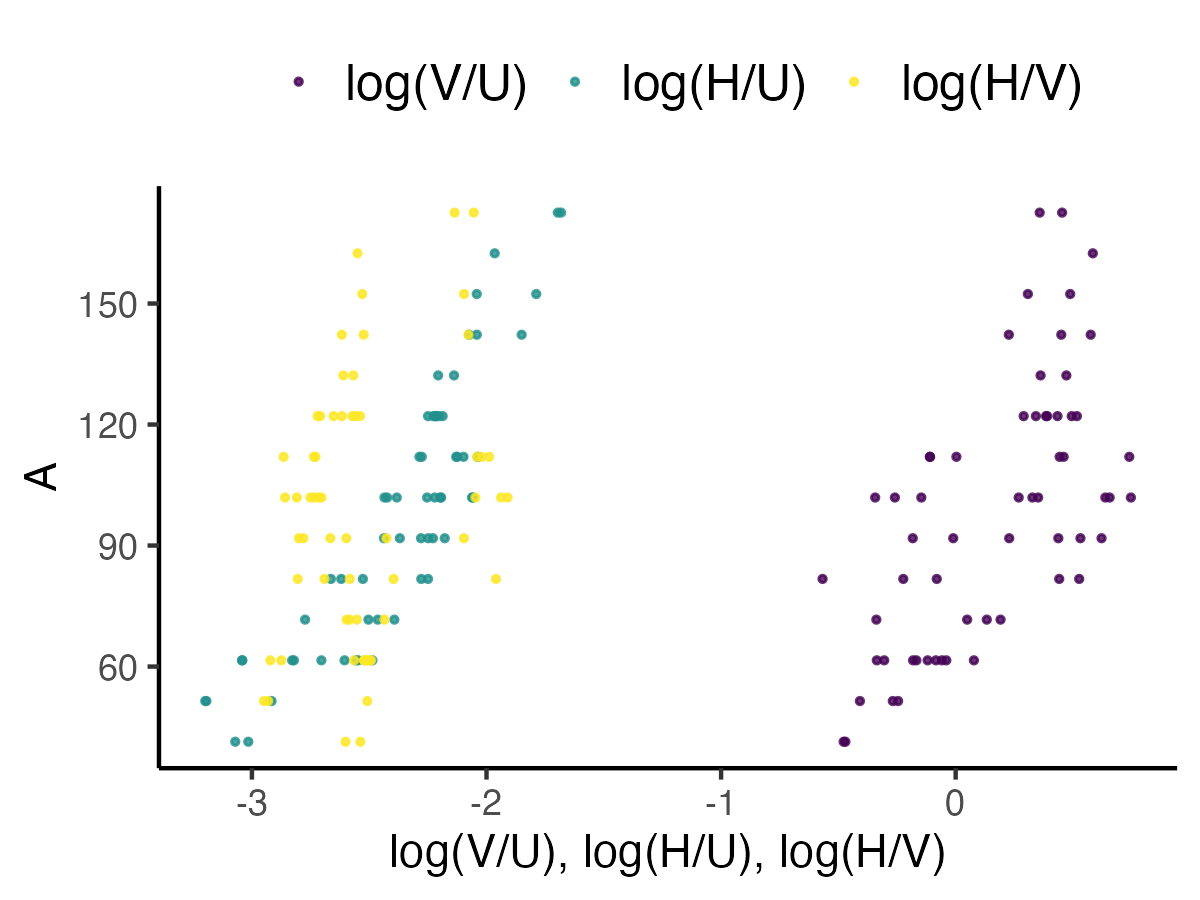}}
  \caption{Year-level results 1963--2023}
  \label{fg:year_results}
  \end{center}
  \footnotesize
  %Note:
\end{figure}

Figures \ref{fg:year_results} provide year-level trends in 1963--2023, the counterpart to Figure \ref{fg:month_part_and_full_time_results}. The findings in the main text remain valid.

\subsubsection{Decomposition of Part-Time and Full-Time Trends in 1972--2023}
\begin{figure}[!ht]
  \begin{center}
  \subfloat[Unemployed ($U$), Vacancy ($V$), and Tightness ($\frac{V}{U}$)]{\includegraphics[width = 0.42\textwidth]
  {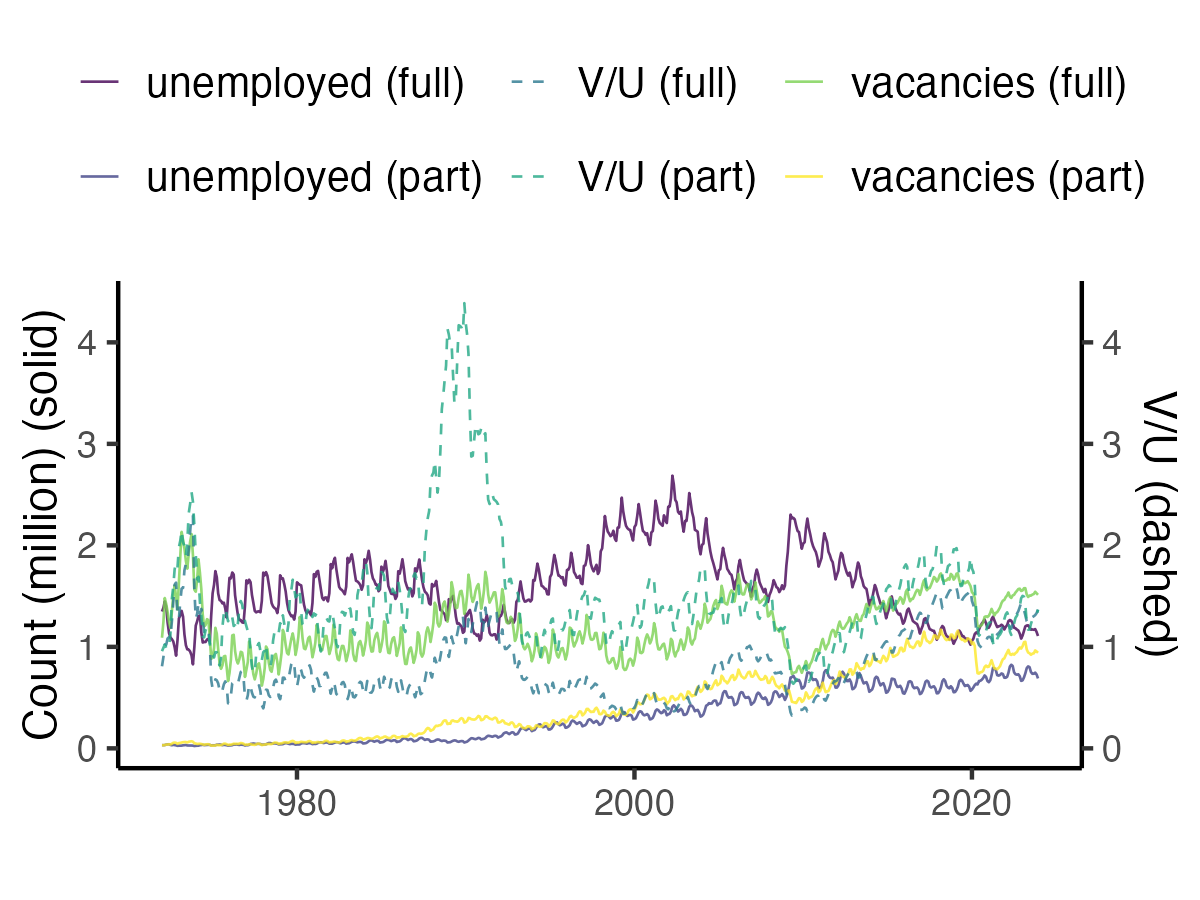}}\hspace{0.08\textwidth}
  \subfloat[Hire ($H$)]{\includegraphics[width = 0.42\textwidth]
  {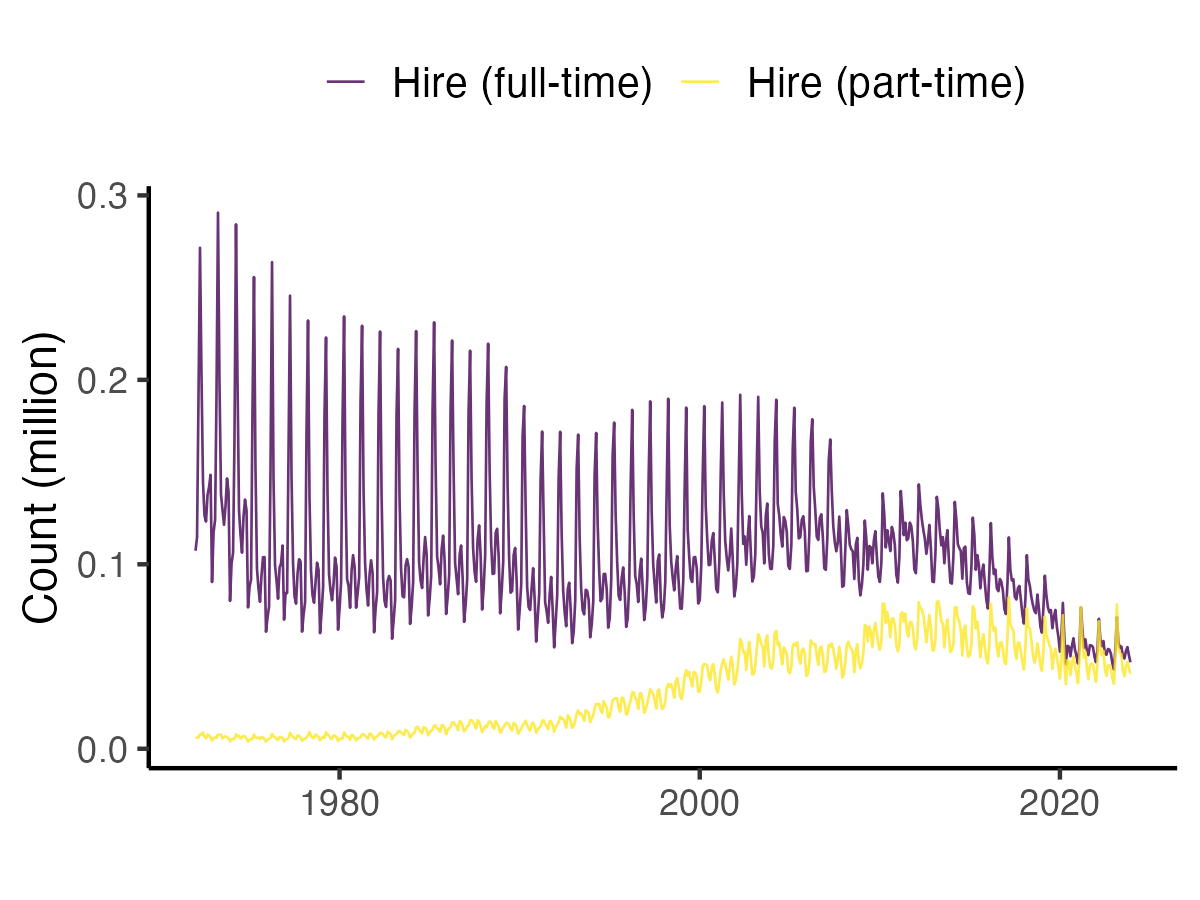}}\\
  \subfloat[Beveridge curve]{\includegraphics[width = 0.42\textwidth]
  {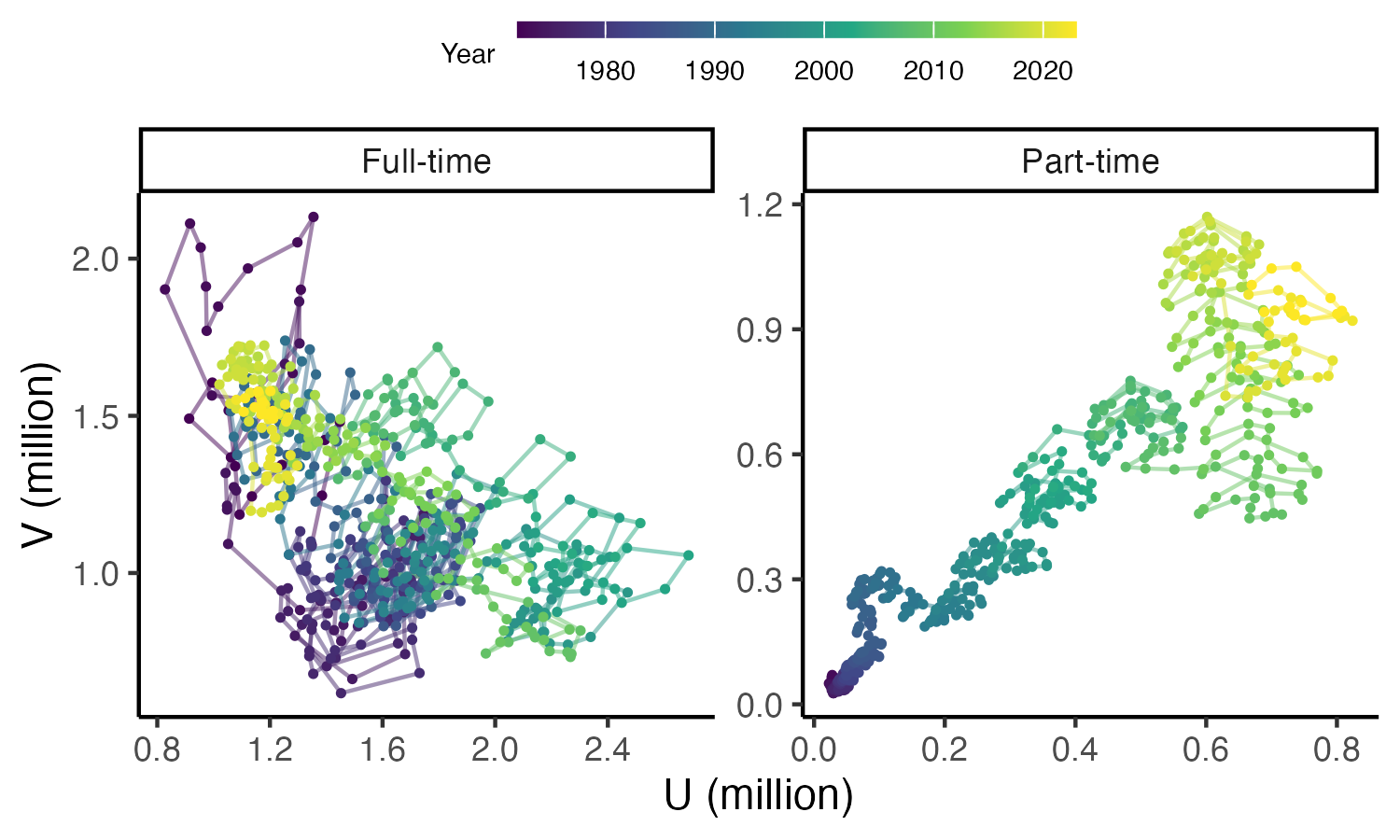}}\hspace{0.08\textwidth}
  \subfloat[Job-finding rate ($\frac{H}{U}$) and vacancy-filling rate ($\frac{H}{V}$)]{\includegraphics[width = 0.42\textwidth]
  {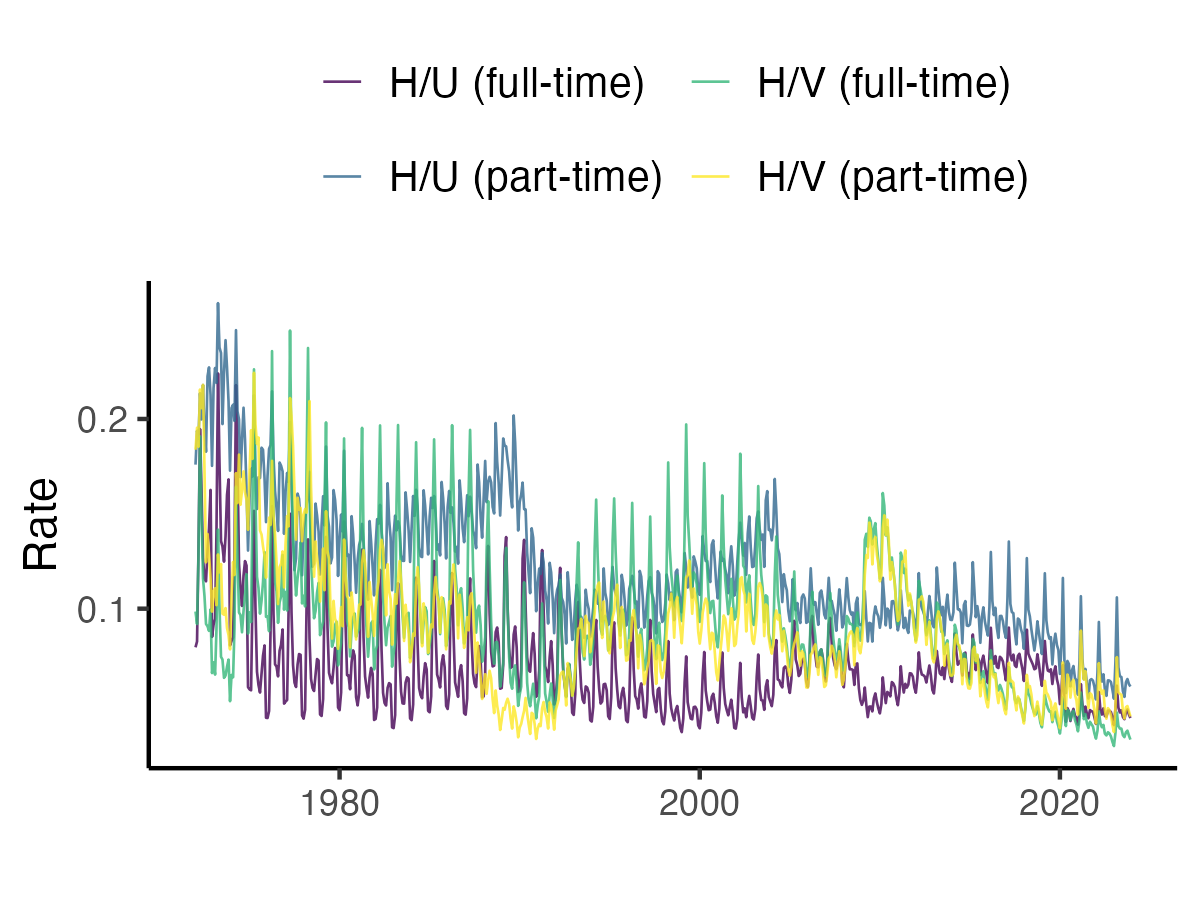}}
  \\
  \subfloat[Matching Efficiency ($A$)]{\includegraphics[width = 0.42\textwidth]
  {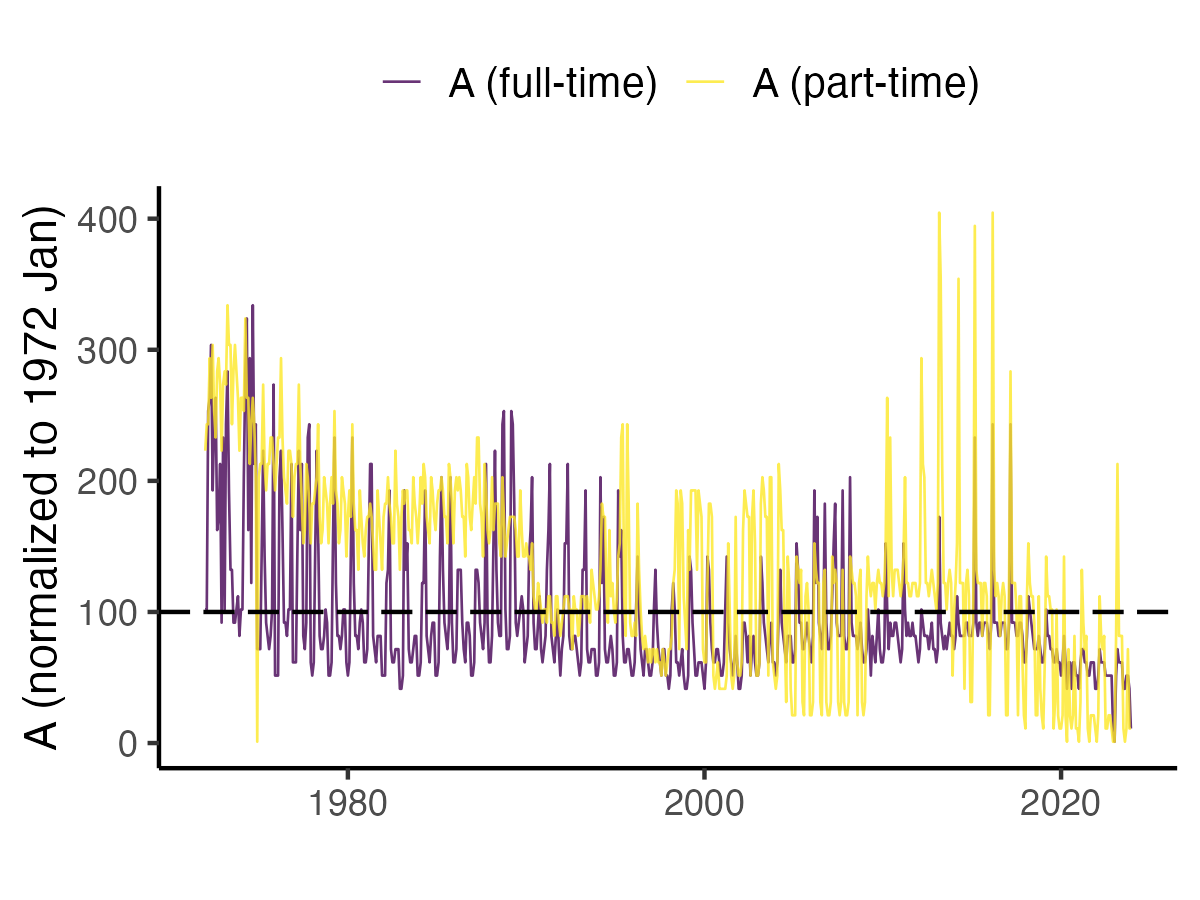}}\hspace{0.08\textwidth}
  \subfloat[Matching Elasticity ($\frac{d\ln m}{d \ln AU}$, $\frac{d\ln m}{d\ln V}$)]{\includegraphics[width = 0.42\textwidth]
  {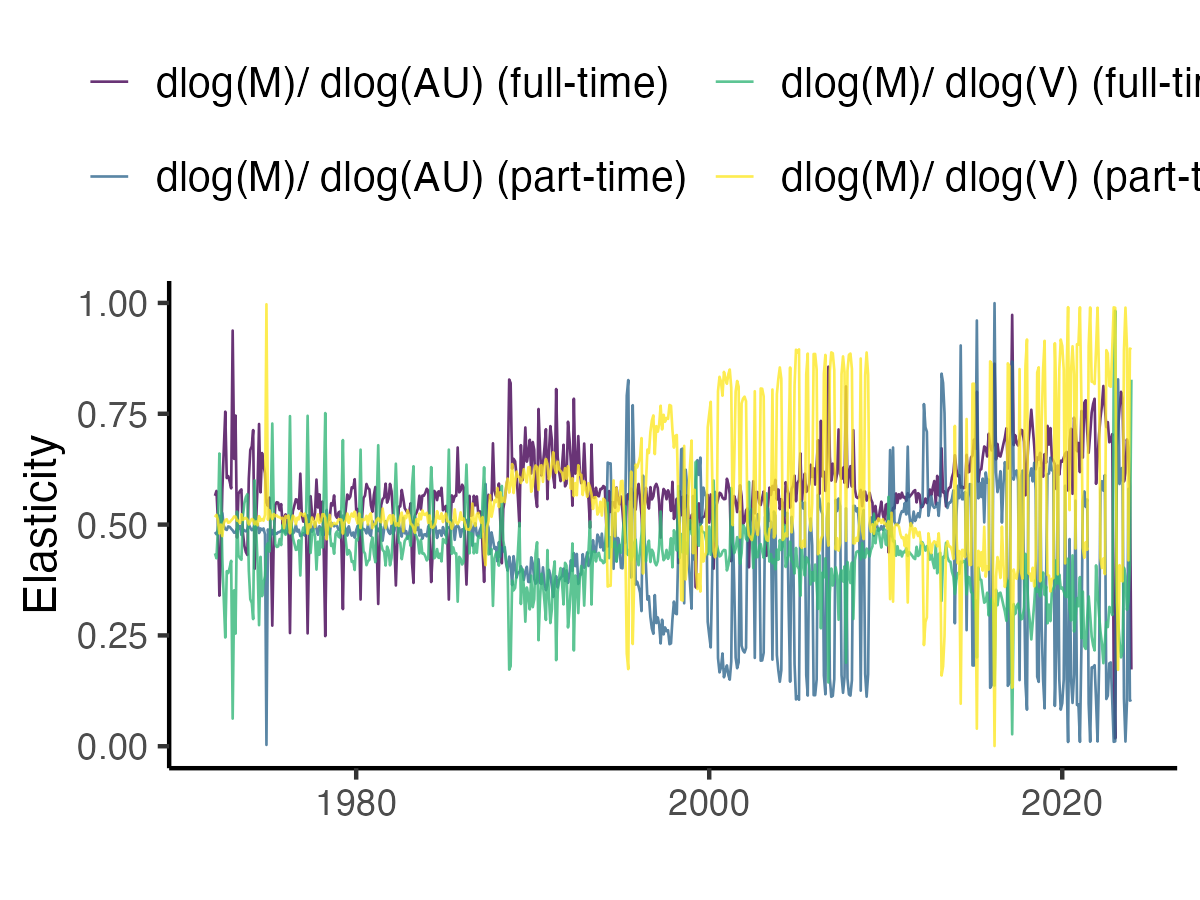}}
  \caption{Month-level full-time and part-time results 1972--2023}
  \label{fg:month_full_time_part_time_results}
  \end{center}
  \footnotesize
  %Note:
\end{figure}

Next, I decompose the aggregate trends into full-time and part-time labor markets' trends.
Figure \ref{fg:month_full_time_part_time_results} provides full-time and part-time labor market trends corresponding to Figures \ref{fg:month_part_and_full_time_results} (a)--(e).
Panel (a) shows that full-time unemployment and vacancies show higher counts and more significant fluctuations compared to part-time data and that part-time unemployment and vacancies exhibit a more gradual and steady trend over the years.
Both full-time and part-time labor markets experience periods of high tightness during the same general periods, notably in the early 1990s.
Part-time labor market tightness shows more pronounced peaks compared to the full-time labor market.
Panel (b) shows that the sharp decline in full-time hires post-2008 suggests greater sensitivity to economic downturns and structural changes in the labor market.
The steady rise in part-time hires indicates increasing reliance on part-time employment, possibly due to greater flexibility and adaptability in the labor market.
Panel (c) illustrates that full-time employment data covers a wider range of both unemployed and vacancies, indicating higher variability in the full-time labor market in contrast to a more stable part-time labor market.
Panel (d) shows that full-time finding rates show higher initial values and more significant declines compared to part-time finding rates.
Part-time finding rates demonstrate a more gradual decline and stable fluctuations over time.
Full-time finding rates exhibit larger fluctuations and more pronounced declines, indicating greater sensitivity to economic changes.
Part-time finding rates, while also showing initial declines, stabilize earlier and display less volatility, suggesting a more resilient part-time labor market.

\begin{figure}[!ht]
  \begin{center}
  \subfloat[Efficiency ($A$) and Tightness ($\ln\frac{V}{U}$), Full-time (left) and Part-time (right)]{\includegraphics[width = 0.4\textwidth]
  {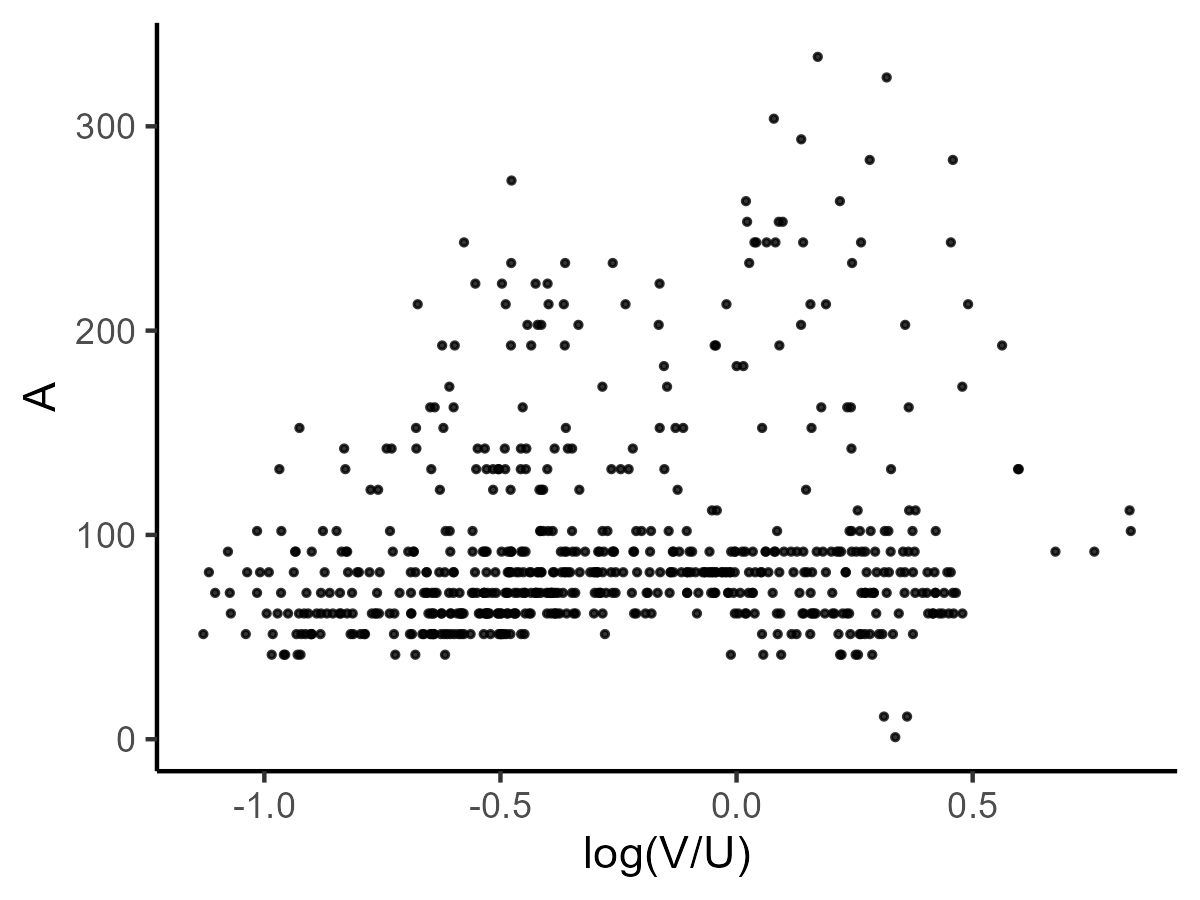}\includegraphics[width = 0.4\textwidth]
  {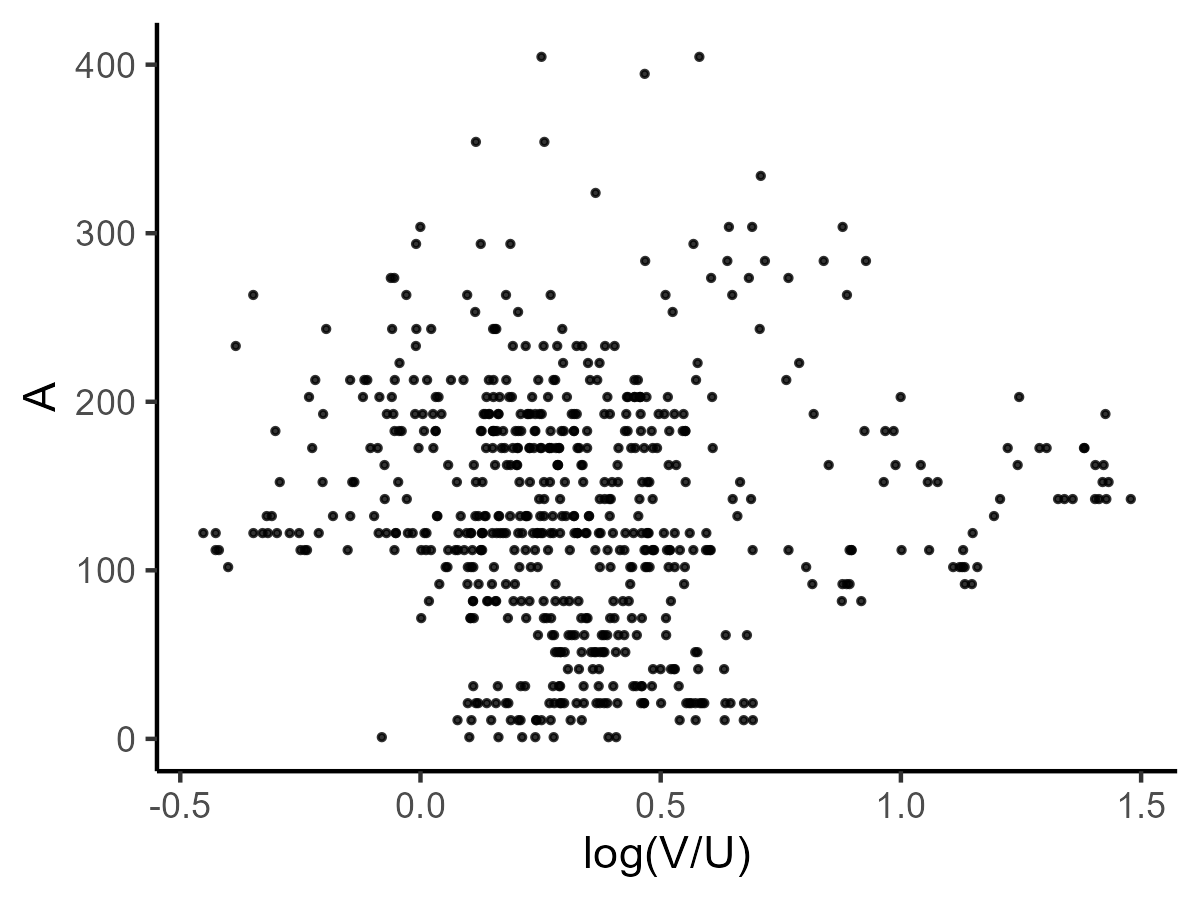}}\\
  \subfloat[Efficiency ($A$) and ($\ln\frac{H}{U}$, $\ln\frac{H}{V}$), Full-time (left) and Part-time (right)]{\includegraphics[width = 0.4\textwidth]
  {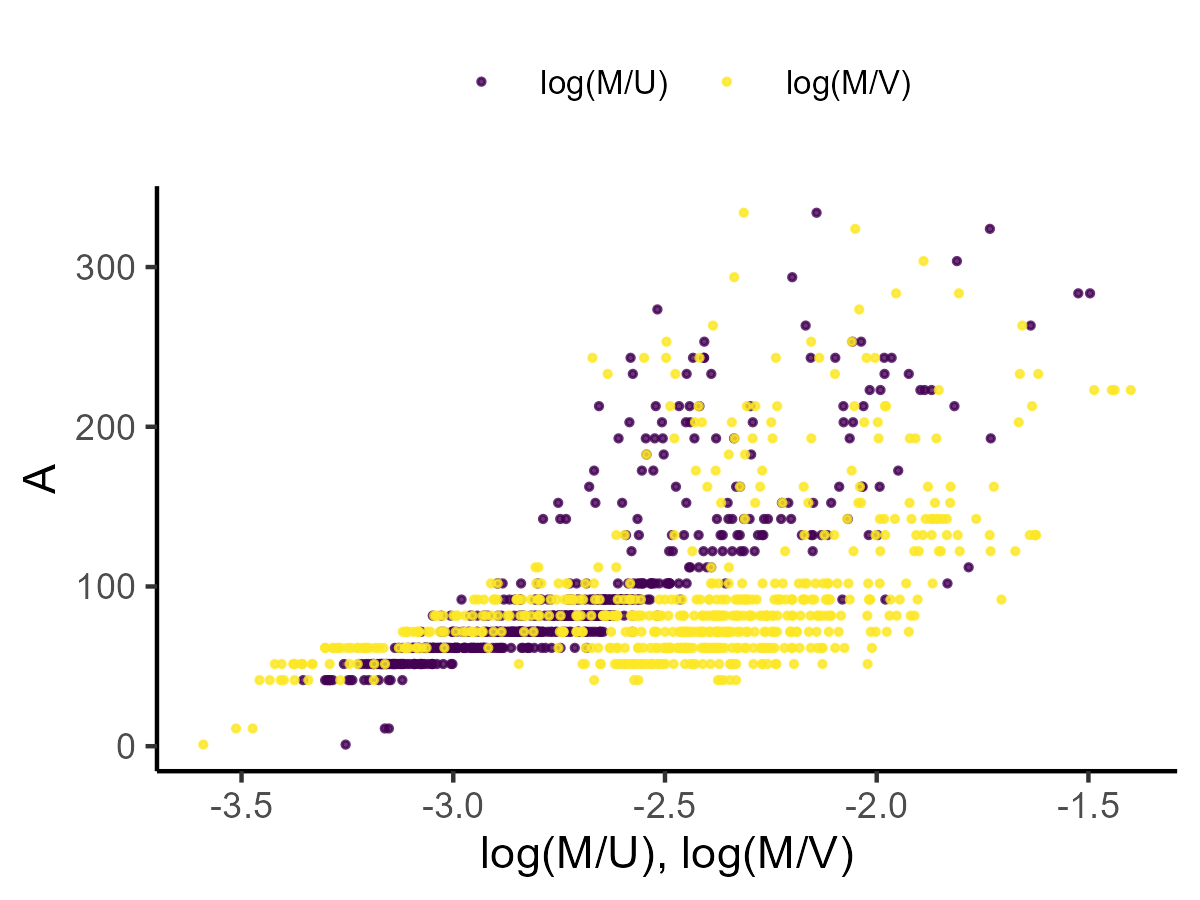}\includegraphics[width = 0.4\textwidth]
  {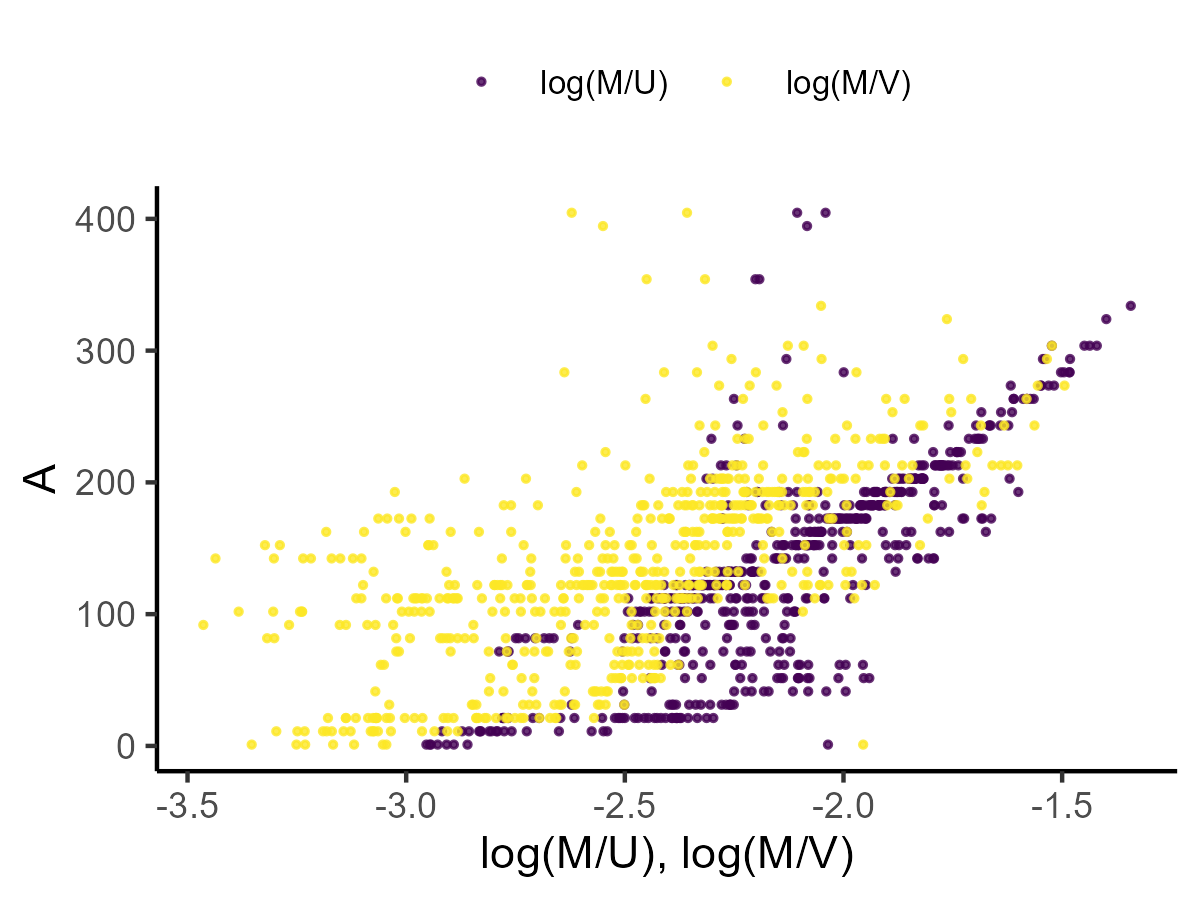}}
  \caption{Month-level full-time and part-time results 1972--2023 (continued)}
  \label{fg:month_full_time_part_time_correlation_results}
  \end{center}
  \footnotesize
  %Note:
\end{figure}

Given estimation results in each dataset, Figure \ref{fg:month_full_time_part_time_correlation_results} (e) depicts the trend of implied matching efficiency.
The downward trends are common.
Full-time matching efficiency fluctuates from 80\% to 300\%, whereas part-time matching efficiency fluctuates from 60\% to 300\% compared to January 1972.
This decomposition shows that the sharp post-COVID decline in aggregate matching efficiency is driven by declining matching efficiency for both full-time and part-time workers.

Figure \ref{fg:month_full_time_part_time_correlation_results} (f) depicts the trends of matching elasticities.
The elasticity with respect to unemployment for full-time workers shows moderate fluctuations, particularly between the 1970s and the early 1980s, when it oscillates between 0.4 and 0.8. After the 1980s, this elasticity stabilizes around 0.5. In contrast, the part-time elasticity with respect to unemployment follows a more unstable pattern, fluctuating between 0.1 and 0.9 with significant variation over time. This instability highlights the vulnerability of part-time labor markets to economic shocks, particularly during recessions such as the COVID-19 pandemic. The larger fluctuations in elasticity suggest that changes in unemployment among part-time workers have had a more uneven impact on the matching process than changes among full-time workers.

Figure \ref{fg:month_full_time_part_time_correlation_results} illustrates some correlation patterns between matching efficiency and market structure variables corresponding with Figure \ref{fg:month_part_and_full_time_results} (f).
I find stronger and more consistent relationships in the full-time labor market.
This confirms that positive correlations between efficiency and market structure are driven mainly by full-time markets.

\subsection{The Cobb-Douglas Mismatch Index and Matching Function}\label{sec:cobb_douglas_mismatch_index}

Under the Cobb-Douglas matching function with additive matching efficiency, i.e., \(H_{r t} = A_{r}A_{t}U_{r t}^{1-\gamma}V_{r t}^{\gamma}\), the equilibrium condition \eqref{eq:equilibrium_condition} can be rewritten as \(\frac{V_{r t}}{U_{r t}^{\star}} = \left(\frac{A_{r'}}{A_{r}}\right)^{\frac{1}{\gamma}}\frac{V_{r' t}}{U_{r' t}^{\star}}\). The optimal nationwide number of new hires is then derived as \(H_{t}^{*} = A_{t}U_{t}^{1-\gamma}V_{t}^{\gamma}\left[\sum_{r=1}^{R}A_{r}\left(\frac{U_{r t}}{V_{t}}\right)^{1-\gamma}\left(\frac{V_{r t}}{V_{t}}\right)^{\gamma}\right]\). Substituting this into the rewritten equilibrium condition yields \(H_{t}^{*} = \bar{A}_{t}A_{t}U_{t}^{1-\gamma}V_{t}^{\gamma}\), where \(\bar{A}_{t} = \left[\sum_{r=1}^{R}A_{r}^{\frac{1}{\gamma}}\left(\frac{V_{r t}}{V_{t}}\right)\right]^{\gamma}\). Using this, the Cobb-Douglas mismatch index takes an analytical form as \(\mathcal{M}_{t} = 1 - \frac{H_{t}}{H_{t}^{*}} = 1 - \sum_{r=1}^{R}\left(\frac{A_{r}}{\bar{A}_{t}}\right)\left(\frac{U_{r t}}{U_{t}}\right)^{1-\gamma}\left(\frac{V_{r t}}{V_{t}}\right)^{\gamma}\). When matching efficiencies are identical across labor markets, the Cobb-Douglas mismatch index is equivalent to the conventional mismatch index proposed by \cite{jackman1987structural}. The detailed property is shown in \cite{csahin2014mismatch}.

Table \ref{tb:estimation_results_cobb_douglas} presents the Cobb-Douglas matching function estimation results for both datasets. Panel (a) uses the nationally representative LFS + SET data: column (1) is the region-year specification with region and year fixed effects, and column (2) is the industry-year specification with industry and year fixed effects. The estimated unemployment elasticity is 0.103 (region) and 0.139 (industry), and the vacancy elasticity is 0.236 (region) and 0.107 (industry), with $R^2$ around 0.97--0.99 in both columns. Panel (b) uses the Hello Work data: column (1) employs prefecture and month fixed effects, while column (2) uses occupation and month fixed effects. The unemployment elasticity is 0.266 in column (1) and 0.510 in column (2), and the vacancy elasticity is 0.319 in column (1) and 0.546 in column (2), indicating greater sensitivity at the occupation level in the Hello Work data. The smaller LFS + SET elasticities reflect the broader cross-sectional aggregation (13 regional blocks, 14 industries) and the inclusion of long-run and recent low-frequency variation, in contrast to the Hello Work month-level panel.

\begin{table}[htbp]
  \begin{center}
      \caption{Cobb-Douglas matching function estimation results}
      \label{tb:estimation_results_cobb_douglas}
      \subfloat[LFS + SET]{
      \begin{tabular}[t]{lcc}
      \toprule
        & (1) & (2)\\
      \midrule
      Dependent var & log(H) & log(H)\\
      log(U) & 0.103 & 0.139\\
       & (0.206) & (0.061)\\
      log(V) & 0.236 & 0.107\\
       & (0.048) & (0.032)\\
      \midrule
      FE & area, year & industry, year\\
      Num.Obs. & 169 & 180\\
      R2 & 0.974 & 0.986\\
      R2 Adj. & 0.969 & 0.984\\
      \bottomrule
      \end{tabular}
      }\\
      \subfloat[Hello Work]{
      \begin{tabular}[t]{lcc}
      \toprule
        & (1) & (2)\\
      \midrule
      Dependent var & log(H) & log(H)\\
      log(U) & 0.266 & 0.510\\
       & (0.019) & (0.013)\\
      log(V) & 0.319 & 0.546\\
       & (0.014) & (0.010)\\
      \midrule
      FE & pref, month & occupation, month\\
      Num.Obs. & 6392 & 8908\\
      R2 & 0.987 & 0.992\\
      R2 Adj. & 0.987 & 0.992\\
      \bottomrule
      \end{tabular}
      }
  \end{center}
  \footnotesize
  \textit{Note}: Panel (a) reports OLS with region and year fixed effects (column 1) and industry and year fixed effects (column 2) using the nationally representative LFS + SET annual panel. Panel (b) reports OLS with prefecture and month fixed effects (column 1) and occupation and month fixed effects (column 2) using Hello Work month-level data. Sample sizes are smaller than the corresponding nonparametric estimation samples because the Cobb-Douglas specification requires $\log H$, $\log V$, and $\log U$ to be defined: panel (a) column (2) drops 2 cells (Compound services in 2021 and 2023, where $U = 0$), reducing $N$ from 182 to 180; panel (b) column (2) drops 3 cells (Railway Operations in August 2015, May 2017, and April 2018, where $H = 0$), reducing $N$ from 8{,}911 to 8{,}908.
\end{table}

\section{Robustness Checks}\label{sec:robustness}

I conduct two robustness checks to assess the sensitivity of the baseline results.

\begin{figure}[!ht]
  \begin{center}
  \subfloat[Matching Efficiency ($A$)]{\includegraphics[width = 0.45\textwidth]
  {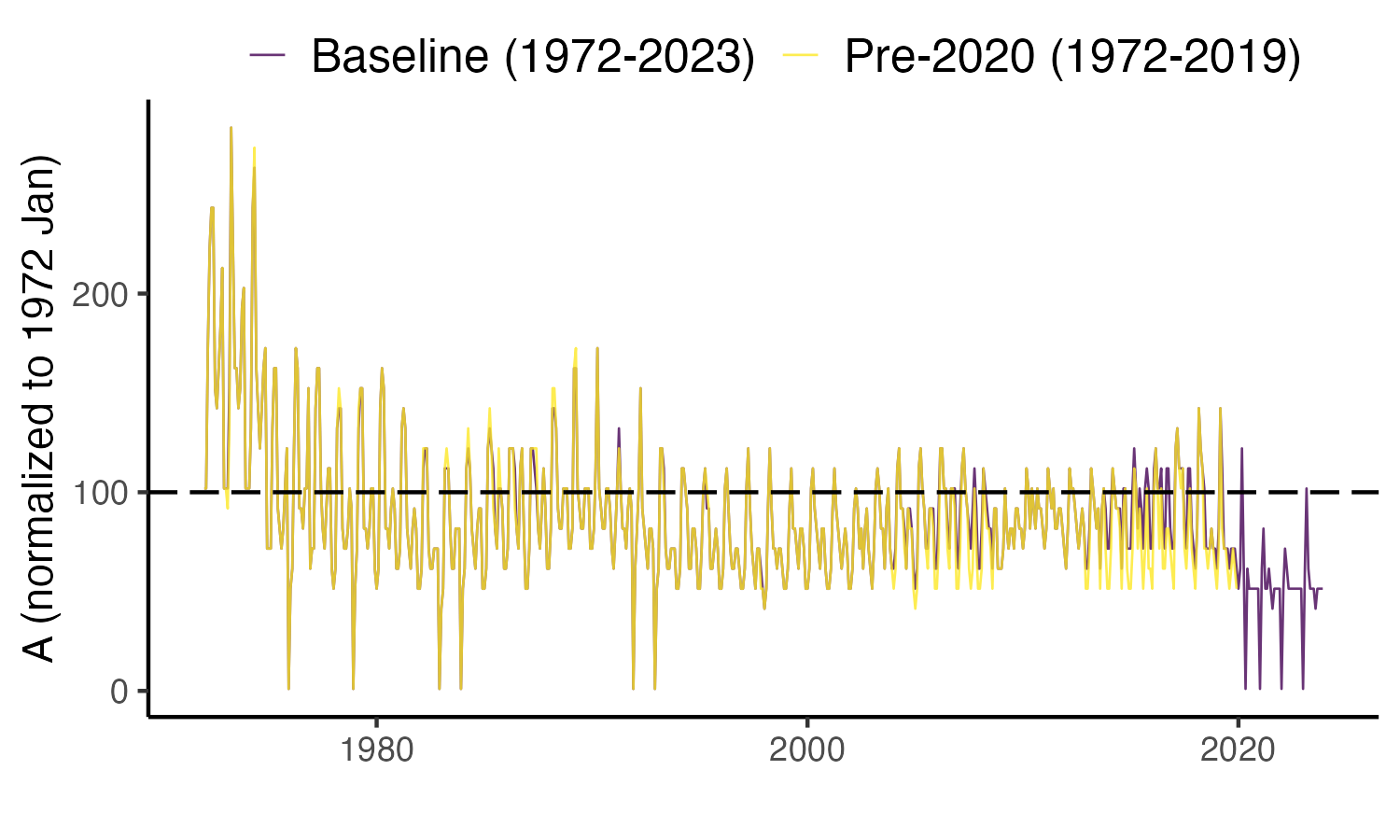}}
  \subfloat[Elasticity]{\includegraphics[width = 0.45\textwidth]
  {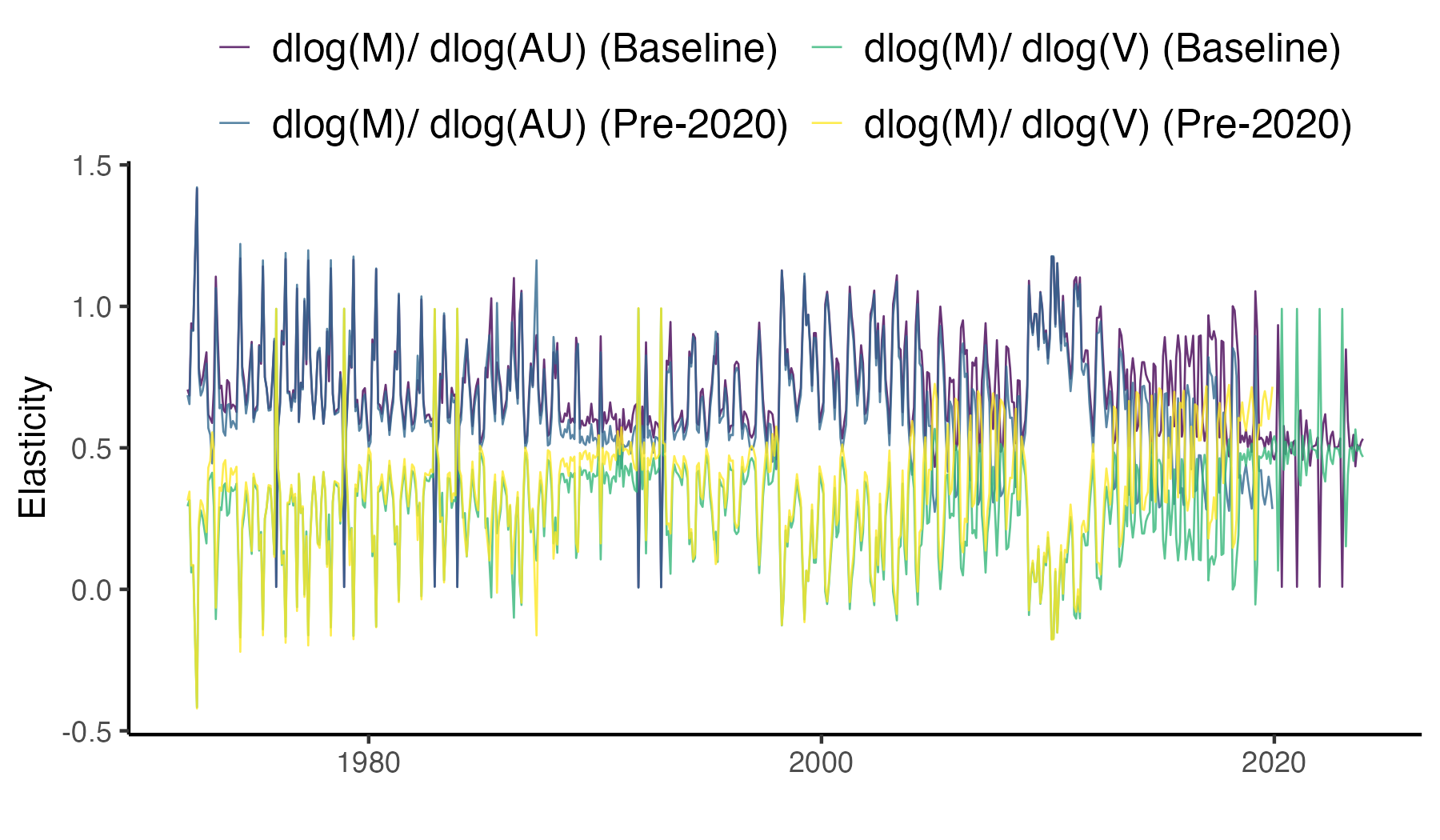}}
  \caption{Pre-COVID robustness: baseline vs.\ pre-2020 sample}
  \label{fg:robustness_pre2020}
  \end{center}
  \footnotesize
  Note: Baseline uses the full sample (1972--2023). Pre-2020 restricts the sample to 1972--2019.
\end{figure}

\paragraph{Pre-COVID sample (1972--2019).}
To examine whether the COVID-19 period drives the baseline estimates, I re-estimate the model excluding post-2019 observations. Figure \ref{fg:robustness_pre2020} compares matching efficiency and elasticities between the baseline (1972--2023) and the pre-2020 sample (1972--2019). The pre-2020 estimates closely track the baseline over the overlapping period, confirming that COVID-era observations do not substantially alter the estimated patterns of declining matching efficiency or the range of matching elasticities.

\begin{figure}[!ht]
  \begin{center}
  \subfloat[Time Series of $\varepsilon_U + \varepsilon_V^{\text{direct}}$]{\includegraphics[width = 0.45\textwidth]
  {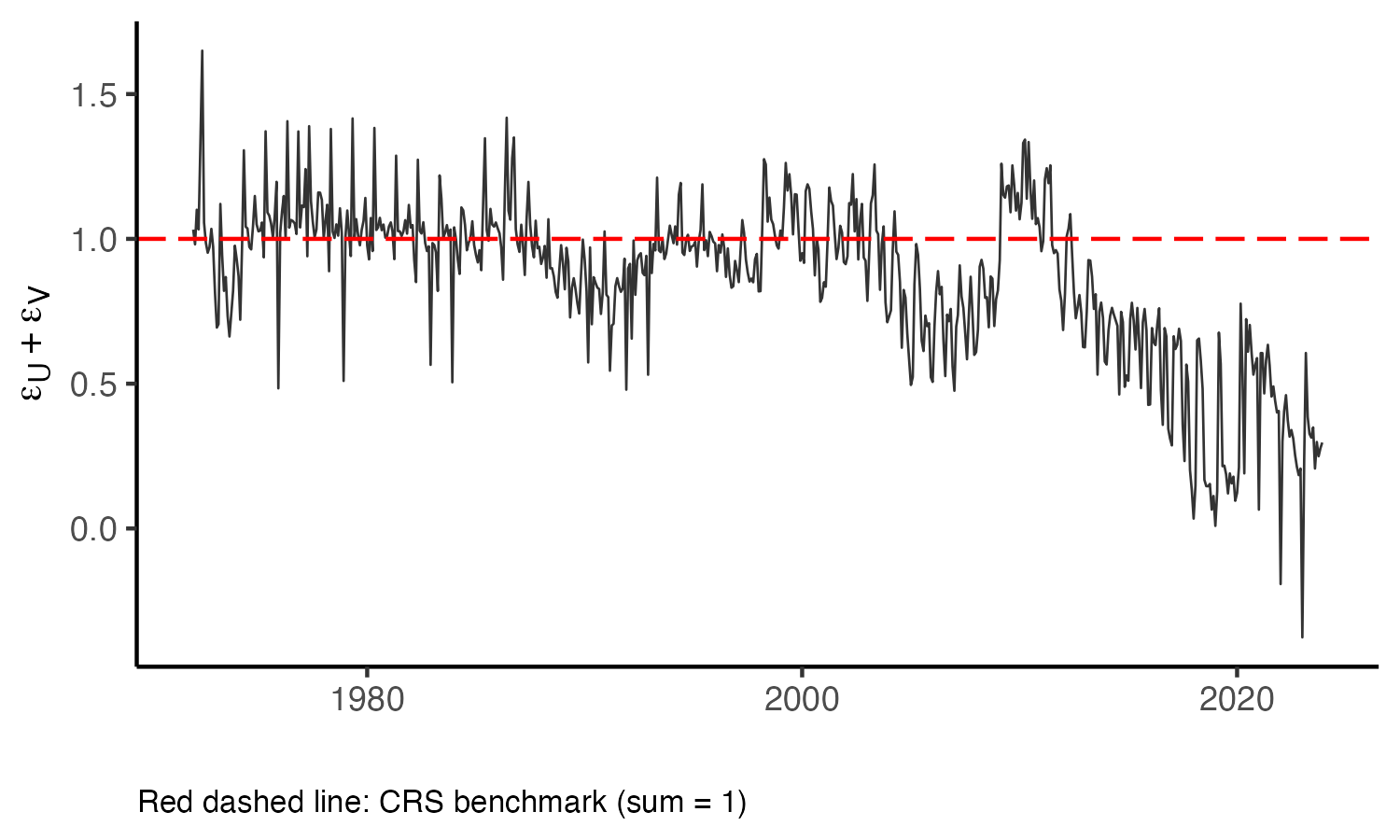}}
  \subfloat[Distribution of $\varepsilon_U + \varepsilon_V^{\text{direct}}$]{\includegraphics[width = 0.45\textwidth]
  {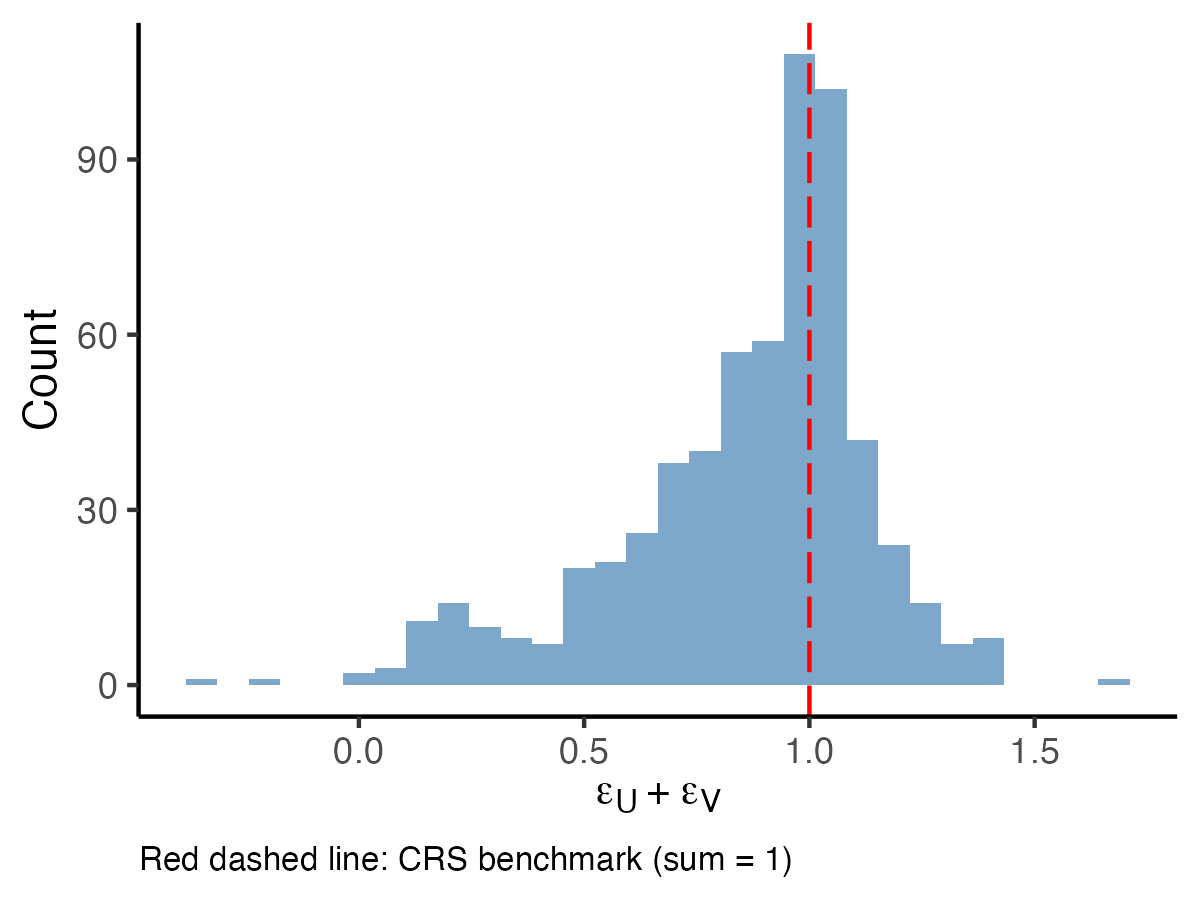}}
  \caption{Illustrative CRS consistency check in the large Hello Work sample}
  \label{fg:robustness_crs}
  \end{center}
  \footnotesize
  Note: Red dashed line indicates the CRS benchmark ($\varepsilon_U + \varepsilon_V^{\text{direct}} = 1$). The main results in the body of the paper instead report $\varepsilon_V = 1 - \varepsilon_U$, which mechanically sums to one. The diagnostic is shown only for the Hello Work monthly aggregate, where the large sample size makes the unconstrained LASSO derivative more stable. The post-COVID period may deviate more from the CRS benchmark and should be interpreted with caution.
\end{figure}

\paragraph{CRS consistency.}
The main results impose CRS by reporting $\varepsilon_V = 1 - \varepsilon_U$. For illustration, I use the large monthly Hello Work aggregate sample to assess whether the unconstrained second-order LASSO approximation produces a direct vacancy derivative that is broadly consistent with this restriction. The unconstrained vacancy elasticity is the direct partial derivative
\[
\varepsilon_V^{\text{direct}} = \left(\beta_2 (AU) + \beta_3 + 2\beta_5 V\right) \frac{V}{H},
\]
where $\beta_1, \dots, \beta_5$ are the LASSO coefficients of the second-order approximation $m=\beta_1(AU)+\beta_2(AU)V+\beta_3 V + \beta_4 (AU)^2 + \beta_5 V^2$ described above. Theoretically, exact CRS would imply $\varepsilon_U + \varepsilon_V^{\text{direct}}=1$ at each point. Because this diagnostic uses an unconstrained local approximation, however, the sum need not equal one mechanically; the relevant check is whether it remains close to the CRS benchmark. Figure \ref{fg:robustness_crs} plots the time series of this sum and its distribution for Hello Work. The main matching-efficiency and elasticity patterns are robust to including or excluding the post-COVID period, as shown in the pre-COVID robustness check above. However, the direct-derivative diagnostic suggests that the post-COVID observations may deviate more visibly from the CRS benchmark, so the post-COVID period should be interpreted with caution when assessing the CRS approximation itself.

\section{Data Construction}\label{sec:data_construction}

This appendix provides a detailed description of the construction of the industry-level and region-level panel datasets used in the nationally representative analysis (Dataset~1). All variables are measured in thousands of persons. The raw data (in Japanese), constructed data, and replication code are available on \texttt{https://github.com/ohtanilson/hellowork}, making the data construction transparent and open to discussion.

\subsection{Data Sources}

I combine two government surveys to construct the matching variables $H_{r t}$ (hires), $U_{r t}$ (unemployment), and $V_{r t}$ (vacancies) for cross-sectional market $r$ and year $t$, covering 2012--2024. Here, $r$ denotes an industry in the industry panel and a region in the region panel.
The raw public data are not provided as a unified dataset but are spread across roughly 60 Excel files with different table layouts, making it difficult to describe every merge and cleaning step fully in prose. The replication code therefore includes the preprocessing scripts used to construct the analysis-ready data, and readers interested in the exact implementation details should refer to those scripts.

\paragraph{Survey on Employment Trends (\textit{Koyou Doukou Chousa}).}
Published by the Ministry of Health, Labour and Welfare (MHLW), this establishment survey covers private-sector establishments with five or more regular employees. I extract two variables:
\begin{itemize}
    \item \textbf{Hires} ($H$): Number of new hires from non-employment by prefecture $\times$ industry, reported as an annual total. Source table: ``Number of Accessions by Prefecture, Industry (Major Group), Enterprise Size, Sex, and Job History.''
    \item \textbf{Unfilled vacancies} ($V$): Number of unfilled positions at the end of June (stock), by region $\times$ industry. This figure is available only from the first-half survey. Source table: ``Number of Unfilled Vacancies by Industry (Major Group), Enterprise Size, Region, and Occupation (Major Group).''
\end{itemize}

\paragraph{Labour Force Survey (\textit{Roudouryoku Chousa}).}
Published by the Ministry of Internal Affairs and Communications (MIC), this household survey provides nationally representative estimates of the unemployed population.
\begin{itemize}
    \item \textbf{Unemployment by industry} ($U$): Annual average of completely unemployed persons classified by industry of previous job (detailed tabulation). Unit: ten thousands (converted to thousands for merging). Available for 21 industry categories, of which 16 match the Survey on Employment Trends classification.
    \item \textbf{Unemployment by prefecture} ($U$): Annual average of completely unemployed persons by prefecture, based on time-series regression model estimates. Unit: thousands. These are model-based estimates because the Labour Force Survey's sample design does not support direct prefecture-level tabulation with sufficient precision.
\end{itemize}

\subsection{Variable Definitions}

\paragraph{Hires ($H$).}
$H_{r t}$ is the annual flow of \emph{new accessions from non-employment} (Japanese: \textit{mishuugyou nyuushokusha}), measured in thousands of persons. I exclude job-to-job transitions because (i) the unemployment stock $U_{r t}$ from the LFS does not include job-changers, so pairing $H$ with $U$ requires a non-employment-only flow to maintain stock--flow conceptual consistency for matching function estimation, and (ii) using total accessions instead inflates $H/U$ to mechanically exceed unity in most years given Japan's high inter-firm mobility. While $H_{r t}$ is an annual flow and $V_{r t}$ and $U_{r t}$ are stocks, this stock--flow pairing is standard in the matching function literature (analogous to JOLTS monthly hires paired with CPS unemployment stocks). However, because the reference periods differ---$H$ covers the full calendar year, $V$ is observed at end-June, and $U$ is an annual average---the timing mismatch may affect the estimates.

\paragraph{Long-run reconstruction of $H$ for the country panel (1972--1990).}
For the country-year panel covering 1972--2024, the SET breakdown of accessions into ``hires from non-employment'' and ``hires from job change'' (Table~15 of the SET annual report) is published only from 1991. For the earlier 1972--1990 segment I therefore recover hires from non-employment as
\[
H^{\text{NE,recovered}}_{t}
= H^{\text{total,recovered}}_{t} \times s^{\text{NE}}_{1991},
\quad t = 1972, \ldots, 1990,
\]
where $H^{\text{total,recovered}}_{t}$ is the calibrated long-run total-accession series obtained from the published SET hire rate multiplied by the LFS employee count and rescaled to match the SET total-accession level over the 1991--2024 overlap, and $s^{\text{NE}}_{1991} = H^{\text{NE}}_{1991}/H^{\text{total}}_{1991} \approx 0.432$ is the boundary-year non-employment share from the SET Table~15 (1991: 2{,}672/6{,}189 thousand). I use the boundary-year share rather than the 1991--2024 average ($\approx 0.382$, range 0.34--0.45) to avoid contaminating the pre-1991 segment with later structural shifts in inter-firm mobility. From 1991 onwards, $H^{\text{NE}}_{t}$ is taken directly from the published SET Table~15 without rescaling. The $H^{\text{NE,recovered}}_{t}$ series is the variable used as $H_t$ in the country-panel matching function estimation; the original $H^{\text{total,recovered}}_{t}$ series and the unscaled SET totals are retained for diagnostic purposes. As a sensitivity check, using the 1991--2024 average non-employment share instead of the boundary-year share changes only the level of the reconstructed pre-1991 hire series and leaves the qualitative aggregate patterns and the paper's main conclusions unchanged.

\paragraph{Vacancy ($V$).}
$V_{r t}$ is the stock of unfilled vacancies at end-June, taken directly from the Survey on Employment Trends. The measure includes both full-time and part-time unfilled positions. Although the same table reports ``part-time workers'' (column~16), this figure refers to the number of employed part-time workers, not the part-time component of unfilled vacancies; hence a part-time-excluded vacancy measure cannot be constructed from this source.

\paragraph{Unemployment ($U$).}
$U_{r t}$ is the annual average of completely unemployed persons from the Labour Force Survey. For the industry panel, unemployment is classified by industry of previous job. For the region panel, prefecture-level estimates are aggregated into 13 regional blocks to match the geographic units of the unfilled vacancy data.

\paragraph{Consistency of $H$, $U$, and $V$.}
Even with $H$ restricted to hires from non-employment, $H_{r t}/U_{r t}$ can still exceed unity at the annual frequency because $H$ is a 12-month cumulated flow whereas $U$ is an annual average stock; over a year, multiple unemployment spells can be turned over from a given average stock. To restore an interpretable ``probability per period'' scale, I report the monthly-equivalent finding rates $H_t/(12 U_t)$ and $H_t/(12 V_t)$ in the country-panel figure, which fall in the 0.05--0.32 and 0.09--0.74 ranges respectively over 1972--2024 and are comparable to the Hello Work monthly ranges (0.04--0.22 and 0.03--0.25). For the industry panel, $H_{r t} > U_{r t}$ can still occur in some cells because annual hires are flows and unemployment is an annual average stock.

\subsection{Panel Construction}

\paragraph{Industry panel.}
I first construct a balanced panel of 16 industries $\times$ 13 years (2012--2024, $N = 208$). For estimation, I drop Mining and Electricity/Gas/Water because unemployment variation is insufficient, leaving 14 industries $\times$ 13 years ($N = 182$). The 16 industries correspond to the major industry groups common to both surveys, as shown in Table~\ref{tab:industry_crosswalk}. Industries present only in the Labour Force Survey are excluded because the Survey on Employment Trends does not report the corresponding hire and vacancy figures.

\begin{table}[htbp]
\centering
\caption{Industry classification crosswalk}\label{tab:industry_crosswalk}
\small
\begin{tabular}{clccc}
\toprule
JSIC & Industry & SET ($H$, $V$) & LFS ($U$) & Used \\
\midrule
C & Mining, quarrying, gravel extraction & \checkmark & \checkmark & Dropped \\
D & Construction & \checkmark & \checkmark & \checkmark \\
E & Manufacturing & \checkmark & \checkmark & \checkmark \\
F & Electricity, gas, heat supply, water & \checkmark & \checkmark & Dropped \\
G & Information and communications & \checkmark & \checkmark & \checkmark \\
H & Transport and postal services & \checkmark & \checkmark & \checkmark \\
I & Wholesale and retail trade & \checkmark & \checkmark & \checkmark \\
J & Finance and insurance & \checkmark & \checkmark & \checkmark \\
K & Real estate, goods rental and leasing & \checkmark & \checkmark & \checkmark \\
L & Scientific research, professional services & \checkmark & \checkmark & \checkmark \\
M & Accommodations, eating and drinking services & \checkmark & \checkmark & \checkmark \\
N & Living-related services, amusement & \checkmark & \checkmark & \checkmark \\
O & Education, learning support & \checkmark & \checkmark & \checkmark \\
P & Medical, health care and welfare & \checkmark & \checkmark & \checkmark \\
Q & Compound services & \checkmark & \checkmark & \checkmark \\
R & Services, N.E.C. & \checkmark & \checkmark & \checkmark \\
\midrule
A & Agriculture, forestry & --- & \checkmark & Excluded \\
B & Fisheries & --- & \checkmark & Excluded \\
S & Government & --- & \checkmark & Excluded \\
--- & Non-agriculture/forestry (aggregate) & --- & \checkmark & Excluded \\
--- & Unclassifiable & --- & \checkmark & Excluded \\
\bottomrule
\end{tabular}
\begin{minipage}{\textwidth}
\vspace{0.5em}
\footnotesize
\textit{Notes:} JSIC = Japan Standard Industrial Classification. SET = Survey on Employment Trends. LFS = Labour Force Survey. Mining and Electricity/Gas/Water are dropped from the estimation sample due to insufficient unemployment variation. Agriculture/forestry (A) and fisheries (B) are outside the SET's target industries. Government (S) is excluded because the SET covers private-sector establishments only. ``Non-agriculture/forestry'' is an aggregate category and ``Unclassifiable'' is a residual category in the LFS, with no corresponding SET data.
\end{minipage}
\end{table}

\paragraph{Region panel.}
I construct a balanced panel of 13 regions $\times$ 13 years (2012--2024, $N = 169$). The regional blocks follow the Survey on Employment Trends classification: Hokkaido, Tohoku, Kita-Kanto, Minami-Kanto, Hokuriku, Tokai, Kinki, Keihanshin, San'in, San'yo, Shikoku, Kita-Kyushu, and Minami-Kyushu. Prefecture-level hires and unemployment are aggregated to these blocks; unfilled vacancies are reported at this level in the original data.

\subsection{Measurement Issues and Caveats}\label{sec:data_caveats}

\paragraph{Timing.}
The three variables are not measured at the same point in time within a year. $V_t$ is a stock measured at end-June, $U_t$ is an annual average, and $H_t$ is an annual flow. This timing mismatch is a common limitation in matching function estimation with annual data. In my baseline specification, I treat $t$ as indexing the calendar year and note that $V_t$ effectively reflects the mid-year labor market state.

\paragraph{Geographic basis: residence vs.\ workplace.}
The Labour Force Survey records unemployment at the place of residence, whereas the Survey on Employment Trends records hires and vacancies at the establishment location (workplace). In the region panel, this discrepancy is mitigated by the use of broad regional blocks (13 blocks rather than 47 prefectures), which encompass most commuting flows within each block. Nevertheless, cross-region commuting---particularly between Keihanshin and Kinki, or within the Kanto area---may introduce measurement error.

\paragraph{Industry classification: previous job vs.\ destination job.}
The Labour Force Survey classifies unemployed persons by the industry of their \emph{previous} job, not by the industry in which they are seeking employment. If inter-industry mobility is substantial, the measured $U_{r t}$ may overstate the effective supply of job seekers in a previous-industry market while understating it in destination industries. This mismatch tends to attenuate the estimated unemployment elasticity of the matching function and may contaminate estimated matching efficiency with industry-switching trends.

\paragraph{Survey coverage.}
The Survey on Employment Trends covers private-sector establishments with five or more regular employees across 16 major industries. The Labour Force Survey, being a household survey, is nationally representative and covers all industries and employment types. The estimated matching function therefore applies to the segment of the labor market covered by the Survey on Employment Trends. Industries and establishment sizes outside this coverage are not represented.

\paragraph{Missing values.}
Small-cell suppression in the Survey on Employment Trends produces missing values (coded as ``$-$'' in the original data) for certain industry $\times$ prefecture and industry $\times$ region combinations. I treat these as missing (NA) rather than replacing them with zero, and estimate on the resulting unbalanced panel. This avoids the systematic downward bias in $H$ and $V$ that would arise from zero-filling suppressed cells.

\end{document}